\documentclass[11pt]{article}
\usepackage[sort&compress,super,numbers,comma]{natbib}
\usepackage{amssymb,amsmath,bm}
\usepackage{graphicx}%
\usepackage{color}
\usepackage{longtable}
\usepackage[dvips]{epsfig}
\usepackage[perpage,symbol*]{footmisc}
\usepackage{multirow}
\usepackage{verbatim}
\usepackage{tablefootnote}
\usepackage{braket}
\usepackage{threeparttable,tabularx}

\usepackage{dcolumn}
\newcolumntype{d}{D{.}{.}{-1}}
\newcolumntype{.}{D{.}{.}{-1}}

\usepackage{hyperref}
\hypersetup{colorlinks,breaklinks,linkcolor=Blue,urlcolor=Blue,anchorcolor=Blue,citecolor=Blue}
\usepackage{color}
\definecolor{DarkBlue}{rgb}{0.0,0.08,0.45}
\definecolor{Blue}{rgb}{0.0,0.0,1.0}
\definecolor{Red}{rgb}{1.0,0.0,0.0}
\definecolor{RedOrange}{rgb}{0.9,0.0,0.2}
\definecolor{dgrn}{RGB}{0,150,0}
\definecolor{dgray}{gray}{0.3}

\usepackage[margin=1.0in,left=1.0in,right=1.0in]{geometry}

\newcommand{\disp}{\text{disp}}

\newcommand{\elst}{\text{elst}}
\newcommand{\exch}{\text{exch}}
\newcommand{\fast}{\text{f}}
\newcommand{\nonelst}{\text{nonelst}}
\newcommand{\rxn}{\text{rxn}}
\newcommand{\slow}{\text{s}}
\newcommand{\vac}{\text{vac}}
\newcommand{\vdw}{\text{vdW}}

\newcommand{\eg}{{\em e.g.}}
\newcommand{\etal}{{\em et al.}}
\newcommand{\ie}{{\em i.e.}}
\newcommand{\fns}{\footnotesize}
\newcommand{\nbd}{\protect\nobreakdash}
\newcommand{\qmmm}{QM\slash MM}
\newcommand{\water}{H$_\text{2}$O}

\newcommand{\apbs}{{\sc apbs}}
\newcommand{\gepol}{{\sc gepol}}
\newcommand{\swig}{SwiG}
\newcommand{\qchem}{{\sc Q-Chem}}
\newcommand{\tinkerHP}{{\sc Tinker-HP}}

\newcommand{\br}{\mathbf{r}}
\newcommand{\bs}{\mathbf{s}}
\newcommand{\cmplxi}{\dot\iota}
\newcommand{\del}{\hat{\bm{\nabla}}}
\newcommand{\realspace}{\mathbb{R}^3}

\newcommand{\cosmoconst}{\zeta}   
\newcommand{\vdwscale}{\alpha}

\newcommand{\Efield}{\mathbf{E}}
\newcommand{\Eperp}{E_\perp}

\newcommand{\diel}{\varepsilon}
\newcommand{\dielw}{\varepsilon} 
\newcommand{\dielst}{\diel_\text{s}}
\newcommand{\dielop}{\diel_\infty}
\newcommand{\dielvac}{\diel_0}
\newcommand{\epsin}{\diel_\text{in}}
\newcommand{\epsout}{\diel_\text{out}}
\newcommand{\felst}[1]{f_{#1}}    
\newcommand{\felsttilde}[1]{\tilde{f}_{#1}}    
\newcommand{\nrefr}{n} 

\newcommand{\E}{\mathcal{E}}
\newcommand{\FreeE}{\mathcal{G}}
\newcommand{\Gelst}{\FreeE_\elst}
\newcommand{\Gnon}{\FreeE_\nonelst}
\newcommand{\work}{\mathcal{W}}

\newcommand{\Ham}{\Hat{\mathcal{H}}}
\newcommand{\Hvac}{\Ham_\vac}

\newcommand{\RxnF}{\Hat{\mathcal{R}}}
\newcommand{\dRxn}{\Delta \mathcal{R}^\text{f}}				

\newcommand{\dGhyd}{\Delta_\text{hyd}\FreeE}
\newcommand{\dGrxn}{\Delta_\text{rxn}\FreeE}
\newcommand{\dGsolv}{\Delta_\text{solv}\FreeE}

\newcommand{\esp}{\varphi}
\newcommand{\kB}{k_\text{B}}
\newcommand{\mathvee}{\upsilon} 

\newcommand{\FLM}{F}
\newcommand{\FOB}{g_1}				

\newcommand{\BornRad}{\PerfRad{}}
\newcommand{\PerfRad}[1]{\bar{R}_{#1}}

\newcommand{\scrA}{\text{\scriptsize A}}
\newcommand{\scrB}{\text{\scriptsize B}}

\newcommand{\cav}{\bm{\Omega}}
\newcommand{\surf}{\bm{\Gamma}}

\newcommand{\CiteN}[1]{\protect\citenum{#1}}
\newcommand{\pcite}[1]{\protect\cite{#1}}

\newcommand{\fig}[2]{\scalebox{#1}{\includegraphics{#2.eps}}}
\newcommand{\mc}[3]{\multicolumn{#1}{#2}{#3}}

\numberwithin{equation}{section} 
\usepackage{titlesec}  

\begin{document}

\title{
	Dielectric continuum methods for quantum chemistry 
	}
	\author{
		John M. Herbert\footnote{\href{mailto:herbert@chemistry.ohio-state.edu}{herbert@chemistry.ohio-state.edu}}  \\
		{\em Department of Chemistry and Biochemistry, The Ohio State University, Columbus, Ohio, USA}
	}

\date{\today}\maketitle
\begin{abstract}\noindent
This review describes the theory and implementation of implicit solvation models based on continuum electrostatics.
Within quantum chemistry this formalism is sometimes synonymous with 
the polarizable continuum model, a particular boundary-element approach to the problem defined by the Poisson
or Poisson-Boltzmann equation, but that 
moniker belies the diversity of available methods.  This work reviews the current state-of-the art, with emphasis on 
theory and methods rather than applications.   The basics of continuum electrostatics are described, 
including the nonequilibrium polarization response upon excitation or ionization of the solute.  Nonelectrostatic interactions, 
which must be included in the model in order to obtain accurate 
solvation energies, are described as well.  Numerical techniques for 
implementing the equations are discussed, including linear-scaling algorithms that can be used in classical or mixed
quantum\slash classical biomolecular electrostatics calculations.
Anisotropic models that can describe interfacial solvation are briefly described.
\end{abstract}

\tableofcontents

\section{Overview}
\label{sec:Intro}
\begin{figure}
	\centering
	\fig{1.0}{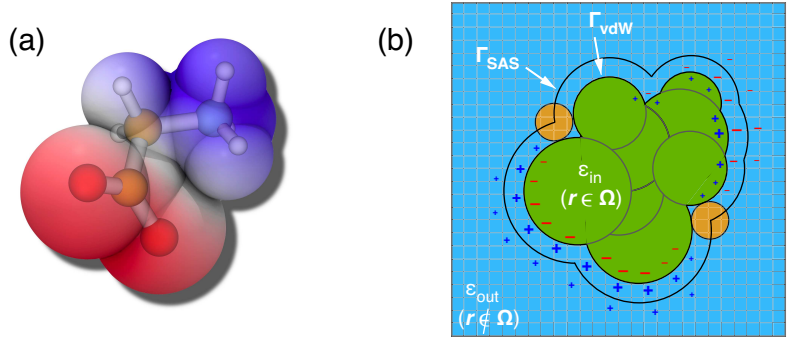}
	\caption{
	(a)
	Zwitterionic tautomer of glycine ($^-$O$_2$CCH$_2$NH$_3^+$) in a molecular cavity constructed from atom-centered spheres.
	Coloring reflects the sign and magnitude of the molecular electrostatic potential 
	evaluated at the cavity surface, $\esp^\rho(\bs)$ for $\bs\in\surf$.
	(b) 
	Schematic illustration of the same molecular cavity (in green) 
	embedded in a dielectric medium (in blue), illustrating how the continuum polarizes in response to the solute's 
	electrostatic potential.
	The orange probe sphere illustrates how the atomic radii that define the van der Waals (vdW) 
	cavity surface might be augmented to afford a ``solvent-accessible surface'' (SAS).
	The region interior to the solute cavity is designated as $\cav$, and for a sharp dielectric interface one sets
	$\diel(\br)\equiv \epsin$ for $\br\in\cav$.   If the solute is described using quantum chemistry then the natural choice is 
	$\epsin=1$.  Outside of the cavity, 
	the permittivity function $\diel(\br)$ takes the value $\epsout$, which is usually the static dielectric constant 
	of the solvent, $\dielst$.
	Panel (b) is adapted from Ref.~\CiteN{GroTry08}; copyright 2008 John Wiley \& Sons.
	}\label{fig:cavity-ESP-schematic}
\end{figure}

The use of dielectric continuum models in quantum chemistry dates to the 
mid-1970s,\cite{RinRiv73,RivRin76,MieScrTom81,MieTom82,BonCimTom83,Tom04}
to when the field itself was still in its infancy.
In their simplest form, these models describe
the solvent in terms of a single parameter $\dielst$, the (static) dielectric constant, a dimensionless 
quantity equal to the electric permittivity relative to vacuum and ranging (for simple liquids) from $\dielst\approx2$
for non-polar solvents such as benzene and hexane, up to $\dielst=78$ for water and $\dielst=110$ for formamide. 
This constant 
describes the solvent's ability to screen a charge, and the Coulomb interaction between charges $Q_1$ and $Q_2$
separated by a distance $r$  
is modified from $V(r) = Q_1 Q_2/4\pi \dielvac r$ in the gas phase, to an attenuated version 
$V(r) = Q_1 Q_2/4\pi \dielvac \dielst r$
within the dielectric medium.   The continuum description represents the ultimate in coarse-graining, reducing the solvent
to a single parameter, which has obvious utility in a quantum chemistry calculation whose cost rises steeply 
with system size.   Within a continuum description, there is no need for sampling over
solvent degrees of freedom (\eg, solvent reorganization in response to an electron transfer event that modifies
the solute's charge distribution), because this averaging is implicitly encoded into the value of $\dielst$.  
While advantageous from the standpoint of cost, limitations of the continuum description are equally apparent:
``specific'' solvation effects such as hydrogen bonding are not captured, and dielectric continuum theory alone does
not describe nonelectrostatic interactions including dispersion and Pauli repulsion.  Without the latter, there is nothing 
to imbue the molecules with finite size, necessitating {\em ad hoc\/} introduction of a ``solute cavity'' to define
the interface between the atomistic solute and the continuum solvent, as depicted in Fig.~\ref{fig:cavity-ESP-schematic}.

Some of the aforementioned limitations can be overcome, in principle, by admission of a small number of explicit solvent
molecules into the atomistic part of the calculation, in what is often called a ``semicontinuum'' or a 
``cluster + continuum'' approach.\cite{PliRiv20} 
As such, the continuum description serves as a flexible and highly useful starting point for the description of solvation 
effects in quantum chemistry.    Introduction of continuum solvation models into quantum chemistry was led by the 
group of Tomasi in 
Pisa,\cite{TomPer94,AmoBarCam99,TomMenCam02,CamMenTom03,TomMenCam05,Tom11,Men12,LipMen16} 
who refer to this approach as the {\em polarizable continuum models\/} (PCM).   That terminology will be introduced 
in Section~\ref{sec:Basics:PCM}, to refer to 
a particular class of continuum solvation models that replace the three-dimensional partial differential equations of continuum
electrostatics with a two-dimensional boundary element problem, defined on the surface of a cavity representing 
the interface between atomistic solute and continuum solvent.    Although PCMs are ubiquitous in quantum chemistry, they are not
the only continuum solvation models that are used in the field and not the only ones discussed herein.   In any case, the combination 
of a quantum-mechanical (QM) description of the solute sets up a {\em self-consistent reaction-field\/} (SCRF) problem 
in which the solute's charge distribution both polarizes, and is polarized by, its environment.   The two effects must be iterated 
to self-consistency.  

The remainder of this review is organized as follows.   Section~\ref{sec:Basics} provides the elementary specification of the 
continuum electrostatics problem, starting from the Poisson and Poisson-Boltzmann equations.  
The mechanics of turning that formalism into a computationally-tractable model are discussed in 
Section~\ref{sec:Smooth}, with an emphasis on the smooth discretization approach developed by this author's  
group.\cite{LanHer10a,LanHer10b,HerLan16,LanHer11c,LanHerAlb20} 
The focus here is on continuum solvation models in quantum chemistry but 
the formalism in Sections~\ref{sec:Basics} and \ref{sec:Smooth} is perfectly applicable to biomolecular implicit solvent
calculations, in which a macromolecular solute is described using a classical force field.\cite{HerLan16} 
Section~\ref{sec:Smooth} also introduces the various flavors of PCM that can be found in the literature and discusses how they can be 
understood in one another's context.   It should be noted that the solution of Poisson's equation, or equivalently the apparent surface 
charge PCM problem, specifies only the {\em electrostatic\/}
contribution to the solvation energy.    Other contributions including cavitation, dispersion, Pauli repulsion, and hydrogen bonding must 
be included in order to predict free energies of solvation ($\dGsolv$) in reasonable agreement with experiment.  
Section~\ref{sec:DeltaG} introduces models for these nonelectrostatic contributions and provides an overview of the accuracy that can
be expected for $\dGsolv$.    Section~\ref{sec:NonEq} introduces several ``nonequilibrium''
formulations of the continuum electrostatics problem, which are used to describe the continuum's response to a sudden change in the solute's
charge density, as in photoexcitation or photoionization.   This provides the machinery to compute solvent effects on vertical excitation, 
ionization, or fluorescence energies.  Finally, Section~\ref{sec:PEQS} discusses modifications to the isotropic continuum model that are 
necessary in order to describe anisotropic solvation environments, such as the air\slash water interface or the 
solid-state\slash aqueous interface.

This review is focused on the theoretical framework and computational mechanics of continuum solvation models, not on applications.
While some limited data to describe the performance of the models are provided, for a complete overview of continuum solvation methods
in action, the reader is directed to several general reviews,\cite{TomMenCam05,Tom11,Men12,Cap16,Car19a,AndFis19,PliRiv20} 
as well as specialized ones concerning the application of PCMs to specific types of 
spectroscopy.\cite{SadPec07,BarCimPav07,Cap07,PecRud07,CamMen07,AgrMik07,Imp12a}
In lieu of a great deal of data, the present work provides copious references to the primary literature.

\section{Continuum Electrostatics}
\label{sec:Basics}
This section reviews the basic electrostatic formalism that underlies continuum solvation theory.
The physical model is defined by Poisson's equation in three-dimensional space (Section~\ref{sec:Basics:Poisson})
but is not fully specified without a surface to demarcate the boundary of the atomistic region
(Section~\ref{sec:Basics:cavity}).    The {\em polarizable continuum models\/} that were introduced into quantum 
chemistry by Tomasi and co-workers\cite{TomPer94,Tom04,TomMenCam05} are introduced in 
Section~\ref{sec:Basics:PCM}, and stem from a reformulation of the Poisson problem as a boundary-element of 
``apparent surface charge" problem.  There are several variants, and these are compared side-by-side
in Section~\ref{sec:Basics:Models}.

\subsection{Poisson's equation}
\label{sec:Basics:Poisson}

The basic tenet of dielectric continuum theory is an assumption that the electric response of matter can be coarse-grained
in the form of a dipole density $\mathbf{P}$ that defines the polarization of the medium.   In the presence of this 
dielectric medium, the role of the 
electric field $\Efield(\br)$ in vacuum is supplanted by the electric displacement field (or electric induction),  
\begin{equation}\label{eq:D(r)}
	\mathbf{D}(\br) = \Efield(\br) + 4\pi\mathbf{P}(\br) = \diel(\br)\, \Efield(\br) \; .
\end{equation}
The electric permittivity $\diel(\br)$ is defined by the manner in which polarization $\mathbf{P}(\br)$ is induced by the 
external field $\Efield(\br)$, and Eq.~\eqref{eq:D(r)} amounts to the definition of a {\em linear\/} dielectric material whose
polarization is proportional to field strength.   Nonlinear susceptibilities are harder to describe within a continuum formalism
and have received less attention.\cite{TomMenCam05}   Whereas a fully general discussion of (linear) dielectric materials 
would allow for a permittivity that is a 
function of frequency also (or even a nonlocal function of space and\slash or time, in some 
formulations),\cite{BasPar96,JenHun03,BasChu07}
the ground-state SCRF problem does not require this generality, and unless otherwise specified, 
$\diel$ will mean the {\em static\/} (zero-frequency) dielectric constant, $\dielst$.   
(The continuum electrostatics community has stubbornly defied the suggestion\cite{Whi79} 
that ``dielectric constant" is obsolete and should be replaced by ``relative electric permittivity".)
For an anisotropic medium, $\diel$ would take the form of a $3\times 3$ tensor rather than scalar,
which could be used to model a liquid crystal in which the electric susceptibility depends on the orientation of
the applied field.\cite{Riz07,Fer07}     
This review will not consider such cases, although a different form of anisotropic solvation is considered
in Section~\ref{sec:PEQS}.  Herein, $\diel$ is a scalar.


That said, Eq.~\eqref{eq:D(r)} does express the permittivity as a scalar-valued {\em function}, $\diel(\br)$, 
rather than simply a dielectric {\em constant}.   This allows for a situation such as that depicted in
Fig.~\ref{fig:cavity-ESP-schematic}(b), wherein a ``solute cavity'' (two-dimensional surface $\surf$) 
defines an interface between the continuum solvent and an atomistic region, the latter of which shall be described using
quantum chemistry.
Within the cavity, inter-particle Coulomb interactions are included explicitly in the Hamiltonian and thus $\diel=1$ in this region.
Outside of the cavity, the permittivity function takes on a value equal to the (static) dielectric constant of the solvent, \eg, $\diel=78$ for water.  
Given a charge density $\rho(\br)$ for the solute,  including both nuclei and electrons, 
Maxwell's equation for the displacement field $\mathbf{D}(\br)$ is 
\begin{equation}\label{eq:Maxwell}
	\del\bm{\cdot}\mathbf{D}(\br) = 4\pi \rho(\br) \; .
\end{equation}
Recognizing that the electric field is $\Efield(\br) = -\del\esp(\br)$, where $\esp(\br)$ is the {\em electrostatic potential}, 
Eq.~\eqref{eq:Maxwell} can be written in the more familiar form of {\em Poisson's equation}, 
\begin{equation}\label{eq:Gen-Poisson}
	\del\bm{\cdot}\left[ \diel(\br) \, \del\esp(\br)\right] = -4\pi \rho(\br) \; .
\end{equation}
All of these equations are expressed in Gaussian electrostatic units, where $4\pi\dielvac =1$.\cite{Wan86}   

Poisson's equation is the mathematical starting point for continuum electrostatics.   Given the charge density
$\rho(\br)$ from an electronic structure calculation, Eq.~\eqref{eq:Gen-Poisson} 
is solved for $\esp(\br)$ throughout space, including both the atomistic region and the 
surrounding dielectric medium.   This potential can be separated into two parts,
\begin{equation}\label{eq:phi(r)}
	\esp(\br) = \esp^\rho(\br) + \esp^{}_\rxn(\br) \; ,
\end{equation}
where the first term is the electrostatic potential generated by the solute's charge density, 
\begin{equation}\label{eq:ESP-rho}
	\esp^\rho(\br) = \int \frac{\rho(\br')}{\|\br - \br'\|} \; d\br' \; ,
\end{equation}
whereas the ``reaction field'' contribution $\esp^{}_\rxn(\br)$ arises from polarization of the continuum,
resulting in an additional charge density $\rho_\text{pol}(\br)$.
Having obtained $\rho(\br)$ from Schr\"odinger's equation
and then $\esp(\br)$ by solving Eq.~\eqref{eq:Gen-Poisson}, the electrostatic solvation energy can be
expressed variously as\cite{LanHer10b,CooHer18}
\begin{equation}\label{eq:G_elst} 
	\Gelst = \frac{1}{2}\int \esp_\rxn(\br) \; \rho(\br) \; d\br 
	= \frac{1}{2}\int \esp^\rho(\br) \; \rho_\text{pol}(\br) \; d\br  \; .
\end{equation}
This is sometimes alternatively called the {\em polarization energy\/} ($\FreeE_\text{pol}$),\cite{LanHer10b} 
but we will use the term {\em electrostatic energy\/} ($\Gelst$) as that is the 
nomenclature typically encountered in the literature on dielectric materials.\cite{Wan86} 
Unlike the theory of intermolecular interactions,\cite{Pat20}
it makes little sense in the present context to consider electro\textit{statics} separate from polarization but the 
reader may, if desired, substitute the phrase ``electrostatics + polarization'' whenever ``electrostatics'' is used herein.
The quantity $\Gelst$ is a {\em free\/} energy insofar as the dielectric formalism implicitly accounts for the averaging over
solvent degrees of freedom, and the factor of $1/2$ in Eq.~\eqref{eq:G_elst} reflects the fact 
that the interaction energy 
is reduced, by precisely half its value, on account of the work required to polarize the 
environment.\cite{Bot76,TomPer94,JacWilHer09,LanHer10b}    

From the point of view of electronic structure theory, 
$\FreeE_\elst[\diel,\rho]$ is a functional of both the permittivity function $\diel(\br)$ and the solute's charge density $\rho(\br)$.
The total (free) energy is 
\begin{equation}\label{eq:E=E0+G}
	\FreeE_0[\Psi] = \big\langle\Psi \big| \Hvac \big|\Psi \big\rangle + \FreeE_\elst[\diel,\rho] \; ,
\end{equation}
where the first term represents the electronic energy {\em in vacuo}.    (It is written here 
in a form that suggests wave function quantum mechanics but could be replaced by a density functional.)
This can equivalently be expressed as a total energy functional 
\begin{equation}\label{eq:G[Psi]}
	\FreeE_0[\Psi] = \big\langle \Psi \big |
		\Hvac + \tfrac{1}{2}\RxnF_0
	\big| \Psi \big\rangle 
\end{equation}
where $\RxnF_0$ is a reaction-field operator that generates the integral in Eq.~\eqref{eq:G_elst}.
Minimization of $\FreeE_0[\Psi]$, or $\FreeE_0[\rho]$ in density functional theory (DFT), 
in conjunction with Poisson's equation to obtain the electrostatic 
potential that defines $\RxnF_0$, defines the SCRF problem.
Provided that the electronic structure model satisfies a variational principle, as it does for self-consistent field (SCF) models, 
then the total energy defined by Eq.~\eqref{eq:E=E0+G} satisfies a variational principle as well,\cite{CarScaFri07,LanHer10b}

Equation~\eqref{eq:Gen-Poisson} 
is sometimes called the ``generalized'' form of Poisson's equation, with the ``ordinary'' form being 
\begin{equation}\label{eq:Poisson}
	\diel\, \hat{\nabla}^2\esp(\br) = -4\pi\rho(\br) \; .
\end{equation}
The distinction is that the permittivity function $\diel(\br)$ in Eq.~\eqref{eq:Gen-Poisson} is replaced 
in Eq.~\eqref{eq:Poisson} by a scalar, $\diel$.   It is the latter equation that is often taken 
to define the continuum electrostatics problem, but this requires additional specification because $\diel=1$ for the atomistic
interactions.   Some sort of molecular surface is needed to delineate the boundary with the continuum, as shown  
in Fig.~\ref{fig:cavity-ESP-schematic} where the cavity is defined by a union of atom-centered spheres.  Given a cavity surface, 
Eq.~\eqref{eq:Poisson} is shorthand for Eq.~\eqref{eq:Gen-Poisson} with the permittivity function 
\begin{equation}\label{eq:eps(r)-sharp}
	\diel(\br) = \begin{cases}
		\epsin, 	& \br \in \cav \\
		\epsout, 	& \br \notin\cav
	\end{cases} \; .
\end{equation}
Note that $\esp(\br)$ is continuous across the cavity surface but its derivative is not.\cite{TomPer94}
In general it makes sense to take $\epsin=1$ although larger values 
(commonly $\epsin=2$--$4$,\cite{Nak96,GroTry08,AleMehBak11} but in some cases 
$\epsin = 10$--$20$\cite{AntMcCGil94,DemWad96,Gry02,AleMehBak11}) 
have sometimes been used in biomolecular electrostatics calculations, in an effort 
to approximate a ``dielectric constant of protein''.   (However, the very concept that such a ``constant''
exists has been vociferously criticized.\cite{WarRus84,Nak96,SchWar01,WarShaKat06,LiLiZha13})
If the solute is described using quantum mechanics, however, then any choice other than $\epsin=1$ represents an 
inconsistent treatment of the Coulomb interactions unless the Coulomb operators that define $\Hvac$ are modified, 
which is seldom done. 

Although we have introduced it as a quantum-mechanical SCRF problem that must be solved in tandem with the 
electronic structure problem, Eq.~\eqref{eq:Gen-Poisson} is also used in classical electrostatics calculations for 
biomolecules,\cite{ShaHon90b,FogBriMol02,Bak05b,BotCaiLuo14} 
where $\rho(\br)$ is comprised of classical point charges (or higher-order 
multipoles\cite{SchPon07,SchBakRen07}) that come from a force field, \eg,
\begin{equation}\label{eq:rho_classical}
	\rho(\br) = \sum_A^\text{atoms} Q_A \, \delta(\br - \mathbf{R}_A) \; .
\end{equation}
For biomolecular applications, one is often interested in an aqueous solvent containing some concentration of 
dissolved ions.   The continuum analogue of that situation is described by the 
{\em Poisson-Boltzmann equation},\cite{ShaHon90a,DesHol01,Lam03,Bak04,GroTry08}
\begin{equation}\label{eq:PBE}
	\del\bm{\cdot}\left[ \diel(\br) \, \del\esp(\br)\right] 
	= -4\pi \big[ \rho(\br) + \rho_\text{ions}(\br)\big]
	\; ,
\end{equation}
in which the right side of Eq.~\eqref{eq:Gen-Poisson} has been augmented with a term that accounts for 
a thermal distribution of dissolved ions.\cite{DesHol01,GroTry08}
Whereas the solute's charge density $\rho(\br)$ reflects atomistic modeling, using either a classical force field or else 
an electronic structure calculation, the density $\rho_\text{ions}(\br)$ of ``mobile'' ions is treated statistically.\cite{MorNet01,DesHol01}   
For an electrolyte with dissolved ion concentrations $\{c_i\}$, 
whose ionic charges are denoted $\{q_i\}$, the statistical charge density for the ions is\cite{GroTry08} 
\begin{equation}\label{eq:rho_mobile_ions}
	\rho_\text{ions}(\br) = \sum_i^\text{ions} q_i c_i \lambda_i(\br) 
	\exp\left(
		\frac{-q_i\esp(\br)}{\kB T}
	\right) \; .
\end{equation}
The {\em ion accessibility function\/} 
$\lambda_i(\br)$ represents some type of step function that serves to exclude the ions from the atomistic region.
With this form for the density of mobile ions, Eq.~\eqref{eq:PBE} 
is sometimes known as the {\em size-modified\/} version of the (nonlinear) Poisson-Boltzmann 
equation.\cite{GroTry08,SteHerHea19}
In the case of a 1:1 electrolyte with monovalent ions ($q_1 = e = -q_2$), Eq.~\eqref{eq:rho_mobile_ions} reduces to 
\begin{equation}
	\rho_\text{ions}(\br) = -2c \lambda(\br) \, \text{sinh}\left(
		\frac{e\,\esp(\br)}{\kB T}
	\right) \; .
\end{equation}
At the physiological ionic strengths, the hyperbolic sine function can 
be linearized without significant error,\cite{Zho93b,FogZucEsp99} resulting in a linearized Poisson-Boltzmann
equation,\cite{Bak04,GroTry08}
\begin{equation}\label{eq:LPBE}	
	\del\bm{\cdot}\left[ \diel(\br) \, \del\esp(\br)\right] 
	= -4\pi \rho(\br) + \kappa^2 \lambda(\br) \, \esp(\br)  \; ,
\end{equation}
where 
\begin{equation}\label{eq:kappa}
	\kappa 
	= \left(
		\frac{8\pi e^2 c}{\kB T}
	\right)^{1/2} \; .
\end{equation}
The dissolved ions screen electrostatic interactions over a length scale $\sim \kappa^{-1}$ 
known as the {\em Debye screening length}, such that the potential that appears in Debye-H\"uckel theory
is the attenuated Coulomb potential 
$e^{-\kappa r}/(\dielst r)$.\cite{DebHuc23b,DesHol01,GroTry08,LanHer11b}


Within the biomolecular electrostatics community there has been significant discussion regarding the accuracy 
of the linearization approximation, with studies noting that the nonlinear form 
[Eq.~\eqref{eq:PBE}, with Eq.~\eqref{eq:rho_mobile_ions} for $\rho_\text{ions}(\br)$]
affords better agreement with explicit solvent simulations when the ionic strength 
is high.\cite{FogZucEsp99,WanRenLuo17}  Deficiencies in the Poisson-Boltzmann model itself---even 
even in its nonlinear form and especially for polyvalent ions---have also been pointed out.\cite{Vla99}
It is therefore worth noting that for the small solutes that characterize most quantum chemistry applications,
the effect of the mobile ions on $\Gelst$ is quite modest,\cite{LanHer11b,SteHerHea19}
although there are effects on activity coefficients.\cite{SteHerHea19,DziBhaAnt20}
These effects are presumably magnified for a solute the size of a protein, but the intermediate size regime has hardly been explored.

Methods for solving the partial differential equations introduced in this section will be introduced below.  
First, however, we need to discuss one 
more aspect of the model problem itself, namely, the definition a surface that establishes the boundary between atomistic 
solute and continuum solvent.

\subsection{Solute cavity}
\label{sec:Basics:cavity}

For the case of a sharp dielectric boundary, Eq.~\eqref{eq:Gen-Poisson} 
has an analytic solution if the cavity surface is spherical and contains the entire charge density $\rho(\br)$.
For a solute consisting of a single point charge, $Q$, 
centered in a spherical cavity of radius $\BornRad$ in a medium with dielectric constant $\diel$, 
this solution affords the well-known ``Born model'' for ion solvation,\cite{Bor20a,RasHon85}
\begin{equation}\label{eq:Born}
	\Delta \FreeE_Q = -\frac{Q^2}{2 \BornRad}\left(\frac{\diel-1}{\diel}\right) \; .
\end{equation}
Here, $\Delta \FreeE$ indicates the change in $\FreeE_\elst$ from its gas-phase value of zero to the solution-phase value
obtained from Eq.~\eqref{eq:G_elst}.    Replacing the point charge $Q$ by a point dipole $\mu$, the solvation energy is 
\begin{equation}\label{eq:Bell}
	\Delta \FreeE_\mu = -\frac{(\diel-1)\mu^2}{(2\diel + 1)\BornRad^3} \; .
\end{equation}
(This result is often attributed to Onsager,\cite{Ons36} although it was derived a few years earlier by Bell.\cite{Bel31})
The dipole solvation energy can alternatively be written 
$\Delta \FreeE_\mu = -\tfrac{1}{2}(\bm{\mu}\bm{\cdot}\Efield_\rxn)$, where
\begin{equation}\label{eq:dipole_rxn_field}
	\Efield_\rxn = 
	\underbrace{
		\frac{1}{\BornRad^3}
		\left(\frac{2(\diel-1)}{2\diel+1}\right)
	}_{
		\FOB(\diel,\BornRad)
	}
	\bm{\mu}
	= \FOB(\diel,\BornRad) \, \bm{\mu}
\end{equation}
is the ``reaction field'' induced by $\bm{\mu}$.\cite{Bot76}   
If $\bm{\mu}$ is polarizable 
($\bm{\mu} = \bm{\mu}_0 + \bm{\alpha}\bm{\cdot}\Efield$), then Eq.~\eqref{eq:dipole_rxn_field}
provides the earliest example of a SCRF model.\cite{Bot76}  
Historically, this model was used 
to formulate a microscopic theory for the dielectric constant of a polar liquid,\cite{Ons36,Bot38,Has72}
going beyond the mean-field description contained in the 
Clausius-Mossotti equation.\cite{HoySte76,Han83}
These attempts were not particularly quantitative,\cite{Has72} and for 
modern purposes a dipolar description of the solute constitutes a needless approximation.

The aforementioned results for $\Delta\FreeE$ were quickly generalized by Kirkwood,\cite{Kir34a,KirWes38,WesKir38} 
and later by others,\cite{Bon51,RinRuiRiv83,MikDalSwa87,MikAgrJen88,KonPon97,SchPon07}
to an arbitrary multipole in a spherical cavity, with the general result
\begin{equation}\label{eq:Gelst-multipole}
	\Gelst = -\frac{1}{2}\sum_{\ell \ge 0} 
	\underbrace{
		\left[\frac{(\ell+1)(\diel-1)}{\ell + (\ell+1)\diel}\right]
		\frac{1}{\BornRad^{2\ell+1}}
	}_{
		g_\ell(\diel,\BornRad) 
	}
	\sum_{m=-\ell}^\ell  
	\big|\big\langle\Psi \big| \hat{T}_{\ell m} \big| \Psi \big\rangle \big|^2
\end{equation}
where the $\hat{T}_{\ell m}$ are spherical multipole operators.\cite{MikJorJen94,MedBudBar17}
Since then, analytic results for ellipsoidal cavities,\cite{WesKir38,RinRuiRiv83}
off-center charges,\cite{KirWes38} and higher-order off-center multipoles\cite{KonPon97} 
have also been derived, along with formulas for the interaction of multipoles contained in 
disjoint spheres with a dielectric medium in between.\cite{LotHea06}
Insofar as an arbitrary charge distribution $\rho(\br)$ has a single-center multipole expansion, these results provide an analytic
solution for a charge density of arbitrary complexity in a spherical or ellipsoidal cavity, 
assuming there is no penetration of $\rho(\br)$ into the continuum
region.   (The latter effect, known as {\em volume polarization},\cite{ZhaBenChi98,Chi97,Chi99,Chi00,Chi02a,Chi06b,Chi02b} 
is discussed below.)  The multipole expansion method 
has been called a ``generalized Kirkwood" solvation model,\cite{SchPon07} and using multipolar formulas for $\RxnF_0$ 
in Eq.~\eqref{eq:G[Psi]}, it can be turned into a generalized Kirkwood SCRF for quantum chemistry. 
Multipolar methods are reviewed elsewhere,\cite{TomPer94,LuqCurMun03}    
but have effectively been rendered obsolete by the polarizable continuum 
methods described in Section~\ref{sec:Basics:PCM}.   In the absence of volume polarization, these approaches 
also afford an exact (albeit numerical) solution to the continuum electrostatics problem, yet 
can used in conjunction with molecule-shaped cavities.

A spherical cavity may make sense 
if a large number of explicit solvent molecules are included in the atomistic solute region, 
and such approaches are known as {\em solvent boundary potential\/} methods.\cite{BegRou94,TirSpeSmi95,ImBerRou01}
These have been developed as replacements for periodic boundary conditions in both 
\qmmm\ simulations\cite{SchRicCui05,BenThi08,BenThi09,BenThi11,AleFie11,ZieCui12,LuCui13}  
and in fully quantum-mechanical 
calculations.\cite{RegBraBar06,BraRegBar06,BraRegBar08a,BraRegBar09}
Spherical cavities make little sense for single molecules, however, and it is clear from Eq.~\eqref{eq:Born} that 
$\Delta\FreeE$ will be quite sensitive to the cavity radius.   This is especially
problematic if that cavity does not correspond closely to the size of the molecule itself, and results 
in Section~\ref{sec:Smooth:SCCS} demonstrate that even for realistic ``molecule-shaped'' cavities, 
solvation energies are quite sensitive to cavity construction.

\begin{figure}
	\centering
	\fig{1.0}{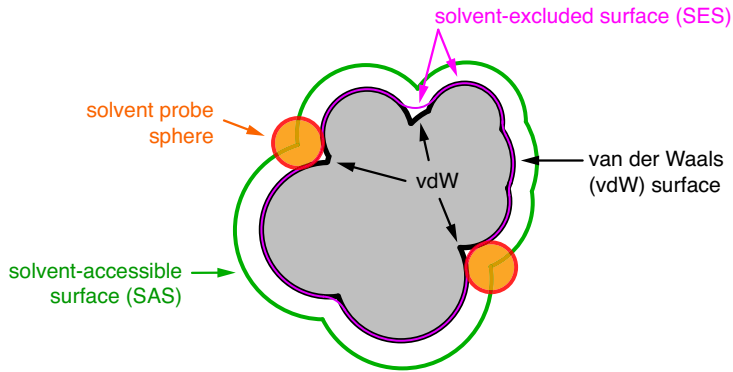}
	\caption{
		Illustration of cavity surface definitions using a set of atomic spheres (in gray) whose envelope defines
		the vdW surface (in black).  The solvent-accessible surface (in green) is defined either
		by augmenting the atomic radii by a probe radius ($R_\text{probe}$, 
		taken here to be smaller than the vdW radii), or equivalently as the center point of the probe sphere as it
		rolls over the vdW surface.  The solvent-excluded surface (SES, in magenta, sometimes called the ``molecular surface")
		is traced out by arcs of the probe that connect points of contact between 
		the probe and the vdW surface.   The SES eliminates the cusps in the vdW surface that otherwise appear along 
		seams of intersection between atom-centered spheres.
	}\label{fig:cavity_types}
\end{figure}

Examples of some molecule-shaped cavity constructions are provided in Fig.~\ref{fig:cavity_types}.
A simple union of atom-centered spheres is often called a {\em van der Waals\/} (vdW) cavity surface.
The atomic radii might simply be empirical parameters of the model,\cite{SwaAdcMcC05} or alternatively 
covalent radii extracted from crystal structures,\cite{Bon64,RowTay96} or based on calculations of 
atomic size.\cite{ManChaVal09}  In the latter two cases these radii represent 
close-contact distances, meaning that an implicit or continuum solvent 
should not be allowed to approach all the way to the vdW radii of the solute atoms.
This exclusion effect has been treated in several ways, most commonly by scaling the atomic vdW radii by a factor
$\vdwscale > 1$.  A scaling factor $\vdwscale = 1.2$ was chosen early on,\cite{BonPalTom84} 
and factors $\vdwscale\approx1.1$--$1.2$ have since become standard choices, 
albeit with little theoretical justification to choose one value over another within a modest range.
Alternatively, and with somewhat better justification,
the atomic vdW radii can be augmented by a ``probe radius'' representing the assumed size of a solute molecule, which can 
be extracted from the liquid structure of the solvent.  For example,  
$R_\text{probe}=1.4$~\AA\ is often used for water, representing 
half the distance to the first peak in the oxygen--oxygen radial distribution function.\cite{BroHea15}
Values for water as small as $R_\text{probe}=0.2$~\AA\ have also been used, however, in an effort to match
solvation energy benchmarks from simulations with explicit solvent.\cite{OnuAgu14}
The cavity surface generated using atomic radii $R_\text{vdW} + R_\text{probe}$ is known as the 
{\em solvent-accessible surface\/} (SAS), which 
was first introduced as a means to measure accessible surface area in proteins and 
is often used to define the ion accessibility function $\lambda(\br)$ in Eq.~\eqref{eq:LPBE}.   
Note that it does not make sense to augment a {\em scaled\/} vdW radius with a probe radius, as this double counts the
size of the exclusion layer.

Both the vdW surface and the SAS exhibit cusps where atomic spheres intersect. 
These cusps are eliminated by {\em solvent-excluded surface\/} (SES) that is generated by the probe sphere as it
rolls over the vdW surface; see Fig.~\ref{fig:cavity_types}.   (In principle, this procedure could be applied to eliminate cusps
in the SAS as well, but that is less often done because cusps in the SAS are less problematic, numerically speaking.\cite{LanHerAlb20})
The center of the probe sphere traces out the SES while its points of contact
with the vdW surface, combined with the concave arcs of the probe sphere that smooth over the cusps, constitute the SES.
(The SES is also known as the Connolly surface,\cite{Con83b} or sometimes simply 
the ``molecular surface",\cite{Ric77} as it is probably the closest approximation to the true shape of the 
molecule.  However, these names have sometimes been used interchangeably or ambiguously in the 
literature, and in particular the term ``Lee-Richards surface''\cite{LeeRic71} 
has been used to mean either the SAS and the SES.\cite{KimWonBha10,OnuAgu14,LanHerAlb20}
(The notation used here is standard in the quantum chemistry literature.\cite{Pom07})
The SES has an analytic construction,\cite{Con83b,LanHerAlb20} although it has 
most often been constructed numerically, for visualization purposes.

To a greater or lesser degree each of these cavity definitions seems physically reasonable, but beyond that 
there is little theoretical justification for any of them, and 
no guarantee that small changes in one or more atomic radii will not have a 
significant impact on computed observables.\cite{ZhaChi98,Chi02b}  
It has been suggested that the ``optimal'' atomic radius for a given atom likely ought to vary as a function of its partial 
atomic charge,\cite{BarCosTom97,GinCamDup08a} but this effect is generally handled empirically, if at all.
``United-atom'' models, in which hydrogen atoms are not given explicit spheres and the radii of the other atoms are increased
to compensate,\cite{BarCosTom97} are also in widespread use.
A more justifiable definition of the solute cavity surface might involve an isocontour of the molecular electron density 
$\rho(\br)$,\cite{ForKeiWib96,Chi00,ChiDup02,CheChi03,Chi06b}
as it is the density that is ultimately responsible for molecular size  and shape.\cite{doCChi10a} 
Isodensity cavity definitions are considered in Section~\ref{sec:Smooth:SCCS}.

With the introduction of a molecule-shaped cavity, one must forego analytic solution of Poisson's equation.  
A variety of numerical algorithms have been introduced, either for classical biomolecular 
applications,\cite{HolSai93,HolSai95,HolBakWan00,LuZhoHol08,WanLuo10,YapHea10,BosFen11,HolMcCYu12,LiLiPet13,GenKra13,mFES} 
or else for quantum electronic structure 
calculations.\cite{FisGenAnd16,RinObeHil16,WomAntDzi18,CooHer18,SteHerHea19} 
Numerically, Eq.~\eqref{eq:Gen-Poisson} and its Poisson-Boltzmann 
analogue are partial differential equations (for $\esp$) that require discretization of three-dimensional space, which 
must extend well into the continuum region 
due to the $(\dielst r)^{-1}$ asymptotic decay of $\esp(\br)$. 
This necessitates the use of multiresolution techniques for efficiency.\cite{HolSai93,HolBakWan00,Bak04,Bak05b,mFES,CooHer18}
Worth mentioning also is the Langevin dipoles model,\cite{LuzWar92,FloWar97,PapWar97a}
which discretizes the continuum solvent using a three-dimensional grid of point dipoles, and is therefore a direct realization
of the conceptual notion that $\mathbf{P}$ is a dipole density.

As compared to these approaches, each of which requires discretization of three-dimensional space, 
a much more efficient two-dimensional formulation of the continuum electrostatics problem is often feasible.  
The transformation of Eq.~\eqref{eq:Gen-Poisson}
from a partial differential equation in three dimensions to an integral equation in two dimensions (over
the cavity surface $\surf$) defines the class of methods known as 
{\em polarizable continuum models\/} (PCMs),\cite{TomMenCam05,Men12,HerLan16} 
which we next describe.

\subsection{Polarizable continuum models}
\label{sec:Basics:PCM}

In quantum chemistry, PCMs are so widely used as to be nearly synonymous
with continuum solvation itself.   This popularity stems from efficiency, which in turn derives from a transformation of the 
volumetric polarization theory of the Poisson or Poisson-Boltzmann equation (Section~\ref{sec:Basics:Poisson}) 
into a surface charge problem that can be solved with numerical methods 
that are described in Section~\ref{sec:Smooth:SWIG}.     That transformation, and the basic working equations of PCMs,
are introduced in this section.   Matrix analogues of these integral equations, which define the various PCMs in practice, 
are introduced in Section~\ref{sec:Smooth:Matrix}.

Physically speaking, polarization of the continuum extends beyond the solute cavity surface, as indicated pictorially in 
Fig.~\ref{fig:cavity-ESP-schematic}(b).
Transformation of the three-dimensional polarization problem into a two-dimensional problem 
defined on the cavity surface $\surf$ relies on a characteristic feature of a sharp dielectric interface, namely, a 
discontinuity in the electric field resulting from a buildup of charge at the interface.   This occurs at any dielectric interface,
including the one that defines the boundary between atomistic solute and continuum solvent. 
Let $\bs\in\surf$ denote
a point on the interface (cavity surface), and let $\mathbf{n}_{\bs}$ be the outward-pointing unit vector normal to the cavity surface 
at the point $\bs$.  At a sharp interface in the permittivity function $\diel(\br)$, as in Eq.~\eqref{eq:eps(r)-sharp},
the normal electric field satisfies a ``jump'' boundary condition,\cite{TomPer94,Chi97}
\begin{equation}\label{eq:jump} 
	\epsout \; \bigl(\mathbf{n}_{\bs}\bm{\cdot}\del\bigr) \, \esp(\bs) \Big|_{\bs = \bs^+} =
	\epsin\; \bigl(\mathbf{n}_{\bs}\bm{\cdot}\del\bigr) \, \esp(\bs) \Big|_{\bs = \bs^-} \; .
\end{equation}
This ensures that the electric displacement $\mathbf{D}(\br)=\diel(\br) \, \Efield(\br)$ is continuous across the 
interface.\cite{Bot76,Wan86}
The notation $\bs^\pm$ indicates that these are one-sided derivatives to be evaluated immediately inside ($\bs^-\in\cav$) or 
outside ($\bs^+\notin\cav$) of the solute cavity, as $\esp$ is only semi-differentiable at the interface.

Polarization of the continuum manifests as a surface charge that accumulates 
at the boundary between the atomistic and the continuum region in order to satisfy Eq.~\eqref{eq:jump}, and whose magnitude is  
proportional to the outward-pointing normal electric field at the dielectric interface.\cite{Wan86}
Let us call that charge $\sigma(\bs)$,
where $\bs\in\surf$, so as to distinguish it from a volume charge such as $\rho(\br)$, where $\br \in \realspace$.   
Introducing the notation $\hat{\partial}_{\,\bs} = \mathbf{n}_{\bs}\bm\cdot\del$ to indicate the normal derivative, 
the surface charge at the dielectric boundary can be expressed in several equivalent ways,\cite{Chi97,Tom11} 
two of which are 
\begin{equation}\label{eq:sigma(s)}
	\sigma(\bs) = 
	\frac{1}{4\pi}	
	\left(\frac{ \epsout-\epsin }{ \epsin }\right)
	\hat{\partial}_{\,\bs} \esp(\bs) 
	\Big|_{\bs = \bs^+} 	
	= \frac{1}{4\pi}
	\left(\frac{ \epsout-\epsin }{ \epsout }\right)
	\hat{\partial}_{\,\bs} \esp(\bs) 		
	\Big|_{\bs = \bs^-}  
\; .
\end{equation}
These differ depending on whether the normal electric field is evaluated immediately inside or outside of the cavity.
In the usual quantum chemistry case where $\epsin=1$ and $\epsout=\dielst$, the first form of this equation (in which the field
is evaluated within the continuum region) is merely the definition of the polarization,
\begin{equation}
	\mathbf{P} = \left(\frac{\dielst-1}{4\pi}\right)\Efield \; ,
\end{equation}
as obtained from Eq.~\eqref{eq:D(r)}.
If the entirety of the solute charge $\rho(\br)$ is confined within the cavity surface, then Eq.~\eqref{eq:sigma(s)} is 
simply a reflection of Gauss' Law.   For cavities 
that are realistic approximations to the size and shape of a molecular solute, however, the
tails of a quantum-mechanical charge distribution will penetrate into the continuum 
region.\cite{ZhaBenChi98,Chi97,Chi99,Chi00,Chi02a,Chi06b,Chi02b} 
This outlying or ``escaped'' charge is discussed in more detail below but will be ignored for now.

It has been argued that the second form of Eq.~\eqref{eq:sigma(s)}, in which the derivative is evaluated
inside of the cavity, should be used in order to avoid ``self-polarization'' of the medium.\cite{TomPer94}   Then  
taking $\epsin=1$ and $\epsout=\dielst$ and recognizing that $\Efield = -\del\esp$, one obtains a model
\begin{equation}\label{eq:D-PCM}
	\sigma(\bs) = -\frac{1}{4\pi}
	\left(\frac{\dielst-1}{\dielst}\right)
	\underbrace{
		\big[\Efield^\rho(\bs) + \Efield^\sigma(\bs)\big]\bm{\cdot}\mathbf{n}_{\bs} 
	}_{
		\Eperp(\bs)
	} 
	= \frac{1}{4\pi}\left(\frac{\dielst-1}{\dielst}\right)
	\left(\frac{\partial \esp}{\partial\mathbf{n}_{\bs}}\right)_{\bs=\bs^-}
\end{equation}
that corresponds to the original PCM introduced by Tomasi and co-workers.\cite{MieScrTom81,MieTom82,BonCimTom83,TomPer94} 
Here, the induced surface charge density $\sigma(\bs)$ is proportional to the outward-directed normal electric field, 
$\Eperp(\bs) = \Efield(\bs)\bm{\cdot}\mathbf{n}_{\bs}$, which is separated into two contributions in Eq.~\eqref{eq:D-PCM}, 
analogous to how $\esp(\br)$ is partitioned in Eq.~\eqref{eq:phi(r)}.   The contribution 
$\Efield^\rho = -\del\esp^\rho$ comes from the solute, whose electrostatic potential $\esp^\rho(\br)$ is defined in 
Eq.~\eqref{eq:phi(r)}, whereas 
$\Efield^\sigma=-\del\esp^\sigma$ can be identified as the reaction-field contribution and originates
in the electrostatic potential 
\begin{equation}\label{eq:esp-sigma}
	\esp^\sigma(\br) = \int_{\bs\in\surf}
	 \frac{\sigma(\bs)} {\|\bs - \br \|} \; 
	d\bs 
\end{equation}
that is generated by the surface charge $\sigma$.    To compute $\esp^\sigma(\br)$, it is only necessary to discretize 
the cavity surface $\surf$ rather than the whole of three-dimensional space.   Historically,  
Eq.~\eqref{eq:D-PCM} was the first example of what has been called an {\em apparent surface charge\/} (ASC) formulation of the 
continuum electrostatics problem.      Nowadays the term ``ASC model'' is essentially synonymous with PCM, as multipolar expansions 
and other simplified treatments see little use in contemporary quantum chemistry because the model defined by Eq.~\eqref{eq:D-PCM}
makes it easy to use the exact solute density $\rho(\br)$ along with a cavity or arbitrary shape.   (Discretization of the surface is
considered in Section~\ref{sec:Smooth}.)  For a spherical cavity, Eq.~\eqref{eq:D-PCM} is 
equivalent to the use of the Kirkwood multipolar expansion formulas if the latter are carried to arbitrary order, as can be proven
explicitly.\cite{LanClaCai88}

In early literature, the model defined by Eq.~\eqref{eq:D-PCM} is often called ``the'' ASC-PCM,\cite{TomPer94}
whereas in contemporary literature it is usually called D-PCM,\cite{TomMenCam05,Men12} 
or sometimes the Miertu\v{s}-Scrocco-Tomasi model.\cite{OroLuq00,LuqCurMun03}   
The somewhat arbitrary decision to use the right side of Eq.~\eqref{eq:sigma(s)} can be avoided, and a somewhat different
version of this model obtained, by noting that the discontinuity in $\Eperp$ at the cavity 
surface can be expressed as\cite{Chi97}
\begin{subequations}
\begin{align}
	\hat{\partial}_{\,\bs} \,\esp^\sigma(\bs)  \Big|_{\bs = \bs^-} 
	&= 2\pi\sigma(\bs) + \hat{\partial}_{\,\bs} \,\esp^\sigma(\bs) 
\\
	\hat{\partial}_{\,\bs} \,\esp^\sigma(\bs)  \Big|_{\bs = \bs^+} 
	&= -2\pi\sigma(\bs) + \hat{\partial}_{\,\bs} \,\esp^\sigma(\bs)  	
\; .
\end{align}
\end{subequations}
Adding these two equations and combining them with Eq.~\eqref{eq:sigma(s)} affords a different expression for the
surface charge,\cite{Con86,LanClaCai88,Chi97}
\begin{equation}\label{eq:sigma(s)-2}
	\sigma(\bs) = \left(\frac{\felst{\diel}}{2\pi}\right) \hat{\partial}_{\,\bs}
	\big[
		\esp^\rho(\bs)  + \esp^\sigma(\bs) 
	\big]\; ,
\end{equation}
in which the normal derivative is evaluated {\em at\/} (rather than {\em near}) the point $\bs\in\surf$.
The permittivity-dependent prefactor in this expression is 
\begin{equation}\label{eq:f(eps)}
	\felst{\diel} = \frac{\epsout - \epsin}{\epsout + \epsin} \; .
\end{equation}
(For a QM solute the only sensible choice is $\epsin=1$, but the more general notation is retained for now, in order 
to accommodate different values of $\epsin$ that are sometimes used in biomolecular electrostatics 
calculations.\cite{AntMcCGil94,DemWad96,Nak96,Gry02,GroTry08,AleMehBak11})

The model of Eq.~\eqref{eq:sigma(s)-2} can be recast in a convenient form by defining an operator $\Hat{D}^\dagger$
that acts on surface functions ($\bs\in\surf$) according to\cite{Chi00,Chi02a,Chi06b} 
\begin{equation}\label{eq:Hat(D)-dagger} 
	\Hat{D}^\dagger\sigma(\bs) 
	= \int_{\surf} d\bs' \sigma(\bs') 
		\frac{\partial}{\partial \mathbf{n}_{\bs}} 
		\left(\frac{1}{\|\bs - \bs'\|}\right) 
	= -\Eperp^\sigma(\bs)
		\; .
\end{equation}
The first equality defines $\hat{D}^\dagger$ and the second follows upon realization that 
$\partial/\partial\mathbf{n}_{\bs} = \mathbf{n}_{\bs}\bm{\cdot}\del$ can be pulled outside of the integral, leaving 
$(\mathbf{n}_{\bs}\bm{\cdot}\del)\esp^\sigma(\bs) = -\Eperp^\sigma(\bs)$.
The operator $\hat{D}^\dagger$ is often called $\Hat{D}^\ast$ in the literature,\cite{Chi00,Chi02a,Chi06b,Can07}
but is the adjoint of the double-layer operator $\hat{D}$ that is introduced below, and the notation used here is selected 
to reflect that fact.
Using $\Hat{D}^\dagger$ to rewrite Eq.~\eqref{eq:sigma(s)-2} affords an alternative to Eq.~\eqref{eq:D-PCM}, namely 
\begin{equation}\label{eq:SPE}
	\left[
		\left(\frac{2\pi}{\felst{\diel}}\right)\hat{1} - \Hat{D}^\dagger
	\right]\sigma(\bs) 
	= -\Eperp^\rho(\bs) \; .
\end{equation}
This equation makes clear that the sole ingredient needed to determine the induced surface charge is 
\begin{equation}
	\Eperp^\rho(\bs) = -(\mathbf{n}_{\bs}\bm{\cdot}\del)\esp^\rho(\bs)
	= -\left(\frac{\partial \esp^\rho}{\partial\mathbf{n}_{\bs}}\right) \;, 
\end{equation}
which is the normal electric field due to the solute, 
evaluated at the cavity surface.  Chipman refers to Eq.~\eqref{eq:SPE} as the {\em surface polarization for
electrostatics\/} (SPE) method,\cite{Chi02a} but others have called it D-PCM,\cite{Can07} or simply PCM.\cite{CosScaReg02b}

The D-PCM approach, which requires explicit calculation of the electric field at the cavity surface, has largely been 
superseded by alternative ASC-PCMs that determines $\sigma(\bs)$ using only 
electrostatic potentials and not their derivatives, 
which might be more sensitive to discretization error.    The modern approach is known as the 
{\em integral equation formalism\/} (IEF-) PCM,\cite{TomMenCan99,Can07}  
and is based on a reformulation of the continuum electrostatics problem as a boundary-value 
problem.\cite{CanMenTom97,MenCanTom97,CanMen98a} 
This reformulation is exact provided that the escaped charge is zero (\eg, for a classical solute), and we continue to defer a
discussion of the escaped charge problem.  IEF-PCM is formulated in terms of integral operators $\hat{S}$ and $\hat{D}$ 
that act on surface functions (defined on $\surf$) to generate the single- and double-layer potentials, respectively.  
These operators are defined by 
\begin{equation}\label{eq:Hat(S)}
	\hat{S} \sigma(\bs) = \int_{\surf} d\bs' 
		\frac{\sigma(\bs')}{\|\bs' - \bs\|}
	= \esp^\sigma(\bs) 
	\; ,
\end{equation}
which generates the electrostatic potential associated with the surface charge distribution $\sigma(\bs)$, and 
\begin{equation}\label{eq:Hat(D)}
	\Hat{D}\sigma(\bs) = 
	\int_{\surf} d\bs' \sigma(\bs') 
	\underbrace{
		\frac{\partial}{\partial \mathbf{n}_{\bs'}} 	
		\left(\frac{1}{\|\bs' - \bs\|}\right)
	}_{
		D(\bs,\bs')
	} \; ,
\end{equation}
which generates the double-layer potential, in the language of integral equations.\cite{Can07} 
[The operator $\Hat{D}$ is the adjoint of $\Hat{D}^\dagger$ in Eq.~\eqref{eq:Hat(D)-dagger},\cite{TomMenCam05}
as becomes clear upon reversing the indices of the kernel $D(\bs,\bs')$.]   Using $\Hat{S}$ and $\Hat{D}$, 
the continuum electrostatics problem can be recast as an integral equation on the surface of the cavity:\cite{TomMenCam05}
\begin{equation}\label{eq:IEF-PCM}
	\left[
		\left(\frac{2\pi}{\felst{\diel}}\right)
		\hat{1} - \hat{D}
	\right] \hat{S}\sigma(\bs)
	= \big(-2\pi\hat{1}+\Hat{D}\big)\,\esp^\rho(\bs) 
\; .
\end{equation}
Equation~\eqref{eq:IEF-PCM} is the basic working equation of IEF-PCM.   In early papers the working equation was formulated
somewhat differently, requiring $\Eperp^\rho$ in addition to $\esp^\rho$.\cite{CanMenTom97,MenCanTom97,CanMen98a}
That form is sometimes called simply ``IEF",\cite{Chi02a,Can07} to distinguish it
from the IEF-PCM that requires $\esp^\rho$ but not its derivative, and should therefore be more stable numerically.
Equivalence of the two forms is demonstrated in Ref.~\citenum{CanMen01a}, and in fact 
Eq.~\eqref{eq:IEF-PCM} can be cast in a variety of equivalent forms,\cite{Chi00,Chi02a,Can07} 
by taking advantage of the fact that $\hat{S}=\hat{S}^\dagger$ and 
\begin{equation}\label{eq:DS=SDt}
	\hat{D}\hat{S} = \hat{S}\hat{D}^\dagger \; .
\end{equation}
However, except for spherical cavities (for which $\hat{D}=\hat{D}^\dagger$),\cite{Chi00}
the operator identity in Eq.~\eqref{eq:DS=SDt} is not generally preserved upon discretization,\cite{CosScaReg02b,LanHer11c}
with the practical result that various forms of Eq.~\eqref{eq:IEF-PCM} are {\em inequivalent\/} as finite-dimensional matrix 
equations.\cite{LanHer11c,HerLan16}  These equations are discussed in Section~\ref{sec:Smooth:Matrix}.

For now, we simply note that Eq.~\eqref{eq:IEF-PCM} is an exact reformulation of the {\em classical\/} 
continuum electrostatics problem, meaning the problem that is 
defined by Poisson's equation with a sharp dielectric boundary 
and where the solute's charge density $\rho(\br)$ is contained entirely within the cavity.   That caveat is satisfied if the solute 
consists of atomic point charges (or higher-order multipoles, including polarizable ones) from a force field.   In that case, 
solution of Eq.~\eqref{eq:IEF-PCM} for $\sigma(\bs)$ constitutes an exact solution to the electrostatics problem, and the 
electrostatic solvation energy is\cite{LanHer10b} 
\begin{equation}\label{eq:Gelst-ASC}
        \Gelst = \frac{1}{2}\int_{\realspace}  \esp^\sigma(\br)  \; \rho(\br) \; d\br
        = \frac{1}{2} \int_{\surf}  \esp^\rho(\bs) \; \sigma(\bs) \; d\bs 
      \; .
\end{equation}
These are analogous to the two forms of $\Gelst$ that are given in Eq.~\eqref{eq:G_elst}, however the second form in 
Eq.~\eqref{eq:Gelst-ASC} requires only surface integration.   
Using this ASC formulation, in lieu of discretizing three-dimensional space, has significant 
advantages over the traditional approach to biomolecular electrostatics, which 
require discretization far enough into the continuum such that $\esp^\rho(\br)$ has decayed to zero.
Most contemporary biomolecular electrostatics calculations are also based on 
finite-difference evaluation of the Laplacian $\hat{\nabla}^2\esp(\br)$,\cite{DavMcC89,LutDavMcC92} 
which leads to problems in obtaining smooth forces for molecular dynamics.\cite{WanCaiXia12,XiaCaiYe13,XiaWanLuo14}
Discretization of the action of $\hat{D}$ and $\hat{S}$ on surface functions can be accomplished in a manner that affords
inherently smooth forces;\cite{LanHer10a,LanHer10b,HerLan16} 
see Section~\ref{sec:Smooth:SWIG}.
Especially for 
biomolecular applications it is worth noting that the IEF-PCM has been adapted to provide a solution to the linearized
Poisson-Boltzmann problem,\cite{CanMenTom97,MenCanTom97,Chi04,LanHer11b}
including the ``size-modified'' version that accounts for finite size of the mobile ions.\cite{LanHer11b}
Large biomolecular solutes have been tackled in this way,\cite{LipStaCan13,LipLagSca14,CapJurLag15,HerLan16}  
but this requires iterative solvers for the matrix equations that define the PCM.  
Linear-scaling implementations that can handle biomolecular solutes are discussed in Section~\ref{sec:Smooth:O(N)}.

For QM solutes, however, there is always escaped charge for realistic cavity sizes, and therefore IEF-PCM is not a fully
equivalent substitute for Poisson's equation.   The extent to which this is a problem
is unclear from the original derivation of IEF-PCM provided by   
Canc\`es \etal,\cite{CanMenTom97,MenCanTom97,CanMen98a,Can07} 
which does not provide much physical insight, nor does it emphasize the assumption (inherent in the
derivation) that there is no outlying charge.  
That issue was addressed directly by Chipman,\cite{Chi97,Chi99,Chi00,Chi02a}  
who assumes from the start that there is outlying charge and that as a consequence the correct reaction-field potential is not 
$\esp^\sigma(\br)$ but rather 
\begin{equation}
	\esp_\rxn(\br) = \esp^\sigma(\br) + \esp^\beta(\br) \; ,
\end{equation}
where $\esp^\sigma(\br)$ arises from the accumulation of charge $\sigma(\bs)$ at the dielectric interface but is 
accompanied by an additional potential $\esp^\beta(\br)$ due to volume polarization.   The latter originates with 
the tail of the solute's 
charge density that extends beyond the cavity surface.    Including an additional term for $\esp^\beta$ on the right side
of Eq.~\eqref{eq:sigma(s)-2}, and recognizing that $\hat{\partial}_{\,\bs}\esp^\sigma(\bs) = \hat{D}^\dagger\sigma(\bs)$, 
an exact equation for the surface charge that includes the effects of volume polarization is\cite{Chi00} 
\begin{align}\label{eq:SVPE}
\begin{aligned}
	\left[\hat{1} - \left(\frac{\felst{\diel}}{2\pi}\right)\hat{D}^\dagger\right]\sigma(\bs)
	&= \frac{\felst{\diel}}{2\pi}\big[
		\hat{\partial}_{\,\bs}\,\esp^\rho(\bs) + \hat{\partial}_{\,\bs}\,\esp^\beta(\bs)
	\big] 
\\
	&= -\frac{\felst{\diel}}{2\pi}\big[
		\Eperp^\rho(\bs) + \Eperp^\beta(\bs)
	\big]	
\; .
\end{aligned}
\end{align}
The potential $\esp^\beta(\br)$ can be modeled as the electrostatic potential generated by a charge density 
\begin{equation}
	\beta(\br) = \begin{cases}
		0 	
			& \text{for $\br \in\cav$}
		\\
		\left(\epsout^{-1} - \epsin^{-1}\right) \rho(\br) 
			& \text{for $\br \notin\cav$}
	\end{cases}
\end{equation}
that satisfies a vacuum-like Poisson equation, 
\begin{equation}
	\hat{\nabla}^2 \esp^\beta(\br) = -4\pi\beta(\br) \; .
\end{equation}
Equation~\eqref{eq:SVPE} can be solved numerically,\cite{ZhaBenChi98}
to afford an exact solution to the continuum electrostatics problem, in a method that Chipman calls 
{\em surface and volume polarization for electrostatics\/} (SVPE).\cite{Chi02a}  
This approach is challenging in practice because the volume charge density $\beta(\br)$
is discontinuous at the cavity surface, however Eq.~\eqref{eq:SVPE} 
can be recast into a form that requires only the normal electric field
$\Eperp^\rho(\bs)$ at the cavity surface, along with the solution of the ASC-PCM that is introduced next.\cite{Chi06b} 
This provides a practical means to access {\em exact\/} electrostatics, even in the presence of outlying charge,
while staying within a two-dimensional surface integral formalism.

A simplified (if approximate) treatment is possible, which eliminates the normal electric field in Eq.~\eqref{eq:SVPE}
and ultimately connects back to IEF-PCM.   This model is obtained 
by introducing an additional surface charge $\alpha(\bs)$, 
distinct from $\sigma(\bs)$,  that is defined by the condition
\begin{equation}
	\Hat{S}\alpha(\bs) = \esp^\beta(\bs) \; ,
\end{equation}
meaning that its electrostatic potential $\esp^\alpha = \Hat{S}\alpha$ must 
reproduce $\esp^\beta$ on the cavity surface.\cite{ZhaBenChi98,Chi97,Chi99,Chi00}
This also ensures that $\esp^\alpha(\br) = \esp^\beta(\br)$ for all interior points $\br\in\cav$, and while the two potentials may differ 
{\em outside\/} of the cavity, those contributions are
scaled by $\dielst^{-1}$ and therefore less important.   (This is confirmed in numerical tests.\cite{ZhaBenChi98,Chi02b})
Assuming that the true surface charge, augmented to reflect volume polarization, is 
$\tilde{\sigma}(\bs) = \alpha(\bs) + \sigma(\bs)$, the term $\hat{\partial}_{\,\bs}\,\esp^\beta(\bs)$ in Eq.~\eqref{eq:SVPE} 
can be manipulated into a form that is consistent with the ASC formalism.\cite{Chi00}  
The result is a model that Chipman has called {\em surface and simulation of volume polarization for electrostatics\/} 
[SS(V)PE],\cite{Chi00,Chi02a,Chi06b}
\begin{equation}\label{eq:SS(V)PE}
	\underbrace{
		\Hat{S}\left(\hat{1} - \frac{\felst{\diel}}{2\pi}\Hat{D}^\dagger\right) 
	}_{
		\Hat{K}_{\diel}
	}
	\tilde\sigma(\bs) =
	\underbrace{
		\felst{\diel} \left(\frac{1}{2\pi}\Hat{D} - \hat{1}\right) 
	}_{
		\Hat{Y}_{\diel}
	}
	\esp^\rho(\bs) 
\; .
\end{equation}
Using the identity in Eq.~\eqref{eq:DS=SDt}, this equation is easily rearranged to afford 
Eq.~\eqref{eq:IEF-PCM} that defines IEF-PCM, and therefore the two models are equivalent at the level of integral 
operators.\cite{Chi00,Chi02a,CanMen01a}  (They differ in practice, as described in Section~\ref{sec:Smooth:Matrix}.)
Importantly, what the derivation of SS(V)PE makes clear is that the surface charge that is determined by solving
Eq.~\eqref{eq:IEF-PCM} implicitly contains the (approximate) effects of volume polarization, which was not evident from the
derivation by Canc\`es and co-workers.\cite{CanMenTom97,MenCanTom97,CanMen98a,Can07} 
Chipman's derivation clarifies that both 
approaches constitute an exact treatment of electrostatic interactions in the limiting case that there is no escaped 
charge ($\alpha \equiv 0$).

\subsection{Comparison of ASC-PCMs}
\label{sec:Basics:Models}

Table~\ref{table:Gelst-Chipman} presents electrostatic solvation energies ($\Gelst$) for several small molecules and ions, in both 
a low-dielectric solvent (toluene, $\dielst=2.4$) and a high-dielectric solvent (water, $\dielst=78$).\cite{Chi02a}
The SVPE method [Eq.~\eqref{eq:SVPE}] affords the exact result but SS(V)PE solvation energies are within 0.1~kcal/mol.
In comparison, the SPE method of Eq.~\eqref{eq:SPE}, which is equivalent to D-PCM,  
exhibits noticeable differences, especially for ions. 
The amount of outlying charge ($Q_\text{out}$) is also quantified in Table~\ref{table:Gelst-Chipman}.   For future reference we note the
obvious definitions
\begin{subequations}
\begin{align}
	Q_\text{in}  & = \int_{\br\in\cav} \rho(\br) \; d\br \\
	Q_\text{out} & = \int_{\realspace} \rho(\br)  \; d\br - Q_\text{in} \; .
\end{align}
\end{subequations}
The charge density $\rho(\br)$ includes both nuclei and electrons, so $Q_\text{in}+Q_\text{out}=0$ for a neutral molecule.
The escaped charge is generally $|Q_\text{in}| \approx 0.1$--$0.2e$ for small solutes.\cite{ZhaBenChi98,Chi02b}

\begin{table}
	\centering
	\caption{ 
		Electrostatic solvation energies in toluene ($\dielst=2.4$) and in water ($\dielst=78.3$), computed with various approaches.$^a$
		The SVPE method [Eq.~\protect\eqref{eq:SVPE}] affords the exact result and SPE is the method in Eq.~\protect\eqref{eq:SPE}.
	}\label{table:Gelst-Chipman}
\begin{threeparttable}
	\begin{tabular}{l . .. .. . @{\hspace{0.5cm}} .}
\hline\hline
		Solute &  \mc{1}{c}{$\dielst$}
		& \mc{5}{c}{$\Gelst$ (kcal/mol)}& \mc{1}{c}{$Q_\text{out}$} 
\\ \cline{3-7}
		&& \mc{1}{c}{SVPE} & \mc{1}{c}{SS(V)PE}
		& \mc{1}{c}{SPE}
		& \mc{2}{c}{COSMO$^b$} 
		& \mc{1}{c}{(a.u.)$^c$}
\\ \cline{6-7}
		&&&&& \mc{1}{c}{$\cosmoconst=0$} & \mc{1}{c}{$\cosmoconst=1/2$} & 
\\ \hline
	H$_2$O				& 2.4 	& -3.9	& -3.9	& -4.0	& -4.8	& -3.9	&
\\ 
	CH$_3$CONH$_2$ 	& 2.4	& -5.3	& -5.0	& -5.2	& -5.9	& -4.8	&
\\	
	NO$^+$ 				& 2.4	& -52.2	& -52.2	& -55.3	& -52.5	& -43.4	&
\\
	CN$^-$ 				& 2.4	& -39.4	& -39.4	& -35.0	& -39.4	& -32.5	&
\\ \hline
	H$_2$O 				& 78.3	& -8.6	& -8.6	& -8.7	& -8.6	& -8.6	& -0.06
\\
	CH$_3$CONH$_2$ 	& 78.3 	& -10.9	& -10.8	& -11.1	& -10.9	& -10.8	& -0.15
\\
	NO$^+$ 				& 78.3	& -89.5	& -89.5	& -94.7	& -89.5	& -88.9	& -0.07
\\
	CN$^-$ 				& 78.3	& -67.4	& -67.3	& -56.8	& -67.3	& -66.9	& -0.17
\\ \hline\hline
	\end{tabular}
\begin{tablenotes}[flushleft]
\fns
	\item 
	$^a$From Ref.~\CiteN{Chi02a}; all calculations used an isodensity cavity with $\rho_0=0.001$~a.u.. 
	$^b$Using a renormalization factor $\felsttilde{\diel}(\cosmoconst)$, Eq.~\protect\eqref{eq:f(eps)-COSMO}.
	$^c$From Ref.~\CiteN{Chi02b}.
\end{tablenotes}
\end{threeparttable}
\end{table}

By arbitrarily dropping the $\Hat{D}$- and $\hat{D}^\dagger$-dependent terms in Eq.~\eqref{eq:SS(V)PE}, one obtains a model 
$\hat{S}\sigma = -\felst{\diel}\esp^\rho$ that we rewrite as 
\begin{equation}\label{eq:COSMO}
	\hat{S} \sigma(\bs) = -\felsttilde{\diel}(\cosmoconst) \; \esp^\rho(\bs) \; .
\end{equation}
This introduces a parameter $\cosmoconst$ into the permittivity factor that was defined in 
Eq.~\eqref{eq:f(eps)}, rewriting it as 
\begin{equation}\label{eq:f(eps)-COSMO}
	\felsttilde{\diel}(\cosmoconst) = \frac{\dielst-1}{\dielst + \cosmoconst} 
\end{equation}
for the usual QM case where $\epsin=1$ and $\epsout=\dielst$.   The model in Eq.~\eqref{eq:COSMO} has a long history and
a variety of names, one of which is the {\em conductor-like screening model\/} (COSMO).\cite{KlaSch93} 
Since the neglected double-layer operator embodies the electric field discontinuity at the cavity surface, this condition is not satisfied
when these terms are neglected to obtain Eq.~\eqref{eq:COSMO},\cite{Can07,LanHer11b} and rescaling the surface
charge by $\felsttilde{\diel}(\cosmoconst)$ can be seen as an attempt to mimic the effect of the jump boundary condition.

The name COSMO hints at the original derivation of the model in Eq.~\eqref{eq:COSMO}:  for a conductor ($\dielst=\infty$), 
the total electrostatic potential vanishes at the cavity surface and the ASC formulation of the Poisson problem is simply 
$\hat{S}\sigma = -\esp^\rho$.\cite{Chi99}
A scaling factor $\felsttilde{\diel}(\cosmoconst)$ is then introduced to account for the fact that $\dielst$ is finite, and 
that there is charge penetration into the medium.   The scaling factor [Eq.~\eqref{eq:f(eps)-COSMO}]
can be justified based on the normalization condition for the total surface charge, which is\cite{Chi97,MenCanTom97,CanMen01b}   
\begin{equation}\label{eq:Gauss}
	\int_{\surf} \sigma(\bs) \; d\bs = 
	-\left( \frac{1}{\epsin} - \frac{1}{\epsout}\right) Q_\text{in} \; .
\end{equation}
For $\epsin=1$ and $\epsout=\dielst$, the total surface charge $Q_\text{surf} = -[(\dielst-1)/\dielst]Q_\text{in}$, and it was originally 
suggested to take $\cosmoconst=1/2$ in order to renormalize for outlying charge,\cite{KlaSch93} 
although the value $\cosmoconst=0$ was later recommended for ions,\cite{KlaMoyPal15}
and values $0 \le \cosmoconst \le 2$ have also been used.\cite{AmoBarCam99} 
The choice $\cosmoconst=0$ has variously been called the ``generalized COSMO''
(GCOSMO) model,\cite{SteTru95,TruSte95a,TruSte95b,TruNguSte96} 
or the {\em conductor-like PCM\/} (C-PCM).\cite{BarCos98,CosRegSca03}
In fact, the normalization condition in 
Eq.~\eqref{eq:Gauss} forms the basis of various {\em ad hoc\/} attempts to rescale the surface 
charge,\cite{MieScrTom81,BonCimTom83,CamTom94a,KlaJon96,MenTom97} 
but tests against exact results from Poisson's equation suggest that none of these schemes is very satisfactory.\cite{ZhaBenChi98}  
Charge rescaling also complicates analytic energy gradients.\cite{CamCosTom96}

At a practical level, 
the choice $\cosmoconst=1/2$ for neutral solutes and $\cosmoconst=0$ for ions works remarkably well 
in comparison to the IEF-PCM and SS(V)PE methods, even in low-dielectric solvents.   This is suggested by the smattering of 
data in Table~\ref{table:Gelst-Chipman} and confirmed by calculations on a much larger data set, for which 
the statistical difference between these methods is 
$\lesssim0.1$~kcal/mol for neutral solutes and $\approx 0.5$~kcal/mol for ions, even at $\dielst=2$.\cite{KlaMoyPal15} 
Some justification for these parameter choices is found in the fact that 
$\felsttilde{\diel}(0)$ looks like the $\diel$-dependent factor in the Born ion model [Eq.~\eqref{eq:Born}],  
whereas $\felsttilde{\diel}(1/2)$ affords the prefactor in the Bell-Onsager model of dipole
solvation [Eq.~\eqref{eq:Bell}].    With appropriate choice of $\cosmoconst$, COSMO is therefore little different from 
SS(V)PE or IEF-PCM in practice, and considerably simpler.
It can be extended in a straightforward way to solvents with nonzero
ionic strength, with or without size exclusion.\cite{LanHer11b} 
For large biomolecular applications in aqueous solvent, there 
seems little reason {\em not\/} to use this approach, in lieu of exact but more complicated   
models, and as an alternative to finite-difference solution of Poisson's equation.

\section{Implementation}
\label{sec:Smooth}
Poisson's equation with a sharp dielectric interface makes for a cute problem in applied
mathematics and there has sometimes been a tendency in that community to analyze its properties, or those of the various
cavity surfaces discussed in Section~\ref{sec:Basics:cavity}, at fixed molecular geometries only.  This is not useful for chemistry, 
which requires a potential energy {\em surface}.  Exploration of that surface requires 
gradients of the total energy functional $\FreeE[\Psi]$, which in turn requires discretization of the integral  
equations introduced above.   In addition, many spectroscopic observables 
can be formulated as analytic energy derivatives,\cite{Gau00,RizCorRuu12,HelCorJor12} 
so a solution-phase theory that encompasses molecular properties requires a differentiable model.
Numerical implementation of the models introduced in 
Section~\ref{sec:Basics} is the topic of this section.    We deal specifically with PCMs in Sections~\ref{sec:Smooth:Matrix}
and \ref{sec:Smooth:SWIG}, but admit more general Poisson equation models in Section~\ref{sec:Smooth:SCCS}.
Linear-scaling implementations, which are useful for hybrid \qmmm\  simulations with PCM boundary conditions, are discussed in
Section~\ref{sec:Smooth:O(N)}.

\subsection{Matrix equations for PCMs}
\label{sec:Smooth:Matrix}
In practice, the integral equation that defines any PCM must be discretized in order to obtain a finite-dimensional matrix equation.   
As such, it is convenient to rewrite the equation that defines SS(V)PE and IEF-PCM as 
\begin{equation}\label{eq:Kq=Rv_oper}
	\Hat{K}_{\diel}\sigma(\bs) = \Hat{Y}_{\diel}\,\esp^\rho(\bs) \; ,
\end{equation}
which is simply a restatement of Eq.~\eqref{eq:SS(V)PE}.  
Using various definitions for $\Hat{K}_{\diel}$ and $\Hat{Y}_{\diel}$, this equation 
can encompass a whole family of ASC-PCMs including the simplified C-PCM and COSMO
methods (Table~\ref{table:matrices}).\cite{Chi02a,ChiDup02,LanHer11c,HerLan16}

\begin{table}[t]
	\centering
	\caption{
		Matrices that define various PCMs according to $\mathbf{K}_{\diel}\mathbf{q}=\mathbf{Y}_{\!\diel}\mathbf{v}^\rho$
		[Eq.~\protect\eqref{eq:Kq=Rv_matrix}]. 
	}\label{table:matrices}
\begin{threeparttable}
{\renewcommand{\arraystretch}{1.5}
	\begin{tabular}{lll}
\hline\hline
		Method & Matrix $\mathbf{K}_{\diel}$ & Matrix $\mathbf{Y}_{\!\diel}$ \\
\hline
		C-PCM$^{a,b}$
					& $\mathbf{S}$ 
					& $-\felsttilde{\diel}(0)\,\bm{1}$  \\	
		COSMO$^b$		
					& $\mathbf{S}$
					& $-\felsttilde{\diel}(\cosmoconst)\,\bm{1}$  \\	
		DESMO$^c$ 	  	
					& $\mathbf{S}$ 
					& $-\bm{1} + (1/\diel)\mathbf{M}$ \\
		SS(V)PE$^d$ 		
					& $\mathbf{S}-( f_{\diel}/4\pi)(\mathbf{DAS}+\mathbf{SAD}^\dagger)$ 
					& $-f_{\diel} \big[\bm{1} - (1/2\pi)\mathbf{DA}\big]$ \\
		IEF-PCM$^d$ 		
					& $\mathbf{S}-( f_{\diel}/2\pi)\mathbf{DAS}$ 
					& $-f_{\diel} \big[\bm{1} - (1/2\pi)\mathbf{DA}\big]$ \\		
\hline\hline
	\end{tabular}
}
	\begin{tablenotes}[flushleft]
	\fns
	\item $^a$Also known as GCOSMO.
	$^b$$\felsttilde{\diel}(\cosmoconst) = (\diel-1)/(\diel+\cosmoconst)$.
	$^c$$M_{ij} = \delta_{ij} \, \esp^\rho_\kappa(\bs_i)/\esp^\rho_0(\bs_i)$. 
	$^d$$\felst{\diel} = (\diel-1)/(\diel+1)$. 
	\end{tablenotes}	
\end{threeparttable}	
\end{table}

To discretize Eq.~\eqref{eq:Kq=Rv_oper}, one must first generate a surface grid of points $\bs_i\in\surf$.   The surface
charge distribution $\sigma(\bs)$ is thereby replaced by 
a set of point charges $\{q_i\}$ at the discretization points $\{\bs_i\}$.  The details of this procedure are discussed in 
the next section but for now it suffices to introduce a matrix notation 
\begin{equation}\label{eq:Kq=Rv_matrix}
	\mathbf{K}_{\diel}\mathbf{q}=\mathbf{Y}_{\!\diel}\mathbf{v}^{\rho}
\end{equation}
for the discretized form ofEq.~\eqref{eq:Kq=Rv_oper}, where $\mathbf{q}$ is a vector of surface charges and 
$\mathbf{v}^\rho$ denotes the molecular electrostatic potential evaluated
at the cavity surface, $\mathvee_i^\rho = \esp^\rho(\bs_i)$.     In this form, surface integrals are replaced by scalar products, \eg, 
$\FreeE_\elst = \tfrac{1}{2}\mathbf{q}\bm{\cdot}\mathbf{v}^{\rho}$ replaces Eq.~\eqref{eq:Gelst-ASC}.   An alternative form 
of Eq.~\eqref{eq:Kq=Rv_matrix} that is sometimes encountered is 
$\mathbf{q} = \mathbf{Q}_{\diel} \mathbf{v}^{\rho}$ where $\mathbf{Q}_{\diel} = \mathbf{K}_{\diel}^{-1} \mathbf{Y}_{\!\diel}$, but
whereas the corresponding operator $\Hat{Q}_{\diel} = \hat{K}_{\diel}^{-1}\hat{Y}_{\!\diel}$ is self-adjoint, this property 
is generally not preserved by discretization except in the special case of C-PCM and COSMO.\cite{LanHer10b}   
This means that the mapping from Eq.~\eqref{eq:Kq=Rv_oper} to Eq.~\eqref{eq:Kq=Rv_matrix} is not unique,
because discretization fails to preserve the condition $\hat{D}\hat{S} = \hat{S}\hat{D}^\dagger$.\cite{LanHer11c} 
In matrix form, this implies that 
$\mathbf{DAS} \neq \mathbf{SAD}^\dagger$ (except for spherical cavities),\cite{CosScaReg02b,LanHer11c} 
where $\mathbf{A}$ is a diagonal matrix 
consisting of the surface area $a_i$ associated with each discretization point $\bs_i$. 
This leads to an ambiguity in the matrix representation of the operator 
$\hat{K}_{\diel} = \hat{S}-(\felst{\diel}/2\pi)\hat{S}\hat{D}^\dagger$ [Eq.~\eqref{eq:SS(V)PE}], 
since it can be argued that any matrix of the form 
\begin{equation}\label{eq:K-matrix}
	\mathbf{K}_{\diel} = \mathbf{S} - \left(\frac{\felst{\diel}}{4\pi}\right)
	\underbrace{
		\big(c_1\mathbf{DAS} + c_2\mathbf{SAD}^\dagger\big) 
	}_{
		\mathbf{X}
	} \; . 
\end{equation}
is an equally valid representation, provided that $c_1+c_2=1$.\cite{LanHer11c}
The various choices for $\mathbf{X}$ are inequivalent in matrix form.
Historically, IEF-PCM has been implemented using $\mathbf{X}=\mathbf{DAS}$ (that is, $c_1=1$ and $c_2=0$), 
whereas SS(V)PE was implemented using the 
symmetrized matrix ($\mathbf{DAS} + \mathbf{SAD}^\dagger)/2$.\cite{Chi02a,HerLan16}  
This is indicated in Table~\ref{table:matrices}, which provides matrix definitions for these and other commonly-encountered PCMs.
Precise definitions of the matrices $\mathbf{S}$ and $\mathbf{D}$ that represent the operators $\hat{S}$ and $\hat{D}$ can be found
elsewhere.\cite{ChiDup02,Pom07,LanHer10b,LanHer11c,HerLan16} 
These depend somewhat upon the discretization algorithm that is 
selected, but generally $S_{ij}$ represents the Coulomb interaction between $q_i$ and $q_j$ (which is straightforward to discretize
except when $i=j$), whereas $D_{ij}$ incorporates the effects of the outward-pointing electric field.

\begin{figure}
	\centering
	\fig{1.0}{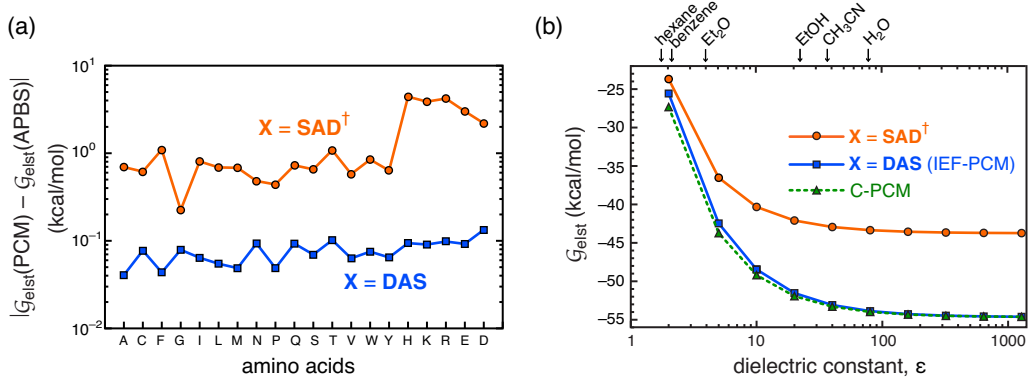}
	\caption{
		(a) Comparison of aqueous solvation energies, $|\FreeE_\elst(\mbox{PCM}) - \FreeE_\elst(\mbox{APBS})|$, 
		computed using either an ASC-PCM or else by 
		numerical solution of Poisson's equation using the \apbs\ software.  
		The data set of solutes consists of amino acids described using atomic partial charges from a force field,
		so that there is no outlying charge.
		The traditional implementation of IEF-PCM corresponds to $\mathbf{X}=\mathbf{DAS}$ (see
		Table~\protect\ref{table:matrices}) but results are also shown for the transpose matrix, $\mathbf{SAD}^\dagger$, and
		SS(V)PE is essentially the average of these two choices.
		(b)  Convergence of $\FreeE_\elst$ for classical histidine as a function of the solvent's dielectric constant, 
		using the \swig\ discretization 
		scheme described in Section~\protect\ref{sec:Smooth:SWIG}.   Dielectric constants for a few common solvents
		are indicated along the top axis.
		Adapted from Ref.~\CiteN{LanHer11c}; copyright 2011 Elsevier.		
	}\label{fig:vs_APBS}
\end{figure}

In the absence of outlying charge, the IEF-PCM and SS(V)PE models are exact up to discretization 
errors, which can be driven to zero in a controlled way.   Although this fact follows from the derivation in Section~\ref{sec:Basics:PCM},
it is worth emphasizing via numerical calculations.
For classical solutes with no outlying charge, which are described using atomic partial charges from a force field, 
Fig.~\ref{fig:vs_APBS}(a) presents a comparison of PCM solvation energies to those obtained by numerical solution of Poisson's
equation,\cite{LanHer11c,HerLan16}    
using a standard multiresolution algorithm as implemented in the \apbs\ software.\cite{APBS}
To examine the 
ambiguity regarding the choice of $\mathbf{X}$, these calculations test both $\mathbf{X}=\mathbf{DAS}$ (\ie, IEF-PCM) 
and also $\mathbf{X}=\mathbf{SAD}^\dagger$; the SS(V)PE model should be essentially the average of these two choices.
Calculations in Fig.~\ref{fig:vs_APBS}
use dense but finite discretization grids, and a small systematic discrepancy is evident between the two choices of $\mathbf{X}$, 
indicating a systematic discrepancy between SS(V)PE and IEF-PCM.   Numerical values of $\FreeE_\elst$
obtained using the IEF-PCM choice of $\mathbf{X}$ agree to within $\lesssim 0.1$~kcal/mol of \apbs\ results, demonstrating the operational
equivalence of the ASC and the volumetric implementation of continuum electrostatics.    This renders the Kirkwood (multipolar
expansion) formulas obsolete, since the requisite formulas are only valid in the absence of escaped charge anyway, and in that
case the ASC-PCM formalism furnishes an exact numerical solution to the continuum electrostatics problem, for cavities of 
arbitrary shape, while also providing a correction for outlying charge.\cite{Chi00,CanMen01b,Chi06b}

The conductor-like model (C-PCM) represents the high-dielectric limit of IEF-PCM and the two models do indeed 
become equivalent as $\diel\rightarrow\infty$,\cite{Chi99,LanHer11b,LanHer11c} 
as demonstrated numerically in Fig.~\ref{fig:vs_APBS}(b).  In practice, there is little difference between the two models 
already for moderately polar solvents.  For $\diel > 10$, there seems little justification for the increased
complexity of IEF-PCM relative to C-PCM.
That said, for non-spherical cavities only the $\mathbf{X}=\mathbf{DAS}$ form of the $\mathbf{K}_{\diel}$ matrix
achieves the correct conductor limit for finite discretization grids, as demonstrated by both formal and numerical 
arguments.\cite{LanHer11c}    
In general, $\hat{D}^\dagger$ proves more challenging to implement as compared to $\hat{D}$,\cite{LanHer11c,YouMewDre15}  
so the absence of $\hat{D}^\dagger$ is one reason to prefer the IEF-PCM form with
$\mathbf{X}=\mathbf{DAS}$.

It is illustrative to note that C-PCM can be derived in an alternative way,\cite{LanHer11b} which also generalizes the method to the 
Poisson-Boltzmann case in which the solvent contains a dissolved electrolyte.  This can be done by introducing the 
following {\em ansatz\/} for the electrostatic potential:\cite{LanHer11b,HerLan16}
\begin{equation}\label{eq:DESMO_ansatz}
	\esp(\br) = \begin{cases}
		\esp_0^\rho(\br) + \esp_0^\sigma(\br) 	& \text{for $\br \in \cav$} \\
		\esp_\kappa^\rho(\br)/\diel			& \text{for $\br\notin \cav$} \\
	\end{cases} \; .
\end{equation}
Within the solute cavity, this looks like the solute's electrostatic potential ($\esp^\rho_0 \equiv \esp^\rho$) plus that arising
from the apparent surface charge, whereas outside of the cavity the potential is modified to include screening by 
dissolved ions:
\begin{equation}
	\esp_\kappa^\rho(\br) = \int \rho(\br') \; 
	\frac{ 
		\exp(-\kappa \| \br - \br' \| ) 
	}{ 
		\| \br - \br' \| 
	} \; d\br'  \; .
\end{equation}
The screened Coulomb potential $e^{-\kappa r}/(\dielst r)$ is the 
form encountered in Debye-H\"uckel theory,\cite{DebHuc23b,DesHol01,GroTry08,LanHer11b}
with screening length $\kappa^{-1}$ that is defined in Eq.~\eqref{eq:kappa}.   C-PCM is immediately recovered by requiring 
$\esp(\br)$ in Eq.~\eqref{eq:DESMO_ansatz} to remain continuous across the solute cavity surface.\cite{LanHer11b}
This simple {\em ansatz\/} cannot, however, be made to satisfy the jump boundary condition in Eq.~\eqref{eq:jump} and incurs
an error of ${\cal O}(1/\diel)$.\cite{LanHer11b,HerLan16}   (This is consistent with the fact that C-PCM can be ``derived'' from IEF-PCM 
simply by dropping the $\hat{D}$- and $\hat{D}^\dagger$-dependent terms from IEF-PCM and rescaling the surface charge 
to compensate,\cite{Can07} as described in Section~\ref{sec:Basics:Models}.)  
The simple {\em ansatz\/} in Eq.~\eqref{eq:DESMO_ansatz}
is easily modifiable to incorporate the effects of an ion exclusion layer (Stern layer) around the solute cavity, as in the size-modified
Poisson-Boltzmann equation of Eq.~\eqref{eq:PBE}, where the ion accessibility function $\lambda(\br)$ serves the same purpose.
In homage to GCOSMO, and in recognition of the fact that this approach generalizes Debye-H\"uckel theory to cavities of arbitrary 
shape, this approach has been named the {\em Debye-H\"uckel-like screening model} (DESMO).\cite{LanHer11b,HerLan16}

\subsection{Discretization}
\label{sec:Smooth:SWIG}
\begin{figure}
	\centering
	\fig{1.0}{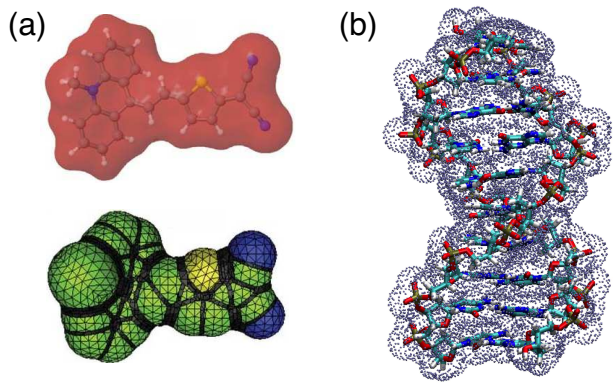}
	\caption{
		(a) Example of a molecular surface (top) and its tessellation using the \gepol\ algorithm (bottom). 
		(b) Surface discretization grid for a double-stranded segment of DNA.
		Panel (a) is reprinted from Ref.~\protect\citenum{Men12}; copyright 2012 John Wiley \& Sons.
	}\label{fig:discretized_cavity}
\end{figure}

Having introduced various PCMs in matrix form, we now turn to the details of discretizing the cavity surface.  
Historically, this has been accomplished using various ``tessellation'' schemes,\cite{Pom07} in which small, flat surface elements 
approximate the curved surface of the cavity.     The ``\gepol'' algorithm\cite{SilVilNil90,PasSil90,SilTunPas91,PasSilTun94}
is a popular version of this finite-element approach, and an example of a molecular cavity discretized in this way is presented in
Fig.~\ref{fig:discretized_cavity}(a).    This approach has several limitations, however, including the fact that the number of tesserae 
per atomic sphere cannot be increased arbitrarily and thus the discretization error cannot be systematically driven to zero.\cite{LanHer11c}
Furthermore, the solid geometry of the tessellation procedure is complicated, leading to very complex formulas for 
surface areas\cite{LioHawLyn95} and analytic energy gradients.\cite{CosMenCam96}    
(In fact, second derivatives of the tesserae areas $a_i$ were considered sufficiently complicated that they were not originally formulated,
and the PCM Hessian was implemented in a semi-analytic way, via finite-difference evaluation of 
$\partial^2 a_i/\partial x \partial y$.\cite{CosScaReg02b})  
More recently, these complexities have been overcome by discretizing the 
surface using atom-centered Lebedev grids,\cite{YorKar99,GreYor05,KhaGreThi05,LanHer10a,LanHer10b,LanHer11c,HerLan16,LanHerAlb20}
which are already widely used in density functional theory,\cite{MurHanLam93,SG1,SG0,DasHer17} 
and therefore widely available in quantum chemistry codes.   An example is depicted in Fig.~\ref{fig:discretized_cavity}(b), 
where the outline of the vdW surface is evident even though only the discretization grid points (and the underlying nuclear
framework) are shown.  Relative to \gepol\ and other tessellation schemes, Lebedev grids have the advantage of being systematically
improvable so that results can be converged to the infinite-grid limit.\cite{LanHer10b,LanHer11c}
Fully analytic Hessians have been formulated and implemented.\cite{LiuLia13}

\begin{figure}[t]
	\centering
	\fig{1.0}{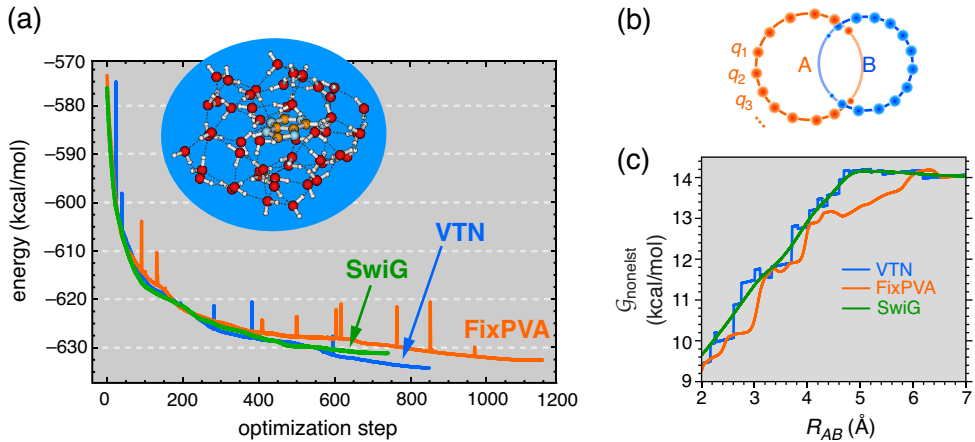}
	\caption{
		Demonstration of the \swig-PCM discretization approach.  
		(a) Geometry optimization trajectories of (adenine)(\water)$_{52}$
		in a C-PCM representation of bulk water, using several different algorithms to discretize the vdW cavity surface.
		(The surface itself is not shown, but the atomistic region appears in the blue inset.)   
		Optimizations are performed in Cartesian coordinates so the total number of steps is large.
		(b) Schematic of the \swig\ discretization algorithm, in which the surface charges $\{q_i\}$ are subject to 
		Gaussian blurring and also to a switching function 
		that attenuates the quadratures weights near the cavity surface.
		(c) Nonelectrostatic solvation energy ($\FreeE_\nonelst$) as spheres $A$ and $B$ are pulled part.
		The value of $\FreeE_\nonelst$ is related to the solvent-exposed surface area 
		and inherits any discontinuities in the surface area function.
		Panels (a) and (c) are adapted from Ref.~\CiteN{LanHer10a}; copyright 2010 American Chemical Society.
	}\label{fig:geom_opt}
\end{figure}

An important issue faced by both tessellation and quadrature is ensuring that the discretization algorithm produces a  
smooth potential energy surface as the atoms are displaced.   The appearance of discontinuities in  
molecular surface area algorithms used in biomolecular electrostatics calculations was pointed out long ago,\cite{WawGibSch94}
and discontinuities are likely the cause of anecodotal complaints about slow convergence of geometry
optimizations using PCMs. 
These discontinuities arise because grid points may disappear into (or emerge from within) the 
interior of the solute cavity, as displacement of the nuclei modifies the extent of interpenetration amongst the atomic spheres that
define the cavity.  
An example is shown in Fig.~\ref{fig:geom_opt}(a), which plots convergence of the energy during geometry 
optimization of (adenine)(\water)$_{52}$ in implicit solvent.   
Two discretization algorithms, the {\em variable tesserae number\/} (VTN) method\cite{LiJen04} and the {\em fixed points with variable 
areas\/} (FixPVA) approach,\cite{SuLi09} are shown to exhibit repeated spikes in the energy.\cite{LanHer10a}    
The VTN algorithm uses a fixed surface grid that unceremoniously
discards surface elements that are swallowed by the cavity, and it is unsurprising that the corresponding potential surface exhibits
discontinuities, although their magnitude ($>20$~kcal/mol in one case) is disturbing.    The FixPVA algorithm, however, 
specifically introduces a switching function to attenuate the surface area of each atomic tesserae within the cavity's interior.
In fact, the sharp changes in energy along the FixPVA optimization in Fig.~\ref{fig:geom_opt}(a) are not discontinuities {\em per se\/} but
rather near-singularities induced by the switching function, which allows surface discretization charges to approach one another
much more closely as compared to the VTN scheme.\cite{LanHer10a,LanHer11c}

To avoid this, one can use a combination of a switching function with Gaussian blurring of the surface charges, 
as illustrated in Fig.~\ref{fig:geom_opt}(b), in what has been called the {\em switching\slash Gaussian\/} (\swig) discretization 
procedure.   This ensures 
that Coulomb interactions between surface discretization elements remain finite even as the distance
between them approaches zero.   Such a procedure was originally developed by York and Karplus 
to obtain a ``smooth COSMO'',\cite{YorKar99} then later extended by Lange and Herbert to the complete family of 
PCMs.\cite{LanHer10a,LanHer10b,LanHer11c,LanHerAlb20}  
This scheme uses Lebedev points to discretize the surface rather than tesserae of finite area.
Nevertheless, the solvent-accessible surface area (SASA) for atom $B$ (whose radius is $R_B$) is easily defined:
\begin{equation}\label{eq:SASA-atomic}
	\text{SASA}(B) = R_B^2 \sum_{i\in B}^\text{grid} w_i F_i \; .
\end{equation}
Here, $w_i$ is the quadrature weight for Lebedev discretization point $\bs_i$, and $F_i$ is the switching function associated
with that point ($0 \leq F_i \leq 1$),
so $a_i=w_i F_i$ is the surface area that is assigned to the point $\bs_i$.\cite{LanHer10a}   
Equation~\eqref{eq:SASA-atomic} is considerably simpler than  
geometric algorithms for determining the exposed surface areas.\cite{LioHawLyn95}
As discussed in Section~\ref{sec:DeltaG:SMx}, models of the nonelectrostatic contributions
to the solvation energy usually include terms proportional to the solvent-exposed surface area so continuity of the potential 
energy surface also demands that the surface area be a continuous function of the nuclear coordinates.  For \swig\ discretization,
it is evident from Fig.~\ref{fig:geom_opt}(c) that this is indeed the case.

\begin{figure}
	\centering
	\fig{1.0}{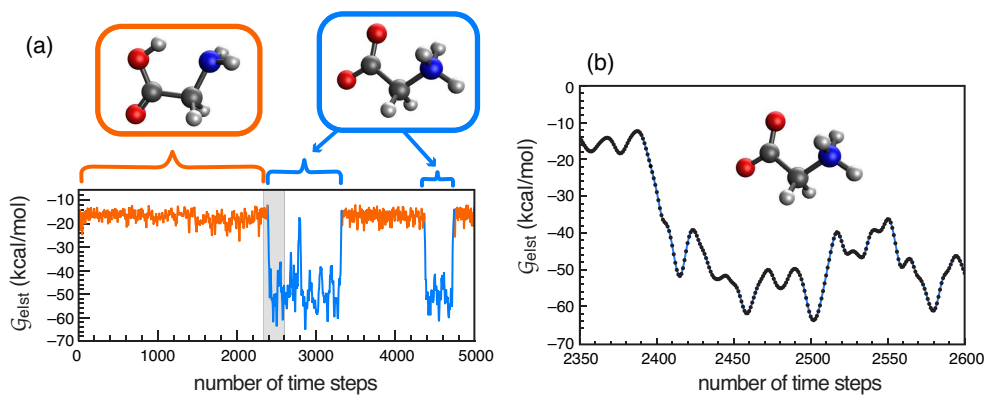}
	\caption{
		(a) Electrostatic solvation energy along an {\em ab initio\/} molecular dynamics trajectory of glycine 
		(PBE0\slash 6-31+G* level) 
		in implicit water (\swig\slash C-PCM).   The simulation starts at $t=0$ from the amino acid
		tautomer (energy data in orange), which is the most stable form of gas-phase glycine, but this species spontaneously
		transfers a proton to form the zwitterionic tautomer (energy data in blue), during the time window highlighted in the gray box
		and shown in detail in (b).  The time step is 0.97~fs.
		(b) Close-up view of $\FreeE_\elst$ in the region where the proton transfer occurs.
		Energy fluctuations are smooth despite the bond-breaking even that occurs around 2.3~ps.
		Data are from Ref.~\CiteN{LanHer10b}.
	}\label{fig:swig-aimd}
\end{figure}

Analytic gradients of \swig-PCMs are greatly simplified relative to those of the corresponding \gepol-discretized models.\cite{LanHer10b}
\swig-PCM potential energy surfaces are provably continuous and differentiable,\cite{YorKar99,LanHer10b} 
and are free of the unwanted oscillations that plague the FixPVA approach
(see Fig.~\ref{fig:geom_opt}).\cite{LanHer10b,LanHer11c}
\swig\ discretization is well-behaved enough to be used for {\em ab initio\/} molecular dynamics simulations involving
bond-breaking, as shown in Fig.~\ref{fig:swig-aimd} for intramolecular proton transfer in glycine.
\swig-PCM provides forces that are stable enough to afford
good energy conservation despite the significant deformation of the solute cavity as it transforms between two
tautomeric forms of glycine.   This is demonstrated by the energy profile in Fig.~\ref{fig:swig-aimd}(b), which provides a closeup 
view of $\approx250$~fs of dynamics during which a bond-breaking event occurs.
On the other hand, it appears that \swig\ discretization exacerbates differences between the 
$\mathbf{X}=\mathbf{DAS}$ and $\mathbf{X} = \mathbf{SAD}^\dagger$ forms of $\mathbf{K}_{\diel}$
[Eq.~\eqref{eq:K-matrix}], at least in comparison to \gepol\ results where these differences are small.\cite{CosScaReg02b}
This is not a major problem insofar as the $\mathbf{X}=\mathbf{DAS}$ form provides correct formal properties and good
numerical agreement with Poisson's equation, as demonstrated in Fig.~\ref{fig:vs_APBS}.

The \swig\ implementation of PCMs first appeared in the \qchem\ program\cite{KryGil13} 
and related discretization schemes have since been adopted in other software.\cite{ScaFri10,GarNee20}
Other discretization approaches have been described\cite{HarRan09,WeiRanHar10,BugDiRMoz15,BugHar19} 
but it is unclear that they have been formulated with gradients in mind.
Given the simplicity and success of the \swig\ approach, it is also unclear what is to be gained from these formulations.

\subsection{Isodensity and self-consistent cavity surfaces}
\label{sec:Smooth:SCCS}

Smooth \swig-PCM discretization schemes been developed for both vdW cavities\cite{LanHer10b} (including the SAS)
and for the ``molecular surface" or SES,\cite{LanHerAlb20} thus providing a solution to the practical problem of discontinuities
in potential energy surfaces.   Unfortunately, these methods do not alter an arguably more fundamental problem, namely, 
that construction of the solute cavity
remains a nontrivial source of arbitrariness in both PCMs and in biomolecular Poisson-Boltzmann electrostatics calculations.
Solvation energies\cite{OnuAgu14} and other properties\cite{ZhaChi99}
can be quite sensitive to the particular atomic radii that are used,
and often the atomic radii that work well for small-molecule solvation energies do not work well for proteins.\cite{TjoZho08}

Both the vdW and SAS cavity constructions consist of atom-centered spheres with radii 
\begin{equation}\label{eq:atomic_radius}
	R_A = \vdwscale R_{\vdw,A} + R_\text{probe} \; .
\end{equation}
The atomic vdW radii $\{R_{\text{vdW},A}\}$ might be taken from crystal structure data
(\eg, Bondi's radii set of radii and its subsequent extensions),\cite{Bon64,RowTay96,ManChaVal09} 
or might simply be parameters of the model.\cite{SwaAdcMcC05}  
For vdW cavities, $R_\text{probe}=0$ and a typical scaling factor is $\vdwscale = 1.2$,\cite{TomPer94} 
whereas one generally does not  scale the vdW radii for SAS cavities.
(As discussed in Ref.~\citenum{TomPer94}, the choice $\vdwscale$ = 1.2 does not result from any kind of elaborate fitting
procedure and is intended only as a rough guide; common choices range from $\vdwscale = 1.1$--$1.4$.)
As an example of just how sensitive $\Gelst$ is to cavity construction, Table~\ref{table:Gelst_vs_radii} reports calculations for
the two tautomers of glycine from Fig.~\ref{fig:swig-aimd}, using various cavity definitions.   A change from $\vdwscale =1.2$
to either $\vdwscale =1.1$ or $\vdwscale =1.3$ results in changes of anywhere from 3--9~kcal/mol in $\FreeE_\elst$.

\begin{table}
	\centering
	\caption{ 
		Aqueous solvation energies computed with the SS(V)PE model for two tautomers of glycine using atomic 
		radii in Eq.~\protect\eqref{eq:atomic_radius}.
		Electronic structure calculations were performed at the B3LYP\slash 6-31+G* level.
	}\label{table:Gelst_vs_radii}
\begin{threeparttable}
	\begin{tabular}{l cc ..}
\hline\hline
		Cavity  & $\vdwscale$ & $R_\text{probe}$
		& \mc{2}{c}{\underline{\hspace{0.65cm} $\FreeE_\elst$ (kcal/mol)\hspace{0.65cm}}} \\ 
		&&\mc{1}{c}{(\AA)} & \mc{1}{c}{amino acid} & \mc{1}{c}{zwitterion} \\
\hline
		vdW$^a$ 		& 1.0	&0.0& -26.1 	& -68.2 \\
		vdW$^a$			& 1.1 	&0.0& -20.2 	& -56.0 \\	
		vdW$^a$			& 1.2 	&0.0& -16.1 	& -46.8 \\	
		vdW$^a$			& 1.3 	&0.0& -12.9 	& -39.0 \\
		vdW$^a$			& 1.4	&0.0& -10.6 	& -32.6 \\
		SAS$^a$			& 1.0	&0.2& -18.0	& -51.4\\
		SAS$^a$			& 1.0	&1.4& -4.3 	& -13.9 \\
		isodensity$^b$ 	&		&      & -16.4 	& -48.1 \\		
\hline\hline
	\end{tabular}
\begin{tablenotes}[flushleft]
\fns
	\item 
	$^a$Using 
	$R_\text{vdW}$ = 1.10~\AA\ (H), 1.70~\AA\ (C), 1.55~\AA\ (N), and 1.52~\AA\ (O),
	discretized using \swig\ with 302 points per atom.
	$^b$Using an isocontour $\rho_0 = 0.001$~a.u.\ and 1,202 grid points.
\end{tablenotes}
\end{threeparttable}
\end{table}

When the solute is described using electronic structure theory, a more satisfying cavity definition uses the solute's own electron
density.\cite{ForKeiWib96,ChiDup02,CheChi03,Chi06b}
It is possible to settle on a numerical isocontour
value (generally $\rho_0\sim 0.001$~a.u.) that appears to have some universal validity,\cite{ZhaChi98,ZhaChi99} 
and results using $\rho_0 = 0.001$~a.u.\ are shown in Table~\ref{table:Gelst_vs_radii}.
It is noteworthy that the 
standard vdW scaling factor $\vdwscale$ = 1.2 affords the best agreement with the isodensity result for both tautomers
of glycine, despite their very different solvation energies.    It is also noteworthy that the two SAS cavities in 
Table~\ref{table:Gelst_vs_radii}, using values of 
$R_\text{probe}$ that are common in biomolecular Poisson-Boltzmann 
calculations,\cite{OnuAgu14} afford very different solvation energies.  The probe radius $R_\text{probe}$ = 1.4~\AA\ is a
realistic measure of the size of a water molecule (based on radial distribution functions for liquid water),\cite{BroHea15} but 
affords rather small values of $\Gelst$.  In contrast, $R_\text{probe}$ = 0.2~\AA\ is much too small to represent an actual water molecule 
but affords solvation energies much closer to values obtained using the 
isodensity and canonical vdW surface definitions.    Values $R_\text{probe} = 0.2$--$0.3$~\AA\ are consistent with typical solvent probe radii
used since the early days of continuum solvation models in quantum chemistry.\cite{MieScrTom81,MieTom82,BonCimTom83}

%

Although the isodensity cavity construction is an appealing choice on physical grounds, existing algorithms
to compute this surface are subject to occasional failure for certain molecular geometries.\cite{ChiDup02,CheChi03}  
In principle, these difficulties could likely be 
overcome using an implementation based on the ``marching cubes'' algorithm that is well known in computer 
graphics.\cite{LorCli87,RajBol03}  
A more fundamental problem is that for an isodensity cavity the surface normal vector $\mathbf{n}_{\bs}$, which is needed
to define the ASC-PCM double-layer operator $\hat{D}$, 
depends on the density:\cite{ForKeiWib96,ChiDup02,Chi06b} 
\begin{equation}
	\mathbf{n}_{\bs} = -\frac{\del\rho(\bs)}{\|\del\rho(\bs)\|} \; .
\end{equation}
The surface area associated with the discretization point $\bs_i$ inherits a density dependence as well, and this 
significantly complicates the formulation of analytic energy gradients.  To date, these have never been
published for the combination of PCM (or related models) with an isodensity cavity construction.    
This limitation could be overcome using an analytically-differentiable pseudo-density,
as has sometimes been used in biomolecular electrostatics calculations,\cite{YuJacFri05,HerLan16}
perhaps using a superposition of frozen atomic densities.
Although this might remove some arbitrariness from the selection of atomic radii, however,  
it would not represent a self-consistent determination 
of the cavity surface that could deform to reflect changes in the molecular electronic structure.  
Conversely, a pseudo-density that is determined in order to reproduce the molecular 
electrostatic potential $\esp^\rho(\br)$ may afford better solvation energies as compared to vdW radii that do not respond
to the electronic structure,\cite{ZhoAgaWon08} but reintroduces the problem of how to compute the analytic gradient.
As such, it is unclear whether such constructions offer advantages over the simplicity of the vdW cavity.    
As a simpler workaround, united-atom radii (in which hydrogen atoms are not given atomic spheres)
have been parameterized in an effort to reproduce results obtained with an 
isodensity cavity.\cite{BarCosTom97}

Within the context of solving Poisson's equation in three-dimensional space, 
this idea has more merit, as one can define ``soft'' atomic spheres that interpolate the permittivity $\diel(\br)$ 
between limiting values inside ($\diel=1$) and outside ($\diel=\dielst$) of the cavity.\cite{FisGenAnd17}   
Related to this, and offering 
a more compelling scheme for self-consistent determination of the solute\slash continuum interface, is a scheme that takes 
$\diel(\br)$ to be a functional of the solute's charge density, $\diel[\rho](\br)$, with 
limiting values $\diel=1$ near the nuclei and $\diel=\dielst$ farther away.
In practice, ``near'' and ``far'' are determined not by distance but by comparison of $\rho(\br)$ to a pair of parameters
$\rho_\text{max}$ and $\rho_\text{min}$, the latter of which establishes what constitutes the ``tail'' of the density.  
This idea was originally developed by Fattebert and Gygi\cite{FatGyg02,FatGyg03,SchFatGyg06}
then later refined by others;\cite{DziHelSky11,AndDabMar12,MatSunLet14,SanSueSch09} 
see Ref.~\CiteN{AndFis19} for a review.    In its modern incarnation,\cite{AndDabMar12}
it uses a permittivity functional 
\begin{equation}\label{eq:eps[rho]}
	\diel[\rho](\br) = \begin{cases}
		1 
			& \rho(\br) > \rho_\text{max} \\
		\exp\big[t(\ln \rho(\br))\big] 	
			& \rho_\text{min} < \rho(\br) < \rho_\text{max} \\
		\dielst				
			& \rho(\br) < \rho_\text{min} \\
	\end{cases}
\end{equation}
in which $t(x)$ is a switching function that interpolates smoothly between values $t(\ln \rho_\text{min}) = \ln\dielst$
and $t(\ln\rho_\text{max}) = 0$, so that $\diel(\br)$ achieves the limits indicated in Eq.~\eqref{eq:eps[rho]}. 
Inserting this {\em ansatz\/} into Poisson's equation [Eq.~\eqref{eq:Gen-Poisson}] affords a model in which the dielectric
interface is smooth, rather than sharp as it is in PCMs, yet one where the definition of the interface is updated self-consistently 
as the density $\rho(\br)$ is iterated to convergence.  For that reason, this approach 
has been called simply the {\em self-consistent continuum solvation\/} (SCCS) model.\cite{AndFis19} 
The dependence of $\dielst(\br)$ on the density does mean that the 
Fock operator $\delta\FreeE/\delta\rho$ acquires an extra term relative to what was discussed in 
Section~\ref{sec:Basics:Poisson}, namely\cite{AndDabMar12}
\begin{equation}
	\mathvee_{\diel}[\rho](\br) = 
	-\frac{1}{8\pi} \big\|\del\esp(\br)\big\|^2
	\left(\frac{\delta \diel[\rho] }{\delta \rho(\br) }\right) \; .
\end{equation}
The SCCS model is increasingly being used in {\em ab initio\/} simulations of materials, 
\eg, to model the aqueous electrolyte\slash solid-state interfaces relevant in 
electrochemistry.\cite{SanSueSch09,NatTruMar19,AndHorNat19,AndFis19,SunSch17,SunLetSch18,SchSun20,
DziBhaAnt20,BraNguGle20,BhaAntDzi20}  
Some of that work points to limitations of the linear dielectric model itself
(\ie, the assumption that $\mathbf{P}\propto\mathbf{E}$), 
because the rotational response of the water molecules saturates at the electrode interface and consequently 
the susceptibility is smaller than in bulk water.\cite{SunLetSch18,SchSun20}   
Limitations in the linearized Poisson-Boltzmann description of electrolyte effects have also been
demonstrated.\cite{SunSch17,SunLetSch18,SchSun20}

\subsection{Linear-scaling algorithms}
\label{sec:Smooth:O(N)}
As noted above, the electrostatic solvation energy obtained from IEF-PCM should be {\em exactly\/} equivalent to that 
obtained by solving Poisson's equation, in the case of a classical solute for which there is no outlying charge.
(This equivalence holds only up to discretization errors, but those are controllable and can be driven to zero if 
systematically-improvable grids are employed.)
It is therefore surprising that the PCM formulation of the classical
continuum problem has seen very little use in biomolecular electrostatics calculations.   Such calculations are almost always
performed in water, meaning that the simpler C-PCM should be essentially exact, and modifications have been proposed to treat
the linearized Poisson-Boltzmann problem,\cite{CanMenTom97,MenCanTom97,Chi04,LanHer11b}
including modifications to simulate an ion exclusion layer.\cite{LanHer11b} 
The Poisson-Boltzmann equation in three dimensions [Eq.~\eqref{eq:PBE}] is typically solved using a finite-difference scheme 
for the Laplacian operator,\cite{DavMcC91,CooHer18} 
and this poses problems for molecular dynamics simulations with implicit solvent, because 
the forces need not be continuous when discretized in this way.  
Although significant effort has gone into obtaining high-quality 
forces,\cite{GilDavLut93,ImBegRou98,LuZhaMcC05,WanTanCha10,CaiYeWan11a,WanCaiXia12,XiaCaiYe13,XiaWanLuo14} 
from a certain point of view this looks like engineering an elaborate means of escape from the very deep hole created by a 
numerical framework that admits discontinuities.  A better strategy is not to get trapped in that hole
in the first place.   The \swig\ discretization
procedure for PCMs (Section~\ref{sec:Smooth:SWIG}) was designed as a starting point that is free of discontinuities.

All of this suggests that \swig-PCMs could make very attractive replacements for the finite-difference Poisson-Boltzmann solvers 
that are commonplace in biomolecular electrostatics calculations.   Efforts to do so will quickly run up against the size  
of the PCM matrix equation, which is equal to the number of discretization grid points, itself
proportional to the number of solvent-exposed atoms.    This means that a straightforward solution of 
Eq.~\eqref{eq:Kq=Rv_matrix}, based on constructing $\mathbf{K}_{\diel}^{-1}\mathbf{Y}_{\diel}$ or its equivalent, incurs a CPU cost of
${\cal O}(N^3_\text{atoms})$ and a memory footprint of ${\cal O}(N^2_\text{atoms})$, albeit with significant prefactors in both 
cases that reflect the number of discretization points per atom.   In \qmmm\ calculations with continuum boundary conditions, 
it is the size of the MM region that dictates the matrix dimension and for small QM regions with large
MM environments, one can easily encounter scenarios wherein the cost of classical electrostatics (PCM part of the calculation)
exceeds the cost of doing the quantum mechanics!\cite{HerLan16}

\begin{figure}
	\centering
	\fig{1.0}{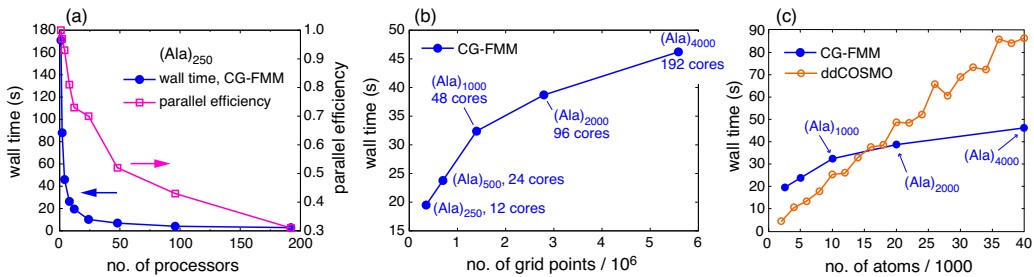}
	\caption{
		Timings and parallel scalability data for linear-scaling 
		PCM solvers applied to polyalanine helices, using a classical force field description of the solute.
		(a) Strong-scaling data for a CG-FMM algorithm applied to (Ala)$_{250}$, 
		running on 12 shared-memory cores per node.  
		(b) Weak-scaling data for (Ala)$_n$ helices of increasing length, 
		versus the number of Lebedev grid points used to discretize the cavity surface. 
		(c) Comparison of timing data for CG-FMM versus ddCOSMO for (Ala)$_n$.  
		The CG-FMM data in (c) are the same as those in (b),
		but all ddCOSMO calculations were run on a single 12-core node.
		Data in (a) and (b) are from Ref.~\protect\citenum{HerLan16} and ddCOSMO 
		data in (c) are from Ref.~\protect\citenum{LipStaCan13}.
	}\label{fig:O(N)}
\end{figure}

A straightforward solution to that problem is to introduce iterative solvers that 
do not require storage or inversion of the matrix $\mathbf{K}_{\diel}$,\cite{RegCosBar98,ScaBarKud04,HerLan16}
such as conjugate gradient (CG) or biconjugate gradient algorithms.\cite{HerLan16}   
The bottleneck in these methods is computing the Coulomb interactions between surface discretization points, and this can
be accelerated using the fast multipole method (FMM),\cite{ScaBarKud04} in either its original formulation or using a simpler 
tree-code approach.\cite{GenKra13,HerLan16}    
Parallelization strategies have also been discussed.\cite{HerLan16}   
Data for polyalanine helices (Fig.~\ref{fig:O(N)}) show that wall times for iterative solution of the PCM equation
(to obtain surface charges) can be reduced to a few seconds, even for (Ala)$_{4000}$ with $5.6\times 10^6$ surface
discretization points.   Proof-of-concept 
MM\slash PCM molecular dynamics simulations have been reported in which the atomistic MM region consists of a 
segment of DNA bound to a histone (21,734 classical atoms from a force field), with $\approx$124,000 point charges used
to discretize the surface.\cite{HerLan16}

An alternative algorithm for fast iterative solution of the PCM equation is based on a domain decomposition 
procedure.\cite{LipStaCan13,LipLagSca14,CapJurLag15,GatLipSta17,StaLagSca19,NotStaSca19} 
Here, the solute cavity is divided into overlapping domains, each consisting of a single atomic sphere, for which the PCM equation can 
be solved analytically.   These single-sphere solutions form the basis for an iterative solution of the PCM equation for the 
full domain, which can be formulated as 
sparse matrix equations that can be solved in ${\cal O}(N_\text{atoms})$ time.
This method was originally developed for C-PCM and is thus named 
``ddCOSMO'',\cite{LipStaCan13,LipLagSca14,CapJurLag15,GatLipSta17,StaLagSca19}
although it has now been extended to arbitrary PCMs (``ddPCM'').\cite{NotStaSca19}
Timing data on just a single compute node [Fig.~\ref{fig:O(N)}(c)] show that the method is competitive with a parallel CG-FMM solver.
The ddCOSMO algorithm has been implemented for MM\slash PCM molecular dynamics in the \tinkerHP\ code,\cite{TINKER-HP}
using a switching function is used to guarantee smooth forces.\cite{LipLagRay15}
Such an approach seems like a practical replacement for
finite-difference Poisson-Boltzmann solvers that is inherently free of problems with discontinuous forces.

\section{Solvation Energies}
\label{sec:DeltaG}

\begin{figure}[t]
	\centering
	\fig{1.0}{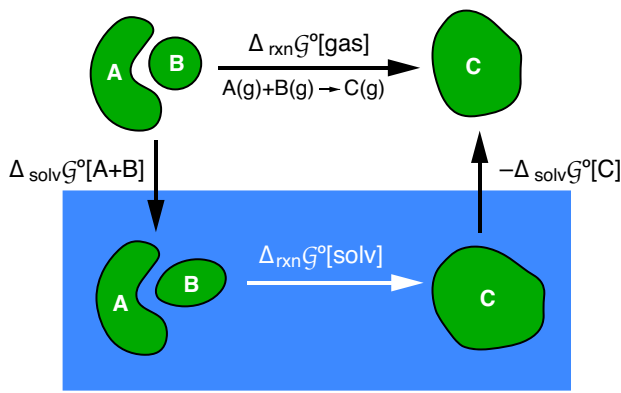}
	\caption{
		Thermodynamic cycle connecting gas- and solution-phase reaction energies for $\rm A + B \rightarrow C$.   
		The gas-phase value of $\dGrxn^\circ$ contains the vibrational entropy change in addition to the reaction enthalpy, and a 
		continuum solvation model is used to compute $\dGsolv^\circ$ for each reactant and product species.    This completes the
		thermodynamic cycle to compute the solution-phase value of $\dGrxn^\circ$.   Changes in shape signify 
		that geometries of A, B, and C may be different in solution than they are in the gas phase, in which case the solvation energies
		$\dGsolv^\circ$ should include a term representing the gas-phase deformation energy.
	}\label{fig:dG_rxn_cycle}
\end{figure}

Perhaps the single most important property afforded by a solvation model is the free energy of solvation, $\dGsolv$.
Using a solvation model to compute $\dGsolv^\circ$ for the reactant and product species in a chemical reaction, 
combined with a gas-phase calculation of the reaction energy $ \Delta_\text{rxn}\mathcal{U}$ and free energy 
\begin{equation}
	\dGrxn^\circ\mbox{[gas]} = \Delta_\text{rxn}\mathcal{U}
	- RT\ln\left(
		\frac{Z_\text{products}}{Z_\text{reactants}}
	\right) \; ,
\end{equation}
where $Z$ represents the partition function, 
one may obtain a value for the solution-phase reaction energy, $\dGrxn^\circ\mbox{[solv]}$.   This procedure is illustrated by 
the thermodynamic cycle shown in Fig.~\ref{fig:dG_rxn_cycle}. 
However, the continuum electrostatics problem discussed heretofore defines only 
the electrostatic contribution to $\dGsolv^\circ$, {\em and this is generally insufficient to predict accurate solvation energies}.   
This point is demonstrated vividly by statistical errors in IEF-PCM solvation energies,\cite{CraTru08}
summarized for aqueous solutes in Table~\ref{table:dG}.
Errors with respect to experimental solvation energies 
average about 6~kcal/mol for charge-neutral solutes, 8~kcal/mol for anions, and 13~kcal/mol for cations, despite
much smaller errors with respect to exact Poisson electrostatics (see Table~\ref{table:Gelst-Chipman}).
In nonaqueous solvents, IEF-PCM errors with respect to experimental values of $\dGsolv^\circ$
are 4.9~kcal/mol for neutral solutes and 12.4~kcal/mol 
for ions.\cite{CraTru08}   Similarly poor results confirmed using other electrostatics-only PCM approaches
including C- and D-PCM.\cite{CraTru06,DziHelSky11}

\begin{table}
	\centering
	\caption{
		Mean unsigned errors (MUEs) for aqueous hydration energies ($\dGhyd^\circ$), using continuum solvation models, 
		for small molecules in the  Minnesota Solvation Database.\protect\cite{ThoCraTru04,SM6,SM12,SMD}
		Estimated uncertainties in the experimental data are $\pm0.2$~kcal/mol for neutral solutes\pcite{ThoCraTru04,SM6}
		and $\pm 3$~kcal/mol for ions.\pcite{SM6}
	}\label{table:dG}
\begin{threeparttable}
	\begin{tabular}{lr .... c ..}
\hline\hline
		Data  & $N_\text{data}$ & 
		\mc{7}{c}{MUE (kcal/mol)} \\ \cline{3-9}
		Set
		&&&& \mc{2}{c}{PCM-based} &&
		\mc{2}{c}{Poisson-based}\\ \cline{5-6}\cline{8-9}
		&&\mc{1}{r}{IEF-PCM$^a$} 
		& \mc{1}{r}{SM12$^b$} & \mc{1}{r}{SMD$^c$} 
		& \mc{1}{r}{CMIRS$^d$}
		&& \mc{1}{r}{SCCS$^e$}  & \mc{1}{r}{soft-sphere$^f$} \\
\hline  			
		neutrals 	& 274	&5.7		& 0.7	& 0.8 	& 0.8 	&& 1.1 	& 1.1 \\
		cations	& 52		&12.7	& 3.5	& 3.2 	& 1.8 	&& 2.3 	& 2.1 \\	
		anions	& 60		&8.0		& 3.8	& 6.2	& 2.8 	&& 5.5 	& 3.0 \\
		all ions	& 112	&9.7		& 3.7	& 4.7 	& 2.4 	&& 4.0 	& 2.6 \\
\hline\hline
	\end{tabular}
	\begin{tablenotes}[flushleft]
	\fns 
	\item $^a$B3LYP\slash 6-31G* level 
		using united-atom radii,\pcite{BarCosTom97} from Ref.~\CiteN{CraTru08}.
	$^b$B3LYP\slash 6-31G* level using ChElPG charges, from Ref.~\CiteN{SM12}, 
	$^c$\textit{m}PW1PW\slash 6-31G* level, from Ref.~\CiteN{MarCraTru09}, 
	$^d$B3LYP\slash 6-31G* level using isodensity cavity with $\rho_0^{} = 0.001$~a.u., 
		from Ref.~\CiteN{YouHer16}.
	$^e$PBE in a plane-wave basis, from Ref.~\CiteN{DupAndMar13}.
	$^f$PBE in a wavelet basis, from Ref.~\CiteN{FisGenAnd17}.
	\end{tablenotes}	
\end{threeparttable}	
\end{table}

To do better, these electrostatic solvation models must be augmented to include nonelectrostatic interactions 
($\FreeE_\nonelst$),\cite{KlaMenTom09} a catch-all term that includes 
%
%
\begin{itemize}
	\item {\em cavitation}, meaning 
	the energy required to carve out a space in the continuum solvent; 
	\item {\em Pauli repulsion}, \ie, short-range repulsive interactions with the solvent molecules; 
	\item {\em dispersion}, which is non-specific but attractive, and finally 
	\item {\em hydrogen-bonding\/} between solute and solvent.
\end{itemize}
Harder-to-define entropic changes to the solvent make a contribution to $\dGsolv^\circ$ as well,\cite{CraTru08}
although this effect could be classified as part of the cavitation (free) energy.   Indeed, the phenomena on the list above are not 
wholly independent, \eg, short-range repulsion is also related to cavitation, and hydrogen bonding has contributions from 
electrostatics, repulsion, and dispersion.  In protic solvents, and for solutes where hydrogen bonding 
is important, one or more explicit solvent molecules could be included as part of the atomistic
solute, thereby removing the largest part of the last item on the list.  Semicontinuum or ``cluster + continuum''
approaches of this kind are common, \eg, for calculating hydration energies of ions.\cite{PliRiv20,TomMucSla20} 
An important special case is p$K_\text{a}$ calculations,\cite{CasOrtFra14,ThaSch15,ThaSch16,PliRiv20} 
corresponding to the ionization reaction 
\begin{equation}\label{eq:pKa_rxn}
	\mbox{HA(aq)} \rightarrow \mbox{H}^+\mbox{(aq)} + \mbox{A}^-\mbox{(aq)}  \; .
\end{equation}
Even with one or more explicit solvent molecules to capture hydrogen bonding, however, the remaining contributions to $\Gnon$ 
must be added to the electrostatic model in order to obtain accurate reaction energetics.
It has occasionally been noted that semicontinuum predictions of hydration energies $\dGhyd^\circ$ may fail to converge as the number 
of explicit water molecules is increased.\cite{BryDiaGod08,RicGuoPar13,DhiEas18,PatEas20} 
In truth, there is no reason why 
such a calculation {\em should\/} converge to $\dGhyd^\circ$, at least in the 
absence of sampling over the explicit water molecules, because such an approach treats
entropic contributions in an unbalanced way.

Nonelectrostatic models appropriate for use with PCMs have been developed over the years on an {\em ad hoc\/} 
basis.\cite{TomPer94,TomMenCam05}    
These have often appeared under the name of ``Miertu\v{s}-Scrocco-Tomasi'' solvation 
model,\cite{LuqCurMun03,SotCurBid05,CurOroLuq06,KlaMenTom09}  
a term that in practice means a PCM used in conjunction with a SASA-type
parameterization of $\Gnon$ along the lines of what is discussed in Section~\ref{sec:DeltaG:SMx}.  
These models have
retained something of a do-it-yourself aesthetic and have not been packaged into ``canned'' or ``black-box'' solvation models,
a fact that has occasionally led to hard feelings
when errors in IEF-PCM solvation energies (such as those in Table~\ref{table:dG}) are pointed out.\cite{KlaMenTom09,CraTru09a} 
Section~\ref{sec:DeltaG:SMx} examines one particular class of black-box solvation model, namely, the SM$x$ models
developed by Cramer, Truhlar, and co-workers.\cite{SM6,SM8,ChaCraTru08b,CraTru08,MarCraTru09,SM12,SMD}   
An even blacker-box approach is ``COSMO-RS'',\cite{Kla11,Kla18} which is 
an extension of the COSMO method\cite{KlaSch93} designed for ``real solvents".   
The formulation and parameters of COSMO-RS are proprietary, however, so the black box cannot be opened and 
this method's performance cannot be independently implemented, verified, or tested.   
As such, COSMO-RS is not discussed any further.

\subsection{SM$x$ and SASA-based models}
\label{sec:DeltaG:SMx}
For computing free energies of solvation, the most successful and popular models within the quantum chemistry 
community are the ``SM$x$" models.\cite{CraTru08}
These have version numbers through $x=12$,\cite{SM12} and later $x = \text{D}$ (for ``density'').\cite{SMD}
Only the most recent versions (SM12 and SMD) are included in Table~\ref{table:dG}.

To understand how these models work, we first take a step back to discuss implicit solvation in classical biomolecular
simulations.    Owing in no small part to difficulties in obtaining stable numerical forces for Poisson-Boltzmann 
electrostatics,\cite{CaiYeWan11a,WanCaiXia12,XiaCaiYe13,XiaWanLuo14}
as discussed in Section~\ref{sec:Smooth:O(N)}, the most popular implicit solvation models in that context are {\em not\/} based
directly on solution of the Poisson-Boltzmann equation but instead are methods known as {\em generalized Born\/} (GB) 
models,\cite{BasCas00,OnuCas19}
whose name is a nod to the Born ion formula of Eq.~\eqref{eq:Born}.   GB models use the Born model 
as a pairwise {\em ansatz\/} to compute the total electrostatic solvation energy according to 
\begin{equation}\label{eq:GB}
	\FreeE^\text{GB}_\elst = -\frac{1}{2}
	\left(
		\frac{1}{\epsin} - \frac{1}{\epsout} 
	\right)
	\sum_{A,B>A}^\text{atoms} \frac{Q_A Q_B}{f_{AB}} \; .
\end{equation}
This expression allows 
for the possibility of a dielectric constant $\epsin$ that is different from unity inside of the solute cavity.
The quantity $f_{AB}^{-1}$ in Eq.~\eqref{eq:GB} is a parameterized, effective Coulomb potential, 
the canonical example of which is\cite{StiTemHaw90,BasCas00,OnuCas19}
\begin{equation}\label{eq:GB-Coulomb}
	f_{AB} = \left[
		R_{AB}^2 + \PerfRad{A} \PerfRad{B}
		\exp\big(- R_{AB}^2/4\PerfRad{A} \PerfRad{B}\big)
	\right]^{1/2} \; .
\end{equation}
Here, $R_{AB} = \big\|\mathbf{R}_A - \mathbf{R}_B\big\|$ is the interatomic distance and the quantity $Q_A^2/f_{AA}$ is 
the Coulomb self-energy in the Born ion model, with an effective cavity radius $f_{AA} \equiv \PerfRad{A}$.  As such, the 
quantities $\{\PerfRad{A}\}$ are a set of effective radii that measure the ``electrostatic size'' of each atom in the 
molecule + continuum environment.\cite{OnuBasCas00,OnuCasBas02,MonSvrOnu07}
For a classical solute, in which $\rho(\br)$ is a collection of 
atom-centered point charges $\{Q_A\}$, these radii can be computed exactly 
by solving either Poisson's equation,\cite{OnuCasBas02} or else its PCM 
analogue,\cite{LanHer12a} once per atomic charge $Q_A$, in a cavity representing the entire molecule.  
This procedure effectively forces the pairwise GB {\em ansatz\/} in Eq.~\eqref{eq:GB} to reproduce the exact electrostatic energy
defined by the continuum electrostatics problem, and these values of $\PerfRad{A}$ have been called the ``perfect'' 
Born radii.\cite{OnuCasBas02,OnuCas19}   The procedure just described is not a practical way to determine $\PerfRad{A}$ but 
can be useful to generate data sets that may 
suggest new analytical forms for GB models.\cite{LeeSalBro02,Sal06,LanHer12a,LanHer12b,HerLan16}

In practice, the radii $\{\PerfRad{K}\}$ are typically determined 
using various surface integration procedures.\cite{MonSvrOnu07}   The most popular of 
these is\cite{OnuBasCas00,BasCas00,MonSvrOnu07,OnuCas19}
\begin{equation}\label{eq:GB_radii}
	\PerfRad{K}  = \left(
		\frac{1}{R_{\vdw,K}} + \frac{1}{4\pi} \int_{\br\in\cav} 
		\frac{
			\theta\big( \|\br - \mathbf{R}_K\| - R_{\vdw,K} \big)
		}{
			\|\br-\mathbf{R}_K\|^4
		} \; d\br
	\right)^{-1} \; .
\end{equation}
Here, $\theta$ is a step function with 
$\theta(x)=1$ for $x>0$ and $\theta(x)=0$ for $x\le0$; this limits the integration domain to the region 
of the solute cavity ($\br\in\cav$) that lies outside of the vdW sphere for atom $K$.   The expression for
$\PerfRad{K}$ in Eq.~\eqref{eq:GB_radii} comes from 
the so-called Coulomb-field approximation,\cite{OnuBasCas00,BasCas00,OnuCas19} 
a charge-in-a-sphere model for the reaction-field potential.   
This is thought to overestimate electrostatic size, especially for charges close to the cavity surface, and 
alternatives have been developed,\cite{MonSvrOnu07,OnuCas19}    
but Eq.~\eqref{eq:GB_radii} is the form used in the SM$x$ models that are described below.   The requisite integral 
can be evaluated numerically using concentric atomic radical shells,\cite{ScaApoCaf97} 
but its gradient with respect to nuclear displacements, which is needed to obtain $d\FreeE^\text{GB}_\elst/dx$, 
is complicated when defined in terms of the solid geometry of interpenetrating vdW spheres.\cite{LioHawLyn95}  
Both the integral in Eq.~\eqref{eq:GB_radii} and its gradient would be straightforward to evaluate using \swig\ 
discretization; see Fig.~\ref{fig:geom_opt}(b). 

The GB {\em ansatz\/} in Eq.~\eqref{eq:GB} specifies only the electrostatic component of the solvation energy.   
The model is completed by adding a nonelectrostatic term $\FreeE_\nonelst$, for which two forms are common.
The simpler of the two consists of a term proportional to the volume of the solute cavity, to model the cavitation energy, 
and another term proportional to the SASA, which serves to model dispersion:
\begin{equation}\label{eq:Gnonelst-simple}
	\FreeE_\nonelst = \beta V_\text{cavity} + \gamma\times(\text{SASA}) \; .
\end{equation}
The quantities $\beta$ and $\gamma$ are empirical parameters.    An alternative form that is also widely used is
\begin{equation}\label{eq:Gnonelst-CDS}
	\Gnon = \sum_K^\text{atoms} \gamma^{}_K \times \text{SASA}(K) \; ,
\end{equation}
where SASA($K$) is the solvent-exposed surface area of atom $K$ and the parameter 
$\{\gamma^{}_K\}$ have units of surface tension.     Note that there is no volumetric term in Eq.~\eqref{eq:Gnonelst-CDS}.
It has been argued, based on the scaled-particle theory of
hard-sphere fluids,\cite{Pie76} that for small molecules the cavitation energy 
ought to be parametrizable in terms of the solvent-exposed surface area,\cite{LanClaCai88,FloSelTan97,LuqCurMun03,CosReg07}
although the form in Eq.~\eqref{eq:Gnonelst-CDS} has been used for macromolecules as well.\cite{GalLev04}
Models that combine GB electrostatics [Eq.~\eqref{eq:GB}] with force-field charges to represent the solute, using 
either Eq.~\eqref{eq:Gnonelst-simple} or Eq.~\eqref{eq:Gnonelst-CDS} for $\Gnon$, 
are known as ``MM\slash GBSA'' methods.\cite{HouWanLi11a,HouWanLi11b,XuSunLi13,SunLiShe14,
SunLiTia14,CheLiuSun16,SunDuaChe18,WenWanChe19,WanWenSun19} 
There is an analogous set of ``MM\slash PBSA'' methods that substitute Poisson-Boltzmann electrostatics in 
place of the GB model.\cite{HouWanLi11a,HouWanLi11b,XuSunLi13,SunLiShe14,
SunLiTia14,CheLiuSun16,SunDuaChe18,WenWanChe19,WanWenSun19,WanGreXia18}
The MM\slash GBSA method is the most widely-used implicit solvation model in biomolecular
simulation,\cite{OnuCas19} but both approaches are popular in drug-discovery applications, \eg, for calculating 
ligand--protein binding affinities.\cite{GenRyd15,WanGreXia18,WanSunWan19,PolGraRiz20}   
In that context, these methods are often used as low-resolution screening tools representing a level of 
sophistication that is intermediate between 
``knowledge-based'' (but largely physics-free) docking or scoring-function procedures, 
and much more expensive atomistic free energy simulations in explicit solvent.

In quantum chemistry contexts, it is not uncommon to use different definitions of the cavity surface for the electrostatic
and nonelectrostatic interactions.   Models for $\Gnon$ introduced above suggest a SAS cavity but 
numerical results for $\Gelst$ in Table~\ref{table:Gelst_vs_radii} suggest that atomic radii 
$R = 1.2 R_\vdw$ correspond more closely to isodensity results 
as compared to SAS radii ($R=R_\vdw + R_\text{probe}$), at least when $R_\text{probe}$ corresponds to a 
realistic estimate of the size of a solvent molecule.  The SM$x$ models,\cite{CraTru08} 
which are the most popular methods for predicting
$\dGsolv$ in quantum chemistry calculations, exemplify this distinction. 
The SM8\cite{SM8} and SM12\cite{SM12} variants, for example, employ 
GB electrostatics based on radii obtained from Eq.~\eqref{eq:GB_radii}, along with 
atom-centered charges derived from the quantum-mechanical charge density by means of certain 
charge models, CM$x$.\cite{CM2,CM3,SM6,CM4M,CM5} 
The latter are empirically-parameterized improvements upon standard atomic partial charge
prescriptions (Mulliken, Hirshfeld, etc.), with additional parameters designed to improve 
the dipole moments obtained from the partial charges, since Mulliken and Hirshfeld charges known to 
do a poor job of reproducing the dipole moment corresponding to the SCF charge density from which they were 
obtained.\cite{DavCha92}   CM$x$ charges inherit the 
basis-set sensitivity of the underlying wave function-derived charges, therefore the SM$x$ models through SM12
are each parameterized for use with particular (small) basis sets, \eg\ 6-31G*.
The SMD model\cite{SMD} was introduced to overcome this limitation, by substituting IEF-PCM electrostatics in 
place of a GB model, with the result that SMD can be used with arbitrary basis sets whereas other 
SM$x$ models should only be used in conjunction with the basis sets for which they were parameterized.

The SM$x$ models are completed by adding a nonelectrostatic term of the form in Eq.~\eqref{eq:Gnonelst-CDS}.
Atomic ``surface-tension'' parameters 
$\{\gamma^{}_K\}$ are themselves empirically-fitted functions of certain ``solvent descriptors'', which include 
the surface tension of the solvent itself, its 
refractive index (which is related to polarizability), and certain Lewis acidity 
parameters.\cite{Abr93}   The result is a ``universal'' model,\cite{CraTru08} 
in the sense that once the fitting procedure is completed, parameters $\{\gamma^{}_K\}$ are available
for any solvent for which the various solvent descriptors are known.   
The SM8 model additionally parameterizes $\FreeE_\nonelst$ as a function of
temperature.\cite{ChaCraTru06,ChaCraTru08a}

Error statistics for small-molecule hydration free energies ($\dGhyd^\circ$), 
obtained using SM12 and SMD, are listed in Table~\ref{table:dG}.
(The SM8 model has slightly larger errors for ions; see Ref.~\citenum{CraTru08}.)
The comparison between IEF-PCM and SMD is especially revealing because these two approaches use the
same approach for $\Gelst$: IEF-PCM is just SMD without $\Gnon$.   Evidently, the nonelectrostatic contribution 
dramatically reduces errors in $\dGhyd^\circ$ for 
both cations and for charge-neutral solutes.   Errors for anions are reduced as well although 
the impact of $\Gnon$ is more muted in that case, which may simply reflect the difficulty of the anion solvation problem.
In any case, it is clear that nonelectrostatic interactions must be included to obtain accurate solvation energies.
For properties that depend on only a single geometry (\eg, for some spectroscopic applications), the nonelectrostatic contributions
may be less important but to compute a reaction profile---reactant, product, and transition state energies in solution---one might
worry that changes in geometry along the reaction coordinate could affect the nonelectrostatic contribution differently for the
various chemical species involved.

Error statistics for solvents other than water are on par with results for aqueous solutes, \eg, mean unsigned errors (MUEs) of 
0.6~kcal/mol for neutral solutes and 4.5~kcal/mol for ions, in the case of SM12,\cite{SM12} versus 
0.7~kcal/mol for neutrals and 4.2~kcal/mol for ions using SMD.\cite{SMD}
Prediction of $\dGsolv^\circ$ in nonaqueous solvents is necessary in order to predict partition coefficients between different solvents.
In particular, the octanol\slash water partition coefficient (equilibrium constant $K_\text{ow}$) 
is a common measure of lipophilicity (or conversely, hydrophobicity),\cite{Dea85,AmeSubFug20,GinVazGil19}
and is related to solvation energies according to 
\begin{equation}\label{eq:Kow}
	 \dGsolv^\circ[\text{octanol}] - \dGsolv^\circ[\text{water}] 
	 = -RT\ln K_\text{ow}  \; .
\end{equation}
The value of $K_\text{ow}$ is widely used in drug-discovery applications,\cite{SahAdhKua16,TsoGiaTsa17,GinVazGil19}   
and atomic decomposition of terms contributing to $\Delta\Delta\FreeE^\circ$ in Eq.~\eqref{eq:Kow} has been used
to determine similarity indices for predicting quantitative structure--activity relationships.\cite{GinVazGil19}
For environmental toxicology purposes, $K_\text{ow}$ is 
an important physical parameter to determine for any new compound.\cite{HerdeBBro13}
The octanol\slash water partition coefficient was 
the subject of a recent blind challenge for theoretical methods,\cite{IsiBerFox20}
with the SM$x$ models emerging as amongst the best performers with a root-mean-square error of 0.44 in 
units of $\log_{10} K_\text{ow}$.\cite{OuiPal20}

Note that by using Eq.~\eqref{eq:Gnonelst-CDS} for $\Gnon$, the SM$x$ models do not contain
any term proportional to the cavity volume, as is often present in MM\slash GBSA and MM\slash PBSA models, and cavitation
effects are therefore included by means of {\em area}-dependent parameters.   Although formally justified by appeal 
to scaled-particle theory,\cite{LuqCurMun03} that argument is probably most convincing for small-molecule solutes.  As such, 
and the success of SM$x$ may partly reflect the fact that it was parameterized using experimental solvation energies 
of mostly small molecules.   (The largest molecules in the training set are $n$-hexadecane and ethyloctadecanoate, at 51 and 
63 atoms, respectively, but most of the solutes are much smaller.\cite{MarCraTru09,SMD})  
Small solutes have limited conformational flexibility and there may not be too much difference in the volumes of different
conformers, therefore a term that explicitly accounts for cavitation may be largely redundant and not required to obtain
good solvation energies.   In contrast, 
MM\slash GBSA and MM\slash PBSA methods are usually parameterized for (or at least tested on) proteins and other 
macromolecular solutes, 
where folded and unfolded structures likely have rather different volumes.  The benchmarks for MM\slash GBSA and MM\slash PBSA
are usually solvation energies obtained from molecular dynamics simulations in explicit solvent.   Since the SM$x$ models have
not been scaled up to macromolecules, it remains unclear how they would perform in that context.

SM$x$ approaches have historically been the go-to models for computing solvation energies in quantum chemistry but 
two newer approaches based directly on Poisson's equation seem very promising.\cite{DupAndMar13,FisGenAnd17}
One of these is the SCCS approach that was described in Section~\ref{sec:Smooth:SCCS}, which uses a density-dependent
permittivity functional $\diel[\rho](\br)$ to determine the solute cavity self-consistently alongside the charge density $\rho(\br)$,
with Poisson's equation solved in three-dimensional space to obtain $\Gelst$.
Combined with the simple two-parameter model for $\Gnon$ in Eq.~\eqref{eq:Gnonelst-simple}, the SCCS approach 
achieves a statistical accuracy for aqueous solvation energies that is comparable to SM12 or SMD
(see Table~\ref{table:dG}).\cite{DupAndMar13} 
This is despite using just two empirical parameters in $\Gnon$, along with two parameters [and an {\em ansatz}, 
Eq.~\eqref{eq:eps[rho]}] for the cavity determination.   
The SM$x$ models have a considerably larger number of empirical parameters, although they are designed to be 
``universal'' models for all solvents, whereas the SCCS model is parameterized only for water.
Improvement upon SCCS results for anions is obtained using a ``soft-sphere'' model,\cite{FisGenAnd17}
which constructs the cavity from vdW spheres but then interpolates the dielectric function 
(from $\diel=1$ to $\diel=78$) over a narrow switching region centered on the vdW radius.\cite{FisGenAnd17}  
When combined with the two-parameter model for $\Gnon$, the soft-sphere solvation model affords a MUE of 3.0~kcal/mol
in $\dGhyd^\circ$ for anions (Table~\ref{table:dG}), which is on par with the estimated
uncertainty (2--3~kcal/mol) in the experimental data themselves.\cite{PliRiv02a,SM6}

\subsection{Physics-based models} 
\label{sec:DeltaG:CMIRS}

Occasionally there have been attempts to put the nonelectrostatic contributions to the solvation energy on a more rigorous 
footing,\cite{Amo94,AmoMen97,WeiMenFre10,CupAmoMen15,
PomThoChi11,PomChi11,PomChi13a,PomChi14,PomChi15,DuiParNin13a,DuiParNin13b} 
or at least to develop parameterized models that are more closely connected to the physics of intermolecular interactions as
compared to the wholly empirical SASA-type approaches.\cite{HurCla72,FloTom89,FloTomAhu91,FloTanTom93,Tru98}
As a simple example of the latter variety, one might borrow from \qmmm\ methodology and assume classical interaction potentials,
\eg, of Lennard-Jones type that are centered on the solute atoms, with an $r^{-6}$ distance dependence for dispersion. 
This affords a model of the form\cite{FloTom89,Tru98} 
\begin{equation}\label{eq:Gdisp-QM/MM}
	\FreeE_\disp = \sum_A^\text{solvent} \bar{\rho}_A 
	\sum_B^\text{solute} \gamma^{}_{\!AB}
	\sum^\text{grid}_{i\in B}
	a_i\left(
	\frac{
		\br^{}_{iB}\bm{\cdot}\mathbf{n}_{\bs_i}
	}{
		3 r_{iB}^6
	}\right)
\; ,
\end{equation}
where $\bar{\rho}_A$ is the average number density of solvent atom $A$, and the parameters 
$\gamma^{}_{\!AB}$ come from a force field.   A similar model, depending for example on $r_{iB}^{-12}$, can be developed
to describe the repulsive contribution, $\FreeE_\text{rep}$.\cite{FloTomAhu91,Tru98}
The quantity $\FreeE_\text{disp} + \FreeE_\text{rep}$ is usually then combined with a cavitation energy of the form 
\begin{equation}\label{eq:Gcav}
	\FreeE_\text{cav} = \sum_B^\text{solute} 
	\left(\frac{\mbox{SASA}(B)}{4\pi R_B^2}\right) \Delta G_\text{HS}(R_B)
\; ,
\end{equation}
in which $\Delta G_\text{HS}(R_B)$ is the solvation energy for a hard sphere of radius $R_B$, obtained from 
scaled-particle theory.\cite{Pie76}   Typically the atomic radii used in these nonelectrostatic terms are SAS radii, \ie, 
they include a probe radius representing the assumed size of a solvent atom; see Eq.~\eqref{eq:atomic_radius}.   
This is the case even if $\FreeE_\elst$ is evaluated using a vdW cavity surface.

This is a classical model for the nonelectrostatic interactions but there have
also been attempts to derive fully quantum-mechanical dispersion--repulsion models within a continuum framework.
One such model, due to Amovilli,\cite{Amo94,AmoMen97}
starts from the generalized Casimir-Polder 	expression for the dispersion energy of a supramolecular complex 
A$\cdots$B,\cite{McL63b,McW84,JasMcW85,McW85,AmoMcW90,McW92} 
\begin{equation}\label{eq:Edisp-FDDS}
%
%
%
	\mathcal{U}_\disp^{\scrA\cdots\scrB}
	= -\frac{\hbar}{2\pi} \left(\frac{e^2}{4\pi\diel_0}\right)^{\!2}
	\int_0^\infty d\omega
	\int_{\realspace} d\br_1 d\br_2 d\br_1' d\br_2' \; 
	\left(\frac{
		\chi^{\scrA}(\br_1,\br_1'|\cmplxi\omega) \; \chi^{\scrB}(\br_2,\br_2'|\cmplxi\omega)
	}{
		\|\br_1 - \br_1'\| ~ \|\br_2 - \br_2'\|
	}\right) \; ,
\end{equation}
in which $\chi(\br,\br'|\omega)$ is the frequency-dependent density susceptibility 
(also known as the polarization propagator),\cite{AmoMcW90,McW92}
evaluated in Eq.~\eqref{eq:Edisp-FDDS} at imaginary frequencies.     
When the monomer separation $R_\text{AB}$ is large, 
second-order perturbation theory affords the ``uncoupled'' 
approximation,\cite{Tan69,LanKar70,SzaOst77,KapRod78,McW84,BuhWel07}   
which was first derived by London:\cite{Lon30,Lon65}    
\begin{equation}\label{eq:LJ}
	\mathcal{U}_\disp^{\scrA\cdots\scrB}(R_\text{AB})
	 = - \left[
	 	\frac{3\hbar}{\pi }\int_0^\infty \bar{\alpha}^{\scrA}(\cmplxi\omega) \; \bar{\alpha}^{\scrB}(\cmplxi\omega) \; d\omega
	\right] \frac{1}{R_\text{AB}^6} \; .
\end{equation}
The quantity $\bar{\alpha}(\omega)$ is the frequency-dependent isotropic polarizability.  Components of the polarizability
tensor, in the spectral representation, are 
\begin{equation}\label{eq:alpha(w)}
	\alpha_{ab}(\omega) = \frac{1}{\hbar}\sum_{n>0} 
	\left[
		\frac{ 
			\langle 0| \hat{\mu}_a | n\rangle \langle n | \hat{\mu}_b | 0 \rangle 
		}{
			\omega - \omega_{n0} 
		}
		+
		\frac{ 
			\langle 0| \hat{\mu}_b | n\rangle \langle n | \hat{\mu}_a | 0 \rangle 
		}{
			\omega + \omega_{n0} 
		}		
	\right] \; ,
\end{equation}
where $\omega_{n0}=(E_n-E_0)/\hbar$.\cite{NorRuu06}
The term in brackets in Eq.~\eqref{eq:LJ} provides a microscopic formula for the Lennard-Jones $C_6$ coefficient.

Returning to the more general expression in Eq.~\eqref{eq:Edisp-FDDS}, a sum-over-states expression (Lehmann representation) 
can be introduced for the requisite density susceptibilities, writing them in terms of transition densities 
$\rho_{0n}(\br)$:\cite{McW85,McW92}
\begin{equation}
	\chi(\br,\br'|\cmplxi\omega) = -\frac{2}{\hbar} \sum_{n>0} 
	\frac{
		\omega_{0n} \; \rho_{0n}(\br) \; \rho_{0n}(\br')
	}{
		\omega_{n0}^2 + \omega^2
	} \; ,
\end{equation}
Inserting this formula for $\chi^{\scrA}(\br_1,\br_1'|\cmplxi\omega)$ in Eq.~\eqref{eq:Edisp-FDDS} affords an expression 
for the dispersion energy of ``A-in-B'', \ie, solute A interacting with an implicit representation of B:\cite{Amo94}
\begin{equation}\label{eq:Gdisp_A-in-B}
	\FreeE_\disp^\text{A-in-B} = \frac{1}{\pi}\int_0^\infty d\omega
	\sum_{n>0} \frac{ 
		\omega_{n0}^{\scrA}
	}{ 
		(\omega_{n0}^{\scrA})^2 + \omega^2 
	} \int_{\realspace} d\br \int_{\surf} d\bs 
	\left(
	\frac{
		\rho^{\scrA}_{0n}(\br) \; 
		\sigma_{\scrB}[\dielw^{\scrB}(\cmplxi\omega),\rho^{\scrA}_{0n}](\bs) 
	}{
		\|\br - \bs\|
	} \right)
	\; .
\end{equation}
In writing this equation, the susceptibility $\chi^{\scrB}$ in Eq.~\eqref{eq:Edisp-FDDS}
has been eliminated by first introducing a polarization density
\begin{equation}
	\rho^{\scrB}_\text{pol}(\br_2) = \int_{\realspace} d\br_1' d\br_2'
	\left(
	\frac{
		\chi^{\scrB}(\br_2,\br_2'|\cmplxi\omega) \; \rho^{\scrA}(\br_1')
	}{
		\|\br_1' - \br_2'\|
	}\right)  \; ,
\end{equation}
induced on B by the presence of A.   This quantity is then replaced with a 
surface charge $\sigma_{\scrB}(\bs)$, in the spirit of the PCM formulation of the continuum electrostatics problem, 
and which depends on the frequency-dependent dielectric function $\dielw(\cmplxi\omega)$ evaluated at imaginary frequencies.
The function $\dielw(\cmplxi\omega)$ is the central quantity in the Lifshitz's general theory of the Casimir force 
and dispersion energies.\cite{DzyLifPit61b,McL63b,ZarKoh76,Par06} 
In a separate approach, Ninham and coworkers\cite{DuiParNin13a,DuiParNin13b} 
use the frequency-dependent dipole- and higher-order (hyper)polarizabilities for the solute, $\bar\alpha(\cmplxi\omega)$ etc., 
in conjunction with models for $\dielw(\omega)$, to model solute--solvent dispersion.

Models for $\FreeE_\disp$ can now be constructed based on Eq.~\eqref{eq:Gdisp_A-in-B} by modeling the  dielectric function
$\dielw(\cmplxi\omega)$ and the surface charge density $\sigma_{\scrB}(\bs)$ that is induced by various excited states of solute~A, via
transitions densities $\rho_{0n}^{\scrA}(\br)$.   These transition densities, along with the corresponding excitation frequencies
$\omega^{\scrA}_{n0}$ that appear in Eq.~\eqref{eq:Gdisp_A-in-B}, 
might be computed explicitly (since the solute~A is described by quantum chemistry),\cite{Amo94}
or perhaps modeled using SCF eigenvalues.\cite{AmoMen97,WeiMenFre10}
A model suggested for the surface charge is\cite{AmoMen97}
\begin{equation}\label{eq:sigma-disp}
	\sigma_{\scrB}[\rho^{\scrA}_{0n}](\bs) =
	-\frac{1}{4\pi}
	\left(
		\frac{\Omega^2}{\Omega^2 + \omega^2} 
	\right)
	\left
		(\frac{\dielop-1}{\dielop} 
	\right)
	\Eperp[\rho^{\scrA}_{0n}](\bs)
 \; ,
\end{equation}
where $\Eperp[\rho^{\scrA}_{0n}](\bs)$ is the normal electric field generated by the density $\rho^{\scrA}_{0n}(\br)$.
Modulo a factor of $\Omega/(\Omega+\omega^2)$, this is precisely the apparent surface charge of the D-PCM approach
[Eq.~\eqref{eq:D-PCM}], albeit generated by the transition density $\rho^{\scrA}_{0n}$ rather than the ground-state density and 
with the ``optical'' dielectric constant $\dielop$ replacing the static dielectric constant $\dielst$.    (As discussed in 
Section~\ref{sec:NonEq}, the value $\dielop$ is the appropriate one for re-polarization upon sudden or vertical excitation, without 
orientational contributions from the solvent.)    Regarding the factor of 
$\Omega^2/(\Omega^2+\omega^2)$, the quantity $\hbar\Omega$ is the characteristic ionization energy of the solvent.   
[This comes from the approximation of setting every $\omega_{n0}$ equal to $\Omega$;\cite{Lon65} 
In practice, $\hbar\Omega = \dielop \times (\text{IE})_\text{solvent}$ has been used.\cite{AmoMen97,WeiMenFre10}]
The integral over $\omega$ in Eq.~\eqref{eq:Gdisp_A-in-B} remains to be evaluated and this factor 
interpolates the apparent surface charge between limits of unity for $\omega=0$, for which the solute sees the full induced polarization
response, and zero as $\omega\rightarrow\infty$ because when $\omega \gg \Omega$ 
the excitation frequency is so large that the response averages to zero.
These models have interesting possibilities for the description of solute--environment dispersion in excited states, which are only
starting to be explored.\cite{WeiMenFre10,CupAmoMen15}   

A rather different formulation of the solute--continuum dispersion interaction has been put forward by Pomogaeva and
Chipman,\cite{PomChi13a} who borrow from the nonlocal correlation energy functional developed by Vydrov and 
Van Voorhis.\cite{VydVan09a,VydVan09b,VydVan10b,VydVan12b} 
This functional, usually known as VV10,\cite{CalOrtSan15b}
represents an attempt to incorporate dispersion interactions into density functional theory in a rigorous way and is 
itself a simplified form of the nonlocal functional developed by Langreth and 
Lundqvist.\cite{LanDioRyd04,DioRydSch04,LanLunCha09,LeeMurKon10}
(In a mildly annoying bit of physicist reductionism, the Langreth-Lundqvist functional is often known as 
 ``the'' van der Waals functional, as if such a designation could possibly be unique.)    These nonlocal correlation functionals are
already based on a pairwise {\em ansatz},\cite{CalOrtSan15b} and to use them in conjunction with a continuum representation of the solvent
one simply replaces the density of one interacting partner with the bulk solvent density, $\bar{\rho}$.  
The functional form is then\cite{PomChi13a,YouHer16}
\begin{equation}\label{eq:Gdisp-CMIRS}
	\FreeE_\text{disp} 
	= A\int_{\br\in\cav}
	\left(\frac{
		\rho(\br) \; I(\br) 
	}{
		w[\rho](\br)\big(w[\rho](\br) + \bar{\rho}^{1/2}_\text{solvent}\big)
	}\right) \; d\br 
\end{equation}
where 
\begin{equation}\label{eq:I(r)}
	I(\br) = \int_{\br'\notin\cav} \left(\frac{d\br'}{\|\br - \br'\|^6 + \delta^6}\right) 
\end{equation}
and 
\begin{equation}\label{eq:w[rho]}
	w[\rho](\br) = \left(
		\rho(\br) + \frac{3C}{4\pi}\frac{\big\| \del \rho(\br) \big\|^4 }{ \rho(\br)^4 }
	\right)^{1/2} \; .
\end{equation}
The parameter $C$ in Eq.~\eqref{eq:w[rho]} is taken without modification from VV10,\cite{VydVan09b,VydVan10b}
but a parameter $\delta$ is introduced in Eq.~\eqref{eq:I(r)} to prevent divergence of the integral.   Aside from that, 
the only additional parameter that has been introduced (beyond those already present in VV10) 
is an overall scaling factor $A$ in Eq.~\eqref{eq:Gdisp-CMIRS}.   The density $\rho(\br)$
is the solute's electron density and the integral in Eq.~\eqref{eq:Gdisp-CMIRS} is evaluated over the solute cavity ($\br\in\cav$).
However, the integration domain in Eq.~\eqref{eq:I(r)} is the region $\br'\notin\cav$ that is {\em outside} of the solute cavity, which 
requires integration of three-dimensional space.   (In practice the discretization need not extend very far beyond the cavity, 
since the integrand decays rapidly.)
It is worth noting that analytic gradients have not been implemented for the 
quantum-mechanical dispersion models described in this section, although they have been implemented for the \qmmm-style 
approach of Eq.~\eqref{eq:Gdisp-QM/MM}.\cite{Tru98}  
However, the functional form for $\FreeE_\disp$ in Eq.~\eqref{eq:Gdisp-CMIRS} may better lend itself to analytic gradients as 
compared to Eq.~\eqref{eq:GB_radii}, 
insofar as the analytic gradient of VV10 has already been reported.\cite{VydVan10b}

In contrast to dispersion, a functional form for Pauli repulsion is rather straightforward.    
This effect arises due to interpenetration of the tails of two non-bonded densities, 
and one functional form that has been suggested is simply
\begin{equation}\label{eq:Gexch}
	\FreeE_\exch = B \int_{\br\notin\cav}  \rho(\br) \; d\br 
\end{equation}
where $B$ is an empirical scaling factor.\cite{AmoMen97} 
(Note also that the integration domain is over the solvent, $\br\notin\cav$.)
An alternative is\cite{PomChi13a}
\begin{equation}\label{eq:Gexch-CMIRS}
	\FreeE_\exch 
	= B\int_{\br\notin\cav} \big\|\del\rho(\br)\big\| \; d\br \; .
\end{equation}
This latter form is suggested by an exact result for the 
exchange-repulsion energy of two hydrogen atoms,\cite{HerFli64}  
but both models have been used in practice.    For example, the model in 
Eq.~\eqref{eq:Gexch} has been used to develop an ``extreme pressure'' (XP-)
PCM,\cite{CamVerMen08,CamCapMen12} which is based on the thermodynamic relation $p = -(\partial \FreeE/\partial V)_T$
and calculation of analytic derivatives of $\FreeE_\exch$ and $\FreeE_\elst$ with respect to the cavity volume.   
XP-PCM has been used to study how pressures $p > 1$~GPa affect both molecular geometries and the equilibrium positions of chemical
reactions.\cite{CheHofCam17,Cam18}   The exchange-repulsion model in Eq.~\eqref{eq:Gexch-CMIRS} has been used as part of 
a black-box solvation model that is described next.

In an attempt to develop a first-principles implicit solvation model that can compete with SM$x$, 
Pomogaeva and Chipman\cite{PomThoChi11,PomChi11,PomChi13a,PomChi14,PomChi15}  
have combined SS(V)PE electrostatics  
(based on an isodensity cavity construction) with a ``minimally parameterized'' nonelectrostatic model of the form 
\begin{equation}\label{eq:Gnonelst-CMIRS}
	\Gnon = \FreeE_\disp + \FreeE_\exch + \FreeE_\text{FESR} \; .
\end{equation}
Here, $\FreeE_\disp$ and $\FreeE_\exch$ are the models in 
Eq.~\eqref{eq:Gdisp-CMIRS} and Eq.~\eqref{eq:Gexch-CMIRS}, respectively, and $\FreeE_\text{FESR}$ is 
``field-effect short-range'' (FESR) term for hydrogen bonding.   The functional form for 
$\FreeE_\text{FESR}$ takes as inputs the maximum and minimum 
values of the normal electric field at the cavity surface, and has the form\cite{PomChi11} 
\begin{equation}\label{eq:FESR}
	\FreeE_\text{FESR}
	= C \big(\min | \Eperp |\big)^\xi + D \big(\max | \Eperp |\big)^\xi \; .
\end{equation}
(The correlation between hydrogen bond strength and local electric fields in water has been made 
before.\cite{FecEavLop03,CorLawSki04,SmiCapWil05}) 
This {\em ansatz\/} is likely only appropriate for small molecules since it would seem to accommodate only a single hydrogen-bond
donor site and a single acceptor site.

The combination of SS(V)PE electrostatics with $\Gnon$ in Eq.~\eqref{eq:Gnonelst-CMIRS} has been called the 
{\em composite method for implicit representation of solvent\/} (CMIRS).\cite{PomChi13a,PomChi14,PomChi15,YouHer16}
This model contains five empirical parameters for a given solvent:  the linear coefficients $A$, $B$, $C$, and $D$,
along with the exponent $\xi$.    (The FESR term is omitted for non-polar solvents.)
Pomogaeva and Chipman determined these parameters, by fitting to experimental solvation energies, 
for benzene,\cite{PomChi13a} cyclohexane,\cite{PomChi13a}
water,\cite{PomChi14} dimethyl sulfoxide,\cite{PomChi15} and acetonitrile,\cite{PomChi15} although all of those parameters
were later adjusted to fix an error in the original implementation.\cite{YouHer16}  Parameters for methanol have also been 
reported based on the original implementation.\cite{SilDegPli16}
For any given solvent, this is is considerably fewer parameters than are contained in any of the SM$x$ models, although the latter are designed
as ``universal'' models in which the nonelectrostatic parameters $\{\gamma^{}_K\}$ in Eq.~\eqref{eq:Gnonelst-CDS} are determined only once,
and then the model is available for any solvent whose macroscopic descriptors are available.\cite{CraTru08}   
For example, the training set for SM12 contains 92 solvents,\cite{SM12}
as compared to the six for which CMIRS parameters are currently available.

\begin{figure}
	\centering
	\fig{1.0}{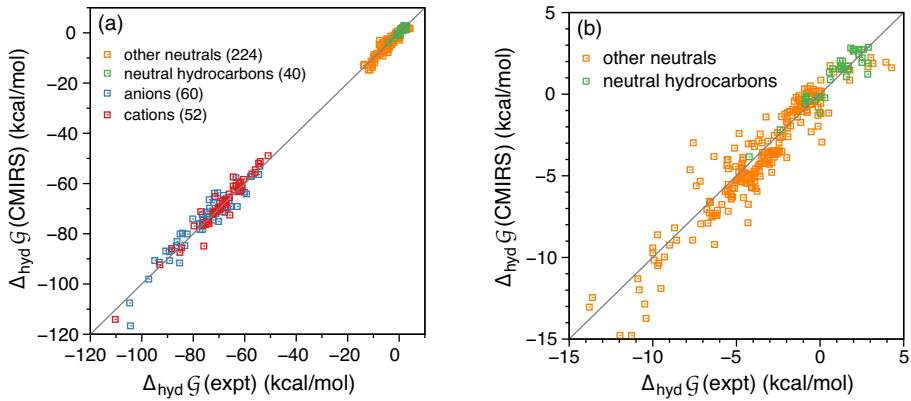}
	\caption{
		Comparison of experimental hydration energies 
		in the Minnesota solvation database\protect\cite{ThoCraTru04,SM6,SM12,SMD} 
		to predictions from the CMIRS model:
		(a) all solvation energies, including both neutral molecules as well as ions
		(with the number of data points indicated in each case), versus
		(b) results for charge-neutral solutes only. 
		Reprinted from Ref.~\protect\citenum{YouHer16}; copyright 2016 American Chemical Society.
	}\label{fig:CMIRS-aq}
\end{figure}

Of course, water is the solvent to have if you're only having one, and error statistics versus experimental hydration energies
(Table~\ref{table:dG}) demonstrate that CMIRS is somewhat more accurate (on average) than SM12 or SMD values, especially
for ions, despite fewer parameters in the model.  The CMIRS error 
of 2.4~kcal/mol for ions is comparable to the estimated uncertainty 
(2--3~kcal/mol) in the experimental solvation energies for aqueous ions,\cite{PliRiv02a,SM6} 
and thus represents the practical lower limit for any solvation model trained on these data.   Correlation between CMIRS and
experimental hydration energies is excellent; see Fig.~\ref{fig:CMIRS-aq}.
That said, despite all of the physical considerations that went into the CMIRS model for $\FreeE_\nonelst$, and despite the limited
number of parameters used per solvent, an error in the original implementation $\FreeE_\disp$ went unnoticed despite the
fact that it modifies dispersion energies in the training set by up to 8~kcal/mol.\cite{YouHer16}   This was able to escape notice 
because the $B$ parameter
in Eq.~\eqref{eq:Gexch-CMIRS} absorbs the discrepancy, making $\FreeE_\exch$ more repulsive in order to offset the 
too-attractive dispersion energy.\cite{YouHer16}   A similar cancellation between the repulsive cavitation term and the attractive
dispersion term has been noted elsewhere,\cite{DuiParNin13b} and in fact $\FreeE_\disp+\FreeE_\exch$ is often 
parameterized together as a single entity in empirical models, including SM$x$ and others that use the atomic surface tension 
{\em ansatz\/} of Eq.~\eqref{eq:Gnonelst-CDS}.    This may hide certain subtleties, such as the fact that cavitation effects
are more important than dispersion to explain binding affinities of rare-gas guest atoms to 
cucurbituril host molecules,\cite{HeBieVan18}
or that the unfavorable hydration energies of small nonpolar polymers are well approximated by the cavitation energy 
($\dGhyd\approx \FreeE_\text{cav}$),\cite{AnsLaiHas19} suggesting near-cancellation of other effects.

\section{Nonequilibrium Solvation}
\label{sec:NonEq}
\subsection{Conceptual overview}
\label{sec:NonEq:Intro}

How does a continuum solvent respond to a sudden change in the solute's charge distribution?
This question must be considered for electronic spectroscopies, including absorption to (or emission from) an
excited electronic state, or photoelectron spectroscopy that removes an electron from the solute.   
The general theory of time-dependent processes in dielectric materials introduces a frequency-dependent 
dielectric function $\dielw(\omega)$, such that the induction field $\mathbf{D}$ responds to a frequency-dependent electric field 
$\Efield$ according to $\mathbf{D}(\br,\omega) = \dielw(\omega)\,\Efield(\br,\omega)$.\cite{BotBor78,Vau79,FelPuzRya06,FelIshPuz15} 
In the presence of a time-dependent field, the permittivity $\dielw(\br,\omega)$ is complex-valued, 
in order to describe the phase lag between $\Efield$ and $\mathbf{D}$.\cite{BotBor78,FelIshPuz15} 
Often, the complex-valued permittivity is denoted as $\hat{\dielw}$ or $\dielw^\ast$ but we will not do so here.   Where needed, 
we will simply indicate the frequency dependence explicitly.

The permittivity is real-valued in two important limits, namely, 
$\dielst \equiv \dielw(0)$, which is the static (zero-frequency) limit, and also in the high-frequency limit.  The limiting value 
$\dielop = \lim_{\omega\rightarrow\infty}\dielw(\omega)$
is known as the {\em optical dielectric constant}, for reasons that are described below, 
and describes the fact that there is always some part of the polarization that is able to remain
in phase with the applied field.  Switching to the time domain and labeling that part of the medium's response as ``fast'' polarization, 
\begin{equation}\label{eq:P_fast}
	\mathbf{P}_\text{fast}(t) = \left(\frac{\dielop-1}{4\pi}\right)\mathbf{E}(t) \; ,
\end{equation}
the remaining (``slow'') component is then defined by the time-dependent 
analogue of Eq.~\eqref{eq:D(r)}:\cite{BotBor78} 
\begin{equation}
	\mathbf{D}(t) = \dielop \Efield(t) + 4\pi \mathbf{P}_\text{slow}(t) \; ,
\end{equation}
The frequency components of $\mathbf{P}_\text{slow}(t)$ depend on $\dielw(\omega)$, 
but generally speaking the slow polarization response consists of vibrational contributions with timescales of 
$10^{-12}$--$10^{-14}$~s, along with orientational components having timescales slower than $10^{-12}$~s.
For example, the primary relaxation timescale for liquid water is 
8--10~ps under ambient conditions.\cite{Has72,Kaa89,Kaa93,Kaa15} 

It is possible to model the frequency dependence of $\dielw(\omega)$ 
directly.\cite{HsuSonMar97,IngMenTom03,CarMenTom06,DinLinMen15,WilDonLip19}
A phenomenological model is
\begin{equation}\label{eq:eps(w)}
	\dielw(\omega) = \dielop + 
	(\dielst - \dielop)\sum_k
	\frac{
		c_k 	
	}{
		1 + \cmplxi\omega\tau_k
	} \; ,
\end{equation}
where the $\tau_k$ are a set of characteristic time constants, representing microscopic relaxation processes, 
and $\sum_k c_k=1$.\cite{BotBor78} 
(The version with only a single timescale was originally introduced by Debye.\cite{Deb29,Has72,FelIshPuz15})
When such a model is used in the context of continuum solvation theory, the polarization charge
becomes explicitly time-dependent and the permittivity for ``fast'' polarization is the real part of $\dielw(\omega)$ for 
frequencies larger than the perturbation of interest. 
Such models have been used to simulate the time-dependent Stokes shift in the 
fluorescence energy,\cite{HsuSonMar97,IngMenTom03,CarMenTom06} 
and to simulate the combined response of the molecule and the medium to an external field that is resonant with an 
excited state of the solute.\cite{WilDonLip19}   The latter application makes the most sense when combined with electronic 
structure methods that simulate time-dependent electron dynamics,\cite{GoiLesLi18,LiGovIsb20}  
but these explicitly time-dependent approaches are not considered in this work.

Instead, this work is focused on the nonequilibrium response to a sudden (vertical or Franck-Condon) 
change in the solute density.  For a vertical absorption, emission, or ionization process, the nuclei are fixed and the 
continuum solvent ``molecules'' cannot vibrate or reorient.   This limits the continuum solvation response to the fast 
component of the polarization, which is electronic in nature and remains in equilibrium with the sudden change in $\rho(\br)$.
The slow component is dictated by the solute's initial electronic state and cannot respond on the timescale
of a vertical excitation or ionization process, and is therefore out of equilibrium with the final electronic state.  
Phenomenologically, this is accomplished by reducing $\dielw(\omega)$ to its limiting values, $\dielst$ and $\dielop$. 

The value of $\dielop$ can be related to the polarizability of the solvent molecules (Lorenz-Lorentz equation),\cite{BotBor78}
but in the present context is simple a measured property of the solvent, determined from the solvent's refractivity
$\nrefr(\omega) = \sqrt{\diel(\omega)}$.\cite{BotBor78,MisKasRud17}  
Formally this should be done in the limit $\omega\rightarrow\infty$ but also needs to be measured
away from any resonances and therefore optical wavelengths are often used, hence the ``optical'' dielectric constant.
Values are often measured at the sodium D-line, $\lambda=589$~nm,\cite{WohWoh96}
and for water the value obtained is $\dielop(\lambda)=1.78$ at $\lambda = 589$~nm, as compared to 
$\dielop(\lambda)=1.95$ at $\lambda = 200$~nm.\cite{HalQue73}
In older literature (and sometimes repeated more recently),\cite{CosBar00b,DinNewMat17}
values of $\dielop = 4.0$--$5.5$ are occasionally reported for water,\cite{ColHasRit48,Has72}
the larger value originally being thought to agree better with predictions from Onsager's reaction-field theory.\cite{ColHasRit48} 
In fact, the larger value of $\dielop$ is based on a false extrapolation using Debye's model of a single relaxation timescale
($\tau_1 = 8$--$10$~ps), whereas permittivity data that extend to terahertz frequencies reveal 
at least two distinct relaxation timescales,\cite{RonThrAst97,RonKei02,NagYadAri06,Kaa15,ZhoRaoLiu19}
including a faster process $\tau_2 < 1$~ps.\cite{RonThrAst97,RonKei02,Kaa15}
(The microscopic explanation for these timescales remains a topic of current debate.\cite{PopIshKha16,Elt17})
The data therefore appear to approach a limiting value $\dielop\approx 5$ in the microwave regime,\cite{Has72,KaaUhl81}  
but further decay to $\dielop\approx 2$ at terahertz frequencies,\cite{RonKei02,ZhoRaoLiu19}
the latter value being 
consistent with $\nrefr^2(\lambda)$ measured at optical wavelengths.\cite{HalQue73}


\begin{table}
	\centering
	\caption{
		Static dielectric constants$^a$ ($\dielst$) and optical dielectric constants$^b$ ($\dielop$)
		for some common solvents.
		(These are given as dimensionless values relative to vacuum permittivity.)
	}\label{table:diel}
\begin{threeparttable}
	\begin{tabular}{l .. l ..}
	\hline\hline
		Solvent  & \mc{1}{c}{$\dielst$} & \mc{1}{c}{$\dielop$}  &
		Solvent  & \mc{1}{c}{$\dielst$} & \mc{1}{c}{$\dielop$}  
\\ \hline
		$n$-hexane			& 1.9 	& 1.89 &    
		ethanol				& 24.3 	& 1.85 \\   											  
		cyclohexane			& 2.0 	& 2.03 & 
		ethylene glycol		& 30.9	& 2.05 \\  
		benzene				& 2.3 	& 2.25 &  
		methanol				& 33.0 	& 1.77 \\   
		toluene				& 2.4 	& 2.24 &  
		nitrobenzene			& 34.7  	& 2.41 \\   
		diethyl ether			& 4.2	& 1.83^c &
		acetonitrile			& 36.0 	& 1.81 \\   
		chloroform			& 4.7	& 2.08 &  
		dimethyl acetamide	& 39.6	& 2.07 \\ 
		dichloromethane		& 8.9	& 2.03 & 
		dimethyl sulfoxide		& 46.6 	& 2.18 \\   
		2-propanol			& 18.2	& 1.92 &  
		water				& 78.4 	& 1.78^d   \\		
		acetone				& 20.8	& 1.85 &  
		formamide			& 109.6	& 2.10 \\
\hline\hline
	\end{tabular}
	\begin{tablenotes}[flushleft]
	\fns
		\item 
		$^a$At 25$^\circ$C, from Ref.~\CiteN{Woh91}.
		$^b$At 20$^\circ$C except where noted, 
		based on refractive indices $\nrefr(\lambda)$ measured at $\lambda=589$~nm, from Ref.~\CiteN{WohWoh96}.
		$^c$At 16.5$^\circ$C.
		$^d$From Ref.~\CiteN{TilTay38}, valid for 20--25$^\circ$C.
	\end{tablenotes}
\end{threeparttable}	
\end{table}

Looking at a modern tabulation of the data for common solvents (Table~\ref{table:diel}),\cite{WohWoh96} one finds 
little variation in refractive indices at visible wavelengths, which generally range from $\nrefr \approx 1.3$--$1.5$.
Despite having {\em static\/} dielectric constants ranging from from $\dielst= 2$--$78$, there is considerable uniformity in $\dielop$, 
which ranges from $\dielop = 1.7$--$2.3$.   This reflects the 
the fact that typical solvents are closed-shell, small-molecule insulators with band gaps in the vacuum ultraviolet,
and therefore roughly similar molecular polarizabilities.  
Inorganic solids may have considerably larger indices of refraction,\cite{Kit05} 
\eg, $\nrefr = 2.43$ at 589~nm (and therefore $\dielop = 5.90$) for BaTiO$_3$(s),\cite{WemDidCam68} 
a material used in nonlinear optical applications.
These larger values are attributable to low-lying excited states that facilitate more substantial electronic polarization and 
therefore significant dispersion of light, but this behavior is simply not found in common solvents.

Before introducing a modern computational formalism for nonequilibrium polarization, we first consider two historical examples.
The first is a well-known result in electronic spectroscopy that relates the Stokes shift
($\Delta \nu = \nu_\text{abs} - \nu_\text{fluor}$) to the change in the solute's dipole moment upon excitation ($\Delta\mu$), and
which is known as the {\em Lippert-Mataga equation},\cite{MatYozKoi56,Lip57,Bot76,MatChoTan05} 
\begin{equation}\label{eq:LM}
	hc\Delta\nu = 
	\text{constant} + \frac{ 2(\Delta\mu)^2}{\BornRad^3} 	
	\underbrace{
		\left[
			\frac{\dielst-1}{\dielst+1}  - \frac{\dielop -1}{2\dielop+1}
		\right]
	}_{
		\FLM(\dielst,\dielop)
	}
 \; .
\end{equation}
In practice this equation is used to determine excited-state dipole moments (assuming that the ground-state 
dipole moment is known) by measuring the Stokes shift in solvents of differing polarity.   
In fact, a variety of alternative formulas for this purpose have been 
suggested,\cite{Bay50,Oos54,McR57,BilKaw62,Bak64,Lip65a,Abe65,ChaVia70,BraCar85} 
along the lines of Eq.~\eqref{eq:LM} but differing somewhat in their treatment of excited-state polarization, which leads
to differences in the form of the ``solvent polarity function'' $\FLM(\dielst,\dielop)$.
These models (including the Lippert-Mataga one) are so crude 
that often experimental data can be fit equally well to any one of 
them.\cite{JozMilHel09,SidHan10,ManKumShi13,DemMenWoy17,KumVarGeo17,RenNadSri18,DivSakWei19} 
More important is the basic molecular physics that underlies this approach.
Comparison to the model of a dipole in a spherical cavity [Eq.~\eqref{eq:Bell}] shows that the physical content of 
of Eq.~\eqref{eq:LM} is to take the difference dipole moment $\Delta \bm{\mu}$ and solvate it using permittivity 
$\dielop$ rather than $\dielst$.   

The solvent parameter $\dielop$ also makes an appearance in Marcus' theory of electron 
transfer,\cite{Mar56a,Mar64,MarSut85,New07,Li15,Hsu20}
in which the ``outer-sphere'' reorganization energy is given by 
\begin{equation}\label{eq:Marcus-lambda}	
	\lambda_\text{outer} = (\Delta Q)^2 \left(\frac{1}{\dielop} - \frac{1}{\dielst}\right)
	\left(\frac{1}{2R_\text{D}} + \frac{1}{2R_\text{A}} - \frac{1}{\|\mathbf{r}_\text{D} - \mathbf{r}_\text{A}\|}\right)
\; .
\end{equation}
This formula is derived from what is essentially a nonequilibrium formulation of the Born ion model [Eq.~\eqref{eq:Born}],
combined with a Coulomb interaction
between charges centered in a donor sphere (radius $R_\text{D}$ centered at $\mathbf{r}_\text{D}$) and an acceptor sphere (radius 
$R_\text{A}$ centered at $\mathbf{r}_\text{A}$).    The electron transfer is assumed to occur instantaneously---before the orientational
degrees of freedom of the solvent molecules can respond---hence the change in $\FreeE_\elst$ involves $\dielop$ in addition to $\dielst$.
The prefactor of $(\dielop^{-1} - \dielst^{-1})$ in Eq.~\eqref{eq:Marcus-lambda}
is sometimes called the {\em Pekar factor},\cite{CraTru99,DinNewMat17} and replaces the 
prefactor of $(1-\dielst^{-1}$) in the equilibrium version of the Born model.
%
%
For water, $\dielst$ is large due to the sizable H$_2$O dipole moment, leading to a large 
orientational component to the dielectric screening effect, but for non-polar solvents most or all of the solvent response is electronic
since $\dielop\approx \dielst$ (see Table~\ref{table:diel}).
The electronic contribution comes from the intrinsic polarizability of the solvent molecules, and for this reason $\dielop$ 
has sometimes been called the ``dielectric constant for induced polarization''.\cite{BotBor78}

\subsection{State-specific approach}
\label{sec:NonEq:SS}
The phenomenology introduced above can be generalized to a rigorous description of electrostatics beyond the Born or 
models,\cite{KimHyn90a,AguOliTom93} 
affording a continuum theory of nonequilibrium solvation.\cite{KimHyn90a,AguOliTom93,CraTru99,CosBar00b,Men07,Li15,GuiCap19}
Several variations and implementations have been formulated for  
PCMs,\cite{MenCamTom98,CosBar00a,Chi09a,MewYouWor15,YouMewDre15,MewHerDre17} 
as well as for continuum solvation based on Poisson's equation.\cite{CooYouHer16,CooHer18}
Operationally, a charge density $\rho_0(\br)$ corresponding to the initial electronic state of the solute is first equilibrated with a continuum
whose dielectric constant is $\dielst$.   The, following excitation (or ionization) from state $|0\rangle$ to state $|k\rangle$, the 
difference density $\Delta\rho_k(\br) = \rho_k(\br) - \rho_0(\br)$ is allowed to polarize a continuum whose dielectric constant is $\dielop$.
We now consider this in detail.


Consistent with the appearance of the zero- and infinite-frequency dielectric constants in 
Eqs.~\eqref{eq:LM} and \eqref{eq:Marcus-lambda}, 
a general nonequilibrium theory of continuum electrostatics is based upon a 
partition $\mathbf{P} = \mathbf{P}_\text{fast} + \mathbf{P}_\text{slow}$ in which the fast component of the polarization remains
in equilibrium with the solute even when the latter experiences a sudden change in its charge density.  The slow polarization is 
frozen on this timescale, fixed at the value determined by equilibrium solvation of the initial state.  This 
is accomplished via a partition of the (linear) electric susceptibility, $\chi(\omega) = [\diel(\omega)-1]/4\pi$.
%
%
Separating the susceptibility $\chi = \chi^{}_\text{f} + \chi^{}_\text{s}$ into fast and slow contributions, the form of 
$\chi^{}_\text{f}$ is suggested by Eq.~\eqref{eq:P_fast}, namely\cite{BraCar85,AguOliTom93,CamTom95c,Agu01,YouMewDre15}  
\begin{subequations}\label{eq:Marcus_partition_chi}
\begin{align}
	\chi^{}_\text{f} & = (\dielop - 1)/4\pi
\\
	\chi^{}_\text{s} & = (\dielst-\dielop)/4\pi \; ,
\end{align}
\end{subequations}
This has been called the ``Marcus partition'' of the of the polarization response,\cite{YouMewDre15} 
although it was actually formalized by Brady and Carr.\cite{BraCar85}
It embodies the phenomenological concepts introduced above, \eg, that $\mathbf{P}_\text{fast}$ originates with 
molecular polarizability while $\mathbf{P}_\text{slow}$ is orientational.

An alternative to Eq.~\eqref{eq:Marcus_partition_chi} is the so-called ``Pekar partition'',\cite{Agu01,YouMewDre15} 
which was originally introduced to describe self-trapped polarons,\cite{Pek54,Pek63}
then subsequently adapted for optical spectroscopy,\cite{BayMcR54a,McR57,Lip57} 
and still later adopted for use in ASC-type continuum solvation 
models.\cite{BonCimTom83,BasChu91,HouSakIno97,CamFreMen02,TomMenCam05}
Its use is prevalent in older literature so the distinction with respect to Eq.~\eqref{eq:Marcus_partition_chi} is worth pointing out.
Within the Pekar approach, the induced surface charge is partitioned into ``inertial'' and ``dynamic'' components
($\mathbf{P} = \mathbf{P}_\text{in} + \mathbf{P}_\text{dyn}$). 
The total surface charge $\sigma(\bs)$ is given by 
Eq.~\eqref{eq:D-PCM} and the dynamical part by the analogous expression\cite{BonCimTom83}
\begin{equation}\label{eq:sigma-partition-Pekar}
	\sigma_\text{dyn}(\bs) = 
	\frac{1}{4\pi}\left(\frac{\dielop-1}{\dielop}\right)
	\left(\frac{\partial \esp}{\partial\mathbf{n}_{\bs}}\right)_{\bs=\bs^-} 
\end{equation}
that is obtained with $\dielop$ replacing $\dielst$.
The inertial charge represents the difference, $\sigma_\text{in}(\bs) = \sigma(\bs)- \sigma_\text{dyn}(\bs)$.   

The difference between these two partitions can readily be understood using the reaction-field for a ground-state dipole $\bm{\mu}_0$ 
in a spherical cavity.  The slow contribution to the reaction field is\cite{Agu01} 
\begin{equation}
	\Efield_\rxn^\slow = 
	\begin{cases}
		\displaystyle 
		\FOB(\dielst,\BornRad) \left(\frac{\dielst - \dielop}{\dielst-1}\right) 
		\bm{\mu}_0
		& \text{(Marcus-Brady-Carr)} 
\\
		\displaystyle
		\big[\FOB(\dielst,\BornRad) - \FOB(\dielop,\BornRad)\big]
		\bm{\mu}_0
		& \text{(Pekar)}
	\end{cases} \; ,
\end{equation}
where $\FOB(\diel,\BornRad)$ is the ``Onsager factor'' defined in Eq.~\eqref{eq:dipole_rxn_field}.
The Marcus-Brady-Carr result follows from the fact that the slow polarization constitutes a fraction 
$\chi^{}_\text{s}/\chi = (\dielst - \dielop)/(\dielst-1)$ of the total polarization, according to Eq.~\eqref{eq:Marcus_partition_chi},
whereas the Pekar result is set by fiat.    Both partitions afford the {\em same\/} total reaction field,\cite{Agu01}
$\Efield_\rxn^\slow + \Efield_\rxn^\fast$, and therefore the same nonequilibrium free energy,\cite{Agu01,MarCraTru11,YouMewDre15} 
up to some minor issues involving discretization along the lines of what was discussed in
Section~\ref{sec:Smooth:Matrix}.\cite{YouMewDre15}   However, the partition into fast and slow components is different.
Noting that $\FOB(\dielst,\BornRad) \approx \BornRad^{-3}$ for high-dielectric solvents, 
Brady and Carr\cite{BraCar85} noted
that the Pekar result for $\Efield_\rxn^\slow$ seems oddly small (and also decoupled from the value of $\dielst$) in this limit.
Indeed, for water one obtains $\Efield_\rxn^\text{s} = 0.97 \bm{\mu}_0/\BornRad^3$ for the Marcus-Brady-Carr case and 
$\Efield_\rxn^\text{s} = 0.63 \bm{\mu}_0/\BornRad^3$ for the Pekar partition.   For that reason, the Marcus-Brady-Carr partition is the 
more common choice in modern literature although the Pekar partition has not vanished.\cite{MarCraTru11}
(The Marcus partition has also been criticized recently, and some alternatives have been suggested.\cite{DinNewMat17})
Side-by-side expressions for the free energy $\Gelst$ in the 
Marcus-Brady-Carr and the Pekar partitions are provided by Tomasi \etal.\cite{TomMenCam05}   As those authors note, there is 
some confusion in the literature regarding the names, \eg, Eq.~\eqref{eq:Marcus_partition_chi} is called Pekar partition by
Chipman.\cite{Chi09a}   As such, Tomasi \etal\ designate these schemes as ``partition~I'', meaning 
Eq.~\eqref{eq:Marcus_partition_chi}, and ``partition~II", meaning Eq.~\eqref{eq:sigma-partition-Pekar}.   Although 
this notation has been adopted in a few places,\cite{Men07,MarCraTru11} the names ``Marcus" (for partition~I) and ``Pekar" 
(for partition~II) remain common.

Having settled on the partition given in Eq.~\eqref{eq:Marcus_partition_chi},  
the basic idea of nonequilibrium polarization is that the susceptibility $\chi^{}_\text{s}$ should be used to induce polarization 
for the initial state (``0''), whose solute charge density is $\rho_0(\br)$, and then $\chi^{}_\text{f}$ should be used in conjunction with the 
difference density $\Delta\rho(\br)$ in order to adjust the polarization to reflect the final state.   
To realize this in practice, one first computes the surface charge $\sigma_0(\bs)$ that is induced by $\rho_0(\br)$ in a medium whose 
dielectric constant is $\dielst$, according to a normal equilibrium solvation calculation.  Next, $\sigma_0(\bs)$ is partitioned into 
fast and slow contributions,\cite{AguOliTom93,CosBar00b,MewYouWor15} 
\begin{subequations}\label{eq:sigma-partition-Marcus}
\begin{align}
	\sigma^\text{f}_0(\bs) &= \left(
		\frac{\dielop - 1}{\dielst - 1}
	\right) \sigma_0(\bs)
\\
	\sigma^\text{s}_0(\bs) &= \left(
		\frac{\dielst - \dielop}{\dielst - 1}	
	\right) \sigma_0(\bs) \; .
\end{align}
\end{subequations}
The quantity $\sigma^\text{s}_0(\bs)$ is retained, whereas 
$\sigma^\text{f}_0(\bs)$ is replaced by a surface charge induced by the excited-state charge distribution, in a medium whose
dielectric constant is $\dielop$.

In order to derive rigorous formulas for the electrostatic free energy of an excited state, 
introduce a Schr\"odinger equation of the form 
\begin{equation}\label{eq:SS-Schrodinger}
	\underbrace{
		\left(\Hvac + \RxnF_0^\text{s} + \RxnF_k^\text{f}\right) 
	}_{
		\Ham_k^\text{SS}
	}
	\big|\Psi_k\big\rangle 
	= \E^\text{SS}_k \, \big|\Psi_k\big\rangle 
\end{equation}
with $k=0$ for the ground state.   
The quantity $\Hvac$ is the vacuum Hamiltonian and the reaction-field contribution 
$\RxnF_k =  \RxnF_0^\text{s} + \RxnF_k^\text{f}$ consists of a slow component $\RxnF_0^\text{s}$ that originates with the 
ground-state density $\rho_0$ and susceptibility $\chi^{}_\text{s}$, along with a fast component $\RxnF_k^\text{f}$ based on the 
final-state density $\rho_k(\br)$ and susceptibility $\chi^{}_\text{f}$.    
Because $\RxnF_k^\text{f}$ depends on the wave function 
$|\Psi_k\rangle$ that is used to compute the excited state's electrostatic potential, the Hamiltonian $\Ham_k^\text{SS} = \Hvac + \RxnF_k$
that is introduced in Eq.~\eqref{eq:SS-Schrodinger} is ``state-specific" (SS), and straightforward attempts to solve this equation encounter
significant complications including convergence difficulties and ambiguous formulas for transition 
moments.\cite{JacHer11c}  These problems can be circumvented by treating $\RxnF_k^\text{f}$ as a perturbation
on top of zeroth-order states that are eigenfunctions of $\Ham_0^\text{SS}$, as discussed below.

First, let us consider an expression for the free energy in an excited state.
Note that Eq.~\eqref{eq:G[Psi]} for the ground-state free energy $\FreeE_0$ is implicitly based on a Hamiltonian 
\begin{equation}
	\Ham_0 \equiv \Ham_0^\text{SS} =\Hvac + \RxnF_0^{\text{s}+\text{f}} \; ,
\end{equation}
however $\FreeE_0$ differs from $\langle\Psi_0|\Ham_0|\Psi_0\rangle$ by an amount equal to the work 
$\work_0 = \tfrac{1}{2}\langle\Psi_0 | \RxnF_0 | \Psi_0\rangle$ that is required to polarize the continuum.
This expression for the work can be generalized to an arbitrary state $|\Psi_k\rangle$: 
\begin{equation}\label{eq:W_k}
	\work_k = \frac{1}{2}\big\langle\Psi_k \big| \RxnF_k \big| \Psi_k \big\rangle
	= \frac{1}{2}\int_{\surf} \sigma_k(\bs) \; \esp^{\rho_k}\!(\bs)  \; .
\end{equation}
Superscripts ``s'' or ``f'' will be added to $\sigma_k(\bs)$, and thus to $\RxnF_k$ and $\work_k$, signifying the partition into slow
or fast charge according to Eq.~\eqref{eq:sigma-partition-Marcus}.   With this notation, the excited-state generalization of $\FreeE_0$ 
is\cite{YouMewDre15}
\begin{equation}\label{eq:G_k}
	\FreeE_k^\text{SS} = \E_k^\text{SS} - \work_0^\text{s} - \work_k^\text{f} + \work_{0,k} \; .
\end{equation}
where
\begin{equation}
	 \E_k^\text{SS} = \big\langle\Psi_k \big| \Ham_k^\text{SS} \big| \Psi_k \big\rangle 
	= \big\langle\Psi_k \big| \Hvac + \RxnF_0^\text{s} + \RxnF_k^\text{f} \big| \Psi_k \big\rangle 
\end{equation}
and
\begin{equation}\label{eq:W_0k}
	\work_{0,k} = \frac{1}{2} \int_{\surf} 
	\esp^{\sigma_0^\text{s}}(\bs)
	\left[
		\sigma_k^\text{f}(\bs) - \sigma_0^\text{f}(\bs)
	\right] d\bs \; .
\end{equation}
Equation~\eqref{eq:G_k} has a straightforward interpretation.  To obtain the {\em free\/} energy $\FreeE_k^\text{SS}$ 
for state $k$, which includes the 
effects of averaging over implicit solvent degrees of freedom, the energy $\E_k^\text{SS}$ that is obtained from the 
Schr\"odinger equation must be reduced by the work $\work_0^\text{s} + \work_k^\text{f}$ that is required for the ground- and 
excited-state polarization processes.  The final term, $\work_{0,k}$, accounts for the Coulomb interaction between initial- and
final-state surface charge.    This term has sometimes been omitted from similar treatments that start from a nonequilibrium 
free energy expression that is otherwise analogous to Eq.~\eqref{eq:G_k},\cite{CarMenTom06,LunKoh13}
however its presence is necessary when the Marcus partition of the polarization is 
used.\cite{Agu01,AguOliTom93,CamTom95c,TomMenCam05,MarCraTru11,YouMewDre15}
The nonequilibrium expression for excitation energies is $\hbar\omega_k = \FreeE_k^\text{SS} - \FreeE_0$, 
which is\cite{YouMewDre15}
\begin{equation}\label{eq:SS_exc_energy}
	\FreeE_k^\text{SS} - \FreeE_0 = \Delta \E_k^\text{SS} - \work_k^\text{f} + \work_0^\text{f} + \work_{0,k}  \; ,
\end{equation}
where $\Delta \E_k^\text{SS} = \E_k^\text{SS}-\E_0$.   Equation~\eqref{eq:SS_exc_energy} also has a straightforward interpretation.
The quantity $\Delta \E_k^\text{SS}$ is the difference between ground- and excited-state eigenvalues of the SS Hamiltonian 
[Eq.~\eqref{eq:SS-Schrodinger}], but must be corrected by the difference in the work required to polarize either state, 
which is restricted to the fast part of the response ($\work_k^\text{f} - \work_0^\text{f}$)
since that is all that changes upon vertical excitation.  


As a simple example of the nonequilibrium formalism, we consider calculations of vertical ionization energies (VIEs) in 
aqueous solution.  These can be measured using liquid microjet photoelectron 
spectroscopy,\cite{WinFau06,SeiThuWin11,SeiWinBra16} and may be quite different from gas-phase values,\cite{WinFau06} 
especially for singly-charged
anions X$^-$(aq) where the initial state is solvated much more strongly than the final state, and while  
equilibrium solvation models might be appropriate for computing {\em adiabatic\/} ionization energies, in which solvent
has the opportunity to re-equilibrate following ionization, these models do a poor job of predicting VIEs.\cite{WinWebHer05,WinFau06}
From a computational perspective, the change in charge upon photoionization means that 
long-range polarization effects are significant, requiring hundreds of explicit solvent molecules
(with concomitant sampling) to obtain a converged 
result.\cite{JacHer10b,GhoIsaSli11,GhoRoySei12,TazGurKim19,TotKubMuc20} 
Convergence is significantly accelerated by continuum
boundary conditions, using an atomistic solute X$^-$(\water)$_n$ that 
contains one or two solvation shells of explicit water molecules.\cite{CooYouHer16,CooHer18} 
The limited size of the atomistic region not only reduces the cost of the quantum chemistry calculation for 
any one structure, but also reduces the sampling that is required in order to obtain converged averages.
(A similar cluster + continuum strategy is often used in 
p$K_\text{a}$ calculations, for essentially the same reasons.\cite{CasOrtFra14,ThaSch15,ThaSch16,PliRiv20}) 
PCM boundary conditions have also been shown to accelerate convergence of absorption spectra with respect to inclusion of 
explicit water molecules,\cite{ProPeeXio16} although the effects (in absolute energy shifts) are not as dramatic as they are for ionization.

\begin{table}
	\centering
	\caption{
		Vertical ionization energies (VIEs) for aqueous ions, comparing experimental results to calculations 
		using nonequilibrium MP2 + PCM calculations.\pcite{PauCooHer19} 
		Each calculation contains $\approx 30$ explicit water molecules and each calculated VIE represents 
		an average over $\approx 100$ snapshots from a simulation.
	}\label{table:VIEs}
\begin{threeparttable}
	\begin{tabular}{l @{\hspace{0.5cm}} . rr c @{\hspace{0.5cm}}  rr @{\hspace{0.5cm}} r}
\hline\hline
		Solute  & \mc{1}{c}{Exptl.\ VIE }
			& \mc{6}{c}{Computed VIE (eV)} 
\\ \cline{3-8}
		& \mc{1}{c}{(eV)}
		&  \mc{2}{c}{Noneq.~PCM} &
		&  \mc{2}{c}{Equil.~PCM} 
		& \mc{1}{c}{No PCM} 
\\ \cline{3-4}\cline{6-7}
		& & \mc{1}{c}{spherical$^a$}& \mc{1}{c}{SAS$^b$} 
		&& \mc{1}{c}{spherical$^a$}& \mc{1}{c}{SAS$^b$} &\\
\hline
	Li$^+$ 	& 60.4^c	& 61.8	& 61.6 	&& 61.3	& 61.0  	& 64.2	\\
	Na$^+$	& 35.4^c	& 36.5	& 36.3 	&& 36.0	& 35.8 	& 38.9 	\\
	H$_2$O	& 11.7^d	& 11.6	& 11.6 	&& 11.1	& 10.9	& 13.8 	\\
	$e^-$	& 3.7^e	& 3.2	& 3.2 	&& 2.6	& 2.6	& 1.8 	\\
	F$^-$	& 11.6^f	& 11.4	& 11.5 	&& 10.8 	& 10.9 	& 10.0 	\\
	Cl$^-$	& 9.6^c	& 9.4	& 9.4 	&& 8.8	& 8.8	& 7.9 	\\
\hline\hline
	\end{tabular}
	\begin{tablenotes}[flushleft]
	\fns
		\item
		$^a$Single spherical cavity for the entire atomistic region, $R=7.525$~\AA.
		$^b$Eq.~\protect\eqref{eq:atomic_radius} with $\vdwscale=1.0$, $R_\text{probe}=1.4$~\AA, 
		and atomic radii $\{R_{\vdw,A}\}$ from Ref.~\CiteN{RowTay96}.
		$^c$Ref.~\CiteN{WinWebHer05}.
		$^d$Ref.~\CiteN{PerZhaNun20}.
		$^e$Ref.~\CiteN{LucYamSuz17}.
		$^f$Ref.~\CiteN{SeiWinBra16}.
	\end{tablenotes}	
\end{threeparttable}	
\end{table}

Table~\ref{table:VIEs} shows aqueous VIEs computed for several small solutes using a cluster + continuum approach with an
atomistic region extending to a radius of 5.5~\AA\ around the solute, containing $\approx 30$ explicit water molecules.  Shown
side-by-side in Table~\ref{table:VIEs} are VIEs computed using 
``gas phase'' (vacuum) boundary conditions that include the explicit water molecules but lack any continuum model, versus 
results using equilibrium and nonequilibrium PCMs.   In the latter case, results are shown both using a SAS cavity but also using
a cavity consisting of a single sphere around atomistic region, which affords VIEs that are essentially identical to the SAS values. 
The nonequilibrium PCM results are $\approx 1$~eV too large for Li$^+$(aq) and Na$^+$(aq) but significantly more accurate for the 
aqueous halide ions.   For the halides, these calculations are also significantly more accurate than previous \qmmm\ or equilibrium
PCM calculations.\cite{WinWebHer05}   For neat liquid water, 
these calculations represent the most accurate VIE to date, in line with new experiments,\cite{PerZhaNun20}    
and are also one of the most accurate VIE calculations to date for the challenging $e^-$(aq) system.\cite{Her19a}

More germane to the present discussion is the comparison of various boundary conditions.   The ``gas phase'' results, meaning 
X$^-$(\water)$_{30}$ and M$^+$(\water)$_{30}$, are in serious error relative to experiment, 
with VIEs that are too small for the anions and too large for the cations.   This is 
consistent in both cases with understabilization 
of the state with larger charge.  Addition of equilibrium PCM boundary conditions modifies VIEs for these systems by up to 3~eV 
for the cations, which is perhaps unsurprising given that the divalent ion M$^{2+}$ incurs
very long-range polarization effects in the final state, but more surprisingly even neutral \water\ as the solute sees a shift of 3~eV
when continuum boundary conditions are activated.  While the application of an equilibrium PCM pushes the VIE substantially in 
the right direction with respect to experiment, results remain
far from quantitative until the nonequilibrium correction is added.  The latter ranges in magnitude up to $\approx 0.6$~eV.
Other calculations for aqueous nucleobases find nonequilibrium corrections to VIEs of $\approx 1$~eV,\cite{MunMarImp15}
and the latter results suggest that for those particular systems at least, the solvent response upon ionization is the most important
difference between vertical and adiabatic ionization energies, more so than geometric relaxation of the ionized solute.\cite{MunMarImp15}


The nonequilibrium formalism is 
relatively straightforward for ionization, assuming that the final state is the ground state of the ionized species, 
but is more complicated for excited states.    For one, the state-specific nature of the Hamiltonian in 
Eq.~\eqref{eq:SS-Schrodinger} 
may cause convergence problems in the presence of (near-)degeneracies.\cite{JacHer11c,KhaKhaHat20} 
Even when the states are well-separated,  
properties such as oscillator strengths are ill-defined because the various final-state wave functions 
are not eigenfunctions of a
common Hamiltonian, and therefore are not mutually orthogonal.\cite{JacHer11c}   
These problems are not unique to continuum solvation methods and arise for any kind of polarizable 
model of the environment, \eg, for \qmmm\ methods that employ polarizable force fields.\cite{ThoSch95,JacHer11c}  
A solution to this conundrum is to treat $\RxnF_k^\text{f}$ in Eq.~\eqref{eq:SS-Schrodinger}
as a perturbation.\cite{CamCorMen05,CarMenTom06,LunKoh13,YouMewDre15,MewYouWor15,MewHerDre17}    
To do so, first introduce a set of zeroth-order states
\begin{equation}\label{eq:0th-order}
	\Ham_0 \left| \Psi_k^{(0)} \right\rangle = \E_k^{(0)} \left| \Psi_k^{(0)} \right\rangle \; ,
\end{equation}
such that the eigenvalue $\E_k^{(0)}$ includes the effects of the ground-state reaction field, $\RxnF_0$.  Eigenfunctions in
Eq.~\eqref{eq:0th-order} are orthonormal, thus a convenient approximation for 
$\FreeE_k^\text{SS}$ in Eq.~\eqref{eq:G_k} uses $\E_k^{(0)}$ in place of 
$\E_k^\text{SS}$ and furthermore uses $\big|\Psi_k^{(0)}\big\rangle$ to evaluate the electrostatic potential for state $k$.
This avoids the complexities of the SS approach, and is equivalent to first-order perturbation theory with respect to a perturbation 
$\hat{W} = \RxnF_k^\text{f} - \RxnF_0^\text{f}$, obtained from a partition 
\begin{equation}
	\Ham_k^\text{SS} = 
	\underbrace{
		\Hvac + \RxnF_0^{\text{s}+\text{f}} 
	}_{
		\Ham_0
	}
	+
	\underbrace{
		 \RxnF_k^\text{f} - \RxnF_0^\text{f}
	}_{
		\Hat{W}
	}
\end{equation}
of the SS Hamiltonian.   
This has been called the {\em perturbation theory state-specific\/} (ptSS) approach to nonequilibrium 
solvation.\cite{YouMewDre15,MewYouWor15,MewHerDre17}  
(In principle, it could be extended to higher-order perturbation theory but it is not clear that is warranted.)
Note that a widely-used ``corrected linear response'' (cLR) procedure,\cite{CarMenTom06} introduced for excited-state
PCM calculations at the level of time-dependent (TD-)DFT 
is really a ptSS approach; the ``linear response'' refers to TDDFT, not to the linear-response
PCM formalism that is discussed in Section~\ref{sec:NonEq:LR}.  In the context of the ptSS or cLR approach, 
it is best to view TDDFT as a form of configuration interaction with single substitutions (CIS), which provides an
eigenvalue equation of the form in Eq.~\eqref{eq:0th-order}.   This is consistent with the idea that the SS version of TDDFT,
as implemented by Improta \etal,\cite{ImpBarSca06,ImpScaFri07} is a fully iterative version of Eq.~\eqref{eq:SS-Schrodinger}
within a CIS-style {\em ansatz}.   Both the ptSS and the fully SS approaches to TDDFT do require construction of the 
``relaxed'' density\cite{FurAhl02,RonAngBel14,MasCamFri18} for the TDDFT excited state in question, 
in order to compute its electrostatic potential.

To obtain practical formulas, \eg, for ASC-PCMs, let us introduce a vector notation for surface integrals.   As an example, we rewrite 
Eq.~\eqref{eq:Gelst-ASC} for $\Gelst$ in the form 
\begin{equation}\label{eq:dot-pdt}
	\Gelst = \frac{1}{2} \int_{\surf}  \sigma(\bs) \; \esp^\rho(\bs) \; d\bs 
	= \tfrac{1}{2}\mathbf{q}\bm{\cdot}\mathbf{v}^{\rho} \; .
\end{equation}
The quantities $\mathbf{q}$ and $\mathbf{v}^\rho$ were introduced in Eq.~\eqref{eq:Kq=Rv_matrix} and the dot-product 
notation represents how ASC-PCM surface integrals are evaluated in practice, upon discretization of the cavity surface.
Using this notation, an expression for the
nonequilibrium free energy for excited state $k$ can be manipulated into the form\cite{CarMenTom06}
\begin{equation}\label{eq:Gneq}
	\FreeE_k^\text{neq} = \E_k^{(0)} + \tfrac{1}{2}\mathbf{v}_0\bm{\cdot}\mathbf{q}_0
	+ \tfrac{1}{2}(\mathbf{v}_k - \mathbf{v}_0)\bm{\cdot} (\Delta\mathbf{q}_k^\text{f}) + \work_{0,k} \; ,
\end{equation}
where $\mathbf{v}_0$ and $\mathbf{v}_k$ represent the electrostatic potentials for states 
$\big|\Psi_0^{(0)}\big\rangle$ and $\big|\Psi_k^{(0)}\big\rangle$, 
and $\Delta\mathbf{q}^\text{f}_k = \mathbf{q}_k^\text{f} - \mathbf{q}_0^\text{f}$ is the 
difference in the fast polarization charges in the two states.  The latter is computed according to 
\begin{equation}\label{eq:q_Delta}
	\Delta\mathbf{q}^\text{f}_k = \mathbf{Q}_{\dielop}(\mathbf{v}_k - \mathbf{v}_0) 
	= \mathbf{Q}_{\dielop}\mathbf{v}^{\Delta\rho}
\end{equation}
for a reaction-field operator ($\mathbf{Q}_{\diel} = \mathbf{K}_{\diel}^{-1}\mathbf{Y}_{\diel}$) involving the optical dielectric constant.
The second equality in Eq.~\eqref{eq:q_Delta} recognizes that $\mathbf{v}^{\Delta\rho} = \mathbf{v}_k - \mathbf{v}_0$ is the 
electrostatic potential corresponding to the difference density 
$\Delta\rho_k(\br) = \rho_k(\br) - \rho_0(\br)$. 
Somewhat similar expressions to Eq.~\eqref{eq:Gneq} can be found elsewhere,\cite{CosBar00a,ImpBarSca06,ImpScaFri07}  
but the connection to the actual free energy of the excited state is most explicit in the work of Caricato \etal.\cite{CarMenTom06}
For cases where the solvent polarization has had time to fully equilibrate with respect to the excited-state density, an
analogous expression for the {\em equilibrium\/} free energy $\FreeE_k^\text{eq}$ is obtained from Eq.~\eqref{eq:Gneq} by replacing
$\Delta\mathbf{q}_{k}^\text{f}$ with $\mathbf{q}_k - \mathbf{q}_0$, where both ground- and excited-state charges are equilibrium ones, 
so that in the equilibrium case $\mathbf{q}_k = \mathbf{Q}_{\dielst}\mathbf{v}_k$.\cite{CarMenTom06}    

The solvent-corrected excitation energy 
is simply the difference between ground- and excited-state free energies,\cite{CarMenTom06,MenCapGui09}
\begin{equation}\label{eq:SS-neq-omega}
	\hbar \omega_{0k}^\text{neq} = 
	\FreeE_k^\text{neq} - \FreeE_0 = \Delta \E_k^{(0)} 
	+ \tfrac{1}{2}(\mathbf{v}_k - \mathbf{v}_0)\bm{\cdot} (\Delta\mathbf{q}^\text{f}_k) + \work_{0,k} \; .
\end{equation}
The quantity $\Delta \E_k^{(0)} = \E_k^{(0)} - \E_0$ is the excitation energy computed in the presence of the ground-state reaction field, 
which is then corrected in Eq.~\eqref{eq:SS-neq-omega} for the change in the fast polarization.
These equations remain valid for the Pekar partition if $\work_{0,k}$ is omitted from Eqs.~\eqref{eq:Gneq} and \eqref{eq:SS-neq-omega}.
As noted by Cammi \etal,\cite{CamCorMen05} Eq.~\eqref{eq:SS-neq-omega} is the detailed analogue of the heuristic theories
of excited-state solvation developed much earlier, \eg, by McRae,\cite{BayMcR54a,McR57,AmoBur73b} 
Lippert,\cite{Lip57} and Mataga.\cite{MatYozKoi56,MatChoTan05}
This becomes clear upon considering the special case of a polarizable dipole in a spherical cavity.\cite{CamCorMen05,ImpBarSca06}

Although presented here in the notation of ASC-PCMs, an analogous theory of nonequilibrium solvation can be developed based
directly on Poisson's equation.\cite{CooYouHer16,CooHer18}  
In that context, the total charge density $\rho_\text{tot}(\br) = \rho(\br) + \rho_\text{pol}(\br)$ 
includes an induced polarization $\rho_\text{pol}(\br)$ [as in Eq.~\eqref{eq:G_elst}], in addition to the charge density
$\rho(\br)$ due to the atomistic solute.
The reaction-field potential is the electrostatic potential arising from $\rho_\text{pol}(\br)$ and surface integrals such as
the ones in Eqs.~\eqref{eq:W_k} and \eqref{eq:W_0k} are replaced by volumetric integration.
Conveniently, the dot product notation introduced in Eq.~\eqref{eq:dot-pdt} is ambivalent to this distinction and a formula analogous
to Eq.~\eqref{eq:Gneq} can be derived,\cite{CooHer18} where the dot product signifies volumetric integration.

In the interest of brevity, the notation introduced above omits a subscript ``elst'' on both 
$\FreeE_k^\text{SS}$ [Eq.~\eqref{eq:G_k}] and $\FreeE_k^\text{neq}$ [Eq.~\eqref{eq:Gneq}], and for that matter on 
$\FreeE_0$ in Eq.~\eqref{eq:G[Psi]} as well.  In each case, these quantities represent only the electrostatic contribution to the
free energy.     Some of the earliest theoretical work on solvatochromatic shifts was concerned not just with changes in the 
chromophore's dipole moment, as in the Onsager-style treatment leading to the Lippert-Mataga equation, 
but also with the role of dispersion effects.\cite{Sup68a,Sup68b,AmoBur73a,AmoBur73b} 
In modern formulations of continuum theory, however, there 
have been only preliminary efforts to incorporate nonelectrostatic interactions (as described in 
Section~\ref{sec:DeltaG}) into excited-state calculations.\cite{WeiMenFre10,MarCraTru13,MarCraTru15,CupAmoMen15}
This is an interesting problem insofar as solute--solvent dispersion is likely more attractive in an excited state, as the 
excited-state wave function is probably 
more polarizable than the ground state, but at the same time Pauli repulsion probably increases in the excited state due to its
larger size.  To a significant extent, the favorable results for solvatochromic shifts that are obtained
with electrostatic-only models\cite{MewYouWor15,MarCraTru11,JacPlaAda12,KhaKhaHat20} 
likely rely on some error cancellation along these lines.
Models that introduce state-specific nonelectrostatic interactions also do well for solvatochromic 
shifts.\cite{MarCraTru15}

\begin{figure}[t]
	\centering
	\fig{1.0}{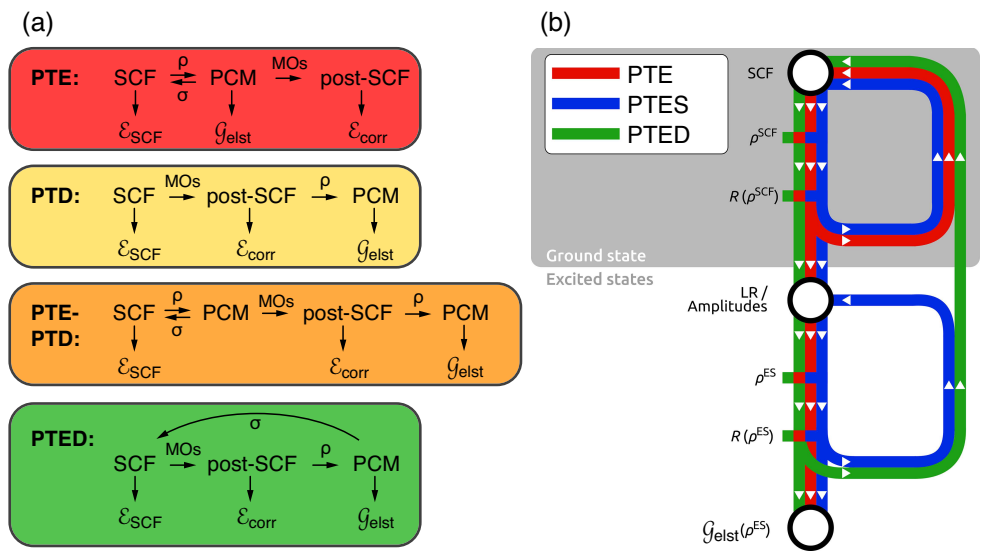}
	\caption{
		Flowcharts of various state-specific pathways for combining a PCM or other SCRF procedure
		with a quantum chemistry method requiring a post-SCF step.
		(a) Illustration of the PTE and PTD schemes and two combinations thereof.   
		Forward-backward arrows ($\rightleftarrows$) indicate where solute density ($\rho$) and polarization charge ($\sigma$) 
		are iterated to self-consistency, and downward arrows
		indicate the point at which various contributions to the energy are extracted.
		(b) Schematic representation of the PTE and PTED procedures for an excited-state (ES) calculation,
		along with the PTES procedure designed as a lower-cost approximation to PTED. 
		Panel (a) is adapted from Ref.~\CiteN{MewHerDre17}; copyright 2017 The PCCP Owner Societies.	
		Panel (b) is adapted from Ref.~\CiteN{GuiCap19}; copyright 2019 John Wiley \& Sons.
	}\label{fig:PTE-PTD}
\end{figure}

There is one remaining source of complexity when the nonequilibrium theory is applied to excited states, in that the 
Schr\"odinger equation in Eq.~\eqref{eq:SS-Schrodinger} leaves open the question of what level of self-consistency should be
sought in obtaining the excited-state reaction-field operator, $\RxnF_k^\text{f}$, which depends on $|\Psi_k\rangle$.
The multiple-choice answer to this question leads to several 
categories of methods that are mapped out in Fig.~\ref{fig:PTE-PTD}, and which are called 
``perturbation to energy'' (PTE), ``perturbation to density'' (PTD), and ``perturbation with self-consistent energy and density'' (PTED).
This nomenclature derives from 
efforts to use perturbation theory to include correlation in the ground-state calculation 
(\eg, MP2 + SCRF),\cite{OliTom91,AguOliTom91,Ang95,LipScaMen09} 
and the question of whether (and how) electron correlation should be
included in the density that is used to polarize the continuum.       The same notation has been adopted for PCM calculations
using non-perturbative models such as coupled-cluster theory,\cite{Cam09,CamTom17} 
and is used here in a discussion that is formulated specifically with excited-state calculations in mind.
As illustrated in Fig.~\ref{fig:PTE-PTD}(a), the PTE scheme involves
self-consistent solution of the SCF + SCRF problem followed by a single-shot post-SCF calculation using solvent-polarized 
molecular orbitals (MOs).   This represents a kind of ``zeroth-order'' inclusion of solvation effects in the correlated calculation,\cite{MewYouWor15} 
and has obvious advantages in terms of cost:  assuming that the post-SCF step dominates the cost of the
gas-phase calculation, then then addition of SCRF boundary conditions adds little to the overall cost of the calculation.  
Alternatively, in the PTD scheme the correlated calculation is performed in the gas phase and then the correlated density (rather than
the SCF density) is used to polarize the solvent.   This introduces solvation effects beyond zeroth order in perturbation theory,\cite{Ang95}
at marginally increased cost:  it is still a single-shot correlation calculation, however the relaxed density is required, which entails 
computational effort along the lines of a gradient calculation at the correlated level of theory.    
A slightly better-performing variant of the traditional PTD approach is the 
PTE-PTD scheme,\cite{MewHerDre17} in which the SCF + SCRF calculation is solved self-consistently and those MOs
are used in the post-SCF calculation, but then the correlated density is used in a final, single-shot PCM calculation to compute the solvation
energy $\FreeE_\elst$.   None of the aforementioned approaches constitutes a fully self-consistent treatment of post-SCF
correlation effects, but this can be accomplished using the PTED scheme 
that is mapped out in Fig.~\ref{fig:PTE-PTD}(a).  Here, the correlated density is used to obtain the PCM surface charge
and this procedure is iterated to self-consistency.   This approach is significantly more expensive because the correlated calculation is
performed at each SCRF iteration.

These ideas have since been extended beyond their perturbation theory origins and represent the various possible levels of 
self-consistency in any calculation that combines an SCRF approach with a quantum chemistry method requiring a post-SCF
calculation,\cite{CarMenSca10,Car11,Car20,MewYouWor15,MewHerDre17,GuiCap19,KhaKhaHat20}
including TDDFT.    An alternative pictorial representation of the simplest method (PTE) and
most complete scheme (PTED) is provided in Fig.~\ref{fig:PTE-PTD}(b), which provides a flowchart for  
an excited-state calculation, indicating which 
densities are used to construct the various reaction-field operators $\RxnF$.   Because of the expense associated with the 
fully self-consistent PTED approach, approximations have been developed 
in which both the ground- and excited-state calculations are iterative, but those two iterative sequences are uncoupled to one 
another.\cite{LipScaMen09,CarMenSca10,Car11,Car20,KhaKhaHat18a}
This scheme, which Caricato calls ``PTES" and has implemented at the coupled-cluster level of theory,\cite{CarMenSca10,Car11,Car20} 
is analogous to a ``vertical excitation model'' introduced for TDDFT.\cite{MarCraTru13}   At the DFT level, 
the PTED scheme in Fig.~\ref{fig:PTE-PTD} is essentially equivalent to the SS-TDDFT + PCM method introduced
by Improta \etal.\cite{ImpBarSca06,ImpScaFri07}  


\begin{figure}[t]
	\centering
	\fig{1.0}{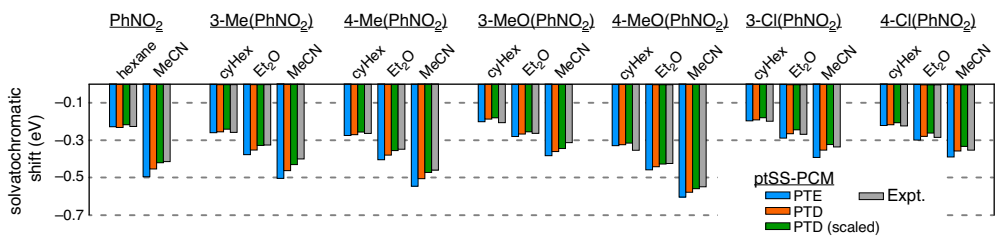}
	\caption{
		Solvatochromic shifts in lowest $^1\pi\pi$ state for derivatives of nitrobenzene (PhNO$_2$)
		in different solvents, comparing experimental values to ADC(2) + C-PCM calculations.   Solvent effects are described  
		using PTE and PTD variants of the ptSS approach.   Also shown are results for an empirically-scaled version of the 
		nonequilibrium PTD correction. 
		Adapted from Ref.~\CiteN{MewYouWor15}; copyright 2015 American Chemical Society.
	}\label{fig:solvent_shifts}
\end{figure}

Figure~\ref{fig:solvent_shifts} presents solvatochromic shifts 
for a set of nitrobenzene derivatives,\cite{MewYouWor15} 
with excitation energies computed at the level of second-order algebraic diagrammatic construction [ADC(2)],\cite{DreWor15}
which is something of an excited-state analogue of MP2.   Solvent contributions in Fig.~\ref{fig:solvent_shifts} are 
incorporated using either the PTE or PTD
variant of the ptSS approach.  Differences between the two variants are negligible, and both
approaches show good agreement with experimental shifts, without the need to invoke the more expensive PTED scheme.   
For many of these molecules, the first-order ptSS contribution to the solvatochromic shift (representing fast polarization)
is 0.10--0.15~eV, in shifts ranging up to 0.6~eV.   The remaining part 
comes from the zeroth-order contribution of simply inserting solvent-polarized MOs into the correlated part
of the calculation.\cite{MewYouWor15} 
Tests on a more diverse set of systems do reveal a small systematic error in the PTE approach,\cite{KhaKhaHat20}
but the mean error with respect to experiment remains $<0.1$~eV and the systematic error can be eliminated by intermediate approaches
that do not require the full self-consistency of the PTED approach.\cite{KhaKhaHat20}    In particular, the PTE-PTD scheme 
[see Fig.~\ref{fig:PTE-PTD}(a)] works well in this regard.\cite{MewHerDre17}    This approach 
requires the correlated density but is not iterative at the correlated level of theory.    Other benchmark studies comparing 
continuum approaches 
to large QM calculations with explicit solvent have suggested that QM\slash PCM excitation energies may agree better with 
full-QM result as compared to \qmmm\ calculations, but explicit water molecules in the QM\slash PCM calculation are required
to obtain good agreement for oscillator strengths.\cite{DeVMenNog18}  Simulation of band shapes requires thermal sampling, 
which cannot be accomplished without at least some explicit solvent molecules.

\subsection{Linear response approach}
\label{sec:NonEq:LR}

Despite its computational complexities, the SS approach to excited-state solvation is conceptually straightforward.  
An alternative to the state-by-state approach, which has fewer moving parts at the computational level, 
is based on linear-response (LR) quantum chemistry methods in which 
excitation energies are computed from the poles of the frequency-dependent 
response to a perturbation, rather than from the Schr\"odinger equation.  Given that the PCM electrostatic contribution 
to the energy is
\begin{equation}
	\FreeE_\elst = \frac{1}{2}\int_{\surf}\int_{\surf} \esp(\bs) \; Q_{\diel}(\bs,\bs') \; \esp(\bs') \; d\bs \; d\bs' \; ,
\end{equation}
where $Q_{\diel}(\bs,\bs')$ is the kernel of the solvent response operator $\hat{Q}_{\diel}=\hat{K}_{\diel}^{-1}\hat{Y}_{\diel}$,
the solvent model contributes only a one-electron potential, $\mathvee^\text{PCM}(\br) = \delta\FreeE_\elst/\delta\rho(\br)$.
The matrix elements of $\mathvee^\text{PCM}(\br)$ are\cite{CosBar01}
\begin{equation}\label{eq:vPCM_mu,nu}
	\mathvee^\text{PCM}_{\mu\nu} = \int_{\surf}\int_{\surf}\esp(\bs) \; Q_{\diel}(\bs,\bs')\; \esp_{\mu\nu}(\bs')   \; d\bs \; d\bs' 
	= (\mathbf{v}^\rho)^\dagger\mathbf{Q}_{\diel}\mathbf{v}^{\mu\nu}
\; ,
\end{equation}
where $\esp_{\mu\nu}(\bs')$ is the electrostatic potential generated by the function pair $\mu\nu$ at the point $\bs'$.
The second equality in Eq.~\eqref{eq:vPCM_mu,nu} demonstrates how the Fock matrix contribution from 
$\mathvee^\text{PCM}(\br)$ is evaluated in practice, and analogous expressions exist for three-dimensional Poisson
approaches.\cite{CooHer18}
The quantities $\mathbf{v}^\rho$ and $\mathbf{v}^{\mu\nu}$ involve
only one-electron integrals, so incorporating the PCM contribution into a LR calculation incurs negligible
overhead with respect to the cost of the gas-phase calculation, and meanwhile this approach is free of the 
iterative complexities of the SS method.   A general LR-PCM formulation has been given by 
Cammi \etal,\cite{CamMen99,CamCorMen05} 
and specific formulations for different excited state methods are available as well, \eg, for 
TDDFT and other single-excitation theories,\cite{CamMen99,CamMenTom00,CosBar01,CorTom02,IozMenTom04}
following on earlier implementations of the coupled-perturbed SCF + PCM procedure for response
properties;\cite{CamCosTom96,CamCosMen96}
for multiconfigurational SCF wave functions;\cite{MikJorJen94} 
for ADC;\cite{LunKoh13,KhaKhaHat18a} 
for the $GW$\slash Bethe-Salpeter equation formalism.\cite{DucGuiJac18}
Finally, it has been implemented for coupled-cluster theory,\cite{Cam10,Cam12a,Car13,Car18,Car19b} 
based on the coupled-cluster response formalism.\cite{CamTom17}

For isolated-molecule quantum chemistry, the LR formalism for excitation energies is generally equivalent to solving the 
corresponding Schr\"odinger equation,\cite{HelJorOls00}  
but the LR- and SS-PCM formalisms are {\em not\/} equivalent.\cite{CamCorMen05,CorCamMen05}
The general form of the LR-PCM result is\cite{CamCorMen05}
\begin{equation}\label{eq:LR-PCM-omega}
	\hbar\omega_{0k}^\text{neq,LR} = \hbar\omega_k^{(0)} + 
	\underbrace{
		\big\langle\Psi_k \big| \hat{\mathcal{V}} \big| \Psi_0 \big\rangle 
		\big\langle\Psi_0 \big| \hat{\mathcal{Q}}^\text{f} \big| \Psi_k \big\rangle 
	}_{
		\dRxn(\bm{\mu}_{0k})
	} \; ,
\end{equation}
where $\big\langle\Psi_k \big| \hat{\mathcal{V}} \big| \Psi_0 \big\rangle$ 
is the electrostatic potential generated by the {\em transition\/} density, $\rho_{k0}(\br)$, and 
$\big\langle\Psi_0 \big| \hat{\mathcal{Q}}^\text{f} \big| \Psi_k \big\rangle$ is the apparent
surface charge induced by $\rho_{k0}(\br)$.
For comparison, the SS-PCM result in Eq.~\eqref{eq:SS-neq-omega} can be rewritten in similar notation: 
\begin{equation}
	\hbar\omega_{0k}^\text{neq,SS} =	\hbar\omega_k^{(0)} + 
	\underbrace{ 
		\tfrac{1}{2} \Big[
			\big\langle\Psi_k \big| \hat{\mathcal{V}} \big| \Psi_k \big\rangle 
			-\big\langle\Psi_0 \big| \hat{\mathcal{V}} \big| \Psi_0 \big\rangle 
		\Big]\bm{\cdot}
		\Big[
			\big\langle\Psi_k \big| \hat{\mathcal{Q}}^\text{f}  \big| \Psi_k \big\rangle 
			-\big\langle\Psi_0 \big| \hat{\mathcal{Q}}^\text{f} \big| \Psi_0 \big\rangle 
		\Big] 	
	}_{
		\dRxn(\Delta\rho_k)
	} \; .
\end{equation}
In both cases, the quantity $\hbar\omega_{0k}^{(0)}\equiv\Delta\E_k^{(0)}$ is the zeroth-order approximation to the 
solution phase excitation energy, calculated in the static reaction field of the ground state.
The quantity $\dRxn{}$ represents the change in the dynamical part of the reaction field, which is a function of the transition
dipole moment $\bm{\mu}_{0k} = \langle \Psi_0 | \hat{\bm{\mu}} | \Psi_k \rangle$
in the LR case but a function of the difference density $\Delta\rho_k(\br)$ in the SS case.
A detailed analysis of the two formalisms suggests that their differences arise from the nonlinear nature of the SS Hamiltonian
combined with the lack of entanglement between the atomistic wave function and its continuum environment.\cite{CorCamMen05}

Whatever the origin of the discrepancy, the 
form of the LR-PCM correction in Eq.~\eqref{eq:LR-PCM-omega} is problematic because the correction vanishes 
for optically-forbidden transitions, as is readily seen from a model of a dipole in a spherical cavity, for which 
$\dRxn = -\FOB(\dielop,\BornRad) \bm{\mu}_{0k}$.\cite{ImpBarSca06} 
For the same reason, the LR-PCM correction $\dRxn(\bm{\mu}_{0k})$ will be rather small for any 
excitation involving significant displacement of charge, whereas intuitively (and in the SS formalism) one expects a significant
solvent effect for a charge-transfer excitation in a polar solvent.
Indeed, SS-PCM results are consistently superior to LR-PCM calculations for excited states with 
charge-transfer character.\cite{Ped13,Min14b,BerZanCal14,BudMacMed15,GuiJacAda15,GuiMenSca18}   
(As discussed in Section~\ref{sec:NonEq:SS}, the cLR formalism encountered in some of these studies is really 
a ptSS-PCM approach.)  Even for states not dominated by charge transfer, 
the SS-PCM approach generally affords smaller errors for solvatochromatic 
shifts in the absorption spectrum as compared to LR-PCM calculations,\cite{MarCraTru11,JacPlaAda12,MewYouWor15}   
although it is worth bearing in mind that the experimental $\lambda_\text{max}$ need not correspond
to the origin (0\nbd--0) transition, due to vibrational structure.\cite{KlaKroSaa10,BloBaiBic16,GarZemPon19}
The ptSS-PCM approach also affords more accurate results 
for emission energies,\cite{ChiBudMed14} although LR-PCM calculations can be used to optimize the excited-state geometries,
which simplifies the procedure.
It has also been argued that the LR correction $\dRxn(\bm{\mu}_{0k})$ constitutes a solute--continuum dispersion 
interaction,\cite{CorCamMen05,Sch16,GuiCap19} insofar as it has the form of the 
solute charge distribution oscillating
at the Bohr frequency $\omega_{0k}^{(0)}$ and coupling to the dynamical response of the environment.
As such, some studies have opted to include both the LR- and ptSS-PCM corrections to $\omega_{0k}^{(0)}$.\cite{MewYouWor15}


\section{Anisotropic Solvation}
\label{sec:PEQS}

Up to this point we have assumed that the continuum environment is isotropic, which is usually the case for a bulk liquid although
there are certain exceptions (notably, liquid crystals) 
where polarization of the medium depends upon the orientation of the electric field vector.  This can be described by 
allowing $\diel$ 
to take the form of a $3\times 3$ matrix, with orientation-dependent permittivities $\diel_{xx}$, $\diel_{yy}$, and $\diel_{zz}$.
The ASC-PCM formalism, and in particular IEF-PCM, has been formulated to handle a permittivity  
tensor.\cite{CanMenTom97,MenCanTom97,CanMen98a,MenCam03}

\begin{figure}
	\centering
	\fig{1.0}{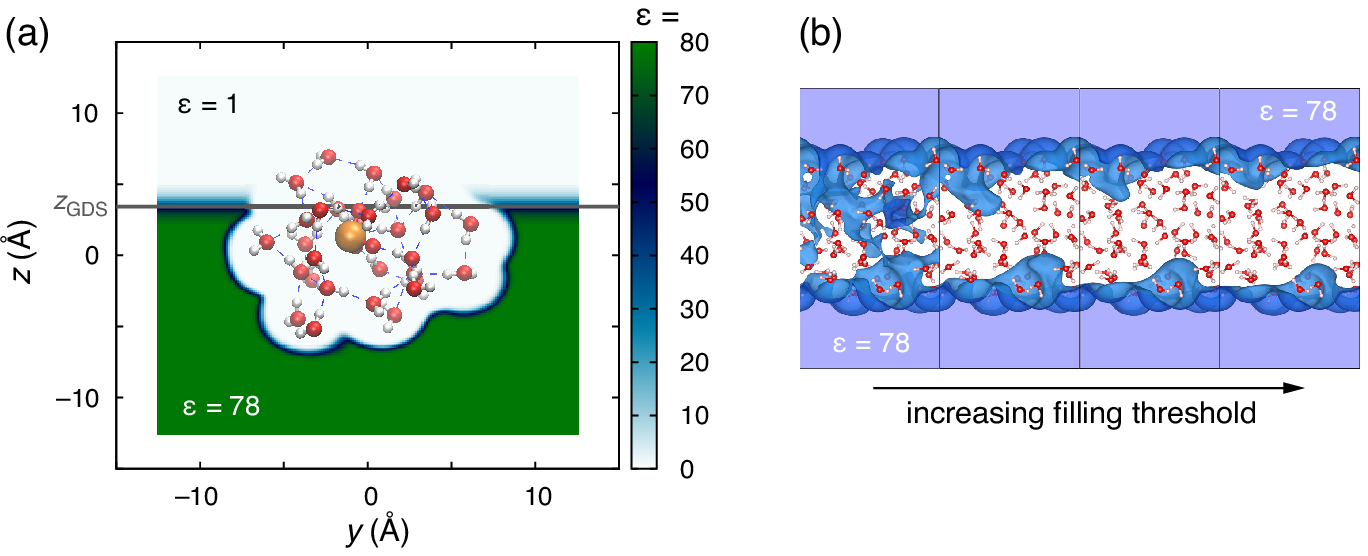}
	\caption{
		Illustrations of anisotropic permittivity functions $\diel(\br)$ for use in Poisson's equation.
		(a) Cluster + continuum description of chlorate ion at the air\slash water interface, 
		in which the atomistic solute is ClO$_3^-$(\water)$_{30}$.  The background color shows the function $\diel(\br)$, 
		interpolating between $\epsout=1$ above the Gibbs dividing surface (GDS) and $\epsout=78$ below it, with $\epsin=1$ inside 
		of the solute cavity.  The horizontal line indicates the position of the dividing surface, $z_\text{GDS}=3.5$~\AA.
		(b) Periodic water slab bounded on either side by continuum water ($\diel=78$, shown in purple), with 
		regions characterized by $\diel>15$ shown in blue.
		From left to right, the interpolating function is modified (using a ``filling threshold'' parameter) in order 
		to exclude pockets of high permittivity 
		that encroach into the interstices between the atomistic water molecules. 
		Panel (b) is adapted from Ref.~\CiteN{AndHorNat19}; copyright 2019 American Chemical Society.
	}\label{fig:eps-interface}
\end{figure}

A more general class of anisotropic solvation problems are interfacial phenomena.   
An example is shown in Fig.~\ref{fig:eps-interface}(a),
in which an atomistic solute consisting for a ClO$_3^-$ ion with approximately two 
solvation shells of explicit water molecules is situated at a dielectric representation of the air\slash water interface. 
Here, the atomistic region is subject to a dielectric environment characterized by $\diel=78$ on one side but $\diel=1$ on the other.
The basic PCM formalism is not equipped to handle such a situation, as it is predicted on a sharp dielectric interface
between $\epsin=1$ inside the cavity and a bulk solvent value ($\epsout$) outside, although it can be accomplished
piecewise if the medium is divided into separate domains, each with its own value of $\diel$.\cite{HosSakIno87} 
Alternatively, to describe the liquid\slash vapor interface
the PCM equations have been to interpolate matrix elements between different values of 
$\diel$.\cite{FreMenCam04,FreCamCor04,BonFreAgr06,IozCosImp06,SiLi09a,WanMaLi10,MozMenFre14,MozFre16}   
Treatment of nonelectrostatic effects proves to be crucial.   
It has been suggested that continuum models cannot describe specific-ion (``Hofmeister'') effects,\cite{JunTob02b,JunTob06}   
in part because the Born model [Eq.~\eqref{eq:Born}] cannot distinguish between cations and anions, although that particular
shortcoming is better ascribed to an overly-simplified cavity shape.\cite{DuiZha20}
Another commonly held view is that continuum models are incapable
of describing the interfacial affinity exhibited by soft ions,\cite{JunTob02b,JunTob06}  
because (so the logic goes) the ion in a continuum solvent ought to be repelled from the
interface by its own image charge.\cite{JunTob06,Cha09}  Despite this conventional wisdom, 
continuum models that include nonelectrostatic interactions have been shown to predict interfacial free energy minima 
for dragging a soft ion through the air\slash water interface.\cite{FreMenCam04,BonFreAgr06,DuiParNin14,DuiParNin15}

Fundamentally, however, the interfacial solvation problem seems to cry out for a permittivity function $\diel(\br)$ 
that can take different values in different regions of space, \ie, a method that solves the generalized Poisson equation with 
an anisotropic permittivity function $\diel(\br)$.
Such a strategy has been pursued to describe the interface between a solid-state electrode and an aqueous 
electrolyte,\cite{SanSueSch09,NatTruMar19,AndHorNat19,AndFis19,SunSch17,SunLetSch18,SchSun20,
DziBhaAnt20,BraNguGle20,BhaAntDzi20}
as well as host\slash guest systems where the guest experiences a low-dielectric environment despite the fact that the host
is dissolved in water.\cite{AksPauHer20}   Finally, anisotropic models have been used 
to compute VIEs of solutes at the air\slash water interface,\cite{CooYouHer16,CooHer18} 
in order to connect with liquid microjet photoelectron spectroscopy. 

The setup for an interfacial calculation of this type is illustrated
in Fig.~\ref{fig:eps-interface}(a), which depicts an atomistic model of ClO$_3^-$(aq) at the air\slash water interface and shows how the
function $\diel(\br)$ is defined.   In this particular example, two solvation shells of explicit water molecules are included in the atomistic region 
in order to account for hydrogen-bonding effects; the continuum model takes care of long-range polarization upon ionization of ClO$_3^-$.   
VIE calculations based on a nonequilibrium formulation of Poisson's equation 
suggest that the VIEs of common inorganic anions are very nearly the 
same at the air\slash water interface as they are in bulk water.\cite{CooHer18} 
Even for an exotic anion like $e^-$(aq),\cite{HerCoo17,Her19a} 
it appears that the interfacial VIE that is no more than 0.2--0.4~eV
different from its bulk value.\cite{CooYouHer16,CooHer18}

In continuum solvation calculations where the atomistic region contains explicit solvent molecules, one must be careful to parameterize
the function $\diel(\br)$ to avoid artificial penetration of high-dielectric regions into the interstices between molecules.   
This includes not just methods that are directly based on Poisson's equation but also PCMs, where explicit solvent molecules might be
also be added, for the purpose of describing 
solute--solvent hydrogen bonds, as in the aforementioned p$K_\text{a}$ calculations.  Oddly, the dielectric penetration problem
in semicontinuum calculations 
has received scant attention in the quantum chemistry literature,\cite{CooHer18,AndHorNat19}
although there is an analogous problem in classical Poisson-Boltzmann electrostatics calculations for biomolecules that is widely 
discussed.   In that context, the problem is that 
standard cavity construction algorithms (based on intersecting atomic spheres) may leave pockets of high-dielectric ``solvent''
in the hydrophobic interior of a protein.\cite{SwaMonMcC05,TjoZho08,ZhoQinTjo08,PanZho13,DecColCat13,LiLiZha13,OnuAgu14} 
This can be mitigated in Poisson-based continuum methods 
by appropriate adjustment of the interpolating function that defines the dielectric boundary.
A spatially-varying permittivity function has been suggested as a solution to the problem that there is no single optimal
value for the dielectric ``constant" inside of a protein.\cite{LiLiZha13}

Figure~\ref{fig:eps-interface}(b) presents an example in which the SCCS model of 
Section~\ref{sec:Smooth:SCCS}, which is based on a functional $\diel[\rho](\br)$, is used to remove artificial high-dielectric regions.
In these simulations, the goal is to perform {\em ab initio\/} molecular dynamics
simulations of explicit water but to use continuum boundary conditions in order to limit the size of the atomistic QM simulation cell that is
required.   The parameters that define $\diel[\rho](\br)$ 
must be chosen carefully, lest high-dielectric regions penetrate between the explicit water molecules,  as they do 
on the left side of Fig.~\ref{fig:eps-interface}(b).   This situation is 
physically incorrect because the QM calculation is based on Coulomb operators that assume vacuum permittivity.
This undesirable situation has been rectified by 
introducing ``solvent awareness'' into the definition of the permittivity function, so that $\diel(\br)$ depends on the coordinates in the
atomistic region {\em directly}, and not just implicitly through the functional $\diel[\rho]$.\cite{AndHorNat19}
Moving from left to right in Fig.~\ref{fig:eps-interface}(b), this solvent awareness is activated and removes the spurious high-dielectric regions.

\section{Closing remarks}
\label{sec:Summary}

%
%
%
%

With the contents of this review serving as a lengthy introduction to what continuum solvation models can do, in closing it
feels {\em a propos\/} to comment briefly on their limitations.    These are very crude models.   That is not inconsistent with being
{\em useful\/} models, but one should not demand too much of something so simple.
It is perhaps best to think of continuum solvation as improved boundary conditions (as compared
to vacuum boundary conditions) for condensed-phase electronic structure calculations.    The primary way in which most users
will encounter the crudeness of these models is in the 
fact that there is considerable arbitrariness in construction of the solute cavity, for which there is no ``right'' choice, although 
there are certainly plenty of wrong ones.   In particular, there is no ``magical'' cavity construction
or scaling factor for the van der Waals radii that will make these quantities universal.   Small tweaks that might 
provide better answers in one system may very well degrade 
the accuracy in other cases.  Isodensity cavity constructions, and smooth interfaces or ``soft" cavities, 
based on permittivity functionals $\diel[\rho(\br)]$, seem like the least arbitrary choices.  

With that in mind, differences between how $\Gelst$ is computed amongst different PCMs seem quite minor in comparison 
to the overall quality of these models, to the point where these differences can likely be parameterized away, or are 
simply washed out, by minor changes in cavity construction.    
The COSMO method, for example, performs well in comparison to more exact formulations of the continuum electrostatics problem, 
even in low-dielectric solvents.\cite{KlaMoyPal15}    
For the SS(V)PE approach, discretization of $\hat{D}^\dagger\sigma(\bs)$ proves to be 
challenging,\cite{LanHer11c,YouMewDre15} leading to a wrong conductor limit in some cases,\cite{LanHer11c}
nevertheless this model has successfully been used 
as the basis of a composite model of solvation.\cite{YouHer16}
Errors in the implementation of certain nonelectrostatic are completely absorbed in the parameterization
of other terms.\cite{YouHer16}

In view of all this, there would seem to be little room to further improve upon the electrostatic part of continuum solvation models.
A corollary is that efforts to make the solvation model fully consistent with correlated wave function electronic
structure methods, either in the ground or excited states, seem misguided.  
The ``zeroth-order'' model, in which solvent-polarized MOs are inserted into a post-SCF correlation calculation, recovers the most
important effects, and a ptSS-style correction for the fast polarization response affords a simple-to-use model for what remains.
Differences with respect to a fully self-consistent model are likely considerably smaller than 
errors introduced by continuum representation of the solvent itself.  
Keeping those caveats in mind, the continuum solvation approach can be highly effective in situations were vacuum 
boundary conditions are dubious, \eg, due to significant charge rearrangement in a polar solvent (including redox chemistry), 
or to modulate the energy levels of the frontier, solvent-exposed orbitals at play in electronic spectroscopy.

Regarding calculation of solvation energies, which is arguably the primary application of continuum solvation models in chemistry, 
the nonelectrostatic contributions that are needed for favorable agreement with experiment can now be modeled (in a variety of
ways) with an accuracy of $< 1$~kcal/mol for charge-neutral solutes.   For ions, the best methods approach the accuracy
($\pm 3$~kcal/mol)\cite{SM6} of the experimental data themselves.\cite{SM12,YouHer16,FisGenAnd17} 
That said, these models are trained on relatively small solutes (for which experimental values of $\dGsolv$ are available); 
it remains unclear whether the aforementioned level of statistical accuracy is transferrable to significantly larger solutes.   

Regarding macromolecular solutes, it is clear that PCMs with linear-scaling solvers
ought to be seriously considered as replacements for biomolecular electrostatics calculations based on finite-difference solution of
the Poisson-Boltzmann equation.  The former approaches provide an {\em exact\/} solution to the classical electrostatics 
problem,\cite{LanHer11c,HerLan16} up to controllable discretization errors, and can be formulated in such a way that potential
energy surfaces are inherently continuous and smooth.\cite{LanHer10a,LanHer10b}   
Indeed, it is this author's opinion that theorists should not accept as their starting point any approach that does not
intrinsically provide a smooth potential energy surface, as the finite-difference approach clearly fails to do.   The ability to explore
the potential energy surface, and thus to have a well-defined ``model chemistry",\cite{HehRadSch86}
is too important to sacrifice.

\section*{Acknowledgements}\noindent
This work was supported by National Science Foundation, grant nos.\ CHE-1665322 and CHE-1955282.
It is a pleasure to acknowledge current and former group members who have worked on these projects:
Dr.\ Adrian Lange, Benjamin Albrecht, Dr.\ Zhi-Qiang You, Dr.\ Marc Coons, and Suranjan Paul.
Collaborations with Dr.\ Jan-Michael Mewes (from the group of Prof.\ Andreas Dreuw) and 
Dr.\ Christopher Stein (from the group of Prof.\ Martin Head-Gordon) are also happily acknowledged.  

\section*{Conflict of Interest}\noindent
J.M.H. serves on the board of directors of Q-Chem Inc.
\addcontentsline{toc}{section}{References}

\begin{thebibliography}{100}

\bibitem{RinRiv73}
D.~Rinaldi and J.-L. Rivail, ``Polarisabilit\'es mol\'eculaires et effet
  di\'electrique de milieu \`a l'\'etat liquide. {\'Etude} th\'eorique de la
  mol\'ecule d'eau et de ses dim\`eres'', {\em Theor.\ Chem.\ Acc.}, {\bf 32},
  57--70 (1973).

\bibitem{RivRin76}
J.-L. Rivail and D.~Rinaldi, ``A quantum chemical approach to dielectric
  solvent effects in molecular liquids'', {\em J.~Chem.\ Phys.}, {\bf 18},
  233--242 (1976).

\bibitem{MieScrTom81}
S.~Miertu\v{s}, E.~Scrocco, and J.~Tomasi, ``Electrostatic interaction of a
  solute with a continuum. {A} direct utilization of ab initio molecular
  potentials for the prevision of solvent effects'', {\em Chem.\ Phys.}, {\bf
  55}, 117--129 (1981).

\bibitem{MieTom82}
S.~Miertu\v{s} and J.~Tomasi, ``Approximate evaluations of the electrostatic
  free energy and internal energy changes in solution processes'', {\em Chem.\
  Phys.}, {\bf 65}, 239--245 (1982).

\bibitem{BonCimTom83}
R.~Bonaccorsi, R.~Cimiraglia, and J.~Tomasi, ``\textit{Ab initio} evaluation of
  absorption and emission transitions for molecular solutes, including separate
  consideration of orientational and inductive solvent effects'', {\em
  J.~Comput.\ Chem.}, {\bf 4}, 567--577 (1983).

\bibitem{Tom04}
J.~Tomasi, ``Thirty years of continuum solvation chemistry: {A} review, and
  prospects for the near future'', {\em Theor.\ Chem.\ Acc.}, {\bf 112},
  184--203 (2004).

\bibitem{GroTry08}
P.~Grochowski and J.~Trylska, ``Continuum molecular electrostatics, salt
  effects, and counterion binding---{A} review of the {Poisson}--{Boltzmann}
  theory and its modifications'', {\em Biopolymers}, {\bf 89}, 93--113 (2008).

\bibitem{PliRiv20}
J.~R. {Pliego Jr.} and J.~M. Riveros, ``Hybrid discrete-continuum solvation
  methods'', {\em WIREs Comput.\ Mol.\ Sci.}, {\bf 10}, e1440:1--25 (2020).

\bibitem{TomPer94}
J.~Tomasi and M.~Persico, ``Molecular interactions in solution: {An} overview
  of methods based on continuous distributions of the solvent'', {\em Chem.\
  Rev.}, {\bf 94}, 2027--2094 (1994).

\bibitem{AmoBarCam99}
C.~Amovilli, V.~Barone, R.~Cammi, E.~Canc\`es, M.~Cossi, B.~Mennucci, C.~S.
  Pomelli, and J.~Tomasi, ``Recent advances in the description of solvent
  effects with the polarizable continuum model'', {\em Adv.\ Quantum Chem.},
  {\bf 32}, 227--261 (1999).

\bibitem{TomMenCam02}
J.~Tomasi, B.~Mennucci, R.~Cammi, and M.~Cossi, ``Quantum mechanical models for
  reactions in solution'', in {\em Computational Approaches to Biochemical
  Reactivity}, G.~{N\^aaray-Szab\^ao} and A.~Warshel, Eds., Vol. ~19 of {\em
  Understanding Chemical Reactivity;}
\newblock Springer: New York, 2002;
\newblock chapter~1, pages 1--102.

\bibitem{CamMenTom03}
R.~Cammi, B.~Mennucci, and J.~Tomasi, ``Computational modelling of the solvent
  effects on molecular properties: {An} overview of the polarizable continuum
  model ({PCM}) approach'', in {\em Computational Chemistry: Reviews of Current
  Trends}, J.~Leszczynski, Ed., Vol. ~8;
\newblock World Scientific: Singapore, 2003;
\newblock chapter~1, pages 1--79.

\bibitem{TomMenCam05}
J.~Tomasi, B.~Mennucci, and R.~Cammi, ``Quantum mechanical continuum solvation
  models'', {\em Chem.\ Rev.}, {\bf 105}, 2999--3093 (2005).

\bibitem{Tom11}
J.~Tomasi, ``Selected features of the polarizable continuum model for the
  representation of solvation'', {\em WIREs Comput.\ Mol.\ Sci.}, {\bf 1},
  855--867 (2011).

\bibitem{Men12}
B.~Mennucci, ``Polarizable continuum model'', {\em WIREs Comput.\ Mol.\ Sci.},
  {\bf 2}, 386--404 (2012).

\bibitem{LipMen16}
F.~Lipparini and B.~Mennucci, ``Perspective: {P}olarizable continuum models for
  quantum-mechanical descriptions'', {\em J.~Chem.\ Phys.}, {\bf 144},
  160901:1--9 (2016).

\bibitem{LanHer10a}
A.~W. Lange and J.~M. Herbert, ``Polarizable continuum reaction-field solvation
  models affording smooth potential energy surfaces'', {\em J.~Phys.\ Chem.\
  Lett.}, {\bf 1}, 556--561 (2010).

\bibitem{LanHer10b}
A.~W. Lange and J.~M. Herbert, ``A smooth, nonsingular, and faithful
  discretization scheme for polarizable continuum models: {The} switching\slash
  {G}aussian approach'', {\em J.~Chem.\ Phys.}, {\bf 133}, 244111:1--18 (2010).

\bibitem{HerLan16}
J.~M. Herbert and A.~W. Lange, ``Polarizable continuum models for
  (bio)molecular electrostatics: {Basic} theory and recent developments for
  macromolecules and simulations'', in {\em Many-Body Effects and
  Electrostatics in Biomolecules}, Q.~Cui, P.~Ren, and M.~Meuwly, Eds.;
\newblock CRC Press: Boca Raton, 2016;
\newblock chapter~11, pages 363--416.

\bibitem{LanHer11c}
A.~W. Lange and J.~M. Herbert, ``Symmetric versus asymmetric discretization of
  the integral equations in polarizable continuum solvation models'', {\em
  Chem.\ Phys.\ Lett.}, {\bf 509}, 77--87 (2011).

\bibitem{LanHerAlb20}
A.~W. Lange, J.~M. Herbert, B.~J. Albrecht, and Z.-Q. You, ``Intrinsically
  smooth discretization of {Connolly's} solvent-excluded molecular surface'',
  {\em Mol.\ Phys.}, {\bf 118}, e1644384:1--18 (2020).

\bibitem{Cap16}
C.~Cappelli, ``Integrated {QM}\slash polarizable {MM}\slash continuum
  approaches to model chiroptical properties of strongly interacting
  solute--solvent systems'', {\em Int.~J. Quantum Chem.}, {\bf 116}, 1532--1542
  (2016).

\bibitem{Car19a}
M.~Caricato, ``Coupled cluster theory with the polarizable continuum model of
  solvation'', {\em Int.~J. Quantum Chem.}, {\bf 119}, e25674:1--19 (2019).

\bibitem{AndFis19}
O.~Andreussi and G.~Fisicaro, ``Continuum embeddings in condensed-matter
  simulations'', {\em Int.~J. Quantum Chem.}, {\bf 119}, e25725:1--17 (2019).

\bibitem{SadPec07}
J.~Sadlej and M.~Pecul, ``Computational modelling of the solvent--solute effect
  on {NMR} parameters by a polarizable continuum model'', in {\em Continuum
  Solvation Models in Chemical Physics}, B.~Mennucci and R.~Cammi, Eds.;
\newblock Wiley: Chichester, UK, 2007;
\newblock pages 125--144.

\bibitem{BarCimPav07}
V.~Barone, P.~Cimino, and M.~Pavone, ``{EPR} spectra of organic free radicals
  in solution from an integrated computational approach'', in {\em Continuum
  Solvation Models in Chemical Physics}, B.~Mennucci and R.~Cammi, Eds.;
\newblock Wiley: Chichester, UK, 2007;
\newblock pages 145--166.

\bibitem{Cap07}
C.~Cappelli, ``Continuum solvation approach to vibrational properties'', in
  {\em Continuum Solvation Models in Chemical Physics}, B.~Mennucci and
  R.~Cammi, Eds.;
\newblock Wiley: Chichester, UK, 2007;
\newblock pages 167--179.

\bibitem{PecRud07}
M.~Pecul and K.~Ruud, ``Solvent effects on natural optical activity'', in {\em
  Continuum Solvation Models in Chemical Physics}, B.~Mennucci and R.~Cammi,
  Eds.;
\newblock Wiley: Chichester, UK, 2007;
\newblock pages 180--205.

\bibitem{CamMen07}
R.~Cammi and B.~Mennucci, ``Macroscopic nonlinear optical properties from
  cavity models'', in {\em Continuum Solvation Models in Chemical Physics},
  B.~Mennucci and R.~Cammi, Eds.;
\newblock Wiley: Chichester, UK, 2007;
\newblock pages 238--251.

\bibitem{AgrMik07}
H.~{\AA}gren and K.~V. Mikkelsen, ``Homogeneous and heterogeneous solvent
  models for nonlinear optical properties'', in {\em Continuum Solvation Models
  in Chemical Physics}, B.~Mennucci and R.~Cammi, Eds.;
\newblock Wiley: Chichester, UK, 2007;
\newblock pages 282--299.

\bibitem{Imp12a}
R.~Improta, ``{UV}--visible absorption and emission energies in condensed phase
  by {PCM}\slash {TD}-{DFT} methods'', in {\em Computational Strategies for
  Spectroscopy: From Small Molecules to Nano Systems}, V.~Barone, Ed.;
\newblock John Wiley \& Sons: Hoboken, NJ, 1st ed., 2012;
\newblock chapter~1, pages 39--76.

\bibitem{BasPar96}
M.~V. Basilevsky and D.~F. Parsons, ``An advanced continuum medium model for
  treating solvation effects: {N}onlocal electrostatics with a cavity'', {\em
  J.~Chem.\ Phys.}, {\bf 105}, 3734--3746 (1996).

\bibitem{JenHun03}
O.~S. Jenkins and K.~L.~C. Hunt, ``Nonlocal dielectric functions on the
  nanoscale: {Screened} forces from unscreened potentials'', {\em J.~Chem.\
  Phys.}, {\bf 119}, 8250--8256 (2003).

\bibitem{BasChu07}
M.~V. Basilevsky and G.~N. Chuev, ``Nonlocal solvation theories'', in {\em
  Continuum Solvation Models in Chemical Physics}, B.~Mennucci and R.~Cammi,
  Eds.;
\newblock Wiley: Chichester, UK, 2007;
\newblock pages 94--109.

\bibitem{Whi79}
D.~H. Whiffen, ``Manual of symbols and terminology for physicochemical
  quantities and units'', {\em Pure Appl.\ Chem.}, {\bf 51}, 1--41 (1979).

\bibitem{Riz07}
A.~Rizzo, ``Birefringences in liquids'', in {\em Continuum Solvation Models in
  Chemical Physics}, B.~Mennucci and R.~Cammi, Eds.;
\newblock Wiley: Chichester, UK, 2007;
\newblock pages 252--264.

\bibitem{Fer07}
A.~Ferrarini, ``Anisotropic fluids'', in {\em Continuum Solvation Models in
  Chemical Physics}, B.~Mennucci and R.~Cammi, Eds.;
\newblock Wiley: Chichester, UK, 2007;
\newblock pages 265--281.

\bibitem{Wan86}
R.~K. Wangsness, {\em Electromagnetic Fields}, John Wiley \& Sons: Hoboken, 2nd
  ed., 1986.

\bibitem{CooHer18}
M.~P. Coons and J.~M. Herbert, ``Quantum chemistry in arbitrary dielectric
  environments: {Theory} and implementation of nonequilibrium {Poisson}
  boundary conditions and application to compute vertical ionization energies
  at the air\slash water interface'', {\em J.~Chem.\ Phys.}, {\bf 148},
  222834:1--21 (2018). Erratum: \textit{J.~Chem.\ Phys}., {\bf 151},
  189901:1--2 (2019).

\bibitem{Pat20}
K.~Patkowski, ``Recent developments in symmetry-adapted perturbation theory'',
  {\em WIREs Comput.\ Mol.\ Sci.}, {\bf 10}, e1452:1--47 (2020).

\bibitem{Bot76}
C.~J.~F. B\"ottcher, {\em Theory of Electric Polarization}, Vol. ~1, Elsevier:
  Amsterdam, 2nd ed., 1976.

\bibitem{JacWilHer09}
L.~D. Jacobson, C.~F. Williams, and J.~M. Herbert, ``The static-exchange
  electron-water pseudopotential, in conjunction with a polarizable water
  model: {A} new {H}amiltonian for hydrated-electron simulations'', {\em
  J.~Chem.\ Phys.}, {\bf 130}, 124115:1--18 (2009).

\bibitem{CarScaFri07}
M.~Caricato, G.~Scalmani, and M.~J. Frisch, ``A {L}agrangian formulation for
  continuum models'', in {\em Continuum Solvation Models in Chemical Physics},
  B.~Mennucci and R.~Cammi, Eds.;
\newblock Wiley: Chichester, UK, 2007;
\newblock pages 64--81.

\bibitem{Nak96}
H.~Nakamura, ``Roles of electrostatic interaction in proteins'', {\em Q.~Rev.\
  Biophys.}, {\bf 29}, 1--90 (1996).

\bibitem{AleMehBak11}
E.~Alexov, E.~L. Mehler, N.~Baker, A.~M. Baptista, Y.~Huang, F.~Milletti, J.~E.
  Nielsen, D.~Farrell, T.~Carstensen, M.~H.~M. Olsson, J.~K. Shen,
  J.~Warwicker, S.~Williams, and J.~M. Word, ``Progress in the prediction of
  p${K}_\text{a}$ values in proteins'', {\em Proteins}, {\bf 79}, 3260--3275
  (2011).

\bibitem{AntMcCGil94}
J.~Antosiewicz, J.~A. {McCammon}, and M.~K. Gilson, ``Prediction of
  {pH}-dependent properties of proteins'', {\em J.~Mol.\ Biol.}, {\bf 238},
  415--436 (1994).

\bibitem{DemWad96}
E.~Demchuk and R.~C. Wade, ``Improving the continuum dielectric approach to
  calculating p${K}_\text{a}$s of ionizable groups in proteins'', {\em
  J.~Phys.\ Chem.}, {\bf 100}, 17373--17387 (1996).

\bibitem{Gry02}
T.~Grycuk, ``Revision of the model system concept for the prediction of
  p${K}_\text{a}$'s in proteins'', {\em J.~Phys.\ Chem.~B}, {\bf 106},
  1434--1445 (2002).

\bibitem{WarRus84}
A.~Warshel and S.~T. Russell, ``Calculations of electrostatic interactions in
  biological systems and in solutions'', {\em Q.~Rev.\ Biophys.}, {\bf 17},
  283--422 (1984).

\bibitem{SchWar01}
C.~N. Schutz and A.~Warshel, ``What are the dielectric ``constants'' of
  proteins and how to validate electrostatic models?'', {\em Proteins}, {\bf
  44}, 400--417 (2001).

\bibitem{WarShaKat06}
A.~Warshel, P.~K. Sharma, M.~Kato, and W.~W. Parson, ``Modeling electrostatic
  effects in proteins'', {\em Biochim.\ Biophys.\ Acta}, {\bf 1764}, 1647--1676
  (2006).

\bibitem{LiLiZha13}
L.~Li, C.~Li, Z.~Zhang, and E.~Alexov, ``On the dielectric ``constant'' of
  proteins: {Smooth} dielectric function for macromolecular modeling and its
  implementation in {DelPhi}'', {\em J.~Chem.\ Theory Comput.}, {\bf 9},
  2126--2136 (2013).

\bibitem{ShaHon90b}
K.~A. Sharp and B.~Honig, ``Electrostatic interactions in macromolecules'',
  {\em Annu.\ Rev.\ Biophys.\ Biophys.\ Chem.}, {\bf 19}, 301--332 (1990).

\bibitem{FogBriMol02}
F.~Fogolari, A.~Brigo, and H.~Molinari, ``The {Poisson}--{Boltzmann} equation
  for biomolecular electrostatics: {A} tool for structural biology'', {\em
  J.~Mol.\ Recognit.}, {\bf 15}, 377--392 (2002).

\bibitem{Bak05b}
N.~A. Baker, ``Biomolecular applications of {Poisson}-{Boltzmann} methods'', in
  {\em Reviews in Computational Chemistry}, K.~Lipkowitz, R.~Larter, and T.~R.
  Cundari, Eds., Vol. ~21;
\newblock John Wiley \& Sons: Hoboken, 2005;
\newblock pages 349--379.

\bibitem{BotCaiLuo14}
W.~M. {Botello-Smith}, Q.~Cai, and R.~Luo, ``Biological applications of
  classical electrostatics methods'', {\em J.~Theor.\ Comput.\ Chem.}, {\bf
  13}, 1440008:1--25 (2014).

\bibitem{SchPon07}
M.~J. Schnieders and J.~W. Ponder, ``Polarizable atomic multipole solutes in a
  generalized {Kirkwood} continuum'', {\em J.~Chem.\ Theory Comput.}, pages
  2083--2097 (2007).

\bibitem{SchBakRen07}
M.~J. Schnieders, N.~A. Baker, P.~Ren, and J.~W. Ponder, ``Polarizable atomic
  multipole solutes in a {Poisson}-{Boltzmann} continuum'', {\em J.~Chem.\
  Phys.}, {\bf 126}, 124114:1--21 (2007).

\bibitem{ShaHon90a}
K.~A. Sharp and B.~Honig, ``Calculating total electrostatic energies with the
  nonlinear {Poisson}--{Boltzmann} equation'', {\em J.~Phys.\ Chem.}, {\bf 94},
  7684--7692 (1990).

\bibitem{DesHol01}
M.~Deserno and C.~Holm, ``Cell model and {Poisson}-{Boltzmann} theory: {A}
  brief introduction'', in {\em Electrostatic Effects in Soft Matter and
  Biophysics}, C.~Holm, P.~K\'ekicheff, and R.~Podgornik, Eds., Vol. ~46 of
  {\em NATO Science Series;}
\newblock Springer Science+Business Media: Dordrecht, 2001;
\newblock pages 27--52.

\bibitem{Lam03}
G.~Lamm, ``The {Poisson}-{Boltzmann} equation'', in {\em Reviews in
  Computational Chemistry}, K.~B. Lipkowitz, R.~Larter, T.~R. Cundari, and
  D.~B. Boyd, Eds., Vol. ~19;
\newblock Wiley-VCH: New York, 2003;
\newblock chapter~4, pages 147--366.

\bibitem{Bak04}
N.~A. Baker, ``Poisson--{Boltzmann} methods for biomolecular electrostatics'',
  {\em Method.\ Enzymol.}, {\bf 383}, 94--118 (2004).

\bibitem{MorNet01}
A.~G. Moreira and R.~R. Netz, ``Field-theoretic approaches to classical charged
  systems'', in {\em Electrostatic Effects in Soft Matter and Biophysics},
  C.~Holm, P.~K\'ekicheff, and R.~Podgornik, Eds., Vol. ~46 of {\em NATO
  Science Series;}
\newblock Springer Science+Business Media: Dordrecht, 2001;
\newblock pages 367--408.

\bibitem{SteHerHea19}
C.~J. Stein, J.~M. Herbert, and M.~{Head-Gordon}, ``The {Poisson}--{B}oltzmann
  model for implicit solvation of electrolyte solutions: {Q}uantum chemical
  implementation and assessment via {S}echenov coefficients'', {\em J.~Chem.\
  Phys.}, {\bf 151}, 224111:1--14 (2019).

\bibitem{Zho93b}
H.-X. Zhou, ``Macromolecular electrostatic energy within the nonlinear
  {Poisson}--{Boltzmann} equation'', {\em J.~Chem.\ Phys.}, {\bf 100},
  3152--3162 (1993).

\bibitem{FogZucEsp99}
F.~Fogolari, P.~Zuccato, G.~Esposito, and P.~Viglino, ``Biomolecular
  electrostatics with the linearized {Poisson}-{Boltzmann} equation'', {\em
  Biophys.~J.}, {\bf 76}, 1--16 (1999).

\bibitem{DebHuc23b}
P.~Debye and E.~H\"uckel, ``On the theory of electrolytes. {I}. {Freezing}
  point depression and related phenomena'', in {\em Collected Papers of Peter
  J. W. Debye;}
\newblock Interscience: New York, 1954;
\newblock pages 217--263.

\bibitem{LanHer11b}
A.~W. Lange and J.~M. Herbert, ``A simple polarizable continuum solvation model
  for electrolyte solutions'', {\em J.~Chem.\ Phys.}, {\bf 134}, 204110:1--15
  (2011).

\bibitem{WanRenLuo17}
C.~Wang, P.~Ren, and R.~Luo, ``Ionic solution: {What} goes right and wrong with
  continuum solvation modeling'', {\em J.~Phys.\ Chem.~B}, {\bf 121},
  11159--11179 (2017).

\bibitem{Vla99}
V.~Vlachy, ``Ionic effects beyond {Poisson}-{Boltzmann} theory'', {\em Annu.\
  Rev.\ Phys.\ Chem.}, {\bf 50}, 145--165 (1999).

\bibitem{DziBhaAnt20}
J.~Dziedzic, A.~Bhandari, L.~Anton, C.~Peng, J.~Womack, M.~Famili, D.~Kramer,
  and C.-K. Skylaris, ``Practical approach to large-scale electronic structure
  calculations in electrolyte solutions via continuum-embedded linear-scaling
  density functional theory'', {\em J.~Phys.\ Chem.~C}, {\bf 124}, 7860--7872
  (2020).

\bibitem{Bor20a}
M.~Born, ``Volumen und {H}ydratationsw\"arme der {Ionen}'', {\em Z.~Phys.},
  {\bf 1}, 45--48 (1920).

\bibitem{RasHon85}
A.~A. Rashin and B.~Honig, ``Reevaluation of the {Born} model of ion
  hydration'', {\em J.~Phys.\ Chem.}, {\bf 89}, 5588--5593 (1985).

\bibitem{Ons36}
L.~Onsager, ``Electric moments of molecules in liquids'', {\em J.~Am.\ Chem.\
  Soc.}, {\bf 58}, 1486--1493 (1936).

\bibitem{Bel31}
R.~P. Bell, ``The electrostatic energy of dipole molecules in different
  media'', {\em Trans.\ Faraday Soc.}, {\bf 27}, 797--802 (1931).

\bibitem{Bot38}
C.~J.~F. B\"ottcher, ``The dielectric constant of dipole liquids'', {\em
  Physica}, {\bf 5}, 635--639 (1938).

\bibitem{Has72}
J.~B. Hasted, ``Liquid water: {D}ielectric properties'', in {\em Water: A
  Comprehensive Treatise}, F.~Franks, Ed., Vol. ~1;
\newblock Plenum Press: New York, 1972;
\newblock pages 255--309.

\bibitem{HoySte76}
J.~S. H{\o}ye and G.~Stell, ``Statistical mechanics of polar systems. {II}'',
  {\em J.~Chem.\ Phys.}, {\bf 64}, 1952--1966 (1976).

\bibitem{Han83}
J.~H. Hannay, ``The {Clausius}--{Mossotti} equation: {An} alternative
  derivation'', {\em Eur.~J. Phys.}, {\bf 4}, 141--143 (1983).

\bibitem{Kir34a}
J.~G. Kirkwood, ``Theory of solutions of molecules containing widely separated
  charges with special application to zwitterions'', {\em J.~Chem.\ Phys.},
  {\bf 2}, 351--361 (1934).

\bibitem{KirWes38}
J.~G. Kirkwood and F.~H. Westheimer, ``The electrostatic influence of
  substituents on the dissociation constants of organic acids. {I}.'', {\em
  J.~Chem.\ Phys.}, {\bf 6}, 506--512 (1938).

\bibitem{WesKir38}
F.~H. Westheimer and J.~G. Kirkwood, ``The electrostatic influence of
  substituents on the dissociation constants of organic acids. {II}.'', {\em
  J.~Chem.\ Phys.}, {\bf 6}, 513--517 (1938).

\bibitem{Bon51}
W.~B. Bonner, ``The electrostatic energy of molecules in solution'', {\em
  Trans.\ Faraday Soc.}, {\bf 47}, 1143--1152 (1951).

\bibitem{RinRuiRiv83}
D.~Rinaldi, M.~F. {Ruiz-Lopez}, and J.-L. Rivail, ``\textit{Ab initio} SCF
  calculations on electrostatically solvated molecules using a deformable three
  axes ellipsoidal cavity'', {\em J.~Chem.\ Phys.}, {\bf 78}, 834--838 (1983).

\bibitem{MikDalSwa87}
K.~V. Mikkelsen, E.~Dalgaard, and P.~Swanstr{\o}m, ``Electron-transfer
  reactions in solution. {An} ab initio approach'', {\em J.~Phys.\ Chem.}, {\bf
  91}, 3081--3092 (1987).

\bibitem{MikAgrJen88}
K.~V. Mikkelsen, H.~{\AA}gren, H.~J.~A. Jensen, and T.~Helgaker, ``A
  multiconfigurational self-consistent reaction-field method'', {\em J.~Chem.\
  Phys.}, {\bf 89}, 3086--3095 (1988).

\bibitem{KonPon97}
Y.~Kong and J.~W. Ponder, ``Calculation of the reaction field due to off-center
  point multipoles'', {\em J.~Chem.\ Phys.}, {\bf 107}, 481--492 (1997).

\bibitem{MikJorJen94}
K.~V. Mikkelsen, P.~J{\o}rgensen, and H.~J.~A. Jensen, ``A multiconfigurational
  self-consistent reaction field response method'', {\em J.~Chem.\ Phys.}, {\bf
  100}, 6597--6607 (1994).

\bibitem{MedBudBar17}
M.~Medved', \v{S}. Budz\'ak, W.~Bartkowiak, and H.~Reis, ``Solvent effects on
  molecular electric properties'', in {\em Handbook of Computational
  Chemistry}, J.~Leszczynski, A.~{Kaczmarek-Kedziera}, T.~Puzyn, M.~G.
  Papadopoulos, H.~Reis, and M.~K. Shukla, Eds.;
\newblock Springer International Publishing: Switzerland, 2nd ed., 2017;
\newblock chapter~17, pages 741--794.

\bibitem{LotHea06}
I.~Lotan and T.~{Head-Gordon}, ``An analytical electrostatic model for salt
  screened interactions between multiple proteins'', {\em J.~Chem.\ Theory
  Comput.}, {\bf 2}, 541--555 (2006).

\bibitem{ZhaBenChi98}
C.-G. Zhan, J.~Bentley, and D.~M. Chipman, ``Volume polarization in reaction
  field theory'', {\em J.~Chem.\ Phys.}, {\bf 108}, 177--192 (1998).

\bibitem{Chi97}
D.~M. Chipman, ``Charge penetration in dielectric models of solvation'', {\em
  J.~Chem.\ Phys.}, {\bf 106}, 10194--10206 (1997).

\bibitem{Chi99}
D.~M. Chipman, ``Simulation of volume polarization in reaction field theory'',
  {\em J.~Chem.\ Phys.}, {\bf 110}, 8012--8018 (1999).

\bibitem{Chi00}
D.~M. Chipman, ``Reaction field treatment of charge penetration'', {\em
  J.~Chem.\ Phys.}, {\bf 112}, 5558--5565 (2000).

\bibitem{Chi02a}
D.~M. Chipman, ``Comparison of solvent reaction field representations'', {\em
  Theor.\ Chem.\ Acc.}, {\bf 107}, 80--89 (2002).

\bibitem{Chi06b}
D.~M. Chipman, ``New formulation and implementation for volume polarization in
  dielectric continuum theory'', {\em J.~Chem.\ Phys.}, {\bf 124}, 224111:1--10
  (2006).

\bibitem{Chi02b}
D.~M. Chipman, ``Energy correction to simulation of volume polarization in
  reaction field theory'', {\em J.~Chem.\ Phys.}, {\bf 116}, 10129--10138
  (2002).

\bibitem{LuqCurMun03}
F.~J. Luque, C.~Curutchet, J.~{Mu\~{n}oz-Muriedas}, A.~{Bidon-Chanal},
  I.~Soteras, A.~Morreale, J.~L. Gelpi, and M.~Orozco, ``Continuum solvation
  models: {D}issecting the free energy of solvation'', {\em Phys.\ Chem.\
  Chem.\ Phys.}, {\bf 5}, 3827--3836 (2003).

\bibitem{BegRou94}
D.~Beglov and B.~Roux, ``Finite representation of an infinite bulk system:
  {S}olvent boundary potential for computer simulations'', {\em J.~Chem.\
  Phys.}, {\bf 100}, 9050--9063 (1994).

\bibitem{TirSpeSmi95}
I.~G. Tironi, R.~Sperb, P.~E. Smith, and W.~F. {van Gunsteren}, ``A generalized
  reaction field method for molecular dynamics simulations'', {\em J.~Chem.\
  Phys.}, {\bf 102}, 5451--5459 (1995).

\bibitem{ImBerRou01}
W.~Im, S.~Bern\`eche, and B.~Roux, ``Generalized solvent boundary potential for
  computer simulations'', {\em J.~Chem.\ Phys.}, {\bf 114}, 2924--2937 (2001).

\bibitem{SchRicCui05}
P.~Schaefer, D.~Riccardi, and Q.~Cui, ``Reliable treatment of electrostatics in
  combined {QM}\slash {MM} simulation of macromolecules'', {\em J.~Chem.\
  Phys.}, {\bf 123}, 014905:1--14 (2005).

\bibitem{BenThi08}
T.~Benighaus and W.~Thiel, ``Efficiency and accuracy of the generalized solvent
  boundary potential for hybrid {QM}\slash {MM} simulations: {I}mplementation
  for semiempirical {H}amiltonians'', {\em J.~Chem.\ Theory Comput.}, {\bf 4},
  1600--1609 (2008).

\bibitem{BenThi09}
T.~Benighaus and W.~Thiel, ``A general boundary potential for hybrid {QM}\slash
  {MM} simulations of solvated biomolecular systems'', {\em J.~Chem.\ Theory
  Comput.}, {\bf 5}, 3114--3128 (2009).

\bibitem{BenThi11}
T.~Benighaus and W.~Thiel, ``Long-range electrostatic effects in {QM}\slash
  {MM} studies of enzymatic reactions: {A}pplication of the solvated
  macromolecule boundary potential'', {\em J.~Chem.\ Theory Comput.}, {\bf 7},
  238--249 (2011).

\bibitem{AleFie11}
A.~Aleksandrov and M.~Field, ``Efficient solvent boundary potential for hybrid
  potential simulations'', {\em Phys.\ Chem.\ Chem.\ Phys.}, {\bf 13},
  10503--10509 (2011).

\bibitem{ZieCui12}
J.~Zienau and Q.~Cui, ``Implementation of the solvent macromolecular boundary
  potential and application to model and realistic enzyme systems'', {\em
  J.~Phys.\ Chem.~B}, {\bf 116}, 12522--12534 (2012).

\bibitem{LuCui13}
X.~Lu and Q.~Cui, ``Charging free energy calculations using the generalized
  solvent boundary potential {(GSBP)} and periodic boundary condition: {A}
  comparative analysis using ion solvation and oxidation free energy in
  proteins'', {\em J.~Phys.\ Chem.~B}, {\bf 117}, 2005--2018 (2013).

\bibitem{RegBraBar06}
N.~Rega, G.~Brancato, and V.~Barone, ``Non-periodic boundary conditions for ab
  initio molecular dynamics in condensed phase using localized basis
  functions'', {\em Chem.\ Phys.\ Lett.}, {\bf 422}, 367--371 (2006).

\bibitem{BraRegBar06}
G.~Brancato, N.~Rega, and V.~Barone, ``Reliable molecular simulations of
  solute-solvent systems with a minimum number of solvent shells'', {\em
  J.~Chem.\ Phys.}, {\bf 124}, 214505:1--9 (2006).

\bibitem{BraRegBar08a}
G.~Brancato, N.~Rega, and V.~Barone, ``A hybrid explicit\slash implicit
  solvation method for first-principle molecular dynamics simulations'', {\em
  J.~Chem.\ Phys.}, {\bf 128}, 144501:1--10 (2008).

\bibitem{BraRegBar09}
G.~Brancato, N.~Rega, and V.~Barone, ``Molecular dynamics simulations in a
  ${N}p{T}$ \textit{ensemble} using non-periodic boundary conditions'', {\em
  Chem.\ Phys.\ Lett.}, {\bf 483}, 177--181 (2009).

\bibitem{SwaAdcMcC05}
J.~M.~J. Swanson, S.~A. Adcock, and J.~A. {McCammon}, ``Optimized radii for
  {Poisson}--{Boltzmann} calculations with the {AMBER} force field'', {\em
  J.~Chem.\ Theory Comput.}, {\bf 1}, 484--493 (2005).

\bibitem{Bon64}
A.~Bondi, ``Van der {Waals} volumes and radii'', {\em J.~Phys.\ Chem.}, {\bf
  68}, 441--451 (1964).

\bibitem{RowTay96}
R.~S. Rowland and R.~Taylor, ``Intermolecular nonbonded contact distances in
  organic crystal structures: {C}omparison with distances expected from van der
  {Waals} radii'', {\em J.~Phys.\ Chem.}, {\bf 100}, 7384--7391 (1996).

\bibitem{ManChaVal09}
M.~Manjeera, A.~C. Chamberlin, R.~Valero, C.~J. Cramer, and D.~G. Truhlar,
  ``Consistent van der {Waals} radii for the whole main group'', {\em J.~Phys.\
  Chem.~A}, {\bf 113}, 5806--5812 (2009).

\bibitem{BonPalTom84}
R.~Bonaccorsi, P.~Palla, and J.~Tomasi, ``Conformational energy of glycine in
  aqueous solutions and relative stability of the zwitterionic and neutral
  forms. {An} ab initio study'', {\em J.~Am.\ Chem.\ Soc.}, {\bf 106},
  1945--1950 (1984).

\bibitem{BroHea15}
D.~H. Brookes and T.~{Head-Gordon}, ``Family of oxygen--oxygen radial
  distribution functions for water'', {\em J.~Phys.\ Chem.\ Lett.}, {\bf 6},
  2938--2943 (2015).

\bibitem{OnuAgu14}
A.~V. Onufriev and B.~Aguilar, ``Accuracy of continuum electrostatic
  calculations based on three common dielectric boundary definitions'', {\em
  J.~Theor.\ Comput.\ Chem.}, {\bf 13}, 1440006:1--25 (2014).

\bibitem{Con83b}
M.~L. Connolly, ``Solvent-accessible surfaces of proteins and nucleic acids'',
  {\em Science}, {\bf 221}, 709--713 (1983).

\bibitem{Ric77}
F.~M. Richards, ``Areas, volumes, packing, and protein structure'', {\em Annu.\
  Rev.\ Biophys.\ Bio.}, {\bf 6}, 151--176 (1977).

\bibitem{LeeRic71}
B.~Lee and F.~M. Richards, ``The interpretation of protein structures:
  {E}stimation of static accessibility'', {\em J.~Mol.\ Biol.}, {\bf 55},
  379--400 (1971).

\bibitem{KimWonBha10}
D.-S. Kim, C.-I. Won, and J.~Bhak, ``A proposal for the revision of molecular
  boundary typology'', {\em J.~Biomol.\ Struct.\ Dyn.}, {\bf 28}, 277--287
  (2010).

\bibitem{Pom07}
C.~S. Pomelli, ``Cavity surfaces and their discretization'', in {\em Continuum
  Solvation Models in Chemical Physics}, B.~Mennucci and R.~Cammi, Eds.;
\newblock Wiley: Chichester, UK, 2007;
\newblock pages 49--63.

\bibitem{ZhaChi98}
C.-G. Zhan and D.~M. Chipman, ``Cavity size in reaction field theory'', {\em
  J.~Chem.\ Phys.}, {\bf 109}, 10543--10558 (1998).

\bibitem{BarCosTom97}
V.~Barone, M.~Cossi, and J.~Tomasi, ``A new definition of cavities for the
  computation of solvation free energies by the polarizable continuum model'',
  {\em J.~Chem.\ Phys.}, {\bf 107}, 3210--3221 (1997).

\bibitem{GinCamDup08a}
B.~Ginovska, D.~M. Camaioni, M.~Dupuis, C.~A. Schwerdtfeger, and Q.~Gil,
  ``Charge-dependent cavity radii for an accurate dielectric continuum model of
  solvation with emphasis on ions: {A}queous solutes with oxo, hydroxo, amino,
  methyl, chloro, bromo, and fluoro functionalities'', {\em J.~Phys.\ Chem.~A},
  {\bf 112}, 10604--10613 (2008).

\bibitem{ForKeiWib96}
J.~B. Foresman, T.~A. Keith, K.~B. Wiberg, J.~Snoonian, and M.~J. Frisch,
  ``Solvent effects. 5. {I}nfluence of cavity shape, truncation of
  electrostatics, and electron correlation on ab initio reaction field
  calculations'', {\em J.~Phys.\ Chem.}, {\bf 100}, 16098--16104 (1996).

\bibitem{ChiDup02}
D.~M. Chipman and M.~Dupuis, ``Implementation of solvent reaction fields for
  electronic structure'', {\em Theor.\ Chem.\ Acc.}, {\bf 107}, 90--102 (2002).

\bibitem{CheChi03}
F.~Chen and D.~M. Chipman, ``Boundary element methods for dielectric cavity
  construction and integration'', {\em J.~Chem.\ Phys.}, {\bf 119},
  10289--10297 (2003).

\bibitem{doCChi10a}
P.~C. {do Couto} and D.~M. Chipman, ``How does dielectric solvation affect the
  size of an ion?'', {\em J.~Phys.\ Chem.~A}, {\bf 114}, 12788--12793 (2010).

\bibitem{HolSai93}
M.~Holst and F.~Saied, ``Multigrid solution of the {Poisson}--{Boltzmann}
  equation'', {\em J.~Comput.\ Chem.}, {\bf 14}, 105--113 (1993).

\bibitem{HolSai95}
M.~J. Holst and F.~Saied, ``Numerical solution of the nonlinear
  {Poisson}--{Boltzmann} equation: {D}eveloping more robust and efficient
  methods'', {\em J.~Comput.\ Chem.}, {\bf 16}, 337--364 (1995).

\bibitem{HolBakWan00}
M.~Holst, N.~Baker, and F.~Wang, ``Adaptive multilevel finite element solution
  of the {Poisson}--{Boltzmann} equation {I}. {A}lgorithms and examples'', 
  {\em J.~Comput.\ Chem.}, {\bf 21}, 1319--1342 (2000). 
  Erratum: {\em J.~Comput.\ Chem.}, {\bf 22}, 45 (2001). 

\bibitem{LuZhoHol08}
B.~Z. Lu, Y.~C. Zhou, M.~J. Holst, and J.~A. {McCammon}, ``Recent progress in
  numerical methods for the {Poisson}-{Boltzmann} equation in biophysical
  applications'', {\em Commun.\ Comput.\ Phys.}, {\bf 3}, 973--1009 (2008).

\bibitem{WanLuo10}
J.~Wang and R.~Luo, ``Assessment of linear finite-difference
  {Poisson}--{Boltzmann} solvers'', {\em J.~Comput.\ Chem.}, {\bf 31},
  1689--1698 (2010).

\bibitem{YapHea10}
E.-H. Yap and T.~{Head-Gordon}, ``New and efficient {Poisson}--{Boltzmann}
  solver for interaction of multiple proteins'', {\em J.~Chem.\ Theory
  Comput.}, {\bf 6}, 2214--2224 (2010).

\bibitem{BosFen11}
A.~H. Boschitsch and M.~O. Fenley, ``A fast and robust Poisson--{Boltzmann}
  solver based on adaptive {C}artesian grids'', {\em J.~Chem.\ Theory Comput.},
  {\bf 7}, 1524--1540 (2011).

\bibitem{HolMcCYu12}
M.~Holst, J.~A. {McCammon}, Z.~Yu, and Y.~C. Zhou, ``Adaptive finite element
  modeling techniques for the {Poisson}-{Boltzmann} equation'', {\em Commun.\
  Comput.\ Phys.}, {\bf 11}, 179--214 (2012).

\bibitem{LiLiPet13}
C.~Li, L.~Li, M.~Petukh, and E.~Alexov, ``Progress in developing
  {Poisson}-{Boltzmann} equation solvers'', {\em Mol.-Based Math.\ Biol.}, {\bf
  1}, 42--62 (2013).

\bibitem{GenKra13}
W.~Geng and R.~Krasny, ``A treecode-accelerated boundary integral
  {Poisson}--{Boltzmann} solver for electrostatics of solvated biomolecules'',
  {\em J.~Comput.\ Phys.}, {\bf 247}, 62--78 (2013).

\bibitem{mFES}
I.~Sakalli, J.~Sch\"oberl, and E.~W. Knapp, ``{mFES}: {A} robust molecular
  finite element solver for electrostatic energy computations'', {\em J.~Chem.\
  Theory Comput.}, {\bf 10}, 5095--5112 (2014).

\bibitem{FisGenAnd16}
G.~Fisicaro, L.~Genovese, O.~Andreussi, N.~Marzari, and S.~Goedecker, ``A
  generalized {Poisson} and {Poisson}-{Boltzmann} solver for electrostatic
  environments'', {\em J.~Chem.\ Phys.}, {\bf 144}, 014103:1--12 (2016).

\bibitem{RinObeHil16}
S.~Ringe, H.~Oberhofer, C.~Hille, S.~Matera, and K.~Reuter,
  ``Function-space-based solution scheme for the size-modified
  {Poisson}--{Boltzmann} equation in full-potential {DFT}'', {\em J.~Chem.\
  Theory Comput.}, {\bf 12}, 4052--4066 (2016).

\bibitem{WomAntDzi18}
J.~C. Womack, L.~Anton, J.~Dziedzic, P.~J. Hasnip, M.~I.~J. Probert, and C.-K.
  Skylaris, ``{DL}\_{MG}: {A} parallel multigrid {Poisson} and
  {Poisson}--{Boltzmann} solver for electronic structure calculations in vacuum
  and solution'', {\em J.~Chem.\ Theory Comput.}, {\bf 14}, 1412--1432 (2018).

\bibitem{LuzWar92}
V.~Luzhkov and A.~Warshel, ``Microscopic models for quantum mechanical
  calculations of chemical processes in solutions: {LD}\slash {AMPAC} and
  {SCAAS}\slash {AMPAC} calculations of solvation energies'', {\em J.~Comput.\
  Chem.}, {\bf 13}, 199--213 (1992).

\bibitem{FloWar97}
J.~Flori\'an and A.~Warshel, ``Langevin dipoles model for ab initio
  calculations of chemical processes in solution: {P}arameterization and
  application to hydration free energies of neutral and ionic solutes and
  conformational analysis in aqueous solution'', {\em J.~Phys.\ Chem.~B}, {\bf
  101}, 5583--5595 (1997).

\bibitem{PapWar97a}
A.~Papazyan and A.~Warshel, ``Continuum and dipole-lattice models of
  solvation'', {\em J.~Phys.\ Chem.~B}, {\bf 101}, 11254--11264 (1997).

\bibitem{LanClaCai88}
J.~Langlet, P.~Claverie, J.~Caillet, and A.~Pullman, ``Improvements of the
  continuum model. 1. {A}pplication to the calculation of the vaporization of
  thermodynamic quantities of nonassociated liquids'', {\em J.~Phys.\ Chem.},
  {\bf 92}, 1617--1631 (1988).

\bibitem{OroLuq00}
M.~Orozco and F.~J. Luque, ``Theoretical methods for the description of the
  solvent effect in biomolecular systems'', {\em Chem.\ Rev.}, {\bf 100},
  4187--4225 (2000). Erratum:{\em Chem.\ Rev.}, {\bf 101}, 203 (2001).

\bibitem{Con86}
R.~Constanciel, ``Theoretical basis of the empirical reaction field
  approximations through continuum model'', {\em Theor.\ Chem.\ Acc.}, {\bf
  69}, 505--523 (1986).

\bibitem{Can07}
E.~Canc\`es, ``Integral equation approaches for continuum models'', in {\em
  Continuum Solvation Models in Chemical Physics}, B.~Mennucci and R.~Cammi,
  Eds.;
\newblock Wiley: Chichester, UK, 2007;
\newblock pages 29--48.

\bibitem{CosScaReg02b}
M.~Cossi, G.~Scalmani, N.~Rega, and V.~Barone, ``New developments in the
  polarizable continuum model for quantum mechanical and classical calculations
  on molecules in solution'', {\em J.~Chem.\ Phys.}, {\bf 117}, 43--54 (2002).

\bibitem{TomMenCan99}
J.~Tomasi, B.~Mennucci, and E.~Canc\`es, ``The {IEF} version of the {PCM}
  solvation method: {An} overview of a new method addressed to study molecular
  solutes at the {QM} ab initio level'', {\em J.~Mol.\ Struct.\ (Theochem)},
  {\bf 464}, 211--226 (1999).

\bibitem{CanMenTom97}
E.~Canc\'es, B.~Mennucci, and J.~Tomasi, ``A new integral equation formalism
  for the polarizable continuum model: {T}heoretical background and
  applications to isotropic and anisotropic dielectrics'', {\em J.~Chem.\
  Phys.}, {\bf 107}, 3032--3041 (1997).

\bibitem{MenCanTom97}
B.~Mennucci, E.~Canc\'es, and J.~Tomasi, ``Evaluation of solvent effects in
  isotropic and anisotropic dielectrics and in ionic solutions with a unified
  integral equation method: {T}heoretical bases, computational implementation,
  and numerical applications'', {\em J.~Phys.\ Chem.~B}, {\bf 101},
  10506--10517 (1997).

\bibitem{CanMen98a}
E.~Canc\`es and B.~Mennucci, ``New applications of integral equations methods
  for solvation continuum models: {Ionic} solutions and liquid crystals'', {\em
  J.~Math.\ Chem.}, {\bf 23}, 309--326 (1998).

\bibitem{CanMen01a}
E.~Canc\`es and B.~Mennucci, ``Comment on ``{R}eaction field treatment of
  charge penetration'''', {\em J.~Chem.\ Phys.}, {\bf 114}, 4744--4745 (2001).

\bibitem{DavMcC89}
M.~E. Davis and J.~A. {McCammon}, ``Solving the finite difference linearized
  {Poisson}-{Boltzmann} equation: {A} comparison of relaxation and conjugate
  gradient methods'', {\em J.~Comput.\ Chem.}, {\bf 10}, 386--391 (1989).

\bibitem{LutDavMcC92}
B.~A. Luty, M.~E. Davis, and J.~A. {McCammon}, ``Solving the finite-difference
  non-linear {Poisson}--{Boltzmann} equation'', {\em J.~Comput.\ Chem.}, {\bf
  13}, 1114--1118 (1992).

\bibitem{WanCaiXia12}
J.~Wang, Q.~Cai, Y.~Xiang, and R.~Luo, ``Reducing grid dependence in
  finite-difference {Poisson}--{Boltzmann} calculations'', {\em J.~Chem.\
  Theory Comput.}, {\bf 8}, 2741--2751 (2012).

\bibitem{XiaCaiYe13}
L.~Xiao, Q.~Cai, X.~Ye, J.~Wang, and R.~Luo, ``Electrostatic forces in the
  {Poisson}-{Boltzmann} systems'', {\em J.~Chem.\ Phys.}, {\bf 139},
  094106:1--12 (2013).

\bibitem{XiaWanLuo14}
L.~Xiao, C.~Wang, and R.~Luo, ``Recent progress in adapting
  {Poisson}--{Boltzmann} methods to molecular simulations'', {\em J.~Theor.\
  Comput.\ Chem.}, {\bf 13}, 1430001:1--19 (2014).

\bibitem{Chi04}
D.~M. Chipman, ``Solution of the linearized {Poisson}--{Boltzmann} equation'',
  {\em J.~Chem.\ Phys.}, {\bf 120}, 5566--5575 (2004).

\bibitem{LipStaCan13}
F.~Lipparini, B.~Stamm, E.~Canc\`es, Y.~Maday, and B.~Mennucci, ``Fast domain
  decomposition algorithm for continuum solvation models: {Energy} and first
  derivatives'', {\em J.~Chem.\ Theory Comput.}, {\bf 9}, 3637--3648 (2013).

\bibitem{LipLagSca14}
F.~Lipparini, L.~Lagard\`ere, G.~Scalmani, B.~Stamm, E.~Canc\`es, Y.~Maday,
  J.-P. Piquemal, M.~J. Frisch, and B.~Mennucci, ``Quantum calculations in
  solution for large to very large molecules: {A} new linear scaling {QM}\slash
  continuum approach'', {\em J.~Phys.\ Chem.\ Lett.}, {\bf 5}, 953--958 (2014).

\bibitem{CapJurLag15}
S.~Caprasecca, S.~Jurinovich, L.~Lagard\`ere, B.~Stamm, and F.~Lipparini,
  ``Achieving linear scaling in computational cost for a fully polarizable
  {MM}\slash continuum embedding'', {\em J.~Chem.\ Theory Comput.}, {\bf 11},
  694--704 (2015).

\bibitem{KlaSch93}
A.~Klamt and G.~Sch\"u\"urmann, ``{COSMO}: {A} new approach to dielectric
  screening in solvents with explicit expressions for the screening energy and
  its gradient'', {\em J.~Chem.\ Soc., Perkin Trans.~2}, pages 799--805 (1993).

\bibitem{CanMen01b}
E.~Canc\`es and B.~Mennucci, ``The escaped charge problem in solvation
  continuum models'', {\em J.~Chem.\ Phys.}, {\bf 115}, 6130--6135 (2001).

\bibitem{KlaMoyPal15}
A.~Klamt, C.~Moya, and J.~Palomar, ``A comprehensive comparison of the {IEFPCM}
  and {SS(V)PE} continuum solvation methods with the {COSMO} approach'', {\em
  J.~Chem.\ Theory Comput.}, {\bf 11}, 4220--4225 (2015).

\bibitem{SteTru95}
E.~V. Stefanovich and T.~N. Truong, ``Optimized atomic radii for quantum
  dielectric continuum solvation models'', {\em Chem.\ Phys.\ Lett.}, {\bf
  244}, 65--74 (1995).

\bibitem{TruSte95a}
T.~N. Truong and E.~V. Stefanovich, ``A new method for incorporating solvent
  effects into the classical, ab initio molecular orbital and density
  functional theory frameworks for arbitrary cavity shape'', {\em Chem.\ Phys.\
  Lett.}, {\bf 240}, 253--260 (1995).

\bibitem{TruSte95b}
T.~N. Truong and E.~V. Stefanovich, ``Analytical first and second energy
  derivatives of the generalized conductorlike screening model for free energy
  of solvation'', {\em J.~Chem.\ Phys.}, {\bf 103}, 3709--3717 (1995).

\bibitem{TruNguSte96}
T.~N. Truong, U.~N. Nguyen, and E.~V. Stefanovich, ``Generalized conductor-like
  screening model ({GCOSMO}) for solvation: {An} assessment of its accuracy and
  applicability'', {\em Int.~J. Quantum Chem.\ Symp.}, {\bf 60}, 1615--1622
  (1996).

\bibitem{BarCos98}
V.~Barone and M.~Cossi, ``Quantum calculation of molecular energies and energy
  gradients in solution by a conductor solvent model'', {\em J.~Phys.\
  Chem.~A}, {\bf 102}, 1995--2001 (1998).

\bibitem{CosRegSca03}
M.~Cossi, N.~Rega, G.~Scalmani, and V.~Barone, ``Energies, structures, and
  electronic properties of molecules in solution with the {C}-{PCM} solvation
  model'', {\em J.~Comput.\ Chem.}, {\bf 24}, 669--681 (2003).

\bibitem{CamTom94a}
R.~Cammi and J.~Tomasi, ``Analytical derivatives for molecular solutes. {I}.
  {Hartree}--{Fock} energy first derivatives with respect to external
  parameters in the polarizable continuum model'', {\em J.~Chem.\ Phys.}, {\bf
  100}, 7495--7502 (1994).

\bibitem{KlaJon96}
A.~Klamt and V.~Jonas, ``Treatment of the outlying charge in continuum
  solvation models'', {\em J.~Chem.\ Phys.}, {\bf 105}, 9972--9981 (1996).

\bibitem{MenTom97}
B.~Mennucci and J.~Tomasi, ``Continuum solvation models: {A} new approach to
  the problem of solute's charge distribution and cavity boundaries'', {\em
  J.~Chem.\ Phys.}, {\bf 106}, 5151--5158 (1997).

\bibitem{CamCosTom96}
R.~Cammi, M.~Cossi, and J.~Tomasi, ``Analytical derivatives for molecular
  solutes. {III}. {Hartree}--{Fock} static polarizabilities in the polarizable
  continuum model'', {\em J.~Chem.\ Phys.}, {\bf 104}, 4611--4620 (1996).

\bibitem{Gau00}
J.~Gauss, ``Molecular properties'', in {\em Modern Methods and Algorithms of
  Quantum Chemistry}, J.~Grotendorst, Ed., Vol. ~3 of {\em NIC Series;}
\newblock John von Neumann Institute for Computing: J\"ulich, 2nd ed., 2000;
\newblock pages 541--592.

\bibitem{RizCorRuu12}
A.~Rizzo, S.~Coriani, and K.~Ruud, ``Response function theory computational
  approaches to linear and nonlinear optical spectroscopy'', in {\em
  Computational Strategies for Spectroscopy: From Small Molecules to Nano
  Systems}, V.~Barone, Ed.;
\newblock John Wiley \& Sons: Hoboken, NJ, 1st ed., 2012;
\newblock chapter~2, pages 77--136.

\bibitem{HelCorJor12}
T.~Helgaker, S.~Coriani, P.~J{\o}rgensen, K.~Kristensen, J.~Olsen, and K.~Ruud,
  ``Recent advances in wave function-based methods of molecular-property
  calculations'', {\em Chem.\ Rev.}, {\bf 112}, 543--631 (2012).

\bibitem{APBS}
E.~Jurrus, D.~Engel, K.~Star, K.~Monson, J.~Brandi, L.~E. Felberg, D.~H.
  Brookes, L.~Wilson, J.~Chen, K.~Liles, M.~Chun, P.~Li, D.~W. Gohara,
  T.~Dolinsky, R.~K. D.~R. Koes, J.~E. Nielsen, T.~{Head-Gordon}, W.~Geng,
  R.~Krasny, G.-W. Wei, M.~J. Holst, J.~A. {McCammon}, and N.~A. Baker,
  ``Improvements to the {APBS} biomolecular solvation software suite'', {\em
  Protein Sci.}, {\bf 27}, 112--128 (2018).

\bibitem{YouMewDre15}
Z.-Q. You, J.-M. Mewes, A.~Dreuw, and J.~M. Herbert, ``Comparison of the
  {Marcus} and {Pekar} partitions in the context of non-equilibrium,
  polarizable-continuum reaction-field solvation models'', {\em J.~Chem.\
  Phys.}, {\bf 143}, 204107:1--14 (2015).

\bibitem{SilVilNil90}
E.~Silla, F.~Villar, O.~Nilsson, J.~L. {Pascual-Ahuir}, and O.~Tapia,
  ``Molecular volumes and surfaces of biomacromolecules via {GEPOL}: {A} fast
  and efficient algorithm'', {\em J.~Mol.\ Graphics}, {\bf 8}, 168--172 (1990).

\bibitem{PasSil90}
J.~L. {Pascual-Ahuir} and E.~Silla, ``{GEPOL}: An improved description of
  molecular surfaces. {I}. {B}uilding the spherical surface set'', {\em
  J.~Comput.\ Chem.}, {\bf 11}, 1047--1060 (1990).

\bibitem{SilTunPas91}
E.~Silla, I.~{Tu\~{n}on}, and J.~L. {Pascual-Ahuir}, ``{GEPOL}: An improved
  description of molecular surfaces. {II}. {C}omputing the molecular area and
  volume'', {\em J.~Comput.\ Chem.}, {\bf 12}, 1077--1088 (1991).

\bibitem{PasSilTun94}
J.~L. {Pascual-Ahuir}, E.~Silla, and I.~{Tu\~{n}on}, ``{GEPOL}: An improved
  description of molecular surfaces. {III}. {A} new algorithm for the
  computation of a solvent-excluding surface'', {\em J.~Comput.\ Chem.}, {\bf
  15}, 1127--1138 (1994).

\bibitem{LioHawLyn95}
D.~A. Liotard, G.~D. Hawkins, G.~C. Lynch, C.~J. Cramer, and D.~G. Truhlar,
  ``Improved methods for semiempirical solvation models'', {\em J.~Comput.\
  Chem.}, {\bf 16}, 422--440 (1995).

\bibitem{CosMenCam96}
M.~Cossi, B.~Mennucci, and R.~Cammi, ``Analytical first derivatives of
  molecular surfaces with respect to nuclear coordinates'', {\em J.~Comput.\
  Chem.}, {\bf 17}, 57--73 (1996).

\bibitem{YorKar99}
D.~M. York and M.~Karplus, ``Smooth solvation potential based on the
  conductor-like screening model'', {\em J.~Phys.\ Chem.~A}, {\bf 103},
  11060--11079 (1999).

\bibitem{GreYor05}
B.~A. Gregersen and D.~M. York, ``High-order discretization schemes for
  biochemical applications of boundary element solvation and variational
  electrostatic projection methods'', {\em J.~Chem.\ Phys.}, {\bf 122},
  194110:1--4 (2005).

\bibitem{KhaGreThi05}
J.~Khandogin, B.~A. Gregersen, W.~Thiel, and D.~M. York, ``Smooth solvation
  method for d-orbital semiemprical calculations of biological reactions. 1.
  {I}mplementation'', {\em J.~Phys.\ Chem.~B}, {\bf 109}, 9799--9809 (2005).

\bibitem{MurHanLam93}
C.~W. Murray, N.~C. Handy, and G.~J. Laming, ``Quadrature schemes for integrals
  of density functional theory'', {\em Mol.\ Phys.}, {\bf 78}, 997--1014
  (1993).

\bibitem{SG1}
P.~M.~W. Gill, B.~G. Johnson, and J.~A. Pople, ``A standard grid for
  density-functional calculations'', {\em Chem.\ Phys.\ Lett.}, {\bf 209},
  506--512 (1993).

\bibitem{SG0}
S.-H. Chien and P.~M.~W. Gill, ``{SG}-0: {A} small standard grid for {DFT}
  quadrature on large systems'', {\em J.~Comput.\ Chem.}, {\bf 27}, 730--739
  (2006).

\bibitem{DasHer17}
S.~Dasgupta and J.~M. Herbert, ``Standard grids for high-precision integration
  of modern density functionals: {SG}-2 and {SG}-3'', {\em J.~Comput.\ Chem.},
  {\bf 38}, 869--882 (2017).

\bibitem{LiuLia13}
J.~Liu and W.~Liang, ``Analytical second derivatives of excited-state energy
  within the time-dependent density functional theory coupled with a
  conductor-like polarizable continuum model'', {\em J.~Chem.\ Phys.}, {\bf
  138}, 024101:1--10 (2013).

\bibitem{WawGibSch94}
R.~J. Wawak, K.~D. Gibson, and H.~A. Scheraga, ``Gradient discontinuities in
  calculations involving molecular surface area'', {\em J.~Math.\ Chem.}, {\bf
  15}, 207--232 (1994).

\bibitem{LiJen04}
H.~Li and J.~H. Jensen, ``Improving the efficiency and convergence of geometry
  optimization with the polarizable continuum model: {New} energy gradients and
  molecular surface tessellation'', {\em J.~Comput.\ Chem.}, {\bf 25},
  1449--1462 (2004).

\bibitem{SuLi09}
P.~Su and H.~Li, ``Continuous and smooth potential energy surface for
  conductor-like screening solvation model using fixed points with variable
  areas'', {\em J.~Chem.\ Phys.}, {\bf 130}, 074109:1--13 (2009).

\bibitem{KryGil13}
A.~I. Krylov and P.~M.~W. Gill, ``Q-{Chem}: {An} engine for innovation'', {\em
  WIREs Comput.\ Mol.\ Sci.}, {\bf 3}, 317--326 (2013).

\bibitem{ScaFri10}
G.~Scalmani and M.~J. Frisch, ``Continuous surface charge polarizable continuum
  models of solvation. {I}. {G}eneral formalism'', {\em J.~Chem.\ Phys.}, {\bf
  132}, 114110:1--15 (2010).

\bibitem{GarNee20}
M.~{Garcia-Rat\`es} and F.~Neese, ``Effect of the solute cavity on the
  solvation energy and its derivatives within the framework of the {G}aussian
  charge scheme'', {\em J.~Comput.\ Chem.}, {\bf 41}, 922--939 (2020).

\bibitem{HarRan09}
H.~Harbrecht and M.~Randrianarivony, ``Wavelent {BEM} on molecular surfaces:
  {P}arameterization and implementation'', {\em Computing}, {\bf 86}, 1--22
  (2009).

\bibitem{WeiRanHar10}
V.~Weijo, M.~Randrianarivony, H.~Harbrecht, and L.~Frediani, ``Wavelet
  formulation of the polarizable continuum model'', {\em J.~Comput.\ Chem.},
  {\bf 31}, 1469--1477 (2010).

\bibitem{BugDiRMoz15}
M.~Bugeanu, R.~{Di Remigio}, K.~Mozgawa, S.~S. Reine, H.~Harbrecht, and
  L.~Frediani, ``Wavelet formulation of the polarizable continuum model. {II}.
  {Use} of piecewise bilinear boundary elements'', {\em Phys.\ Chem.\ Chem.\
  Phys.}, {\bf 17}, 31566--31581 (2015).

\bibitem{BugHar19}
M.~Bugeanu and H.~Harbrecht, ``Parametric representation of molecular
  surfaces'', {\em Int.~J. Quantum Chem.}, {\bf 119}, e25695:1--12 (2019).

\bibitem{ZhaChi99}
C.-G. Zhan and D.~M. Chipman, ``Reaction field effects on nitrogen shielding'',
  {\em J.~Chem.\ Phys.}, {\bf 110}, 1611--1622 (1999).

\bibitem{TjoZho08}
H.~Tjong and H.-X. Zhou, ``On the dielectric boundary in {Poisson}--{Boltzmann}
  calculations'', {\em J.~Chem.\ Theory Comput.}, {\bf 4}, 507--514 (2008).

\bibitem{LorCli87}
W.~E. Lorensen and H.~E. Cline, ``Marching cubes: {A} high resolution {3D}
  surface construction algorithm'', {\em Comp.\ Graph.}, {\bf 21}, 163--169
  (1987).

\bibitem{RajBol03}
D.~A. Rajon and W.~E. Bolch, ``Marching cubes algorithm: {Review} and trilinear
  interpolation adaptation for image-based dosimetric models'', {\em Comput.\
  Med.\ Imag.\ Grap.}, {\bf 27}, 411--435 (2003).

\bibitem{YuJacFri05}
Z.~Yu, M.~P. Jacobson, and R.~Friesner, ``What role do surfaces play in {GB}
  models? {A} new-generation of surface-generalized {Born} model based on a
  novel {Gaussian} surface for biomolecules'', {\em J.~Comput.\ Chem.}, {\bf
  27}, 72--89 (2005).

\bibitem{ZhoAgaWon08}
B.~Zhou, M.~Agarwal, and C.~F. Wong, ``Variable atomic radii for
  continuum-solvent electrostatics calculation'', {\em J.~Chem.\ Phys.}, {\bf
  129}, 014509:1--9 (2008).

\bibitem{FisGenAnd17}
G.~Fisicaro, L.~Genovese, O.~Andreussi, S.~Mandai, N.~N. Nair, N.~Marzari, and
  S.~Goedecker, ``Soft-sphere continuum solvation in electronic-structure
  calculations'', {\em J.~Chem.\ Theory Comput.}, {\bf 13}, 3829--3845 (2017).

\bibitem{FatGyg02}
J.-L. Fattebert and F.~Gygi, ``Density functional theory for efficieint
  \textit{ab initio} molecular dynamics simulations in solution'', {\em
  J.~Comput.\ Chem.}, {\bf 23}, 662--666 (2002).

\bibitem{FatGyg03}
J.-L. Fattebert and F.~Gygi, ``First-principles molecular dynamics simulations
  in a continuum solvent'', {\em Int.~J. Quantum Chem.}, {\bf 93}, 139--147
  (2003).

\bibitem{SchFatGyg06}
D.~A. Scherlis, J.-L. Fattebert, F.~Gygi, M.~Cococcioni, and N.~Marzari, ``A
  unified electrostatic and cavitation model for first-principles molecular
  dynamics in solution'', {\em J.~Chem.\ Phys.}, {\bf 124}, 074103:1--12
  (2006).

\bibitem{DziHelSky11}
J.~Dziedzic, H.~H. Helal, C.-K. Skylaris, A.~A. Mostofi, and M.~C. Payne,
  ``Minimal parameter implicit solvent model for ab initio electronic-structure
  calculations'', {\em Europhys.\ Lett.}, {\bf 95}, 43001:1--6 (2011).

\bibitem{AndDabMar12}
O.~Andreussi, I.~Dabo, and N.~Marzari, ``Revised self-consistent continuum
  solvation in electronic-structure calculations'', {\em J.~Chem.\ Phys.}, {\bf
  136}, 064102:1--20 (2012).

\bibitem{MatSunLet14}
K.~Mathew, R.~Sundararaman, K.~{Letchworth-Weaver}, T.~A. Arias, and R.~G.
  Hennig, ``Implicit solvation model for density-functional study of
  nanocrystal surfaces and reaction pathways'', {\em J.~Chem.\ Phys.}, {\bf
  140}, 084106:1--8 (2014).

\bibitem{SanSueSch09}
V.~M. S\'anchez, M.~Sued, and D.~A. Scherlis, ``First-principles molecular
  dynamics simulations at solid-liquid interfaces with a continuum solvent'',
  {\em J.~Chem.\ Phys.}, {\bf 131}, 174108:1--9 (2009).

\bibitem{NatTruMar19}
F.~Nattino, M.~Truscott, N.~Marzari, and O.~Andreussi, ``Continuum models of
  the electrochemical diffuse layer in electronic-structure calculations'',
  {\em J.~Chem.\ Phys.}, {\bf 150}, 041722:1--17 (2019).

\bibitem{AndHorNat19}
O.~Andreussi, N.~G. H\"ormann, F.~Nattino, G.~Fisicaro, S.~Goedecker, and
  N.~Marzari, ``Solvent-aware interfaces in continuum solvation'', {\em
  J.~Chem.\ Theory Comput.}, {\bf 15}, 1996--2009 (2019).

\bibitem{SunSch17}
R.~Sundararaman and K.~Schwarz, ``Evaluating continuum solvation models for the
  electrode-electrolyte interface: {C}hallenges and strategies for
  improvement'', {\em J.~Chem.\ Phys.}, {\bf 146}, 084111:1--5 (2017).

\bibitem{SunLetSch18}
R.~Sundararaman, K.~{Letchworth-Weaver}, and K.~A. Schwarz, ``Improving
  accuracy of electrochemical capacitance and solvation energetics in
  first-principles calculations'', {\em J.~Chem.\ Phys.}, {\bf 148},
  144105:1--7 (2018).

\bibitem{SchSun20}
K.~Schwarz and R.~Sundararaman, ``The electrochemical interface in
  first-principles calculations'', {\em Surf.\ Sci.\ Rep.}, {\bf 75},
  100492:1--22 (2020).

\bibitem{BraNguGle20}
G.~Bramley, M.-T. Nguyen, V.-A. Glezakou, R.~Rousseau, and C.-K. Sylaris,
  ``Reconciling work functions and adsorption enthalpies for implicit solvent
  models: {A} {Pt(111)}\slash water interface case study'', {\em J.~Chem.\
  Theory Comput.}, {\bf 16}, 2703--2715 (2020).

\bibitem{BhaAntDzi20}
A.~Bhandari, L.~Anton, J.~Dziedzic, C.~Peng, D.~Kramer, and C.-K. Skylaris,
  ``Electronic structure calculations in electrolyte solutions: {Methods} for
  neutralization of extended charged interfaces'', {\em J.~Chem.\ Phys.}, {\bf
  153}, 124101:1--12 (2020).

\bibitem{DavMcC91}
M.~E. Davis and J.~A. {McCammon}, ``Dielectric boundary smoothing in finite
  difference solutions of the {Poisson} equation: {An} approach to improve
  accuracy and convergence'', {\em J.~Comput.\ Chem.}, {\bf 12}, 909--912
  (1991).

\bibitem{GilDavLut93}
M.~K. Gilson, M.~E. Davis, B.~A. Luty, and J.~A. {McCammon}, ``Computation of
  electrostatic forces on solvated molecules using the {Poisson}-{Boltzmann}
  equation'', {\em J.~Phys.\ Chem.}, {\bf 97}, 3591--3600 (1993).

\bibitem{ImBegRou98}
W.~Im, D.~Beglov, and B.~Roux, ``Continuum solvation model: {C}omputation of
  electrostatic forces from numerical solutions to the {Poisson}--{Boltzmann}
  equation'', {\em Comput.\ Phys.\ Commun.}, {\bf 111}, 59--75 (1998).

\bibitem{LuZhaMcC05}
B.~Lu, D.~Zhang, and J.~A. {McCammon}, ``Computation of electrostatic forces
  between solvated molecules determined by the {Poisson}--{Boltzmann} equation
  using a boundary element method'', {\em J.~Chem.\ Phys.}, {\bf 122},
  214102:1--7 (2005).

\bibitem{WanTanCha10}
J.~Wang, C.~Tan, E.~Chanco, and R.~Luo, ``Quantitative analysis of
  {Poisson}--{Boltzmann} implicit solvent in molecular dynamics'', {\em Phys.\
  Chem.\ Chem.\ Phys.}, {\bf 12}, 1194--1202 (2010).

\bibitem{CaiYeWan11a}
Q.~Cai, X.~Ye, J.~Wang, and R.~Luo, ``Dielectric boundary force in numerical
  {Poisson}--{Boltzmann} methods: {Theory} and numerical strategies'', {\em
  Chem.\ Phys.\ Lett.}, {\bf 514}, 368--373 (2011).

\bibitem{RegCosBar98}
N.~Rega, M.~Cossi, and V.~Barone, ``Towards linear scaling in continuum solvent
  models. {A} new iterative procedure for energies and geometry
  optimizations'', {\em Chem.\ Phys.\ Lett.}, {\bf 293}, 221--229 (1998).

\bibitem{ScaBarKud04}
G.~Scalmani, V.~Barone, K.~N. Kudin, C.~S. Pomelli, G.~E. Scuseria, and M.~J.
  Frisch, ``Achieving linear-scaling computational cost for the polarizable
  continuum model of solvation'', {\em Theor.\ Chem.\ Acc.}, {\bf 111}, 90--100
  (2004).

\bibitem{GatLipSta17}
P.~Gatto, F.~Lipparini, and B.~Stamm, ``Computation of forces arising from the
  polarizable continuum model within the domain-decomposition paradigm'', {\em
  J.~Chem.\ Phys.}, {\bf 147}, 224108:1--11 (2017).

\bibitem{StaLagSca19}
B.~Stamm, L.~Lagard\`ere, G.~Scalmani, P.~Gatto, E.~Canc\`es, J.-P. Piquemal,
  Y.~Maday, B.~Mennucci, and F.~Lipparini, ``How to make continuum solvation
  incredibly fast in a few simple steps: {A} practical guide to the domain
  decomposition paradigm for the conductor-like screening model'', {\em Int.~J.
  Quantum Chem.}, {\bf 119}, e25669:1--15 (2019).

\bibitem{NotStaSca19}
M.~Nottoli, B.~Stamm, G.~Scalmani, and F.~Lipparini, ``Quantum calculations in
  solution of energies, structures, and properties with a domain decomposition
  polarizable continuum model'', {\em J.~Chem.\ Theory Comput.}, {\bf 15},
  6061--6073 (2019).

\bibitem{TINKER-HP}
L.~Lagard\`ere, L.-H. Jolly, F.~Lipparini, F.~Aviat, B.~Stamm, Z.~F. Jing,
  M.~Harger, H.~Torabifard, G.~A. Cisneros, M.~J. Schneiders, N.~Gresh,
  Y.~Maday, P.~Y. Ren, J.~W. Ponder, and J.-P. Piquemal, ``Tinker-{HP}: {A}
  massively parallel molecular dynamics package for multiscale simulations of
  large complex systems with advanced point dipole polarizable force fields'',
  {\em Chem.\ Sci.}, {\bf 9}, 956--972 (2018).

\bibitem{LipLagRay15}
F.~Lipparini, L.~Lagard\`ere, C.~Raynaud, B.~Stamm, E.~Canc\`es, B.~Mennucci,
  M.~Schnieders, P.~Ren, Y.~Maday, and J.-P. Piquemal, ``Polarizable molecular
  dynamics in a polarizable continuum solvent'', {\em J.~Chem.\ Theory
  Comput.}, {\bf 11}, 623--634 (2014).

\bibitem{CraTru08}
C.~J. Cramer and D.~G. Truhlar, ``A universal approach to solvation modeling'',
  {\em Acc.\ Chem.\ Res.}, {\bf 41}, 760--768 (2008).

\bibitem{CraTru06}
C.~J. Cramer and D.~G. Truhlar, ``{SM}$x$ continuum models for condensed
  phases'', in {\em Trends and Perspectives in Modern Computational Science},
  G.~Maroulis and T.~E. Simos, Eds., Vol. ~6 of {\em Lecture Series on Computer
  and Computational Sciences;}
\newblock Brill/VSP: Leiden, 2006;
\newblock pages 112--140.

\bibitem{ThoCraTru04}
J.~D. Thompson, C.~J. Cramer, and D.~G. Truhlar, ``New universal solvation
  model and comparison of the accuracy of the {SM}5.42{R}, {SM}5.43{R},
  {C}-{PCM}, {D}-{PCM}, and {IEF}-{PCM} continuum solvation models for aqueous
  and organic solvation free energies and for vapor pressures'', {\em J.~Phys.\
  Chem.~A}, {\bf 108}, 6532--6542 (2004).

\bibitem{SM6}
C.~P. Kelly, C.~J. Cramer, and D.~G. Truhlar, ``{SM6}: A density functional
  theory continuum solvation model for calculating aqueous solvation free
  energies of neutrals, ions, and solute--water clusters'', {\em J.~Chem.\
  Theory Comput.}, {\bf 1}, 1133--1152 (2005).

\bibitem{SM12}
A.~V. Marenich, C.~J. Cramer, and D.~G. Truhlar, ``Generalized {Born} solvation
  model {SM12}'', {\em J.~Chem.\ Theory Comput.}, {\bf 9}, 609--620 (2013).

\bibitem{SMD}
A.~V. Marenich, C.~J. Cramer, and D.~G. Truhlar, ``Universal solvation model
  based on solute electron density and on a continuum model of the solvent
  defined by the bulk dielectric constant and atomic surface tensions'', {\em
  J.~Phys.\ Chem.~B}, {\bf 113}, 6378--6396 (2009).

\bibitem{MarCraTru09}
A.~V. Marenich, C.~J. Cramer, and D.~G. Truhlar, ``Performance of {SM6}, {SM8},
  and {SMD} on the {SAMPL1} test set for the prediction of small-molecule
  solvation free energies'', {\em J.~Phys.\ Chem.~B}, {\bf 113}, 4538--4543
  (2009).

\bibitem{YouHer16}
Z.-Q. You and J.~M. Herbert, ``Reparameterization of an accurate, few-parameter
  implicit solvation model for quantum chemistry: {C}omposite method for
  implicit representation of solvent, {CMIRS} v.~1.1'', {\em J.~Chem.\ Theory
  Comput.}, {\bf 12}, 4338--4346 (2016).

\bibitem{DupAndMar13}
C.~Dupont, O.~Andreussi, and N.~Marzari, ``Self-consistent continuum solvation
  ({SCCS}): {The} case of charged systems'', {\em J.~Chem.\ Phys.}, {\bf 139},
  214110:1--8 (2013).

\bibitem{KlaMenTom09}
A.~Klamt, B.~Mennucci, J.~Tomasi, V.~Barone, C.~Curutchet, M.~Orozco, and F.~J.
  Luque, ``On the performance of continuum solvation methods. {A} comment on
  ``{U}niversal approaches to solvation modeling'''', {\em Acc.\ Chem.\ Res.},
  {\bf 42}, 489--492 (2009).

\bibitem{TomMucSla20}
L.~Toman\'{\i}k, E.~Muchov\'a, and P.~Slav\'{\i}\v{c}ek, ``Solvation energies
  of ions with ensemble cluster-continuum approach'', {\em Phys.\ Chem.\ Chem.\
  Phys.}, {\bf 22}, 22357--22368 (2020).

\bibitem{CasOrtFra14}
R.~Casasnovas, J.~{Ortega-Castro}, J.~Frau, J.~Donoso, and F.~{Mu\~{n}oz},
  ``Theoretical p${K}_\text{a}$ calculations with continuum model solvents,
  alternative protocols to thermodynamic cycles'', {\em Int.~J. Quantum Chem.},
  {\bf 114}, 1350--1363 (2014).

\bibitem{ThaSch15}
B.~Thapa and H.~B. Schlegel, ``Calculations of p${K}_{\text{a}}$'s and redox
  potentials of nucleobases with explicit waters and polarizable continuum
  solvation'', {\em J.~Phys.\ Chem.~A}, {\bf 119}, 5134--5144 (2015).

\bibitem{ThaSch16}
B.~Thapa and H.~B. Schlegel, ``Density functional theory calculation of
  p$K_\text{a}$'s of thiols in aqueous solution using explicit water molecules
  and the polarizable continuum model'', {\em J.~Phys.\ Chem.~A}, {\bf 120},
  5726--5735 (2016).

\bibitem{BryDiaGod08}
V.~S. Bryantsev, M.~S. Diallo, and W.~A. {Goddard III}, ``Calculation of
  solvation free energies of charged solutes using mixed cluster\slash
  continuum models'', {\em J.~Phys.\ Chem.~B}, {\bf 112}, 9709--9719 (2008).

\bibitem{RicGuoPar13}
D.~Riccardi, H.-B. Guo, J.~M. Parks, B.~Gu, L.~Liang, and J.~C. Smith,
  ``Cluster-continuum calculations of hydration free energies of anions and
  group~12 divalent cations'', {\em J.~Chem.\ Theory Comput.}, {\bf 9},
  555--569 (2013).

\bibitem{DhiEas18}
S.~Dhillon and A.~L. East, ``Challenges in predicting $\Delta_\text{rxn}G$ in
  solution: {H}ydronium, hydroxide, and water autoionization'', {\em Int.~J.
  Quantum Chem.}, {\bf 118}, e25703:1--7 (2018).

\bibitem{PatEas20}
D.~H. Patel and A.~L.~L. East, ``Semicontinuum (cluster-continuum) modeling of
  acid-catalyzed aqueous reactions: {Alkene} hydration'', {\em J.~Phys.\
  Chem.~A}, {\bf 124}, 9088--9104 (2020).

\bibitem{SotCurBid05}
I.~Soteras, C.~Curutchet, A.~{Bidon-Chanal}, M.~Orozco, and F.~J. Luque,
  ``Extension of the {MST} model to the {IEF} formalism: {HF} and {B3LYP}
  parameterizations'', {\em J.~Mol.\ Struct.\ (Theochem)}, {\bf 727}, 29--40
  (2005).

\bibitem{CurOroLuq06}
C.~Curutchet, M.~Orozco, F.~J. Luque, B.~Mennucci, and J.~Tomasi, ``Dispersion
  and repulsion contributions to the solvation free energy: {C}omparison of
  quantum mechanical and classical approaches in the polarizable continuum
  model'', {\em J.~Comput.\ Chem.}, {\bf 27}, 1769--1780 (2006).

\bibitem{CraTru09a}
C.~J. Cramer and D.~G. Truhlar, ``Reply to comment on ``{A} universal approach
  to solvation modeling'''', {\em Acc.\ Chem.\ Res.}, {\bf 42}, 493--497
  (2009).

\bibitem{SM8}
A.~V. Marenich, R.~M. Olson, C.~P. Kelly, C.~J. Cramer, and D.~G. Truhlar,
  ``Self-consistent reaction field model for aqueous and nonaqueous solutions
  based on accurate polarized partial charges'', {\em J.~Chem.\ Theory
  Comput.}, {\bf 3}, 2011--2033 (2007).

\bibitem{ChaCraTru08b}
A.~C. Chamberlin, C.~J. Cramer, and D.~G. Truhlar, ``Performance of {SM8} on a
  test to predict small-molecule solvation free energies'', {\em J.~Phys.\
  Chem.~B}, {\bf 112}, 8651--8655 (2008).

\bibitem{Kla11}
A.~Klamt, ``The {COSMO} and {COSMO}-{RS} solvation models'', {\em WIREs
  Comput.\ Mol.\ Sci.}, {\bf 1}, 699--709 (2011).

\bibitem{Kla18}
A.~Klamt, ``The {COSMO} and {COSMO}-{RS} solvation models'', {\em WIREs
  Comput.\ Mol.\ Sci.}, {\bf 8}, e1338:1--11 (2018).

\bibitem{BasCas00}
D.~Bashford and D.~A. Case, ``Generalized {Born} models of macromolecular
  solvation effects'', {\em Annu.\ Rev.\ Phys.\ Chem.}, {\bf 51}, 129--152
  (2000).

\bibitem{OnuCas19}
A.~V. Onufriev and D.~A. Case, ``Generalized {Born} implicit solvent models for
  biomolecules'', {\em Annu.\ Rev.\ Biophys.}, {\bf 48}, 275--296 (2019).

\bibitem{StiTemHaw90}
W.~C. Still, A.~Tempczyk, R.~C. Hawley, and T.~Hendrickson, ``Semianalytical
  treatment of solvation for molecular mechanics and dynamics'', {\em J.~Am.\
  Chem.\ Soc.}, {\bf 112}, 6127--6129 (1990).

\bibitem{OnuBasCas00}
A.~Onufriev, D.~Bashford, and D.~A. Case, ``Modification of the generalized
  {Born} models suitable for macromolecules'', {\em J.~Phys.\ Chem.~B}, {\bf
  104}, 3712--3720 (2000).

\bibitem{OnuCasBas02}
A.~Onufriev, D.~A. Case, and D.~Bashford, ``Effective {Born} radii in the
  generalized {Born} model approximation: {The} importance of being perfect'',
  {\em J.~Comput.\ Chem.}, {\bf 23}, 1297--1304 (2002).

\bibitem{MonSvrOnu07}
J.~Mongan, W.~A. Svrcek-Seiler, and A.~Onufriev, ``Analysis of integral
  expressions for effective {Born} radii'', {\em J.~Chem.\ Phys.}, {\bf 127},
  185101:1--10 (2007).

\bibitem{LanHer12a}
A.~W. Lange and J.~M. Herbert, ``Improving generalized {Born} models by
  exploiting connections to polarizable continuum models. {I}. {An} improved
  effective {Coulomb} operator'', {\em J.~Chem.\ Theory Comput.}, {\bf 8},
  1999--2011 (2012).

\bibitem{LeeSalBro02}
M.~S. Lee, F.~R. {Salsbury, Jr.}, and C.~L. {Brooks III}, ``Novel generalized
  {Born} methods'', {\em J.~Chem.\ Phys.}, {\bf 116}, 10606--10614 (2002).

\bibitem{Sal06}
F.~R. {Salsbury, Jr.}, ``Analysis of errors in {Still's} equation for
  macromolecular electrostatic solvation energies'', {\em Mol.\ Phys.}, {\bf
  104}, 1299--1309 (2006).

\bibitem{LanHer12b}
A.~W. Lange and J.~M. Herbert, ``Improving generalized {Born} models by
  exploiting connections to polarizable continuum models. {II}. {C}orrections
  for salt effects'', {\em J.~Chem.\ Theory Comput.}, {\bf 8}, 4381--4392
  (2012).

\bibitem{ScaApoCaf97}
M.~Scarsi, J.~Apostolakis, and A.~Caflisch, ``Continuum electrostatics energies
  of macromolecules in aqueous solutions'', {\em J.~Phys.\ Chem.~A}, {\bf 101},
  8098--8106 (1997).

\bibitem{Pie76}
R.~A. Pierotti, ``A scaled particle theory of aqueous and nonaqueous
  solutions'', {\em Chem.\ Rev.}, {\bf 76}, 717--726 (1976).

\bibitem{FloSelTan97}
F.~M. Floris, M.~Selmi, A.~Tani, and J.~Tomasi, ``Free energy and entropy for
  inserting cavities in water: {C}omparison of {Monte} {Carlo} simulation and
  scaled particle theory results'', {\em J.~Chem.\ Phys.}, {\bf 107},
  6353--6365 (1997).

\bibitem{CosReg07}
M.~Cossi and N.~Rega, ``First and second derivatives of the free energy in
  solution'', in {\em Continuum Solvation Models in Chemical Physics},
  B.~Mennucci and R.~Cammi, Eds.;
\newblock Wiley: Chichester, UK, 2007;
\newblock pages 313--322.

\bibitem{GalLev04}
E.~Gallicchio and R.~M. Levy, ``{AGBNP}: {An} analytic implicit solvent model
  suitable for molecular dynamics simulations and high-resolution modeling'',
  {\em J.~Comput.\ Chem.}, {\bf 25}, 479--499 (2004).

\bibitem{HouWanLi11a}
T.~Hou, J.~Wang, Y.~Li, and W.~Wang, ``Assessing the performance of the
  {MM}\slash {PBSA} and {MM}\slash {GBSA} methods. 1. {The} accuracy of binding
  free energy calculations based on molecular dynamics simulations'', {\em
  J.~Chem.\ Inf.\ Model.}, {\bf 51}, 69--82 (2011).

\bibitem{HouWanLi11b}
T.~Hou, J.~Wang, Y.~Li, and W.~Wang, ``Assessing the performance of the
  molecular mechanics\slash {Poisson} {Boltzmann} surface area and molecular
  mechanics\slash generalized {Born} surface area methods. {II}. {The} accuracy
  of ranking poses generated from docking'', {\em J.~Comput.\ Chem.}, {\bf 32},
  866--877 (2011).

\bibitem{XuSunLi13}
L.~Xu, H.~Sun, Y.~Li, J.~Wang, and T.~Hou, ``Assessing the performance of
  {MM}\slash {PBSA} and {MM}\slash {GBSA} methods. 3. {The} impact of force
  fields and ligand charge models'', {\em J.~Phys.\ Chem.~B}, {\bf 117},
  8408--8421 (2013).

\bibitem{SunLiShe14}
H.~Sun, Y.~Li, M.~Shen, S.~Tian, L.~Xu, P.~Pan, Y.~Guan, and T.~Hou,
  ``Assessing the performance of {MM}\slash {PBSA} and {MM}\slash {GBSA}
  methods. 5. {I}mproved docking performance using high solute dielectric
  constant {MM}\slash {GBSA} and {MM}\slash {PBSA} rescoring'', {\em Phys.\
  Chem.\ Chem.\ Phys.}, {\bf 16}, 22035--22045 (2014).

\bibitem{SunLiTia14}
H.~Sun, Y.~Li, S.~Tian, L.~Xu, and T.~Hou, ``Assessing the performance of
  {MM}\slash {PBSA} and {MM}\slash {GBSA} methods. 4. {A}ccuracies of
  {MM}\slash {PBSA} and {MM}\slash {GBSA} methodologies evaluated by various
  simulation protocols using {PDBbind} data set'', {\em Phys.\ Chem.\ Chem.\
  Phys.}, {\bf 16}, 15719--16729 (2014).

\bibitem{CheLiuSun16}
F.~Chen, H.~Liu, H.~Sun, P.~Pan, Y.~Li, D.~Li, and T.~Hou, ``Assessing the
  performance of the {MM}\slash {PBSA} and {MM}\slash {GBSA} methods. 6.
  {C}apability to predict protein--protein binding free energies and re-rank
  binding poses generated by protein--protein docking'', {\em Phys.\ Chem.\
  Chem.\ Phys.}, {\bf 18}, 22129--22139 (2016).

\bibitem{SunDuaChe18}
H.~Sun, L.~Duan, F.~Chen, H.~Liu, Z.~Wang, P.~Pan, F.~Zhu, J.~Z.~H. Zhang, and
  T.~Hou, ``Assessing the performance of {MM}\slash {PBSA} and {MM}\slash
  {GBSA} methods. 7. {Entropy} effects on the performance of end-point binding
  free energy calculation approaches'', {\em Phys.\ Chem.\ Chem.\ Phys.}, {\bf
  20}, 14450--14460 (2018).

\bibitem{WenWanChe19}
G.~Weng, E.~Wang, F.~Chen, H.~Sun, Z.~Wang, and T.~Hou, ``Assessing the
  performance of {MM}\slash {PBSA} and {MM}\slash {GBSA} methods. 9.
  {P}rediction of reliability of binding affinities and binding poses for
  protein--peptide complexes'', {\em Phys.\ Chem.\ Chem.\ Phys.}, {\bf 21},
  10135--10145 (2019).

\bibitem{WanWenSun19}
E.~Wang, G.~Weng, H.~Sun, H.~Du, F.~Zhu, F.~Chen, Z.~Wang, and T.~Hou,
  ``Assessing the performance of the {MM}\slash {PBSA} and {MM}\slash {GBSA}
  methods. 10. {Impacts} of enhanced sampling and variable dielectric model on
  protein--protein interactions'', {\em Phys.\ Chem.\ Chem.\ Phys.}, {\bf 21},
  18958--18969 (2019).

\bibitem{WanGreXia18}
C.~Wang, D.~Greene, L.~Xiao, R.~Qi, and R.~Luo, ``Recent developments and
  applications of the {MMPBSA} method'', {\em Front.\ Mol.\ Biosci.}, {\bf 4},
  87:1--18 (2018).

\bibitem{GenRyd15}
S.~Genheden and U.~Ryde, ``The {MM}\slash {PBSA} and {MM}\slash {GBSA} methods
  to estimate ligand-binding affinities'', {\em Expert Opin.\ Drug Dis.}, {\bf
  10}, 449--461 (2015).

\bibitem{WanSunWan19}
E.~Wang, H.~Sun, J.~Wang, Z.~Wang, H.~Liu, J.~Z.~H. Zhang, and T.~Hou,
  ``End-point binding free energy calculation with {MM}\slash {PBSA} and
  {MM}\slash {GBSA}: {S}trategies and applications in drug design'', {\em
  Chem.\ Rev.}, {\bf 119}, 9478--9508 (2019).

\bibitem{PolGraRiz20}
G.~Poli, C.~Granchi, F.~Rizzolio, and T.~Tuccinardi, ``Applications of
  {MM}-{PBSA} methods in virtual screening'', {\em Molecules}, {\bf 25},
  1971:1--19 (2020).

\bibitem{CM2}
J.~Li, T.~Zhu, C.~J. Cramer, and D.~G. Truhlar, ``New class~{IV} charge model
  for extracting accurate partial charges from wave functions'', {\em J.~Phys.\
  Chem.~A}, {\bf 102}, 1820--1831 (1998).

\bibitem{CM3}
P.~Winget, J.~D. Thompson, J.~D. Xidos, C.~J. Cramer, and D.~G. Truhlar,
  ``Charge model~3: {A} class~{IV} charge model based on hybrid density
  functional theory with variable exchange'', {\em J.~Phys.\ Chem.~A}, {\bf
  106}, 10707--10717 (2002).

\bibitem{CM4M}
R.~M. Olson, A.~V. Marenich, C.~J. Cramer, and D.~G. Truhlar, ``Charge model~4
  and intramolecular charge polarization'', {\em J.~Chem.\ Theory Comput.},
  {\bf 3}, 2046--2054 (2007).

\bibitem{CM5}
A.~V. Marenich, S.~V. Jerome, C.~J. Cramer, and D.~G. Truhlar, ``Charge
  model~5: {An} extension of {H}irshfeld population analysis for the accurate
  description of molecular interactions in gaseous and condensed phases'', {\em
  J.~Chem.\ Theory Comput.}, {\bf 8}, 527--541 (2012).

\bibitem{DavCha92}
E.~R. Davidson and S.~Chakravorty, ``A test of the {Hirshfeld} definition of
  atomic charges and moments'', {\em Theor.\ Chem.\ Acc.}, {\bf 83}, 319--330
  (1992).

\bibitem{Abr93}
M.~H. Abraham, ``Scales of solute hydrogen-bonding: {Their} construction and
  application to physicochemical and biochemical processes'', {\em Chem.\ Soc.\
  Rev.}, {\bf 22}, 73--83 (1993).

\bibitem{ChaCraTru06}
A.~C. Chamberlin, C.~J. Cramer, and D.~G. Truhlar, ``Predictiing aqueous free
  energies of solvation as functions of temperature'', {\em J.~Phys.\ Chem.~B},
  {\bf 110}, 5665--5675 (2006).

\bibitem{ChaCraTru08a}
A.~C. Chamberlin, C.~J. Cramer, and D.~G. Truhlar, ``Extension of a
  temperature-dependent aqueous solvation model to compounds containing
  nitrogen, fluorine, chlorine, bromine, and sulfur'', {\em J.~Phys.\ Chem.~B},
  {\bf 112}, 3024--3039 (2008).

\bibitem{Dea85}
J.~C. Dearden, ``Partitioning and lipophilicity in quantitative
  structure--activity relationships'', {\em Environ.\ Health Persp.}, {\bf 61},
  203--228 (1985).

\bibitem{AmeSubFug20}
S.~Am\'ezqueta, X.~Subirats, E.~Fuguet, M.~Ros\'es, and C.~R\`afols,
  ``Octanol-water partition constant'', in {\em Liquid-Phase Extraction}, C.~F.
  Poole, Ed.;
\newblock Elsevier: Amsterdam, 2020;
\newblock chapter~6, pages 183--208.

\bibitem{GinVazGil19}
T.~Ginex, J.~Vazquez, E.~Gilbert, E.~Herrero, and F.~J. Luque, ``Lipophilicity
  in drug design: {An} overview of lipophilicity descriptors in {3D}-{QSAR}
  studies'', {\em Future Med.\ Chem.}, {\bf 11}, 1177--1193 (2019).

\bibitem{SahAdhKua16}
S.~Sahoo, C.~Adhikari, M.~Kuanar, and B.~K. Mishra, ``A short review of the
  generation of molecular descriptors and their applications in quantitative
  structure property\slash activity relationships'', {\em Curr.\ Comput.\ Aided
  Drug Des.}, {\bf 12}, 181--205 (2016).

\bibitem{TsoGiaTsa17}
F.~Tsopelas, C.~Giaginis, and A.~{Tsantili-Kakoulidou}, ``Lipophilicity and
  biomimetic properties to support drug discovery'', {\em Expert Opin.\ Drug
  Dis.}, {\bf 12}, 885--896 (2017).

\bibitem{HerdeBBro13}
J.~L.~M. Hermens, J.~H.~M. {de Bruijn}, and D.~N. Brooke, ``The octanol--water
  partition coefficient: {Stengths} and limitations'', {\em Environ.\ Toxicol.\
  Chem.}, {\bf 32}, 732--733 (2013).

\bibitem{IsiBerFox20}
M.~I\c{s}ik, T.~D. Bergazin, T.~Fox, A.~Rizzi, J.~D. Chodera, and D.~L. Mobley,
  ``Assessing the accuracy of octanol--water partition coefficient predictions
  in the {SAMPL6} part~{II} log~${P}$ challenge'', {\em J.~Comput.-Aided Mol.\
  Des.}, {\bf 34}, 335--370 (2020).

\bibitem{OuiPal20}
J.~A. Ouimet and A.~S. Paluch, ``Predicting octanol\slash water partition
  coefficients for the {SAMPL6} challenge using the {SM12}, {SM8}, and {SMD}
  solvation models'', {\em J.~Comput.-Aided Mol.\ Des.}, {\bf 34}, 575--588
  (2020).

\bibitem{PliRiv02a}
J.~R. {Pliego Jr.} and J.~M. Riveros, ``Gibbs energy of solvation of organic
  ions in aqueous and dimethyl sulfoxide solutions'', {\em Phys.\ Chem.\ Chem.\
  Phys.}, {\bf 4}, 1622--1627 (2002).

\bibitem{Amo94}
C.~Amovilli, ``Calculation of the dispersion energy contribution to the
  solvation free energy'', {\em Chem.\ Phys.\ Lett.}, {\bf 229}, 244--249
  (1994).

\bibitem{AmoMen97}
C.~Amovilli and B.~Mennucci, ``Self-consistent-field calculation of {Pauli}
  repulsion and dispersion contributions to the solvation free energy in the
  polarizable continuum model'', {\em J.~Phys.\ Chem.~B}, {\bf 101}, 1051--1057
  (1997).

\bibitem{WeiMenFre10}
V.~Weijo, B.~Mennucci, and L.~Frediani, ``Toward a general formulation of
  dispersion effects for solvation continuum models'', {\em J.~Chem.\ Theory
  Comput.}, {\bf 6}, 3358--3364 (2010).

\bibitem{CupAmoMen15}
L.~Cupellini, C.~Amovilli, and B.~Mennucci, ``Electronic excitations in
  nonpolar solvents: {Can} the polarizable continuum model accurately reproduce
  solvent effects?'', {\em J.~Phys.\ Chem.~B}, {\bf 119}, 8984--8991 (2015).

\bibitem{PomThoChi11}
A.~Pomogaeva, D.~W. Thompson, and D.~M. Chipman, ``Modeling short-range
  contributions to hydration energies with minimal parameterization'', {\em
  Chem.\ Phys.\ Lett.}, {\bf 511}, 161--165 (2011).

\bibitem{PomChi11}
A.~Pomogaeva and D.~M. Chipman, ``Field-extremum model for short-range
  contributions to hydration free energy'', {\em J.~Chem.\ Theory Comput.},
  {\bf 7}, 3952--3960 (2011).

\bibitem{PomChi13a}
A.~Pomogaeva and D.~M. Chipman, ``New implicit solvation models for dispersion
  and exchange energies'', {\em J.~Phys.\ Chem.~A}, {\bf 117}, 5812--5820
  (2013).

\bibitem{PomChi14}
A.~Pomogaeva and D.~M. Chipman, ``Hydration energy from a composite method for
  implicit representation of the solvent'', {\em J.~Chem.\ Theory Comput.},
  {\bf 10}, 211--219 (2014).

\bibitem{PomChi15}
A.~Pomogaeva and D.~M. Chipman, ``Composite method for implicit representation
  of solvent in dimethyl sulfoxide and acetonitrile'', {\em J.~Phys.\ Chem.~A},
  {\bf 119}, 5173--5180 (2015).

\bibitem{DuiParNin13a}
T.~T. Duignan, D.~F. Parsons, and B.~W. Ninham, ``A continuum solvent model of
  the multipolar dispersion solvation energy'', {\em J.~Phys.\ Chem.~B}, {\bf
  117}, 9412--9420 (2013).

\bibitem{DuiParNin13b}
T.~T. Duignan, D.~F. Parsons, and B.~W. Ninham, ``A continuum model of
  solvation energies including electrostatic, dispersion, and cavity
  contributions'', {\em J.~Phys.\ Chem.~B}, {\bf 117}, 9421--9429 (2013).

\bibitem{HurCla72}
M.-J. Huron and P.~Claverie, ``Calculation of the interaction energy of one
  molecule with its whole surrounding. {I}. {Method} and application to pure
  nonpolar compounds'', {\em J.~Phys.\ Chem.}, {\bf 76}, 2123--2133 (1972).

\bibitem{FloTom89}
F.~Floris and J.~Tomasi, ``Evaluation of the dispersion contribution to the
  solvation energy. {A} simple computational model in the continuum
  approximation'', {\em J.~Comput.\ Chem.}, {\bf 10}, 616--627 (1989).

\bibitem{FloTomAhu91}
F.~M. Floris, J.~Tomasi, and J.~L.~P. Ahuir, ``Dispersion and repulsion
  contributions to the solvation energy: {R}efinements to a simple
  computational model in the continuum approximation'', {\em J.~Comput.\
  Chem.}, {\bf 12}, 784--791 (1991).

\bibitem{FloTanTom93}
F.~M. Floris, A.~Tani, and J.~Tomasi, ``Evaluation of dispersion--repulsion
  contributions to the solvation energy. {C}alibration of the uniform
  approximation with the aid of {RISM} calculations'', {\em Chem.\ Phys.}, {\bf
  169}, 11--20 (1993).

\bibitem{Tru98}
T.~N. Truong, ``Quantum modelling of reactions in solution: {An} overview of
  the dielectric continuum methodology'', {\em Int.\ Rev.\ Phys.\ Chem.}, {\bf
  17}, 525--546 (1998).

\bibitem{McL63b}
A.~D. {McLachlan}, ``Retarded dispersion forces between molecules'', {\em
  Proc.~R. Soc.\ Lond.~A}, {\bf 271}, 387--401 (1963).

\bibitem{McW84}
R.~{McWeeny}, ``Weak interactions between molecules'', {\em Croatica Chem.\
  Acta}, {\bf 57}, 865--878 (1984).

\bibitem{JasMcW85}
M.~Jaszunski and R.~{McWeeny}, ``Time-dependent {Hartree}--{Fock} calculations
  of dispersion energy'', {\em Mol.\ Phys.}, {\bf 55}, 1275--1286 (1985).

\bibitem{McW85}
R.~{McWeeny}, ``Electron density and response theory'', {\em J.~Mol.\ Struct.\
  (Theochem)}, {\bf 123}, 231--242 (1985).

\bibitem{AmoMcW90}
C.~Amovilli and M.~{McWeeny}, ``A matrix partitioning approach to the
  calculation of intermolecular potentials. {General} theory and some
  examples'', {\em Chem.\ Phys.}, {\bf 140}, 343--361 (1990).

\bibitem{McW92}
R.~{McWeeny}, {\em Methods of Molecular Quantum Mechanics}, Academic Press: New
  York, 2nd ed., 1992.

\bibitem{Tan69}
K.~T. Tang, ``Dynamic polarizabilities and van der {Waals} coefficients'', {\em
  Phys.\ Rev.}, {\bf 177}, 108--114 (1969).

\bibitem{LanKar70}
P.~W. Langhoff and M.~Karplus, ``Application of {Pad\'e} approximants to
  dispersion force and optical polarizability computations'', in {\em The
  Pad\'e Approximant in Theoretical Physics}, G.~A. {Baker Jr.} and J.~L.
  Gammel, Eds., Vol. ~71 of {\em Mathematics in Science and Engineering;}
\newblock Academic Press: New York, 1970;
\newblock chapter~2, pages 41--97.

\bibitem{SzaOst77}
A.~Szabo and N.~S. Ostlund, ``The correlation energy in the random phase
  approximation: {I}ntermolecular forces between closed-shell systems'', {\em
  J.~Chem.\ Phys.}, {\bf 67}, 4351--4360 (1977).

\bibitem{KapRod78}
I.~G. Kaplan and O.~B. Rodimova, ``Intermolecular interactions'', {\em Sov.\
  Phys.\ Usp.}, {\bf 21}, 918--943 (1978).

\bibitem{BuhWel07}
S.~Y. Buhmann and D.-G. Welsch, ``Dispersion forces in macroscopic quantum
  electrodynamics'', {\em Prog.\ Quantum Electron.}, {\bf 31}, 51--130 (2007).

\bibitem{Lon30}
F.~London, ``Zur {Theorie} und {S}ystematik der {M}olekularkr\"afte'', {\em
  Z.~Phys.}, {\bf 63}, 245--279 (1930).

\bibitem{Lon65}
H.~C. {Longuet-Higgins}, ``Intermolecular forces'', {\em Discuss.\ Faraday
  Soc.}, {\bf 40}, 7--18 (1965).

\bibitem{NorRuu06}
P.~Norman and K.~Ruud, ``Microscopic theory of nonlinear optics'', in {\em
  Non-Linear Optical Properties of Matter}, M.~G. Papadopoulos, A.~J. Sadlej,
  and J.~Leszczynski, Eds., Vol. ~1 of {\em Challenges and Advances in
  Computational Chemistry and Physics;}
\newblock Springer: Dordrecht, 2006;
\newblock chapter~1, pages 1--49.

\bibitem{DzyLifPit61b}
I.~E. Dyzaloshinskii, E.~M. Lifshitz, and L.~P. Pitaevskii, ``The general
  theory of van der {Waals} forces'', {\em Adv.\ Phys.}, {\bf 10}, 165--209
  (1961).

\bibitem{ZarKoh76}
E.~Zaremba and W.~Kohn, ``Van der {Waals} interaction between an atom and a
  solid surface'', {\em Phys.\ Rev.~B}, {\bf 13}, 2270--2285 (1976).

\bibitem{Par06}
V.~A. Parsegian, {\em Van der Waals Forces: A Handbook for Biologists,
  Chemists, Engineers, and Physicists}, Cambridge University Press: New York,
  2006.

\bibitem{VydVan09a}
O.~A. Vydrov and T.~{Van Voorhis}, ``Improving the accuracy of the nonlocal van
  der {Waals} density functional with minimal empiricism'', {\em J.~Chem.\
  Phys.}, {\bf 130}, 104105:1--7 (2009).

\bibitem{VydVan09b}
O.~A. Vydrov and T.~{Van Voorhis}, ``Nonlocal van der {Waals} density
  functional theory made simple'', {\em Phys.\ Rev.\ Lett.}, {\bf 103},
  063004:1--4 (2009).

\bibitem{VydVan10b}
O.~A. Vydrov and T.~{Van Voorhis}, ``Nonlocal van der {Waals} density
  functional: {The} simpler the better'', {\em J.~Chem.\ Phys.}, {\bf 133},
  244103:1--9 (2010).

\bibitem{VydVan12b}
O.~A. Vydrov and T.~{Van Voorhis}, ``Nonlocal van der {Waals} density
  functionals based on local response models'', in {\em Fundamentals of
  Time-Dependent Density Functional Theory}, M.~A.~L. Marques, N.~T. Maitra,
  F.~M.~S. Nogueira, E.~K.~U. Gross, and A.~Rubio, Eds., Vol.  837 of {\em
  Lecture Notes in Physics;}
\newblock Springer-Verlag: Berlin, 2012;
\newblock chapter~23, pages 443--456.

\bibitem{CalOrtSan15b}
J.~Calbo, E.~Ort\'{\i}, J.~C. {Sancho-Garc\'{\i}a}, and J.~Arag\'o, ``The
  nonlocal correlation density function {VV10}: {A} successful attempt to
  accurately capture noncovalent interactions'', {\em Annu.\ Rep.\ Comput.\
  Chem.}, {\bf 11}, 37--102 (2015).

\bibitem{LanDioRyd04}
D.~C. Langreth, M.~Dion, H.~Rydberg, E.~Schr\"oder, P.~Hyldgaard, and B.~I.
  Lundqvist, ``Van der {Waals} density functional theory with applications'',
  {\em Int.~J. Quantum Chem.}, {\bf 101}, 599--610 (2005).

\bibitem{DioRydSch04}
M.~Dion, H.~Rydberg, E.~Schr\"oder, D.~C. Langreth, and B.~I. Lundqvist, ``Van
  der {Waals} density functional for general geometries'', {\em Phys.\ Rev.\
  Lett.}, {\bf 92}, 246401:1--4 (2004).

\bibitem{LanLunCha09}
D.~C. Langreth, B.~I. Lundqvist, S.~D. {Chakarova-K\"ack}, V.~R. Cooper,
  M.~Dion, P.~Hyldgaard, A.~Kelkkanen, J.~Kleis, L.~Kong, S.~Li, P.~G. Moses,
  E.~Murray, A.~Puzder, H.~Rydberg, E.~Schr\"oder, and T.~Thonhauser, ``A
  density functional for sparse matter'', {\em J.~Phys.: Condens.\ Matt.}, {\bf
  21}, 084203:1--15 (2009).

\bibitem{LeeMurKon10}
K.~Lee, {\'{E}}.~D. Murray, L.~Kong, B.~I. Lundqvist, and D.~C. Langreth,
  ``Higher-accuracy van der {Waals} density functional'', {\em Phys.\ Rev.~B},
  {\bf 82}, 081101:1--4 (2010).

\bibitem{HerFli64}
C.~Herring and M.~Flicker, ``Asymptotic exchange coupling of two hydrogen
  atoms'', {\em Phys.\ Rev.}, {\bf 134}, A362--A366 (1964).

\bibitem{CamVerMen08}
R.~Cammi, V.~Verdolino, B.~Mennucci, and J.~Tomasi, ``Towards the elaboration
  of a {QM} method to describe molecular solutes under the effect of a very
  high pressure'', {\em Chem.\ Phys.}, {\bf 344}, 135--141 (2008).

\bibitem{CamCapMen12}
R.~Cammi, C.~Cappelli, B.~Mennucci, and J.~Tomasi, ``Calculation and analysis
  of the harmonic vibrational frequencies in molecules at extreme pressure:
  {M}ethodology and diborane as a test case'', {\em J.~Chem.\ Phys.}, {\bf
  137}, 154112:1--16 (2012).

\bibitem{CheHofCam17}
B.~Chen, R.~Hoffmann, and R.~Cammi, ``The effect of pressure on organic
  reactions in fluids---a new theoretical perspective'', {\em Angew.\ Chem.\
  Int.\ Ed.\ Engl.}, {\bf 56}, 11126--11142 (2017).

\bibitem{Cam18}
R.~Cammi, ``Quantum chemistry at the high pressures: {The} {eXtreme} pressure
  polarizable continuum model ({XP}-{PCM})'', in {\em Frontiers of Quantum
  Chemistry}, M.~J. W\'ojcik, H.~Nakatsuji, B.~Kirtman, and Y.~Ozaki, Eds.;
\newblock Springer Nature: Singapore, 2018;
\newblock chapter~12, pages 273--288.

\bibitem{FecEavLop03}
C.~J. Fecko, J.~D. Eaves, J.~J. Loparo, A.~Tokmakoff, and P.~L. Geissler,
  ``Ultrafast hydrogen-bond dynamics in the infrared spectroscopy of water'',
  {\em Science}, {\bf 301}, 1698--1702 (2003).

\bibitem{CorLawSki04}
S.~A. Corcelli, C.~P. Lawrence, and J.~L. Skinner, ``Combined electronic
  structure\slash molecular dynamics approach for ultrafast infrared
  spectroscopy of dilute {HOD} in liquid {H}$_2${O} and {D}$_2${O}'', {\em
  J.~Chem.\ Phys.}, {\bf 120}, 8107--8117 (2004).

\bibitem{SmiCapWil05}
J.~D. Smith, C.~D. Cappa, K.~R. Wilson, R.~C. Cohen, P.~L. Geissler, and R.~J.
  Saykally, ``Unified description of temperature-dependent hydrogen-bond
  rearrangements in liquid water'', {\em Proc.\ Natl.\ Acad.\ Sci.\ USA}, {\bf
  102}, 14171--14174 (2005).

\bibitem{SilDegPli16}
N.~M. Silva, P.~Deglmann, and J.~R. {Pliego, Jr.}, ``{CMIRS} solvation model
  for methanol: {P}arameterization, testing, and comparison with {SMD}, {SM8},
  and {COSMO}-{RS}'', {\em J.~Phys.\ Chem.~B}, {\bf 120}, 12660--12668 (2016).

\bibitem{HeBieVan18}
S.~He, F.~Biedermann, N.~Vankova, L.~Zhechkov, T.~Heine, R.~E. Hoffman, A.~{De
  Simone}, T.~T. Duignan, and W.~M. Nau, ``Cavitation energies can outperform
  dispersion interactions'', {\em Nat.\ Chem.}, {\bf 10}, 1252--1257 (2018).

\bibitem{AnsLaiHas19}
N.~Ansari, A.~Laio, and A.~Hassanali, ``Spontaneously forming dendritic voids
  in liquid water can host small polymers'', {\em J.~Phys.\ Chem.\ Lett.}, {\bf
  10}, 5585--5591 (2019).

\bibitem{BotBor78}
C.~J.~F. B\"ottcher and P.~Bordewijk, {\em Theory of Electric Polarization},
  Vol. ~2, Elsevier: Amsterdam, 1978.

\bibitem{Vau79}
W.~E. Vaughan, ``Dielectric relaxation'', {\em Annu.\ Rev.\ Phys.\ Chem.}, {\bf
  30}, 103--124 (1979).

\bibitem{FelPuzRya06}
Y.~Feldman, A.~Puzenko, and Y.~Ryabov, ``Dielectric relaxation phenomena in
  complex materials'', {\em Adv.\ Chem.\ Phys.}, {\bf 133}, 1--125 (2006).

\bibitem{FelIshPuz15}
Y.~Feldman, P.~B. Ishai, A.~Puzenko, and V.~Raicu, ``Elementary theory of the
  interaction of electromagnetic fields with dielectric materials'', in {\em
  Dielectric Relaxation in Biological Systems}, V.~Raicu and Y.~Feldman, Eds.;
\newblock Oxford University Press: Oxford, first ed., 2015;
\newblock pages 33--59.

\bibitem{Kaa89}
U.~Kaatze, ``Complex permittivity of water as a function of frequency and
  temperature'', {\em J.~Chem.\ Eng.\ Data}, {\bf 34}, 371--374 (1989).

\bibitem{Kaa93}
U.~Kaatze, ``Dielectric spectroscopy of aqueous solutions. {H}ydration
  phenomena and hydrogen-bonded networks'', {\em J.~Mol.\ Liq.}, {\bf 56},
  95--115 (1993).

\bibitem{Kaa15}
U.~Kaatze, ``Dielectric relaxation of water'', in {\em Dielectric Relaxation in
  Biological Systems}, V.~Raicu and Y.~Feldman, Eds.;
\newblock Oxford University Press: Oxford, first ed., 2015;
\newblock pages 189--227.

\bibitem{HsuSonMar97}
C.-P. Hsu, X.~Song, and R.~A. Marcus, ``Time-dependent {Stokes} shift and its
  calculation from solvent dielectric dispersion data'', {\em J.~Phys.\
  Chem.~B}, {\bf 101}, 2546--2551 (1997).

\bibitem{IngMenTom03}
F.~Ingrosso, B.~Mennucci, and J.~Tomasi, ``Quantum mechanical calculations
  coupled with a dynamical continuum model for the descriptiion of dielectric
  relaxation: {Time} dependent {Stokes} shift of coumarin {C153} in polar
  solvents'', {\em J.~Mol.\ Liq.}, {\bf 108}, 21--46 (2003).

\bibitem{CarMenTom06}
M.~Caricato, B.~Mennucci, J.~Tomasi, F.~Ingrosso, R.~Cammi, S.~Corni, and
  G.~Scalmani, ``Formation and relaxation of excited states in solution: {A}
  new time dependent polarizable continuum model based on time dependent
  density functional theory'', {\em J.~Chem.\ Phys.}, {\bf 124}, 124520:1--13
  (2006).

\bibitem{DinLinMen15}
F.~Ding, D.~B. Lingerfelt, B.~Mennucci, and X.~Li, ``Time-dependent
  non-equilibrium dielectric response in {QM}\slash continuum approaches'',
  {\em J.~Chem.\ Phys.}, {\bf 142}, 034120:1--8 (2015).

\bibitem{WilDonLip19}
A.~Wildman, G.~Donati, F.~Lipparini, B.~Mennucci, and X.~Li, ``Nonequilibrium
  environment dynamics in a frequency-dependent polarizable embedding model'',
  {\em J.~Chem.\ Theory Comput.}, {\bf 15}, 43--51 (2019).

\bibitem{Deb29}
P.~J.~W. Debye, {\em Polar Molecules}, The Chemical Catalog Company: New York,
  1929.

\bibitem{GoiLesLi18}
J.~J. Goings, P.~J. Lestrange, and X.~Li, ``Real-time time-dependent electronic
  structure theory'', {\em WIREs Comput.\ Mol.\ Sci.}, {\bf 8}, e1341:1--19
  (2018).

\bibitem{LiGovIsb20}
X.~Li, N.~Govind, C.~Isborn, A.~E. {DePrince III}, and K.~Lopata, ``Real-time
  time-dependent electronic structure theory'', {\em Chem.\ Rev.}, {\bf 120},
  9951--9993 (2020).

\bibitem{MisKasRud17}
J.~Mistrik, S.~Kasap, H.~E. Ruda, C.~Koughia, and J.~Singh, ``Optical
  properties of electronic materials: {F}undamentals and characterization'', in
  {\em Springer Handbook of Electronic and Photonic Materials}, S.~Kasap and
  P.~Capper, Eds.;
\newblock Springer International Publishing: Cham, Switzerland, 2017;
\newblock chapter~3, pages 47--83.

\bibitem{WohWoh96}
C.~Wohlfarth and B.~Wohlfarth, {\em Refractive Indices of Organic Liquids},
  Vol.  38B of {\em Group~III Condensed Matter}, Landolt-B\"ornstein,
  Springer-Verlag: Berlin, 1996.

\bibitem{HalQue73}
G.~M. Hale and M.~R. Querry, ``Optical constants of water in the 200-nm to
  200-$\mu$m wavelength region'', {\em Appl.\ Opt.}, {\bf 12}, 555--563 (1973).

\bibitem{CosBar00b}
M.~Cossi and V.~Barone, ``Separation between fast and slow polarizations in
  continuum solvation models'', {\em J.~Phys.\ Chem.~A}, {\bf 104},
  10614--10622 (2000).

\bibitem{DinNewMat17}
M.~Dinpajooh, M.~D. Newton, and D.~V. Matyushov, ``Free energy functionals for
  polarization fluctuations: {Pekar} factor revisited'', {\em J.~Chem.\ Phys.},
  {\bf 146}, 064504:1--18 (2017).

\bibitem{ColHasRit48}
C.~H. Collie, J.~B. Hasted, and D.~M. Ritson, ``The dielectric properties of
  water and heavy water'', {\em Proc.\ Phys.\ Soc.}, {\bf 60}, 145--160 (1948).

\bibitem{RonThrAst97}
C.~R{\o}nne, L.~Thrane, P.-O. {\AA}strand, A.~Wallqvist, K.~V. Mikkelsen, and
  S.~R. Keiding, ``Investigation of the temperature dependence of dielectric
  relaxation in liquid water by {THz} reflection spectroscopy and molecular
  dynamics simulation'', {\em J.~Chem.\ Phys.}, {\bf 107}, 5319--5331 (1997).

\bibitem{RonKei02}
C.~R{\o}nne and S.~R. Keiding, ``Low frequency spectroscopy of liquid water
  using {THz}-time domain spectroscopy'', {\em J.~Mol.\ Liq.}, {\bf 101},
  199--218 (2002).

\bibitem{NagYadAri06}
M.~Nagai, H.~Yada, T.~Arikawa, and K.~Tanaka, ``Terahertz time-domain
  attenuated total reflection spectroscopy in water and biological solution'',
  {\em Int.~J. Infrared Milli.}, {\bf 27}, 505--515 (2006).

\bibitem{ZhoRaoLiu19}
J.~Zhou, X.~Rao, X.~Liu, T.~Li, L.~Zhou, Y.~Zheng, and Z.~Zhu, ``Temperature
  dependent optical and dielectric properties of liquid water studied by
  terahertz time-domain spectroscopy'', {\em AIP Adv.}, {\bf 9}, 035346:1--7
  (2019).

\bibitem{PopIshKha16}
I.~Popov, P.~B. Ishai, A.~Khamzin, and Y.~Feldman, ``The mechanism of the
  dielectric relaxation of water'', {\em Phys.\ Chem.\ Chem.\ Phys.}, {\bf 18},
  13941--13953 (2016).

\bibitem{Elt17}
D.~C. Elton, ``The origin of the {Debye} relaxation in liquid water and fitting
  the high frequency excess response'', {\em Phys.\ Chem.\ Chem.\ Phys.}, {\bf
  19}, 18739--18749 (2017).

\bibitem{KaaUhl81}
U.~Kaatze and V.~Uhlendorf, ``The dielectric properties of water at microwave
  frequencies'', {\em Z.~Phys.\ Chem.\ Neue Folge}, {\bf 126}, 151--165 (1981).

\bibitem{Woh91}
C.~Wohlfarth, {\em Static Dielectric Constants of Pure Liquids and Binary
  Liquid Mixtures}, Vol. ~6 of {\em Landolt-B\"ornstein, New Series~IV},
  Springer Science + Business Media, 1991.

\bibitem{TilTay38}
L.~W. Tilton and J.~K. Taylor, ``Refractive index and dispersion of distilled
  water for visible radiation, at temperatures 0 to 60$^\circ$~{C}'', {\em
  J.~Res.\ Nat.\ Bur.\ Stand.}, {\bf 20}, 419--477 (1938).

\bibitem{Kit05}
C.~Kittel, {\em Introduction to Solid State Physics}, John Wiley \& Sons:
  Hoboken, 8th ed., 2005.

\bibitem{WemDidCam68}
S.~H. Wemple, M.~{Didomenico Jr.}, and I.~Camlibel, ``Dielectric and optical
  properties of melt-grown {BaTiO}$_3$'', {\em J.~Phys.\ Chem.\ Solids}, {\bf
  29}, 1797--1803 (1968).

\bibitem{MatYozKoi56}
N.~Mataga, Y.~Kaifu, and M.~Koizumi, ``Solvent effects upon fluorescence
  spectra and the dipole moments of excited molecules'', {\em Bull.\ Chem.\
  Soc.\ Jpn.}, {\bf 29}, 465--470 (1956).

\bibitem{Lip57}
E.~Lippert, ``Spektroskopische {B}estimmung des {D}ipolmomentes aromatischer
  {V}erbindungen im ersten angeregten {S}ingulettzustand'', {\em
  Z.~Electrochem.}, {\bf 61}, 962--975 (1957).

\bibitem{MatChoTan05}
N.~Mataga, H.~Chosrowjan, and S.~Taniguchi, ``Ultrafast charge transfer in
  excited electronic states and investigations into fundamental problems of
  exciplex chemistry: {Our} early studies and recent developments'', {\em
  J.~Photoch.\ Photobio.~C}, {\bf 6}, 37--79 (2005).

\bibitem{Bay50}
N.~S. Bayliss, ``The effect of the electrostatic polarization of the solvent on
  electronic absorption spectra in solution'', {\em J.~Chem.\ Phys.}, {\bf 18},
  292--296 (1950).

\bibitem{Oos54}
Y.~Ooshika, ``Absorption spectra of dyes in solution'', {\em J.~Phys.\ Soc.\
  Jpn.}, {\bf 9}, 594--602 (1954).

\bibitem{McR57}
E.~G. {McRae}, ``Theory of solvent effects on molecular electronic spectra.
  {F}requency shifts'', {\em J.~Phys.\ Chem.}, {\bf 61}, 562--572 (1957).

\bibitem{BilKaw62}
L.~Bilot and A.~Kawski, ``Zue {Theorie} des {E}influsses von {L}\"osungsmitteln
  auf die {E}lectronenspektren der {M}olek\"ule'', {\em Z.~Naturforsch.~A},
  {\bf 17}, 621--627 (1962).

\bibitem{Bak64}
N.~G. Bakshiev, ``Universal intermolecular interactions and their effect on the
  position of the electronic spectra of molecules in two component solutions'',
  {\em Opt.\ Spectrosc.}, {\bf 16}, 821--832 (1964).

\bibitem{Lip65a}
W.~Liptay, ``Dipole moments of molecules in excited states and the effect of
  external electric fields on the optical absorption of molecules in
  solution'', in {\em Modern Quantum Chemistry: Istanbul Lectures. Part III.
  Action of Light and Organic Crystals}, O.~Sinano\v{g}lu, Ed.;
\newblock Academic Press: New York, 1965;
\newblock pages 45--66.

\bibitem{Abe65}
T.~Abe, ``Theory of solvent effects on molecular electronic spectra.
  {F}requency shifts'', {\em Bull.\ Chem.\ Soc.\ Jpn.}, {\bf 38}, 1314--1318
  (1965).

\bibitem{ChaVia70}
A.~Chamma and P.~Viallet, ``Determination du moment dipolaire d'une molecule
  dans un etat excite singulet'', {\em Sci.\ Paris Ser.~C}, {\bf 270},
  1901--1904 (1970).

\bibitem{BraCar85}
J.~E. Brady and P.~W. Carr, ``An analysis of dielectric models of
  solvatochromism'', {\em J.~Phys.\ Chem.}, {\bf 89}, 5759--5766 (1985).

\bibitem{JozMilHel09}
M.~J\'ozefowicz, P.~Milart, and J.~R. Heldt, ``Determination of ground and
  excited state dipole moments of
  $4,5'$-diamino[$1,1'$:$3',1''$-terphenyl]-$4',6'$-dicarbonitrile using
  solvatochromic method and quantum-chemical calculations'', {\em Spectrochim.\
  Acta~A}, {\bf 74}, 959--963 (2009).

\bibitem{SidHan10}
R.~Siddlingeshwar and S.~M. Hanagodimath, ``Estimation of the ground and the
  first excited singlet-state dipole moments of 1,4-disubstituted anthraquinone
  dyes by the solvatochromic method'', {\em Spectrochim.\ Acta~A}, {\bf 75},
  1203--1210 (2010).

\bibitem{ManKumShi13}
S.~R. Manohara, V.~U. Kumar, Shivakumaraiah, and G.~Gerward, ``Estimation of
  ground and excited-state dipole moments of 1,2-diazines by solvatochromic
  method and quantum-chemical calculation'', {\em J.~Mol.\ Liq.}, {\bf 181},
  97--104 (2013).

\bibitem{DemMenWoy17}
E.~G. Demissie, E.~T. Mengesha, and G.~W. Woyessa, ``Modified solvatochromic
  equations for better determination of ground and excited state dipole moments
  of $p$-aminobenzoicacid ({PABA}): {A}ccounting for real shape over
  hypothetical spherical solvent shell'', {\em J.~Photochem.\ Photobiol.~A},
  {\bf 337}, 184--191 (2017).

\bibitem{KumVarGeo17}
R.~Kumari, A.~Varghese, L.~George, and Y.~N. Sudhaker, ``Effect of solvent
  polarity on the photophysical properties of chalcone derivatives'', {\em RSC
  Adv.}, {\bf 7}, 24204:1--11 (2017).

\bibitem{RenNadSri18}
C.~G. Renuka, Y.~F. Nadaf, G.~Sriprakash, and S.~R. Prasad, ``Solvent
  dependence on structure and electronic properties of
  7-(diethylamino)-{2H}-1-benzopyran-2-one ({C}-466) laser dye'', {\em
  J.~Fluoresc.}, {\bf 28}, 839--854 (2018).

\bibitem{DivSakWei19}
V.~M. Divac, D.~\v{S}aki\'c, T.~Weitner, and M.~Gabri\v{c}evi\'c, ``Solvent
  effects on the absorption and fluorescence spectra of {Zaleplon}:
  {D}etermination of ground and excited state dipole moments'', {\em
  Spectrochim.\ Acta~A}, {\bf 212}, 356--362 (2019).

\bibitem{Mar56a}
R.~A. Marcus, ``On the theory of oxidation-reduction reactions involving
  electron transfer.~{I}.'', {\em J.~Chem.\ Phys.}, {\bf 24}, 966--978 (1956).

\bibitem{Mar64}
R.~A. Marcus, ``Chemical and electrochemical electron-transfer theory'', {\em
  Annu.\ Rev.\ Phys.\ Chem.}, {\bf 15}, 155--196 (1964).

\bibitem{MarSut85}
R.~A. Marcus and N.~Sutin, ``Electron transfers in chemistry and biology'',
  {\em Biochim.\ Biophys.\ Acta}, {\bf 811}, 265--322 (1985).

\bibitem{New07}
M.~D. Newton, ``The role of solvation in electron transfer: {T}heoretica and
  computational aspects'', in {\em Continuum Solvation Models in Chemical
  Physics}, B.~Mennucci and R.~Cammi, Eds.;
\newblock Wiley: Chichester, UK, 2007;
\newblock pages 389--413.

\bibitem{Li15}
X.-Y. Li, ``An overview of continuum models for nonequilibrium solvation:
  {Popular} theories and new challenge'', {\em Int.~J. Quantum Chem.}, {\bf
  115}, 700--721 (2015).

\bibitem{Hsu20}
C.-P. Hsu, ``Reorganization energies and spectral densities for electron
  transfer problems in charge transporting materials'', {\em Phys.\ Chem.\
  Chem.\ Phys.}, {\bf 22}, 21630--21641 (2020).

\bibitem{CraTru99}
C.~J. Cramer and D.~G. Truhlar, ``Implicit solvation models: {E}quilibria,
  structure, spectra, and dynamics'', {\em Chem.\ Rev.}, {\bf 99}, 2161--2200
  (1999).

\bibitem{KimHyn90a}
H.~J. Kim and J.~T. Hynes, ``Equilibrium and nonequilibrium solvation and
  solute electronic structure. {I}. {F}ormulation'', {\em J.~Chem.\ Phys.},
  {\bf 93}, 5194--5210 (1990).

\bibitem{AguOliTom93}
M.~A. Aguilar, F.~J. {Olivares del Valle}, and J.~Tomasi, ``Nonequilibrium
  solvation: {An} \textit{ab initio} quantum-mechanical method in the continuum
  cavity model approximation'', {\em J.~Chem.\ Phys.}, {\bf 98}, 7375--7384
  (1993).

\bibitem{Men07}
B.~Mennucci, ``Continuum models for excited states'', in {\em Continuum
  Solvation Models in Chemical Physics}, B.~Mennucci and R.~Cammi, Eds.;
\newblock Wiley: Chichester, UK, 2007;
\newblock pages 110--123.

\bibitem{GuiCap19}
C.~A. Guido and S.~Caprasecca, ``On the description of the environment
  polarization response to electronic transitions'', {\em Int.~J. Quantum
  Chem.}, {\bf 119}, e25711:1--11 (2019).

\bibitem{MenCamTom98}
B.~Mennucci, R.~Cammi, and J.~Tomasi, ``Excited states and solvatochromatic
  shifts within a nonequilibrium solvation approach: {A} new formulation of the
  integral equation formalism method at the self-consistent field,
  configuration interaction, and multiconfiguration self-consistent field
  level'', {\em J.~Chem.\ Phys.}, {\bf 109}, 2798--2807 (1998).

\bibitem{CosBar00a}
M.~Cossi and V.~Barone, ``Solvent effect on vertical electronic transitions by
  the polarizable continuum model'', {\em J.~Chem.\ Phys.}, {\bf 112},
  2427--2435 (2000).

\bibitem{Chi09a}
D.~M. Chipman, ``Vertical electronic excitation with a dielectric continuum
  model of solvation including volume polarization. {I}. {Theory}'', {\em
  J.~Chem.\ Phys.}, {\bf 131}, 014103:1--10 (2009).

\bibitem{MewYouWor15}
J.-M. Mewes, Z.-Q. You, M.~Wormit, T.~Kriesche, J.~M. Herbert, and A.~Dreuw,
  ``Experimental benchmark data and systematic evaluation of two \textit{a
  posteriori}, polarizable-continuum corrections for vertical excitation
  energies in solution'', {\em J.~Phys.\ Chem.~A}, {\bf 119}, 5446--5464
  (2015).

\bibitem{MewHerDre17}
J.-M. Mewes, J.~M. Herbert, and A.~Dreuw, ``On the accuracy of the general,
  state-specific polarizable-continuum model for the description of correlated
  ground- and excited states in solution'', {\em Phys.\ Chem.\ Chem.\ Phys.},
  {\bf 19}, 1644--1654 (2017).

\bibitem{CooYouHer16}
M.~P. Coons, Z.-Q. You, and J.~M. Herbert, ``The hydrated electron at the
  surface of neat liquid water appears to be indistinguishable from the bulk
  species'', {\em J.~Am.\ Chem.\ Soc.}, {\bf 138}, 10879--10886 (2016).

\bibitem{CamTom95c}
R.~Cammi and J.~Tomasi, ``Nonequilibrium solvation theory for the polarizable
  continuum model: {A} new formulation at the {SCF} level with application to
  the case of the frequency-dependent linear electric response function'', {\em
  Int.~J. Quantum Chem.\ Symp.}, {\bf 29}, 465--474 (1995).

\bibitem{Agu01}
M.~A. Aguilar, ``Separation of the electric polarization into fast and slow
  components: {A} comparison of two partition schemes'', {\em J.~Phys.\
  Chem.~A}, {\bf 105}, 10393--10396 (2001).

\bibitem{Pek54}
S.~I. Pekar, {\em Untersuchungen \"uber die {E}lektronentheorie der
  {K}ristalle}, Akademie-Verlag: Berlin, 1954.

\bibitem{Pek63}
S.~I. Pekar ``Research in Electron Theory of Crystals'' Technical Report
  AEC--tr--5575, U.S. Atomic Energy Commission, Division of Technical
  Information, (1963).

\bibitem{BayMcR54a}
N.~S. Bayliss and E.~G. {McRae}, ``Solvent effects in organic spectra: {Dipole}
  forces and the {Franck}-{Condon} principle'', {\em J.~Phys.\ Chem.}, {\bf
  58}, 1002--1006 (1954).

\bibitem{BasChu91}
M.~V. Basilevsky and G.~E. Chudinov, ``Dynamics of charge transfer chemical
  reactions in a polar medium within the scope of the
  {Born}-{Kirkwood}-{Onsager} model'', {\em Chem.\ Phys.}, {\bf 157}, 327--344
  (1991).

\bibitem{HouSakIno97}
H.~Houjou, M.~Sakurai, and Y.~Inoue, ``Theoretical evaluation of medium effects
  on absorption maxima of molecular solutes. {I}. {F}ormulation of a new method
  based on the self-consistent reaction field theory'', {\em J.~Chem.\ Phys.},
  {\bf 107}, 5652--5660 (1997).

\bibitem{CamFreMen02}
R.~Cammi, L.~Frediani, B.~Mennucci, J.~Tomasi, K.~Ruud, and K.~V. Mikkelsen,
  ``A second-order, quadratically convergent multiconfigurational
  self-consistent field polarizable continuum model for equilibrium and
  nonequilibrium solvation'', {\em J.~Chem.\ Phys.}, {\bf 117}, 13--26 (2002).

\bibitem{MarCraTru11}
A.~V. Marenich, C.~J. Cramer, D.~G. Truhlar, C.~A. Guido, B.~Mennucci,
  G.~Scalmani, and M.~J. Frisch, ``Practical computation of electronic
  excitation in solution: {Vertical} excitation model'', {\em Chem.\ Sci.},
  {\bf 2}, 2143--2161 (2011).

\bibitem{JacHer11c}
L.~D. Jacobson and J.~M. Herbert, ``A simple algorithm for determining
  orthogonal, self-consistent excited-state wave functions for a state-specific
  {H}amiltonian: {A}pplication to the optical spectrum of the aqueous
  electron'', {\em J.~Chem.\ Theory Comput.}, {\bf 7}, 2085--2093 (2011).

\bibitem{LunKoh13}
B.~Lunkenheimer and A.~K\"ohn, ``Solvent effects on electronically excited
  states using the conductor-like screening model and the second-order
  correlated method {ADC(2)}'', {\em J.~Chem.\ Theory Comput.}, {\bf 9},
  977--994 (2013).

\bibitem{WinFau06}
B.~Winter and M.~Faubel, ``Photoemission from liquid aqueous solutions'', {\em
  Chem.\ Rev.}, {\bf 106}, 1176--1211 (2006).

\bibitem{SeiThuWin11}
R.~Seidel, S.~Th\"urmer, and B.~Winter, ``Photoelectron spectroscopy meets
  aqueous solution: {S}tudies from a vacuum liquid microjet'', {\em J.~Phys.\
  Chem.\ Lett.}, {\bf 2}, 633--641 (2011).

\bibitem{SeiWinBra16}
R.~Seidel, B.~Winter, and S.~E. Bradforth, ``Valence electronic structure of
  aqueous solutions: {Insights} from photoelectron spectroscopy'', {\em Annu.\
  Rev.\ Phys.\ Chem.}, {\bf 67}, 283--305 (2016).

\bibitem{WinWebHer05}
B.~Winter, R.~Weber, I.~V. Hertel, M.~Faubel, P.~Jungwirth, E.~C. Brown, and
  S.~E. Bradforth, ``Electron binding energies of aqueous alkali and halide
  ions: {EUV} photoelectron spectroscopy of liquid solutions and combined ab
  initio and molecular dynamics calculations'', {\em J.~Am.\ Chem.\ Soc.}, {\bf
  127}, 7203--7214 (2005).

\bibitem{JacHer10b}
L.~D. Jacobson and J.~M. Herbert, ``A one-electron model for the aqueous
  electron that includes many-body electron-water polarization: {Bulk}
  equilibrium structure, vertical electron binding energy, and optical
  absorption spectrum'', {\em J.~Chem.\ Phys.}, {\bf 133}, 154506:1--19 (2010).

\bibitem{GhoIsaSli11}
D.~Ghosh, O.~Isayev, L.~V. Slipchenko, and A.~I. Krylov, ``Effect of solvation
  on the vertical ionization energy of thymine: {From} microhydration to
  bulk'', {\em J.~Phys.\ Chem.~A}, {\bf 115}, 6028--6038 (2011).

\bibitem{GhoRoySei12}
D.~Ghosh, A.~Roy, R.~Seidel, B.~Winter, S.~Bradforth, and A.~I. Krylov,
  ``First-principle protocol for calculating ionization energies and redox
  potentials of solvated molecules and ions: {Theory} and application to
  aqueous phenol and phenolate'', {\em J.~Phys.\ Chem.~B}, {\bf 116},
  7269--7280 (2012).

\bibitem{TazGurKim19}
R.~N. Tazhigulov, P.~K. Gurunathan, Y.~Kim, L.~V. Slipchenko, and K.~B.
  Bravaya, ``Polarizable embedding for simulating redox potentials of
  biomolecules'', {\em Phys.\ Chem.\ Chem.\ Phys.}, {\bf 21}, 11642--11650
  (2019).

\bibitem{TotKubMuc20}
Z.~T\'oth, J.~Kube\v{c}ka, E.~Muchov\'a, and P.~Slavi\v{c}ek, ``Ionization
  energies in solution with the {QM}:{QM} approach'', {\em Phys.\ Chem.\ Chem.\
  Phys.}, {\bf 22}, 10550--10560 (2020).

\bibitem{ProPeeXio16}
M.~R. Provorse, T.~Peev, C.~Xiong, and C.~M. Isborn, ``Convergence of
  excitation energies in mixed quantum and classical solvent: {C}omparison of
  continuum and point charge models'', {\em J.~Phys.\ Chem.~B}, {\bf 120},
  12148--12159 (2016). Erratum: {\em J.~Phys.\ Chem.~B}, {\bf 121}, 2372 (2017).

\bibitem{PauCooHer19}
S.~K. Paul, M.~P. Coons, and J.~M. Herbert, ``Erratum: `{Quantum} chemistry in
  arbitrary dielectric environments: {Theory} and implementation of
  nonequilibrium {Poisson} boundary conditions and application to compute
  vertical ionization energies at the air\slash water interface''', {\em
  J.~Chem.\ Phys.}, {\bf 151}, 189901:1--2 (2019).

\bibitem{PerZhaNun20}
C.~F. Perry, P.~Zhang, F.~B. Nunes, I.~Jordan, A.~{von Conta}, and H.~J.
  W\"orner, ``Ionization energy of liquid water revisited'', {\em J.~Phys.\
  Chem.\ Lett.}, {\bf 11}, 1789--1794 (2020).

\bibitem{LucYamSuz17}
D.~Luckhaus, Y.~Yamamoto, T.~Suzuki, and R.~Signorell, ``Genuine binding energy
  of the hydrated electron'', {\em Sci.\ Adv.}, {\bf 3}, e1603224:1--5 (2017).

\bibitem{Her19a}
J.~M. Herbert, ``Structure of the aqueous electron'', {\em Phys.\ Chem.\ Chem.\
  Phys.}, {\bf 21}, 20538--20565 (2019).

\bibitem{MunMarImp15}
A.~{Mu\~{n}oz-Losa}, D.~Markovitsi, and R.~Improta, ``A state-specific
  {PCM}-{DFT} method to include dynamic solvent effects in the calculation of
  ionization energies: {A}pplication to {DNA} bases'', {\em Chem.\ Phys.\
  Lett.}, {\bf 634}, 20--24 (2015).

\bibitem{KhaKhaHat20}
S.~K. Khani, A.~M. Khah, and C.~H\"attig, ``Comparison of reaction field
  schemes for coupling continuum solvation models with wave function methods
  for excitation energies'', {\em J.~Chem.\ Theory Comput.}, {\bf 16},
  4554--4564 (2020).

\bibitem{ThoSch95}
M.~A. Thompson and G.~K. Schenter, ``Excited states of the bacteriochlorophyll
  $b$ dimer of {\em {R}hodopseudomonas viridis}: {A} {QM}\slash {MM} study of
  the photosynthetic reaction center that includes {MM} polarization'', {\em
  J.~Phys.\ Chem.}, {\bf 99}, 6374--6386 (1995).

\bibitem{CamCorMen05}
R.~Cammi, S.~Corni, B.~Mennucci, and J.~Tomasi, ``Electronic excitation
  energies of molecules in solution: {State} specific and linear response
  methods for nonequilibrium continuum solvation models'', {\em J.~Chem.\
  Phys.}, {\bf 122}, 104513:1--12 (2005).

\bibitem{ImpBarSca06}
R.~Improta, V.~Barone, G.~Scalmani, and M.~J. Frisch, ``A state-specific
  polarizable continuum model time dependent density functional method for
  excited state calculations in solution'', {\em J.~Chem.\ Phys.}, {\bf 125},
  054103:1--9 (2006).

\bibitem{ImpScaFri07}
R.~Improta, G.~Scalmani, M.~J. Frisch, and V.~Barone, ``Toward effective and
  reliable fluorescence energies in solution by a new state specific
  polarizable continuum model time dependent density functional theory
  approach'', {\em J.~Chem.\ Phys.}, {\bf 127}, 074504:1--9 (2007).

\bibitem{FurAhl02}
F.~Furche and R.~Ahlrichs, ``Adiabatic time-dependent density functional
  methods for excited state properties'', {\em J.~Chem.\ Phys.}, {\bf 117},
  7433--7447 (2002). Erratum, {\em J.~Chem.\ Phys.}, {\bf 121}, 12772--12773
  (2004).

\bibitem{RonAngBel14}
E.~Ronca, C.~Angeli, L.~Belpassi, F.~{De Angelis}, F.~Tarantelli, and
  M.~Pastore, ``Density relaxation in time-dependent density functional theory:
  {C}ombining relaxed density natural orbitals and multireference perturbation
  theories for an improved description of excited states'', {\em J.~Chem.\
  Theory Comput.}, {\bf 10}, 4014--4024 (2014).

\bibitem{MasCamFri18}
F.~Maschietto, M.~Campetella, M.~J. Frisch, G.~Scalmani, C.~Adamo, and
  I.~Ciofini, ``How are the charge transfer descriptors affected by the quality
  of the underpinning electronic density?'', {\em J.~Comput.\ Chem.}, {\bf 39},
  735--742 (2018).

\bibitem{MenCapGui09}
B.~Mennucci, C.~Cappelli, C.~A. Guido, R.~Cammi, and J.~Tomasi, ``Structures
  and properties of electronically excited chromophores in solution from the
  polarizable continuum model coupled to the time-dependent density functional
  theory'', {\em J.~Phys.\ Chem.~A}, {\bf 113}, 3009--3020 (2009).

\bibitem{AmoBur73b}
A.~T. Amos and B.~L. Burrows, ``Solvent-shift effects on electronic spectra and
  excited-state dipole moments and polarizabilities'', {\em Adv.\ Quantum
  Chem.}, {\bf 7}, 289--313 (1973).

\bibitem{Sup68a}
P.~Suppan, ``Solvent effects on the energy of electronic transitions:
  {E}xperimental observations and applications to structural problems of
  excited molecules'', {\em J.~Chem.\ Soc.~A}, pages 3125--3133 (1968).

\bibitem{Sup68b}
P.~Suppan, ``Polarizability of excited molecules from spectroscopic studies'',
  {\em Spectrochim.\ Acta~A}, {\bf 24}, 1161--1165 (1968).

\bibitem{AmoBur73a}
A.~T. Amos and B.~L. Burrows, ``Dispersion interactions and solvent-shift
  effects'', {\em Theor.\ Chem.\ Acc.}, {\bf 29}, 139--150 (1973).

\bibitem{MarCraTru13}
A.~V. Marenich, C.~J. Cramer, and D.~G. Truhlar, ``Uniform treatment of
  solute--solvent dispersion in the ground and excited electronic states of the
  solute based on a solvation model with state-specific polarizability'', {\em
  J.~Chem.\ Theory Comput.}, {\bf 9}, 3649--3659 (2013).

\bibitem{MarCraTru15}
A.~V. Marenich, C.~J. Cramer, and D.~G. Truhlar, ``Electronic absorption
  spectra and solvatochromic shifts by the vertical excitation model:
  {S}olvated clusters and molecular dynamics sampling'', {\em J.~Phys.\
  Chem.~B}, {\bf 119}, 958--967 (2015).

\bibitem{JacPlaAda12}
D.~Jacquemin, A.~Planchat, C.~Adamo, and B.~Mennucci, ``{TD}-{DFT} assessment
  of functionals for optical 0--0 transitions in solvated dyes'', {\em
  J.~Chem.\ Theory Comput.}, {\bf 8}, 2359--2372 (2012).

\bibitem{OliTom91}
F.~J. {Olivares del Valle} and J.~Tomasi, ``Electron correlation and solvation
  effects. {I}. {Basic} formulation and preliminary attempt to include the
  electron correlation in the quantum mechanical polarizable continuum model so
  as to study solvation phenomena'', {\em Chem.\ Phys.}, {\bf 150}, 139--150
  (1991).

\bibitem{AguOliTom91}
M.~A. Aguilar, F.~J. {Olivares del Valle}, and J.~Tomasi, ``Electron
  correlation and solvation effects. {II}. {T}he description of the vibrational
  properties of a water molecule in a dielectric given by the application of
  the polarizable continuum model with inclusion of correlation effects'', {\em
  Chem.\ Phys.}, {\bf 150}, 151--161 (1991).

\bibitem{Ang95}
J.~\'Angy\'an, ``Choosing between alternative {MP2} algorithms in the
  self-consistent reaction field theory of solvent effects'', {\em Chem.\
  Phys.\ Lett.}, {\bf 241}, 51--56 (1995).

\bibitem{LipScaMen09}
F.~Lipparini, G.~Scalmani, and B.~Mennucci, ``Non covalent interactions in
  {RNA} and {DNA} base pairs: {A} quantum-mechanical study of the coupling
  between solvent and electronic density'', {\em Phys.\ Chem.\ Chem.\ Phys.},
  {\bf 11}, 11617--11623 (2009).

\bibitem{Cam09}
R.~Cammi, ``Quantum cluster theory for the polarizable continuum model. {I}.
  {The} {CCSD} level with analytical first and second derivatives'', {\em
  J.~Chem.\ Phys.}, {\bf 131}, 164104:1--14 (2009).

\bibitem{CamTom17}
R.~Cammi and J.~Tomasi, ``Quantum cluster theory for the polarizable continuum
  model {(PCM)}'', in {\em Handbook of Computational Chemistry},
  J.~Leszczynski, A.~{Kaczmarek-Kedziera}, T.~Puzyn, M.~G. Papadopoulos,
  H.~Reis, and M.~K. Shukla, Eds.;
\newblock Springer International Publishing: Switzerland, 2nd ed., 2017;
\newblock chapter~34, pages 1517--1556.

\bibitem{CarMenSca10}
M.~Caricato, B.~Mennucci, G.~Scalmani, G.~W. Trucks, and M.~J. Frisch,
  ``Electronic excitation energies in solution at equation of motion {CCSD}
  level within a state specific polarizable continuum model approach'', {\em
  J.~Chem.\ Phys.}, {\bf 132}, 084102:1--7 (2010).

\bibitem{Car11}
M.Caricato, ``{CCSD}-{PCM}: {I}mproving upon the reference reaction field
  approximation at no cost'', {\em J.~Chem.\ Phys.}, {\bf 135}, 074113:1--11
  (2011).

\bibitem{Car20}
M.~Caricato, ``Coupled cluster theory in the condensed phase within the
  singles-{T} density scheme for the environment response'', {\em WIREs
  Comput.\ Mol.\ Sci.}, {\bf 10}, e1463:1--27 (2020).

\bibitem{KhaKhaHat18a}
S.~K. Khani, A.~M. Khah, and C.~H\"attig, ``{COSMO}-{RI}-{ADC(2)} excitation
  energies and excited state gradients'', {\em Phys.\ Chem.\ Chem.\ Phys.},
  {\bf 20}, 16354--16363 (2018).

\bibitem{DreWor15}
A.~Dreuw and M.~Wormit, ``The algebraic diagrammatic construction scheme for
  the polarization propagator for the calculation of excited states'', {\em
  WIREs Comput.\ Mol.\ Sci.}, {\bf 5}, 82--95 (2015).

\bibitem{DeVMenNog18}
M.~{De Vetta}, M.~F. S.~J. Menger, J.~J. Nogueira, and L.~Gonz\'alez, ``Solvent
  effects on electronically excited states: {QM}\slash continuum versus
  {QM}\slash explicit models'', {\em J.~Phys.\ Chem.~B}, {\bf 122}, 2975--2984
  (2018).

\bibitem{CosBar01}
M.~Cossi and V.~Barone, ``Time-dependent density functional theory for
  molecules in liquid solutions'', {\em J.~Chem.\ Phys.}, {\bf 115}, 4708--4717
  (2001).

\bibitem{CamMen99}
R.~Cammi and B.~Mennucci, ``Linear response theory for the polarizable
  continuum model'', {\em J.~Chem.\ Phys.}, {\bf 110}, 9877--9886 (1999).

\bibitem{CamMenTom00}
R.~Cammi, B.~Mennucci, and J.~Tomasi, ``Fast evaluation of geometries and
  properties of excited molecules in solution: {A} {Tamm}-{Dancoff} model with
  application to 4-dimethylaminobenzonitrile'', {\em J.~Phys.\ Chem.~A}, {\bf
  104}, 5631--5637 (2000).

\bibitem{CorTom02}
S.~Corni and J.~Tomasi, ``Excitation energies of a molecule close to a metal
  surface'', {\em J.~Chem.\ Phys.}, {\bf 117}, 7266--7278 (2002).

\bibitem{IozMenTom04}
M.~F. Iozzi, B.~Mennucci, J.~Tomasi, and R.~Cammi, ``Excitation energy transfer
  {(EET)} between molecules in condensed matter: {A} novel application of the
  polarizable continuum model {(PCM)}'', {\em J.~Chem.\ Phys.}, {\bf 120},
  7029--7040 (2004).

\bibitem{CamCosMen96}
R.~Cammi, M.~Cossi, B.~Mennucci, and J.~Tomasi, ``Analytical {Hartree}--{Fock}
  calculation of the dynamical polarizabilities $\alpha$, $\beta$, and $\gamma$
  of molecules in solution'', {\em J.~Chem.\ Phys.}, {\bf 105}, 10556--10564
  (1996).

\bibitem{DucGuiJac18}
I.~Duchemin, C.~A. Guido, D.~Jacquemin, and X.~Blase, ``The {Bethe}--{Salpeter}
  formalism with polarisable continuum embedding: {R}econciling linear-response
  and state-specific features'', {\em Chem.\ Sci.}, {\bf 9}, 4430--4443 (2018).

\bibitem{Cam10}
R.~Cammi, ``Coupled-cluster theories for the polarizable continuum model. {II}.
  {A}nalytical gradients for excited states of molecular solutes by the
  equation of motion coupled-cluster method'', {\em Int.~J. Quantum Chem.},
  {\bf 110}, 3040--3052 (2010).

\bibitem{Cam12a}
R.~Cammi, ``Coupled-cluster theory for the polarizable continuum model. {III}.
  {A} response theory for molecules in solution'', {\em Int.~J. Quantum Chem.},
  {\bf 112}, 2547--2560 (2012).

\bibitem{Car13}
M.~Caricato, ``A comparison between state-specific and linear-response
  formalisms for the calculation of vertical electronic transition energy in
  solution with the {CCSD}-{PCM} method'', {\em J.~Chem.\ Phys.}, {\bf 139},
  044116:1--9 (2013).

\bibitem{Car18}
M.~Caricato, ``Linear response coupled cluster theory with the polarizable
  continuum model within the singles approximation for the solvent response'',
  {\em J.~Chem.\ Phys.}, {\bf 148}, 134113:1--9 (2018).

\bibitem{Car19b}
M.~Caricato, ``{CCSD}-{PCM} excited state energy gradients with the linear
  response singles approximation to study the photochemistry of molecules in
  solution'', {\em ChemPhotoChem}, {\bf 3}, 747--754 (2019).

\bibitem{HelJorOls00}
T.~Helgaker, P.~J{\o}rgensen, and J.~Olsen, {\em Molecular Electronic-Structure
  Theory}, Wiley: New York, 2000.

\bibitem{CorCamMen05}
S.~Corni, R.~Cammi, B.~Mennucci, and J.~Tomasi, ``Electronic excitation
  energies of molecules in solution within continuum solvation models:
  {I}nvestigating the discrepancy between state-specific and linear-response
  methods'', {\em J.~Chem.\ Phys.}, {\bf 123}, 134512:1--10 (2005).

\bibitem{Ped13}
A.~Pedone, ``Role of solvent on charge transfer in 7-aminocoumarin dyes: {New}
  hints from {TD}-{CAM}-{B3LYP} and state specific {PCM} calculations'', {\em
  J.~Chem.\ Theory Comput.}, {\bf 9}, 4087--4096 (2013).

\bibitem{Min14b}
N.~Minezawa, ``State-specific solvation effect on the intramolecular charge
  transfer reaction in solution: {A} linear-response free energy {TDDFT}
  method'', {\em Chem.\ Phys.\ Lett.}, {\bf 608}, 140--144 (2014).

\bibitem{BerZanCal14}
C.~Bernini, L.~Zani, M.~Calamante, G.~Reginato, A.~Mordini, M.~Taddei,
  R.~Basosi, and A.~Sinicropi, ``Excited state geometries and vertical emission
  energies of solvated dyes for {DSSC}: {A} {PCM}\slash {TD}-{DFT} benchmark
  study'', {\em J.~Chem.\ Theory Comput.}, {\bf 10}, 3925--3933 (2014).

\bibitem{BudMacMed15}
\v{S}. Budz\'ak, P.~Mach, M.~{Medved'}, and O.~{Kysel'}, ``Critical analysis of
  spectral solvent shifts calculated by the contemporary {PCM} approaches of a
  representative series of charge-transfer methylated benzenes'', {\em Phys.\
  Chem.\ Chem.\ Phys.}, {\bf 17}, 17618--17627 (2015).

\bibitem{GuiJacAda15}
C.~A. Guido, D.~Jacquemin, C.~Adamo, and B.~Mennucci, ``Electronic excitations
  in solution: {The} interplay between state specific approaches and a
  time-dependent density functional theory description'', {\em J.~Chem.\ Theory
  Comput.}, {\bf 11}, 5782--5790 (2015).

\bibitem{GuiMenSca18}
C.~A. Guido, B.~Mennucci, G.~Scalmani, and D.~Jacquemin, ``Excited state dipole
  moments in solution: {C}omparison between state-specific and linear-response
  {TD}-{DFT} values'', {\em J.~Chem.\ Theory Comput.}, {\bf 14}, 1544--1553
  (2018).

\bibitem{KlaKroSaa10}
B.~Klaum\"unzer, D.~Kr\"oner, and P.~Saalfrank, ``({TD}-){DFT} calculation of
  vibrational and vibronic spectra of riboflavin in solution'', {\em J.~Phys.\
  Chem.~B}, {\bf 114}, 10826--10834 (2010).

\bibitem{BloBaiBic16}
J.~Bloino, A.~Baiardi, and M.~Biczysko, ``Aiming at an accurate prediction of
  vibrational and electronic spectra for medium-to-large molecules: {An}
  overview'', {\em Int.~J. Quantum Chem.}, {\bf 116}, 1543--1574 (2016).

\bibitem{GarZemPon19}
C.~{Garc\'{\i}a-Iriepa}, M.~Zemmouche, M.~{Ponce-Vargas}, and I.~Navizet, ``The
  role of solvation models on the computed absorption and emission spectra:
  {The} case of fireflies oxyluciferin'', {\em Phys.\ Chem.\ Chem.\ Phys.},
  {\bf 21}, 4613--4623 (2019).

\bibitem{ChiBudMed14}
S.~Chibani, \v{S}. Budz\'ak, M.~{Medved'}, B.~Mennucci, and D.~Jacquemin,
  ``Full {cLR}-{PCM} calculations of the solvatochromic effects on emission
  energies'', {\em Phys.\ Chem.\ Chem.\ Phys.}, {\bf 16}, 26024--26029 (2014).

\bibitem{Sch16}
T.~Schwabe, ``General theory for environmental effects on (vertical) electron
  excitation energies'', {\em J.~Chem.\ Phys.}, {\bf 145}, 154105:1--7 (2016).

\bibitem{MenCam03}
B.~Mennucci and R.~Cammi, ``Ab initio model to predict {NMR} shielding tensors
  for solutes in liquid crystals'', {\em Int.~J. Quantum Chem.}, {\bf 93},
  121--130 (2003).

\bibitem{HosSakIno87}
H.~Hoshi, M.~Sakurai, Y.~Inoue, and R.~Ch\^{u}j\^{o}, ``Medium effects on the
  molecular electronic structure. {I}. {T}he formulation of a theory for the
  estimation of a molecular electronic structure surrounded by an anisotropic
  medium'', {\em J.~Chem.\ Phys.}, {\bf 87}, 1107--1115 (1987).

\bibitem{FreMenCam04}
L.~Frediani, B.~Mennucci, and R.~Cammi, ``Quantum-mechanical continuum
  solvation study of the polarizability of halides at the water\slash air
  interface'', {\em J.~Phys.\ Chem.~B}, {\bf 108}, 13796--13806 (2004).

\bibitem{FreCamCor04}
L.~Frediani, R.~Cammi, S.~Corni, and J.~Tomasi, ``A polarizable continuum model
  for molecules at diffuse interfaces'', {\em J.~Chem.\ Phys.}, {\bf 120},
  3893--3907 (2004).

\bibitem{BonFreAgr06}
L.~Bondesson, L.~Frediani, H.~{\AA}gren, and B.~Mennucci, ``Solvation of
  {N}$_3^-$ at the water surface: {The} polarizable continuum model approach'',
  {\em J.~Phys.\ Chem.~B}, {\bf 110}, 11361--11368 (2006).

\bibitem{IozCosImp06}
M.~F. Iozzi, M.~Cossi, R.~Improta, N.~Rega, and V.~Barone, ``A polarizable
  continuum approach for the study of heterogeneous dielectric environments'',
  {\em J.~Chem.\ Phys.}, {\bf 124}, 184103 (2006).

\bibitem{SiLi09a}
D.~Si and H.~Li, ``Heterogeneous conductorlike solvation model'', {\em
  J.~Chem.\ Phys.}, {\bf 131}, 044123:1--8 (2009).

\bibitem{WanMaLi10}
J.-B. Wang, J.-Y. Ma, and X.-Y. Li, ``Polarizable continuum model associated
  with the self-consistent-reaction field for molecular adsorbates at the
  interface'', {\em Phys.\ Chem.\ Chem.\ Phys.}, {\bf 12}, 207--214 (2010).

\bibitem{MozMenFre14}
K.~Mozgawa, B.~Mennucci, and L.~Frediani, ``Solvation at surfaces and
  interfaces: {A} quantum-mechanical\slash continuum approach including
  nonelectrostatic contributions'', {\em J.~Phys.\ Chem.~C}, {\bf 118},
  4715--4725 (2014).

\bibitem{MozFre16}
K.~Mozgawa and L.~Frediani, ``Electronic structure of small surfactants: {A}
  continuum solvation study'', {\em J.~Phys.\ Chem.~C}, {\bf 120}, 17501--17513
  (2016).

\bibitem{JunTob02b}
P.~Jungwirth and D.~J. Tobias, ``Ions at the air\slash water interface'', {\em
  J.~Phys.\ Chem.~B}, {\bf 106}, 6361--6373 (2002).

\bibitem{JunTob06}
P.~Jungwirth and D.~J. Tobias, ``Specific ion effects at the air\slash water
  interface'', {\em Chem.\ Rev.}, {\bf 106}, 1259--1281 (2006).

\bibitem{DuiZha20}
T.~T. Duignan and X.~S. Zhao, ``The {Born} model can accurately describe
  electrostatic ion solvation'', {\em Phys.\ Chem.\ Chem.\ Phys.}, {\bf 22},
  25126--25135 (2020).

\bibitem{Cha09}
M.~Chaplin, ``Theory vs experiment: {What} is the surface charge of water?'',
  {\em Water}, {\bf 1}, 1--28 (2009).

\bibitem{DuiParNin14}
T.~T. Duignan, D.~F. Parsons, and B.~W. Ninham, ``Ion interactions with the
  air--water interface using a continuum solvent model'', {\em J.~Phys.\
  Chem.~B}, {\bf 118}, 8700--8710 (2014).

\bibitem{DuiParNin15}
T.~T. Duignan, D.~F. Parsons, and B.~W. Ninham, ``Hydronium and hydroxide at
  the air--water interface with a continuum solvent model'', {\em Chem.\ Phys.\
  Lett.}, {\bf 635}, 1--12 (2015).

\bibitem{AksPauHer20}
H.~Aksu, S.~K. Paul, J.~M. Herbert, and B.~D. Dunietz, ``How well does a
  solvated octa-acid capsule shield the embedded chromophore? {A} computational
  analysis based on an anisotropic dielectric continuum model'', {\em J.~Phys.\
  Chem.~B}, {\bf 124}, 6998--7004 (2020).

\bibitem{HerCoo17}
J.~M. Herbert and M.~P. Coons, ``The hydrated electron'', {\em Annu.\ Rev.\
  Phys.\ Chem.}, {\bf 68}, 447--472 (2017).

\bibitem{SwaMonMcC05}
J.~M.~J. Swanson, J.~Mongan, and J.~A. {McCammon}, ``Limitations of
  atom-centered dielectric functions in implicit solvent models'', {\em
  J.~Phys.\ Chem.~B}, {\bf 109}, 14769--14772 (2005).

\bibitem{ZhoQinTjo08}
H.-X. Zhou, S.~Qin, and H.~Tjong, ``Modeling protein--protein and
  protein--nucleic acid interactions: {Structure}, thermodynamics, and
  kinetics'', {\em Annu.\ Rep.\ Comput.\ Chem.}, {\bf 4}, 67--87 (2008).

\bibitem{PanZho13}
X.~Pang and H.-X. Zhou, ``Poisson-{Boltzmann} calculations: van der {Waals} or
  molecular surface?'', {\em Commun.\ Comput.\ Phys.}, {\bf 13}, 1--12 (2013).

\bibitem{DecColCat13}
S.~Decherchi, J.~Colmenares, C.~E. Catalano, M.~Spagnuolo, E.~Alexov, and
  W.~Rocchia, ``Between algorithm and model: {D}ifferent molecular surface
  definitions for the {Poisson}-{Boltzmann} based electrostatic
  characterization of biomolecules in solution'', {\em Commun.\ Comput.\
  Phys.}, {\bf 13}, 61--89 (2013).

\bibitem{HehRadSch86}
W.~J. Hehre, L.~Radom, P.~v.~R.~Schleyer, and J.~A. Pople, {\em Ab Initio
  Molecular Orbital Theory}, Wiley-Interscience: New York, 1986.

\end{thebibliography}

\end{document}